\documentclass[10pt, oneside]{article}
\pdfoutput=1
\usepackage{hyperref}
\usepackage{enumitem} 
\usepackage{easylist}  	
\usepackage{graphicx}							
\usepackage{amssymb}
\usepackage[usenames]{color}
\usepackage{epsf}
\usepackage[usenames,dvipsnames]{pstricks}
\usepackage{epsfig}
\usepackage{pst-grad} 
\usepackage{url}
\usepackage{wrapfig}
\usepackage{float} 
\newenvironment{s-itemize}{
\begin{itemize}
  \setlength{\itemsep}{1pt}
  \setlength{\parskip}{0pt}
  \setlength{\parsep}{0pt}
}{\end{itemize}}

\newenvironment{s-enumerate}{
\begin{enumerate}
  \setlength{\itemsep}{1pt}
  \setlength{\parskip}{0pt}
  \setlength{\parsep}{0pt}
}{\end{enumerate}}

\let\OLDthebibliography\thebibliography
\renewcommand\thebibliography[1]{
  \OLDthebibliography{#1}
  \setlength{\parskip}{.5ex}
  \setlength{\itemsep}{0pt plus 0.3ex}
}
\title{The Variant-Rule\\Another Logically Universal Rule}
\author{Jos\'e Manuel G\'omez Soto%
\thanks{jmgomez@uaz.edu.mx, http://matematicas.reduaz.mx/$\sim$jmgomez}%
\hspace{2ex}{\it \small Universidad Aut\'onoma de Zacatecas.}\\
\hspace{2ex}{\it \small  Unidad Acad\'emica de Matem\'aticas. Zacatecas, Zac. M\'exico.}\\
Andrew Wuensche%
\thanks{andy@ddlab.org,  http://www.ddlab.org}%
\hspace{2ex}{\it \small Discrete Dynamics Lab.}\\
}

\begin{document}

\maketitle

\vspace{-3ex}
\begin{abstract}
\noindent The Variant-rule derives from the
\mbox{Precursor-rule}\cite{Gomez2017} by interchanging two classes of
its 28 isotropic mappings.  Although this small mutation conserves
most glider types and stable blocks, glider-gun engines are
changed, as are most large scale pattern behaviors, illustrating both
the robustness and fragility of evolution. We demonstrate these newly
discovered structures and dynamics, and utilising two different
glider types, build the logical gates
required for universality in the logical sence. 

\end{abstract}

\begin{center}
{\it keywords: universality, cellular automata, glider-guns, logical gates.}
\end{center}

\section{Introduction}
\label{Introduction}

The idea of Cellular Automata (CA) was conceived by von~Neumann in the
1940s, applying the earliest computers to construct
an abstract self-reproducing machine. That a CA itself might become a
computer by its dynamic patterns can be traced to back to 1970 with Conway's 
Game-of-Life\cite{Gardner1970} and the first glider-gun, discovered by Gosper,
which lead to a demonstration of universal computation based on memory, transmission
and processing\cite{Berlekamp1982}, and the proof\cite{Randall2002} was based
on the Turing Machine.
 
Here we consider one aspect of universal computation,
``universality in the logical sense'' ---  that logical gates can be built within
the CA dynamics by means of a glider-gun.
By this definition, several logically universal CA have been created, some called
\mbox{``Life-like''} because they are variations on the Game-of-Life
birth-survival scheme\cite{Eppstein2010}, and others were built on schemes
different from birth/survival. Under this last approach, Sapin found a
universal CA in 2004 \cite{Sapin2004}, and G\'omez-Soto/Wuensche
published three more: the \mbox{the X-Rule} in 2015\cite{Gomez2015},
the \mbox{Precursor-Rule} in 2017\cite{Gomez2017}, and the \mbox{Sayab-Rule}
in 2018\cite{Gomez2018}.

In this paper we present the ``Variant-rule'', another logically
universal CA, made from a chance mutation of the
Precursor-rule\cite{Gomez2017}.` The Precursor-rule is part of the
family of CA discovered by G\'omez-Soto/Wuensche using the
input-entropy search method\cite{Wuensche99,Wuensche2005}. The rule
is isotropic --- patterns and mechanisms operate equivalently in any
direction.  A 2D isotropic CA with a $3\times3$ neighborhood can be
defined by 102 symmetric groups that map to either of two states, 0 or
1.  The \mbox{Precursor-rule} has 28 symmetric groups mapping to 1.
The \mbox{Variant-rule} has the same number, but one of these groups
has been interchanged for another.

Essential properties to support CA logical universality are dynamic
periodic patterns: gliders and glider-guns, and stable blocks that
can destroy gliders called ``eaters''. Useful dynamical
interaction between these and other objects include
reflection, transformation and oscillation.

\begin{figure}[htb]
\begin{center}
\begin{minipage}[c]{.95\linewidth}
\begin{minipage}[c]{.30\linewidth} 
\fbox{\includegraphics[width=1\linewidth,bb=155 191 408 481, clip=]{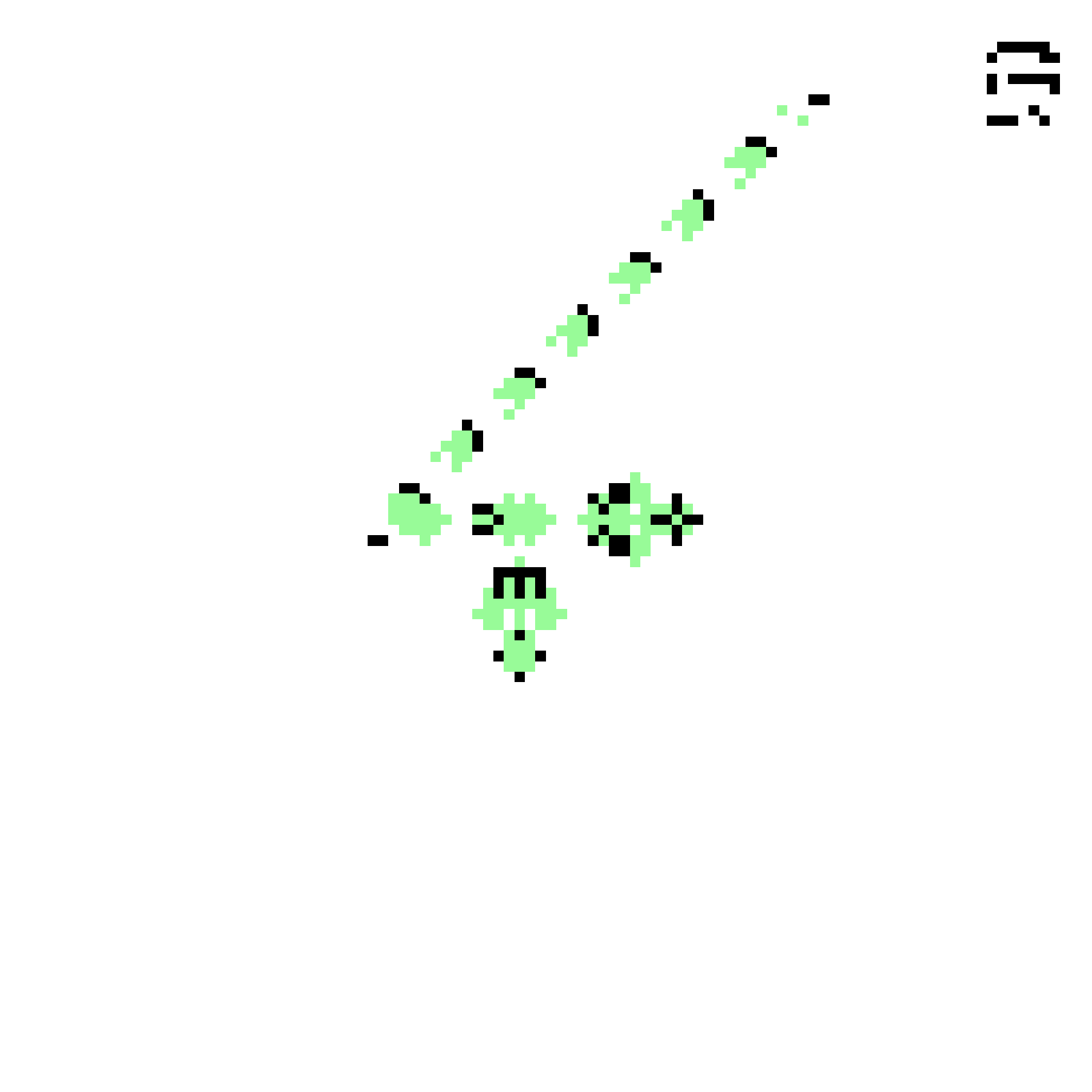}}\\ 
\textsf{\small{Variant: $p$=22}}
\end{minipage}
\hfill
\begin{minipage}[c]{.354\linewidth} 
\fbox{\includegraphics[width=1\linewidth,bb=100 100 415 413, clip=]{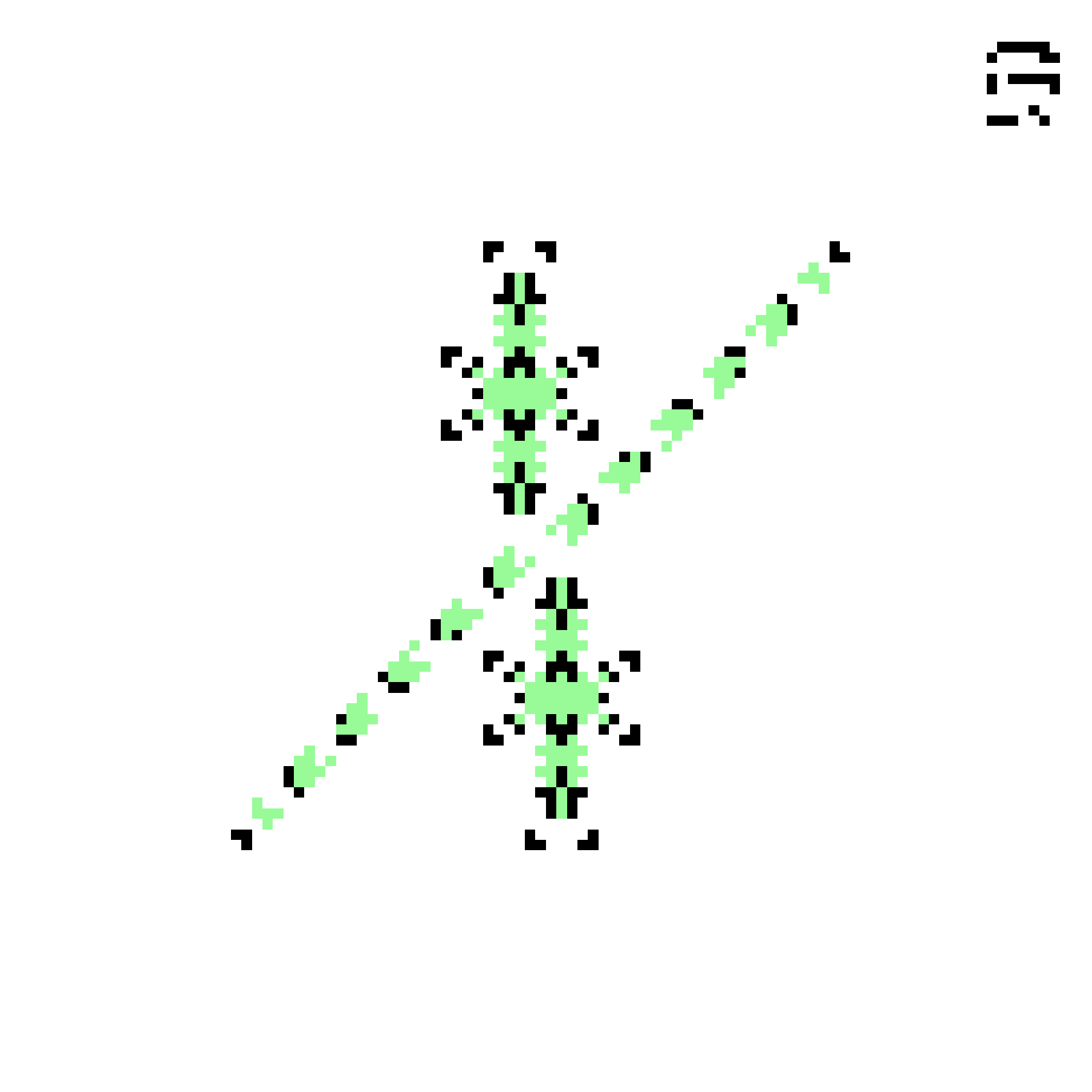}}\\ 
\textsf{\small{Precursor: $p$=19}}
\end{minipage}
\hfill
\begin{minipage}[c]{.236\linewidth} 
\fbox{\includegraphics[width=1\linewidth,bb=45 118 246 415, clip=]{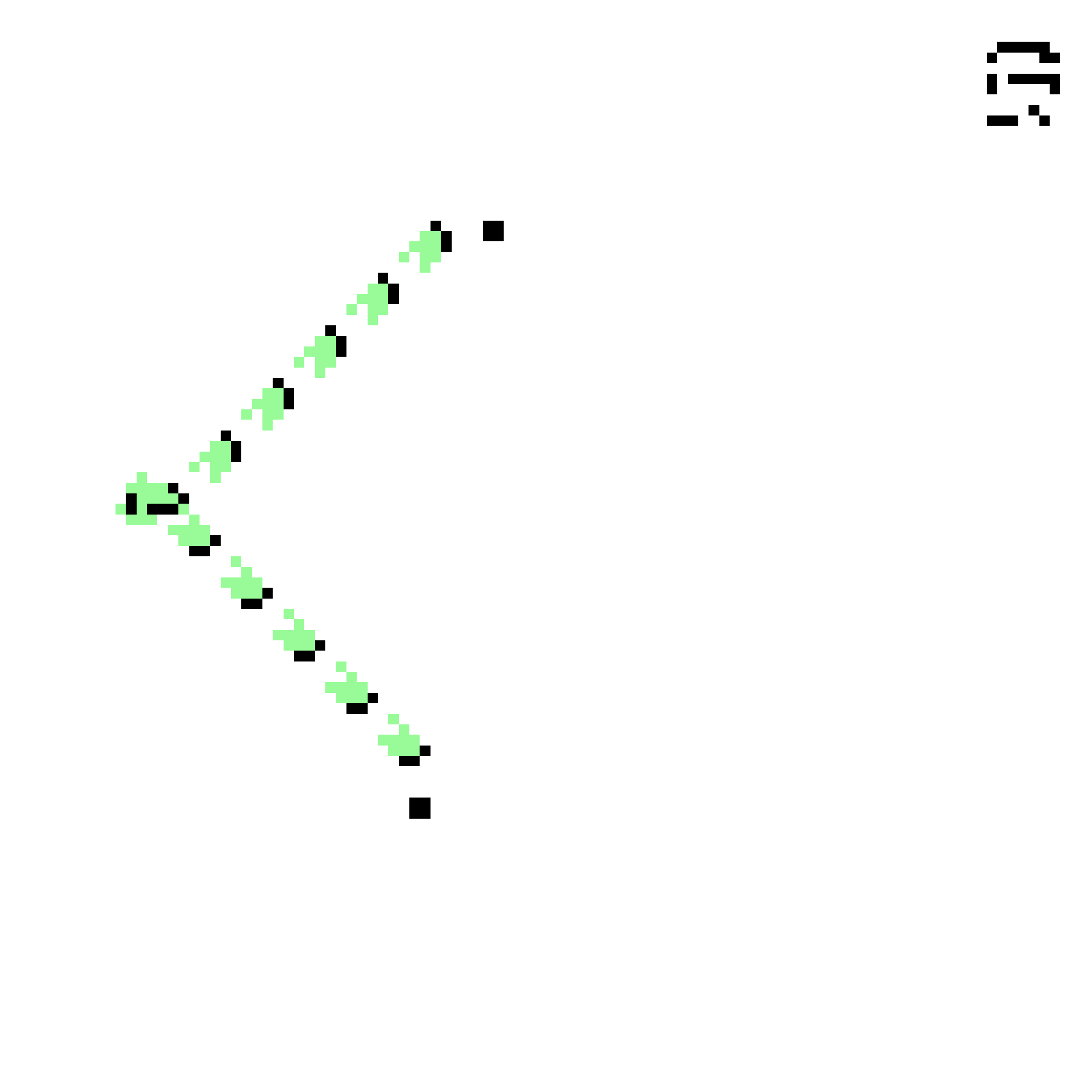}}\\ 
\textsf{\small{Sayab: $p$=20}}
\end{minipage}
\end{minipage}
\end{center}
\vspace{-4ex}
\caption[Comparing three glider-guns]
{\textsf{
Comparing the three GGa glider-guns of the Variant, Precursor, and
Sayab rules firing Ga gliders, with the period $p$ (firing frequency)  indicated.
Note the different mix of glider-stream phases, where Sayab has just one phase
per time-step, the other rules have two, but a different mix. 
Gliders streams are stopped by eaters. Green denotes motion.
\label{three glider-guns}
}}
\vspace{-4ex}
\end{figure}

\begin{figure}[h]
\begin{center}
\textsf{\small
\begin{tabular}[t]{ @{}c@{} @{}c@{}  @{}c@{}  @{}c@{}  @{}c@{}   @{}c@{}   }
& \multicolumn{4}{c}{$\leftarrow$---------------- Ga ----------------$\rightarrow$} &\\
  \includegraphics[height=.08\linewidth,bb= 1 1 30 30, clip=]{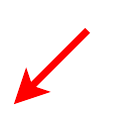}
&  \includegraphics[height=.08\linewidth,bb=4 3  47 35, clip=]{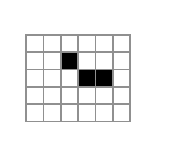}%
& \includegraphics[height=.08\linewidth,bb=2 3  47 35,  clip=]{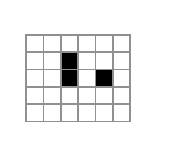}%
& \includegraphics[height=.08\linewidth,bb=-7 3  47 35,  clip=]{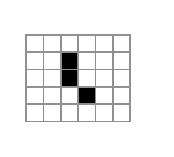}%
& \includegraphics[height=.08\linewidth,bb=-7 3  47 35,  clip=]{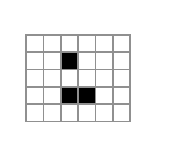}%
& \includegraphics[height=.08\linewidth,bb=2 3  47 35,  clip=]{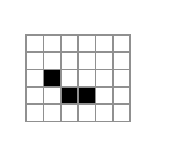}\\[-1ex]  
& 1 & 2 & 3 & 4 & 5 
\end{tabular}}

\textsf{\small
\begin{tabular}[t]{ @{}c@{} @{}c@{}  @{}c@{}  @{}c@{}  @{}c@{}   @{}c@{}   }
 \multicolumn{4}{c}{$\leftarrow$-------------------------- Gc --------------------------$\rightarrow$} & &\\
  \includegraphics[height=.1\linewidth,bb=2 3  67 45, clip=]{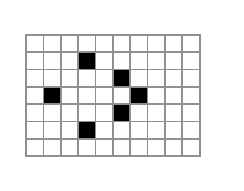}%
& \includegraphics[height=.1\linewidth,bb=2 3  67 45,  clip=]{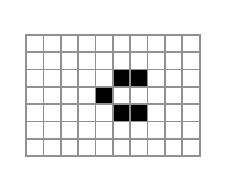}%
& \includegraphics[height=.1\linewidth,bb=2 3  67 45,  clip=]{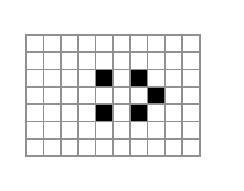}%
& \includegraphics[height=.1\linewidth,bb=2 3  67 45,  clip=]{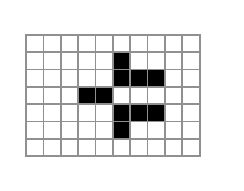}%
& \includegraphics[height=.1\linewidth,bb=2 3  67 45,  clip=]{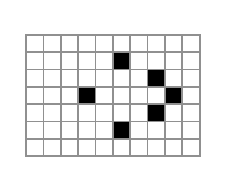}%
& \includegraphics[width=.09\linewidth,bb= 10 2 32 23, clip=]{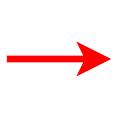}\\[-1ex]  
1 & 2 & 3 & 4 & 5 & 
\end{tabular}}
\end{center}
\vspace{-4ex}
\caption[glidera Ga and Gc]%
{\textsf{The 4 phases of the diagonal glider Ga and the orthogonal glider Gc,
moving as indicated by arrows. The speed of Ga=$c$/4, Gc=$c/2$.
\label{glider-Ga-Gc}
}}
\end{figure}

The glider-gun, a periodic structure ejecting gliders into space, is
the key and most elusive mechanism.  Figure~\ref{three glider-guns}
compares the three very different GGa glider-guns, of the Variant, 
Precursor\cite{Gomez2017} and Sayab\cite{Gomez2018} rules, all 
ejecting the same 4-phase diagonal glider
Ga (figure~\ref{glider-Ga-Gc}), which is used to construct logical
gates. However, the ejected GGa gliders-streams differ in spacing and
mix of phases.

\enlargethispage{3ex}
\clearpage

\enlargethispage{3ex}
\begin{figure}
\begin{minipage}[c]{1\linewidth} 
\includegraphics[width=1\textwidth,bb=177 -10 851 651, clip=]{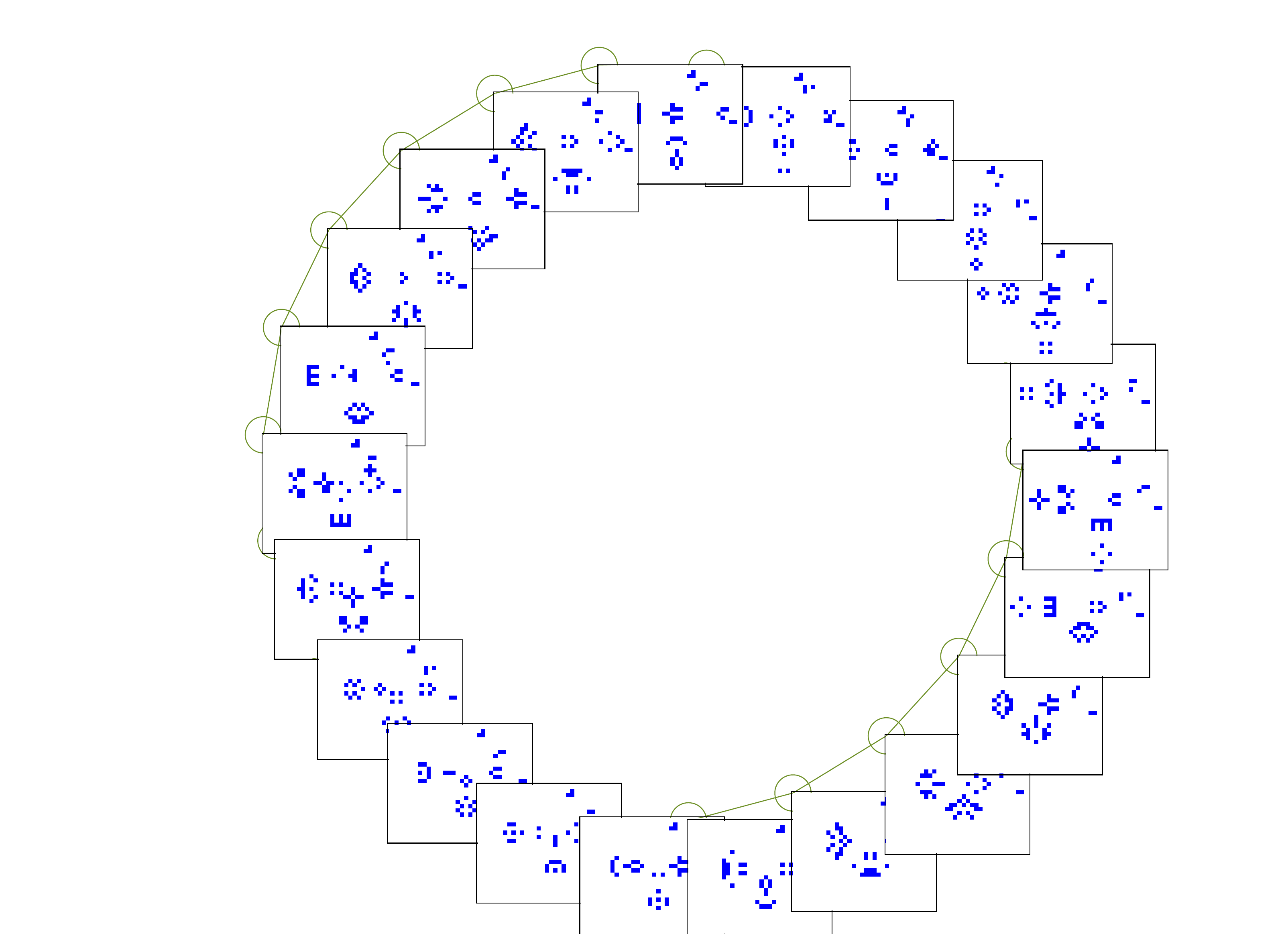}
\end{minipage}
\begin{minipage}[c]{1\linewidth} 
\vspace{-73ex}\hspace{33ex}\includegraphics[width=.34\textwidth,bb=12 11 441 363 ]{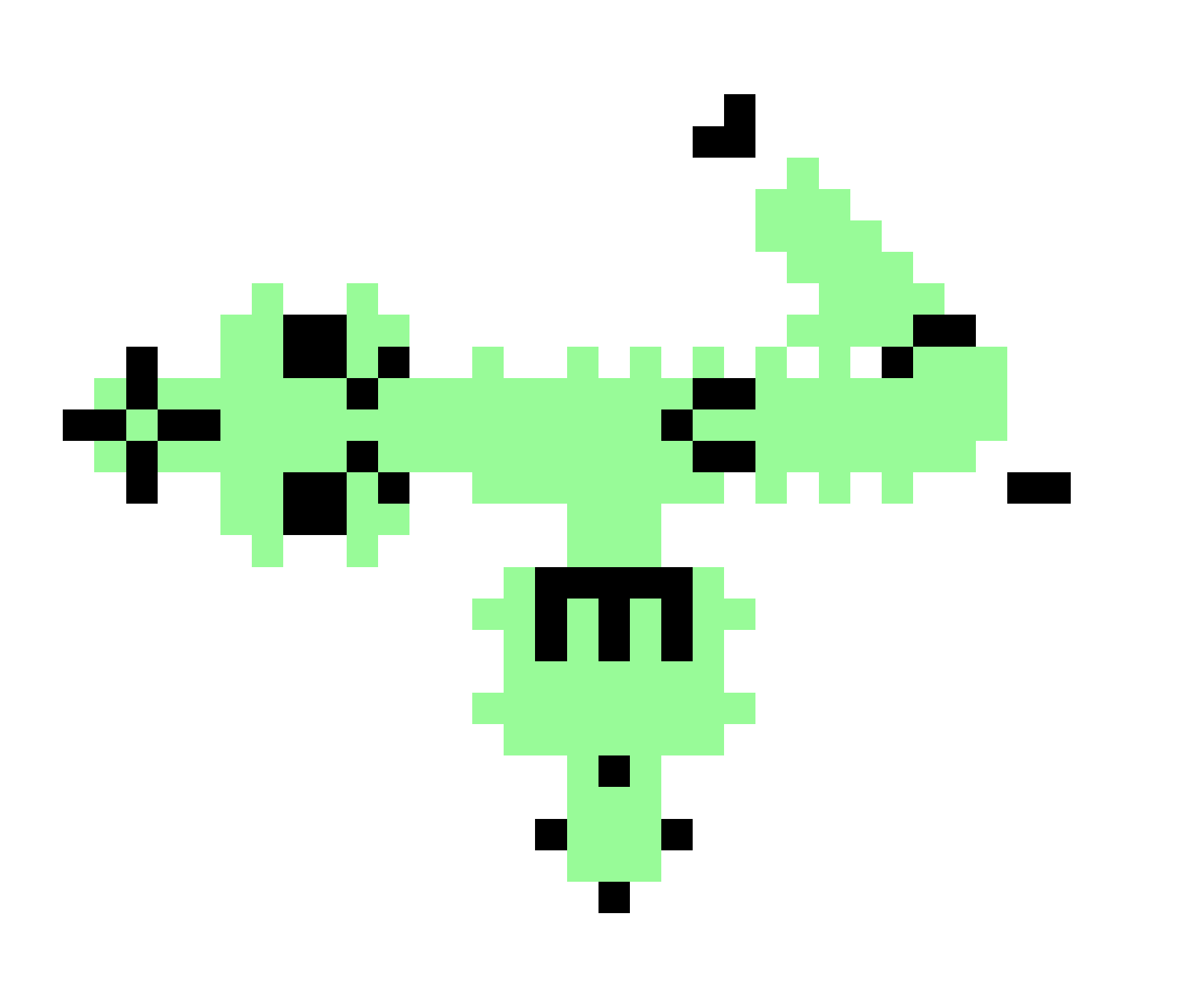}
\end{minipage}
\vspace{-5ex}
\caption[GGa attractor]
{\textsf{The Variant-rule glider-gun GGa attractor
    cycle\cite{Wuensche92,Wuensche2016} incorporates the sub-glider-gun GGc. 
    The period is 22 time-steps showing all
    phases/patterns of the GGa. The direction of time is clockwise.
    $Inset$: A glider-gun phase shown at a larger scale (green denotes
    motion) alongside the same phase on the attractor cycle.  Note the
    the glider Gc is shot to the East then reflected/transformed to glider
    Ga travelling NW, which is stopped by an Eater.
}}
\label{GGa attractor}
\end{figure}

The Precursor and Variant rules both also feature GGc glider-guns
(figure~\ref{three glider-guns}) which act as the initial, intermediate,
components for GGa glider-guns, where Gc gliders
bounce/transform to make Ga gliders, whereas the Sayab-rule GGa
glider-gun ejects Ga gliders directly.

\begin{figure}[htb]
\begin{center}
\fbox{\includegraphics[width=.8\linewidth,bb=97 170 337 313, clip=]{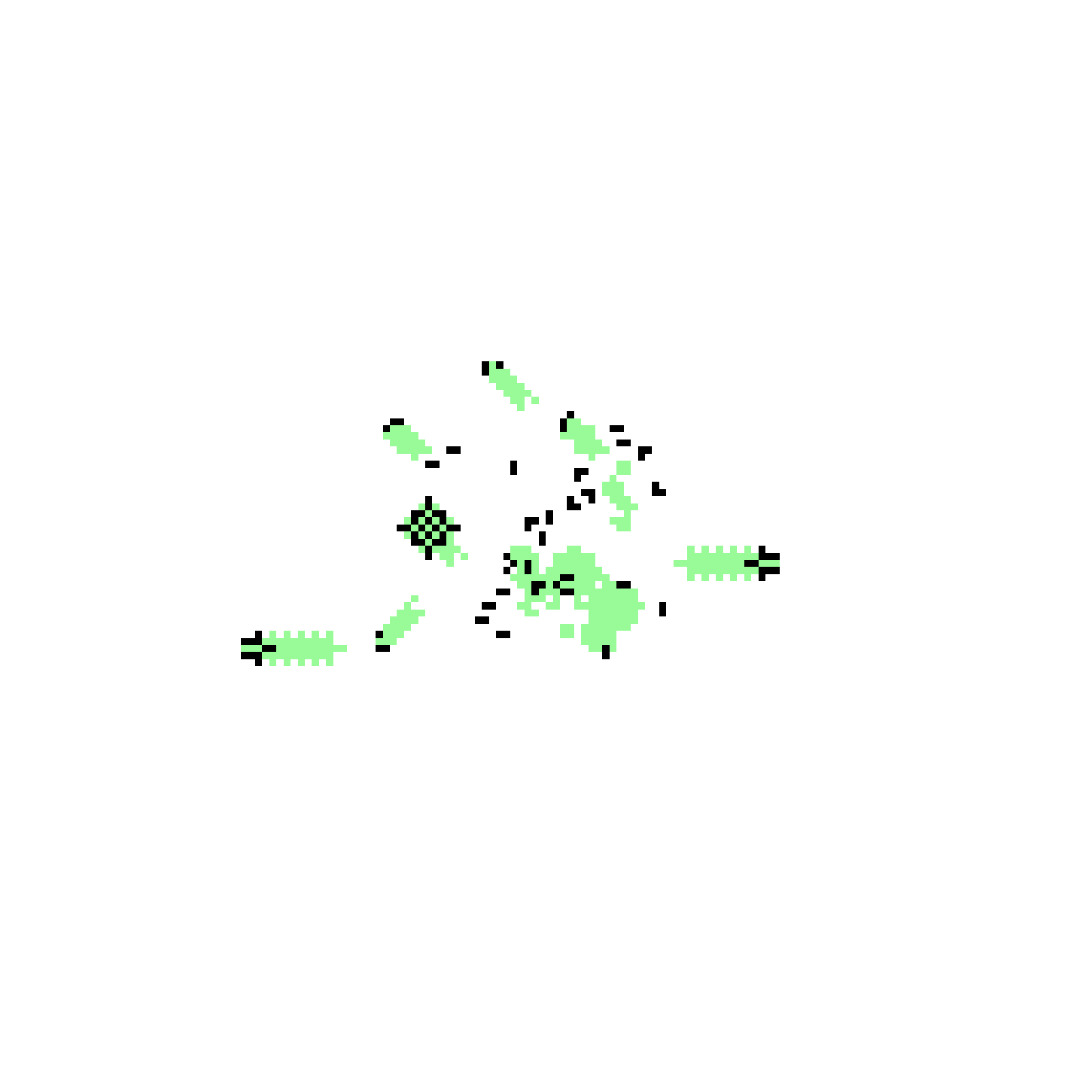}}
\end{center}
\vspace{-3ex}
\caption[Typical evolution from a random zone]
{\textsf{
A typical evolution emerging after 61 time-steps from a 30x30 30\%
density random zone.  Gliders and  still-lives have emerged
spontaneously.  Green denotes dynamics or motion for a given
number of time-steps,  and shows glider direction by their trails
--- this applies in all similar images in the paper.
\label{Typical evolution}
}}
\end{figure}

Given the genetic closeness of the Precursor and Variant rules, it is
not unexpected that both share glider types, small oscillators, and
small stable blocks which act as eaters or reflectors. However,
despite this closeness, their glider-guns and larger scale pattern
behaviors are very different. This perhaps illustrates both the
robustness and fragility of evolution, and suggests a direction for
further study.

The Variant-rule features spontaneously emergent gliders, stable
blocks and oscillators (figure~\ref{Typical evolution}), but the
complex patterns of its glider-guns GGc and GGa, shown in
figure~\ref{GGa attractor} as an attractor
cycle\cite{Wuensche92}\footnote{A ``bare'' attractor cycle, free of
  transients as in figure~\ref{GGa attractor} and similar figures, can
  be generated in DDLab\cite{Wuensche-DDLab} from a seed state by
  setting the number of transient levels to zero\cite{Wuensche2016}.}
of all 22 phases, are unlikely to emerge spontaneously.  Glider-guns
need to be constructed, and this has been achieved from the
interaction of two independent oscillating structures.  There are
other glider-guns types built from combinations of these basic
glider-guns, including glider-guns with variable periods, and surely
glider-guns yet to be discovered.  These structures can combine with
each other and with eaters, reflectors, oscillators and collisions to
build ever increasing complexity\footnote{Some of these, oscillators,
  glider-guns, puffers, rakes, etc. owe their discovery to
  contributions of the ConwayLife forum \cite{ConwayLife-forum}.}  by
multiple assemblies of sub-components, including the logical gates,
NOT, AND and OR, by GGa or GGc glider-guns, demonstrating logical
universality.

\enlargethispage{4ex}
The paper is structured in the following further sections, 
(\ref{The Variant-rule definition}) the
Variant-rule definition, 
(\ref{Gliders}) a description of gliders,
(\ref{Collisions}) collisions, 
(\ref{Glider-Guns, oscillators and reflectors}) gliders-guns, oscillators and reflectors, 
(\ref{Glider stream circuits}) glider stream circuits, 
(\ref{Variable period glider-guns}) variable period glider-guns,
(\ref{Logical Universality}) logical universality, 
(\ref{Spaceships, puffers and rakes}) spaceships, puffers and rakes, and 
(\ref{Concluding remarks}) concluding remarks.

\section{The Variant-rule definition}
\label{The Variant-rule definition}

\begin{figure}[htb]
\begin{center}
\includegraphics[width=1\linewidth]{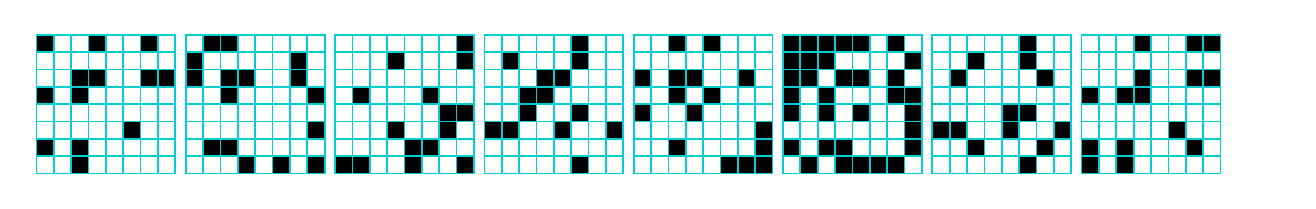} 
\includegraphics[width=1\linewidth]{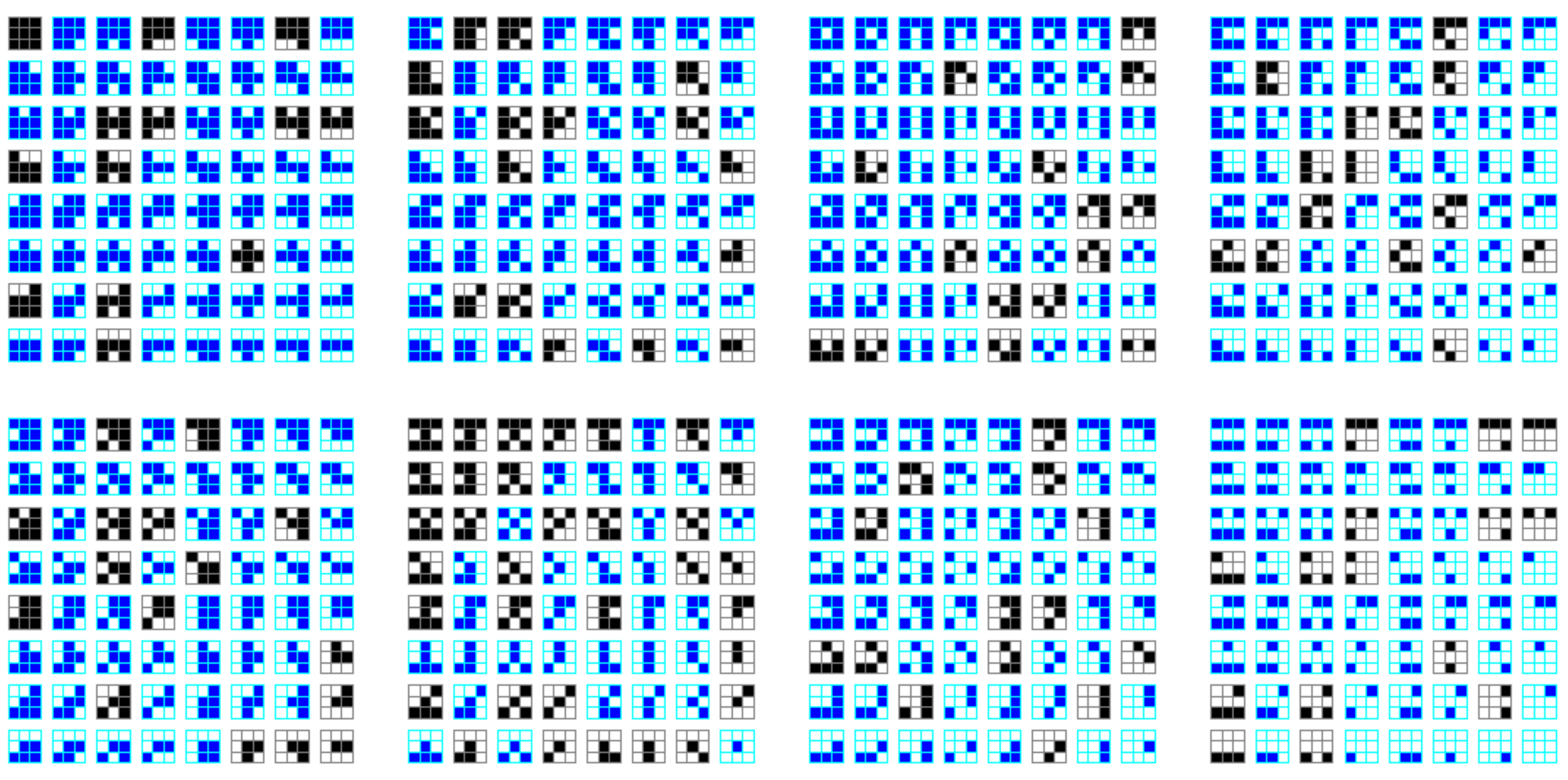} 
\end{center} 
\vspace{-4ex}
\caption[Variant rule's 512 neighborhoods]
{\textsf{\underline{\it Top}: The Variant rule-table based on all 512 neighborhoods, and 
\underline{\it Below}: expanded to show each neighborhood pattern.
136 black neighborhoods map to~1, 386 blue neighborhoods map to 0. Because the rule
is isotropic, only 102 symmetry classes are significant
(figure~\ref{isotropic neighborhoods with output=1}).}}
\label{512-nhoods}
\end{figure}

\begin{figure}[htb]
\begin{center}
\includegraphics[width=1\linewidth]{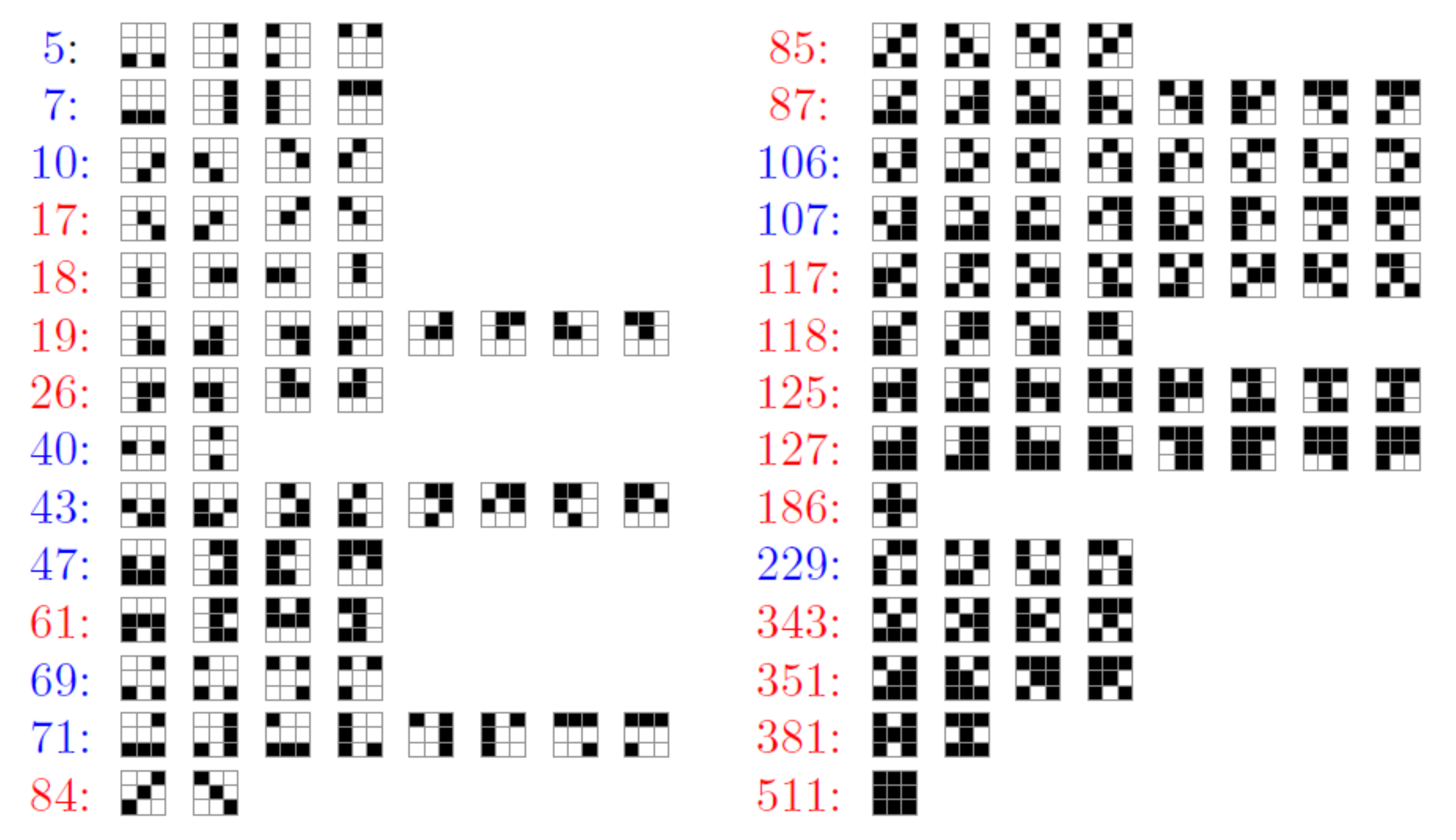}   
\end{center} 
\vspace{-4ex}
\caption[isotropic neighborhoods with output=1]%
{\textsf{The Variant-rule's 28 isotropic neighborhood symmetry classes that map to~1
(the remaining 74 symmetry classes map to~0, making 102 in total).
Each class is identified by the smallest decimal equivalent of the class, where the 3$\times3$
pattern is taken as a string in the order 
\begin{minipage}[c]{4ex}\scriptsize 876\\[-1ex]543\\[-1ex]210\end{minipage} --- for example, 
the pattern 
\raisebox{-.5ex}{\includegraphics[width=3ex,bb=6 8 25 26 2, clip=]{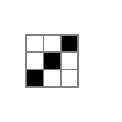}} 
is the string 001010100 representing the symmetry class 84.  The class
numbers are colored depending on the value of the central cell to distinguish
birth (blue) from survival (red), but no clear ÒLife-likeÓ birth/survival logic is
discernible.}}
\label{isotropic neighborhoods with output=1}
\end{figure}

Definitions of CA can be found from many sources, so we will skip the
details here. We just note that this paper deals with binary 2D
classical synchronous CA, comparable to the Game-of-Life (GoL) with a
Moore neighborhood, but not based on birth/survival, and with periodic
(or null) boundary conditions.
The Moore neighborhood has $3 \times 3=9$ cells giving a full
lookup-table with $2^9$ outputs, a rule-space of 512 
(figure~\ref{512-nhoods}), but we consider isotropic rules only, equal
outputs for any neighborhood rotation, reflection, or vertical
flip. If rules are classified by isotropy the number of effective
outputs, one for each symmetry class, reduces rule-space to
102\cite{Sapin2004}.  Within this isotropic rules-space, both the
Precursor and Variant rules have 28 symmetry classes with
an output of 1 (figure~\ref{isotropic neighborhoods with output=1}).
When the Precursor-rule was announced in the ConwayLife forum, an active member
with the handle ``Wildmyron'',
misstranscribed the rule into Golly's software format. The symmetry class\\ 

56:
\raisebox{-.5ex}{\epsfig{file=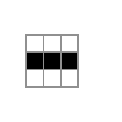,width=3ex,bb=6 8 25 26, clip=}}
\raisebox{-.5ex}{\epsfig{file=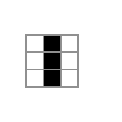,width=3ex,bb=6 8 25 26, clip=}} 
was replaced by the symmetry class\\

85:
\raisebox{-.5ex}{\epsfig{file=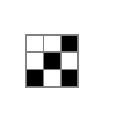,width=3ex,bb=6 8 25 26, clip=}}
\raisebox{-.5ex}{\epsfig{file=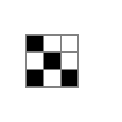,width=3ex,bb=6 8 25 26, clip=}}
\raisebox{-.5ex}{\epsfig{file=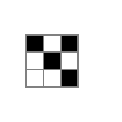,width=3ex,bb=6 8 25 26, clip=}}
\raisebox{-.5ex}{\epsfig{file=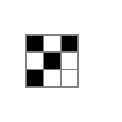,width=3ex,bb=6 8 25 26, clip=}}\\

with the happy consequence that the forum
was able to discover many interesting dynamical properties of the mutated rule,
which we named the ``Variant rule''. The rule, and seed files
for DDLab and Golly, as well as other rules in this family,
can be found in the ``Logical Universality in 2D Cellular Automata'' website\cite{UC2DCA-webpage},
so experiments can be repeated or new ones initiated.

\section{Gliders}
\label{Gliders}

Travelling patterns, and their collisions with each other and with
stationary patterns, can be used to simulate logical
processing in CA.  A travelling pattern is often periodic, translating
through space via a number of phases --- the distance travelled by a
particular phase to a new position gives the velocity measured in
time-steps.  Because of the nearest neighbor Moore neighborhood, the
fastest velocity, the ``speed of light'' $c$, is one lattice cell,
orthogonal or diagonal, per time-step, but there may be slower
velocities --- a stationary pattern has zero velocity. 
In the lexicon of the Game-of-Life played on an
orthogonal lattice, such a pattern moving diagonally is a ``glider''
whereas a pattern moving orthogonally is a ``space-ship'', but in this
paper we use the term ``glider'' for both.

\begin{figure}[htb]
\begin{center}
\begin{minipage}[c]{.85\linewidth} 
\begin{minipage}[c]{.25\linewidth}
\fbox{\includegraphics[width=1\linewidth,bb=55 356 125 386, clip=]{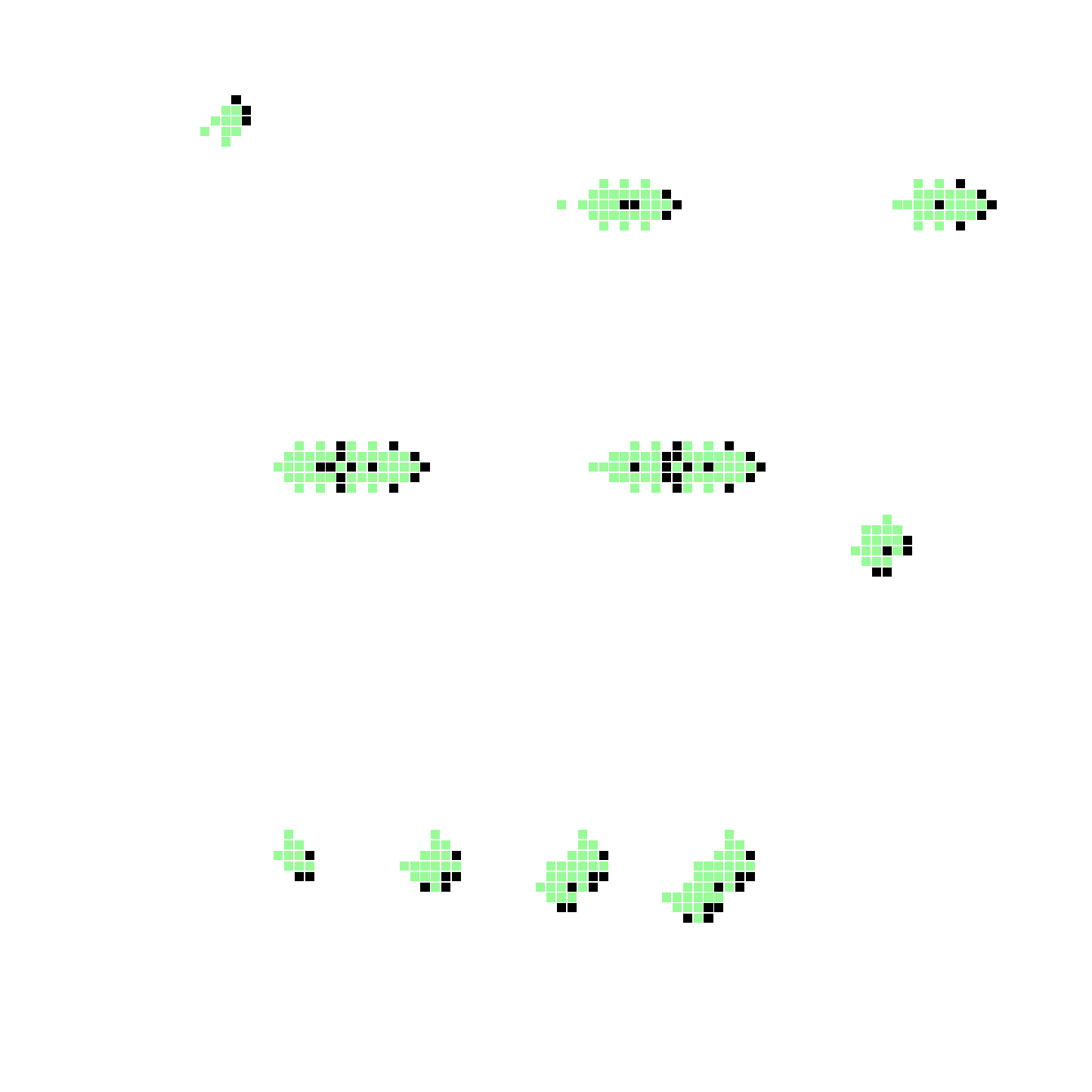}} 
\textsf{\small{Ga, $c$/4}}
\end{minipage}
\hfill
\begin{minipage}[c]{.25\linewidth}
\fbox{\includegraphics[width=1\linewidth,bb=207 323  272 354, clip=]{pdf-figs/gl-all}} 
\textsf{\small{Gb, $c$/2}}
\end{minipage}
\hfill
\begin{minipage}[c]{.25\linewidth}
\fbox{\includegraphics[width=1\linewidth,bb=330 323 395 354, clip=]{pdf-figs/gl-all}} 
\textsf{\small{Gc, $c$/2}}
\end{minipage}\\[1ex]
\begin{minipage}[c]{.25\linewidth}
\fbox{\includegraphics[width=1\linewidth,bb=305 193 370 222, clip=]{pdf-figs/gl-all}} 
\textsf{\small{G2a, $c$/4}}
\end{minipage}
\hfill
\begin{minipage}[c]{.25\linewidth}
\fbox{\includegraphics[width=1\linewidth,bb=97 220 175 255, clip=]{pdf-figs/gl-all}} 
\textsf{\small{Gd, $c$/2}}
\end{minipage}
\hfill
\begin{minipage}[c]{.25\linewidth}
\fbox{\includegraphics[width=1\linewidth,bb=218 220 295 255, clip=]{pdf-figs/gl-all}} 
\textsf{\small{Ge, $c$/2}}
\end{minipage}\\
\begin{center}
\begin{minipage}[c]{.85\linewidth}
\fbox{\includegraphics[width=1\linewidth,bb=90 61 311 103, clip=]{pdf-figs/gl-all}} 
\textsf{\small{Ga compound gliders of increasing size, $c$/4}}
\end{minipage}
\end{center}
\end{minipage}
\end{center}
\vspace{-3ex}
\caption[glider types]
{\textsf{
Glider types in the Variant-rule --- one representative phase for each glider.
Glider-guns have been created for Ga, Gc and G2a in both the Variant and Precursor rules.
All gliders except for Gd and Ge also operate in the Precursor rule. All have 4 phases,
and the velocity is $c$/4 for diagonal gliders, and $c$/2 for
orthogonal.
\label{glider types}
}}
\vspace{-2ex}
\end{figure}

We have seen in figure \ref{glider-Ga-Gc} all 4 phases of the two
gliders, Ga and Gc --- Variant (and Precursor) glider-guns have been
found for these and also for G2a.  Figure~\ref{glider types} is a
summary showing just one phase of these other gliders in the
Variant-rule.  All have 4 phases, and the velocity is $c$/4 for
diagonal gliders, and $c$/2 for orthogonal.  All but Gd and Ge gliders
also operate in the Precursor rule.
  
\clearpage

\section{Collisions}
\label{Collisions}

Collisions are fundamental in the research of logical universality in
CA to manipulate gliders streams shot from a glider-gun, and control
other logical artifacts. There are many possible collision scenarios
between gliders, stable blocks, and oscillators, where
collision outcomes depend on the exact point of impact and phase. As
these are deterministic systems, a theory of collision behavior should
be possible but is beyond the present state of the art, so for a given
CA that supports complex glider dynamics one must resort to experiment
and compile a catalogue of useful collisions that may serve as logical
data transmission components. The Variant rule is rich in useful
collision outcomes, with a degree of overlap with the Precursor rule.

\begin{figure}[htb]
\begin{center}
\begin{minipage}[c]{1\linewidth}
\fbox{\begin{minipage}[c]{.25\linewidth} 
\vspace{1ex}
\includegraphics[width=1\linewidth,bb=140 192 223 220, clip=]{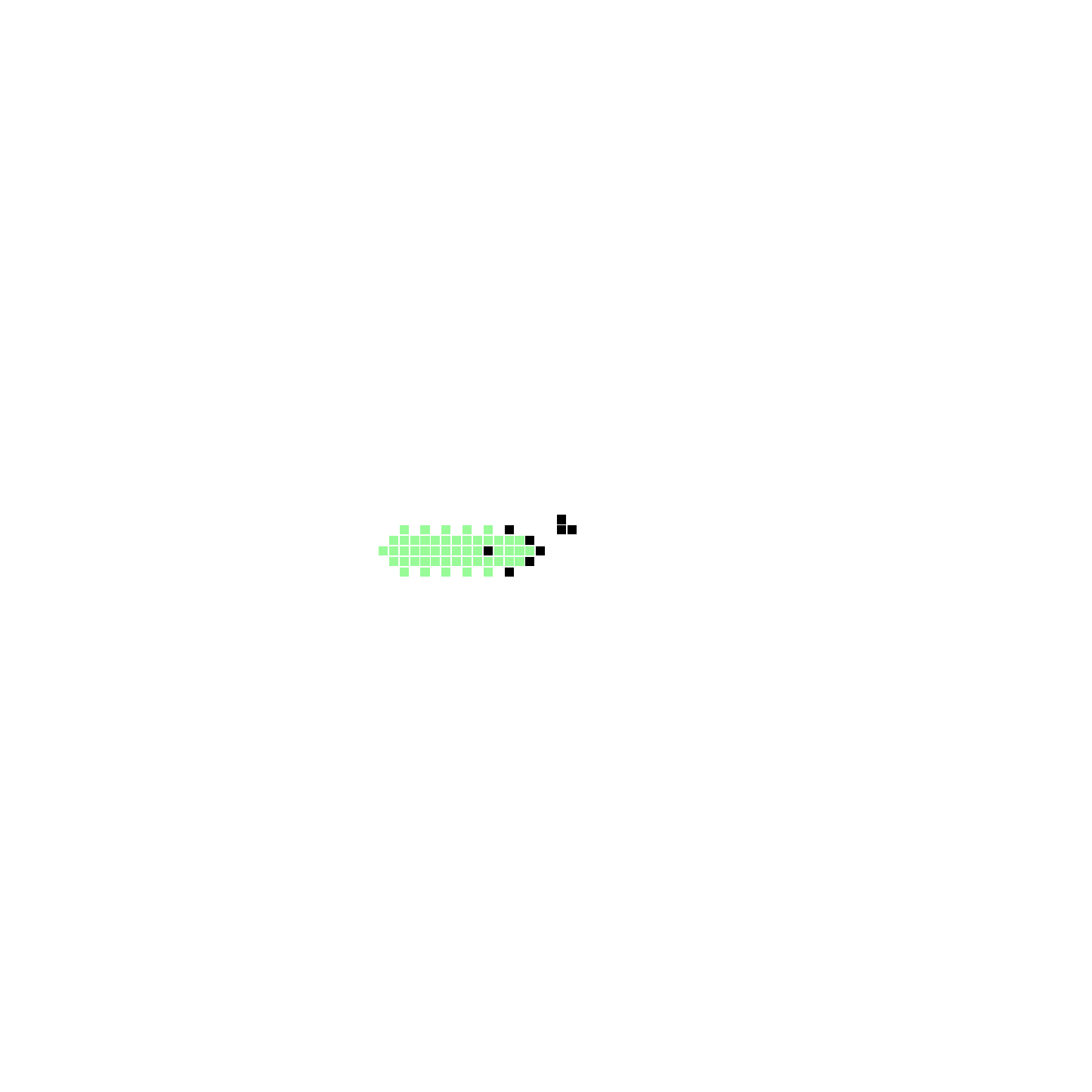}\\
\includegraphics[width=1\linewidth,bb=140 192 223 220, clip=]{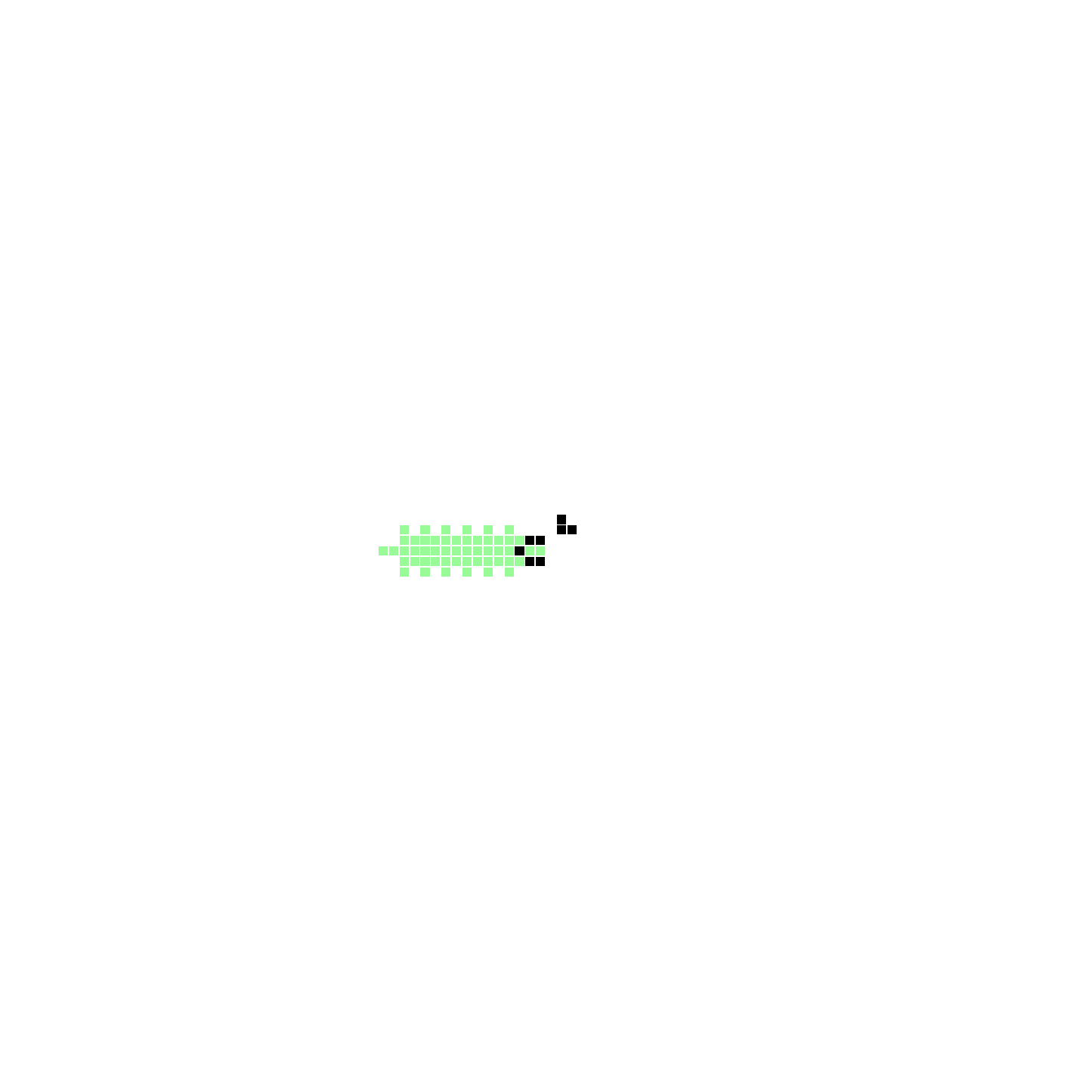}\\
\includegraphics[width=1\linewidth,bb=140 192 223 220, clip=]{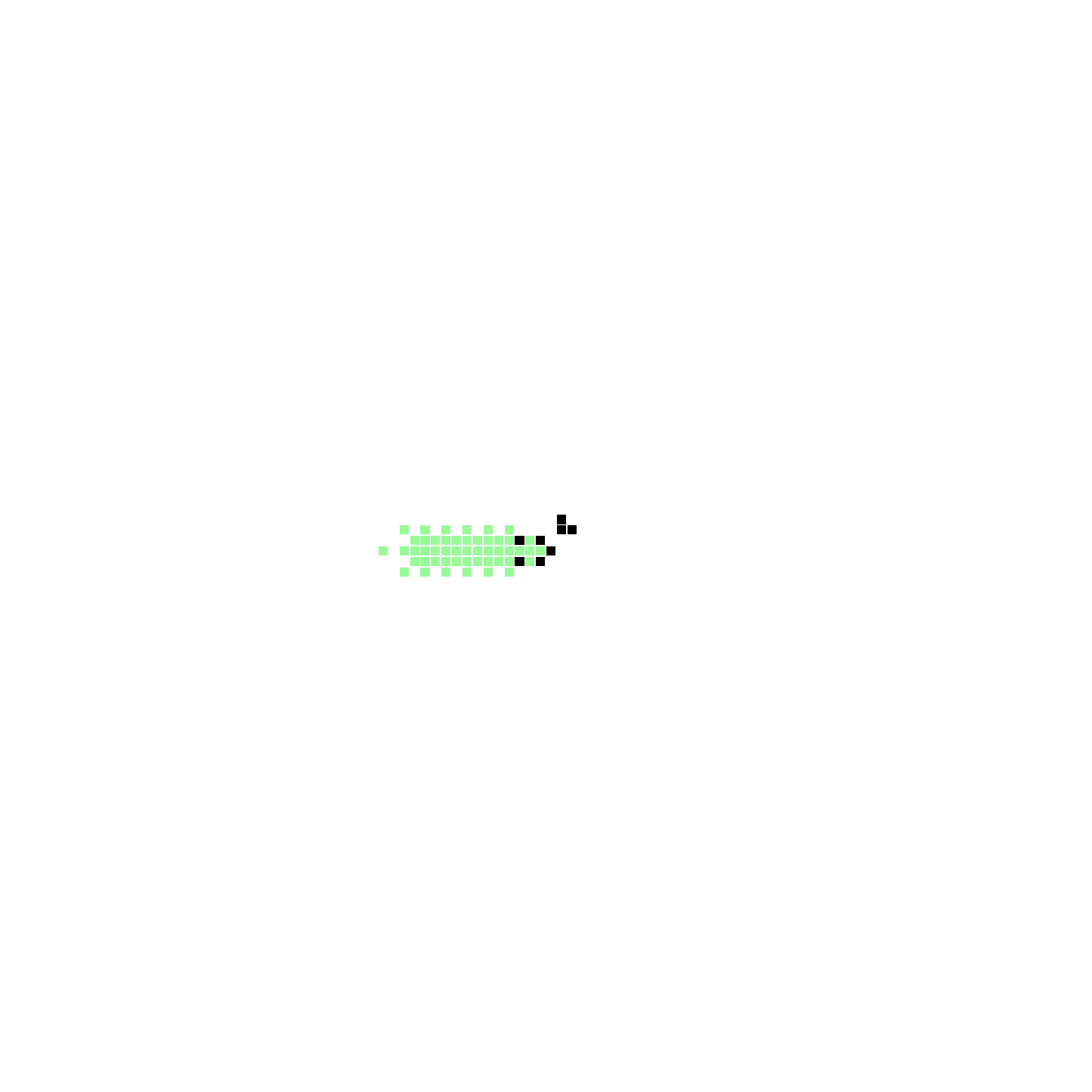}\\
\includegraphics[width=1\linewidth,bb=140 192 223 220, clip=]{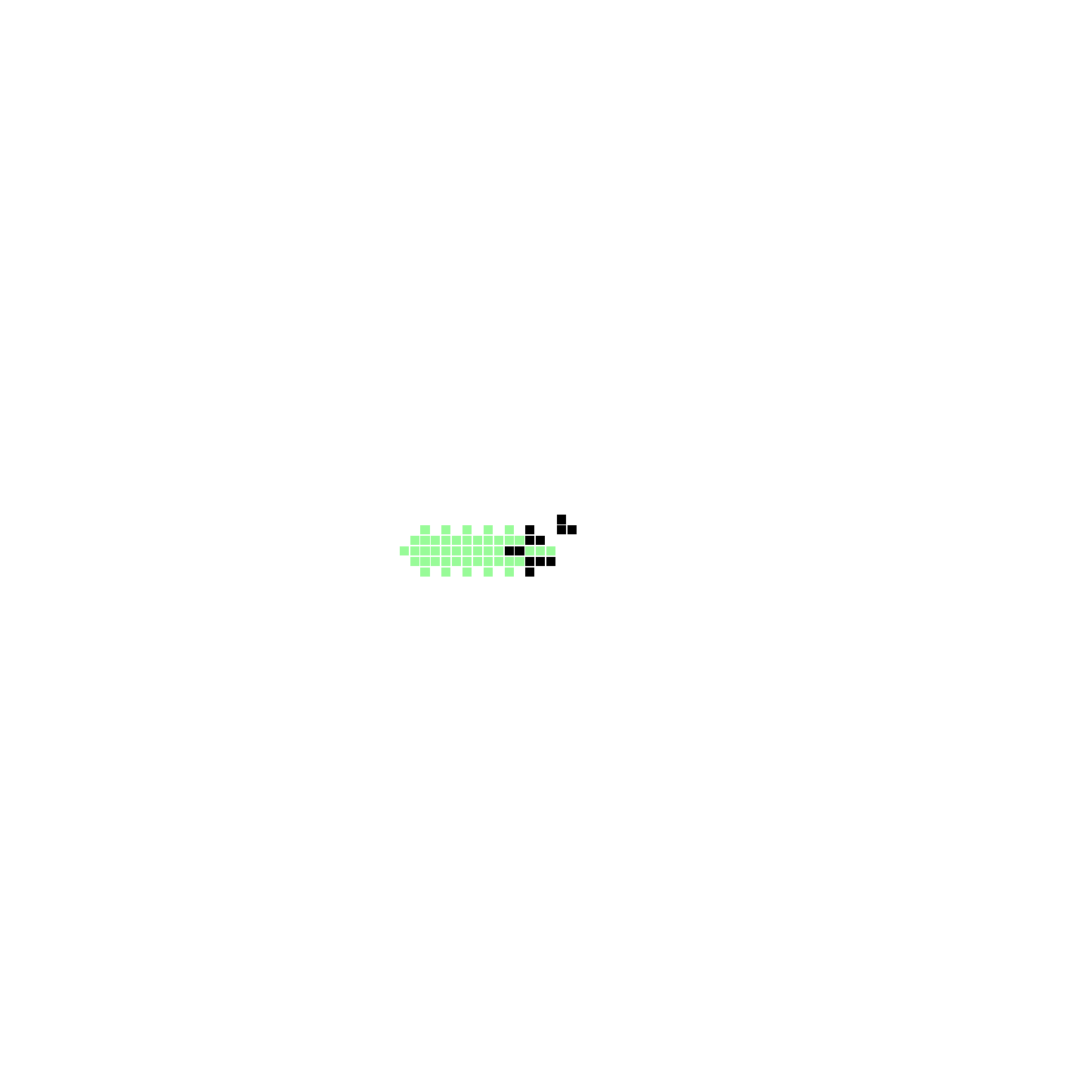}\\
\includegraphics[width=1\linewidth,bb=140 192 223 220, clip=]{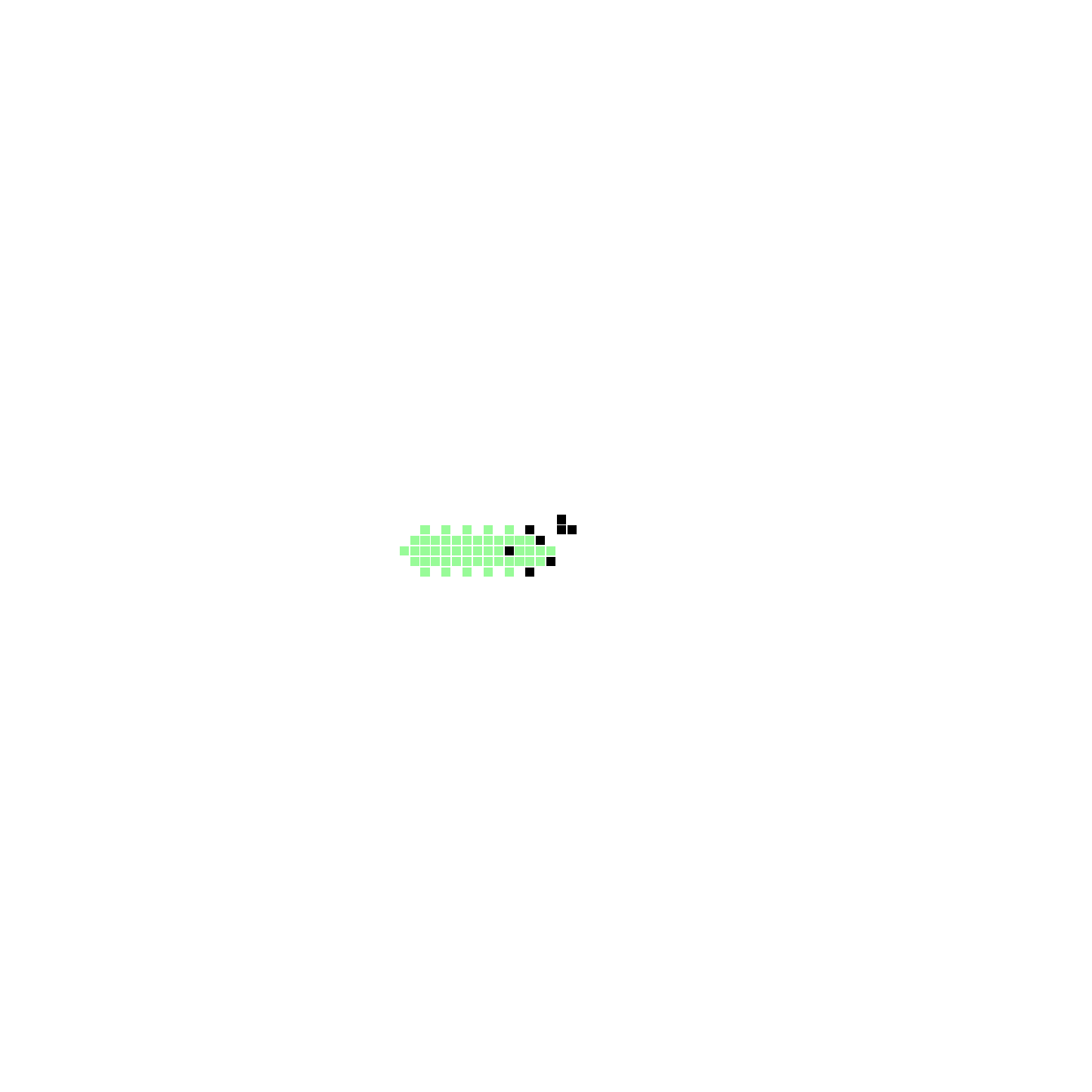}\\
\includegraphics[width=1\linewidth,bb=140 192 223 220, clip=]{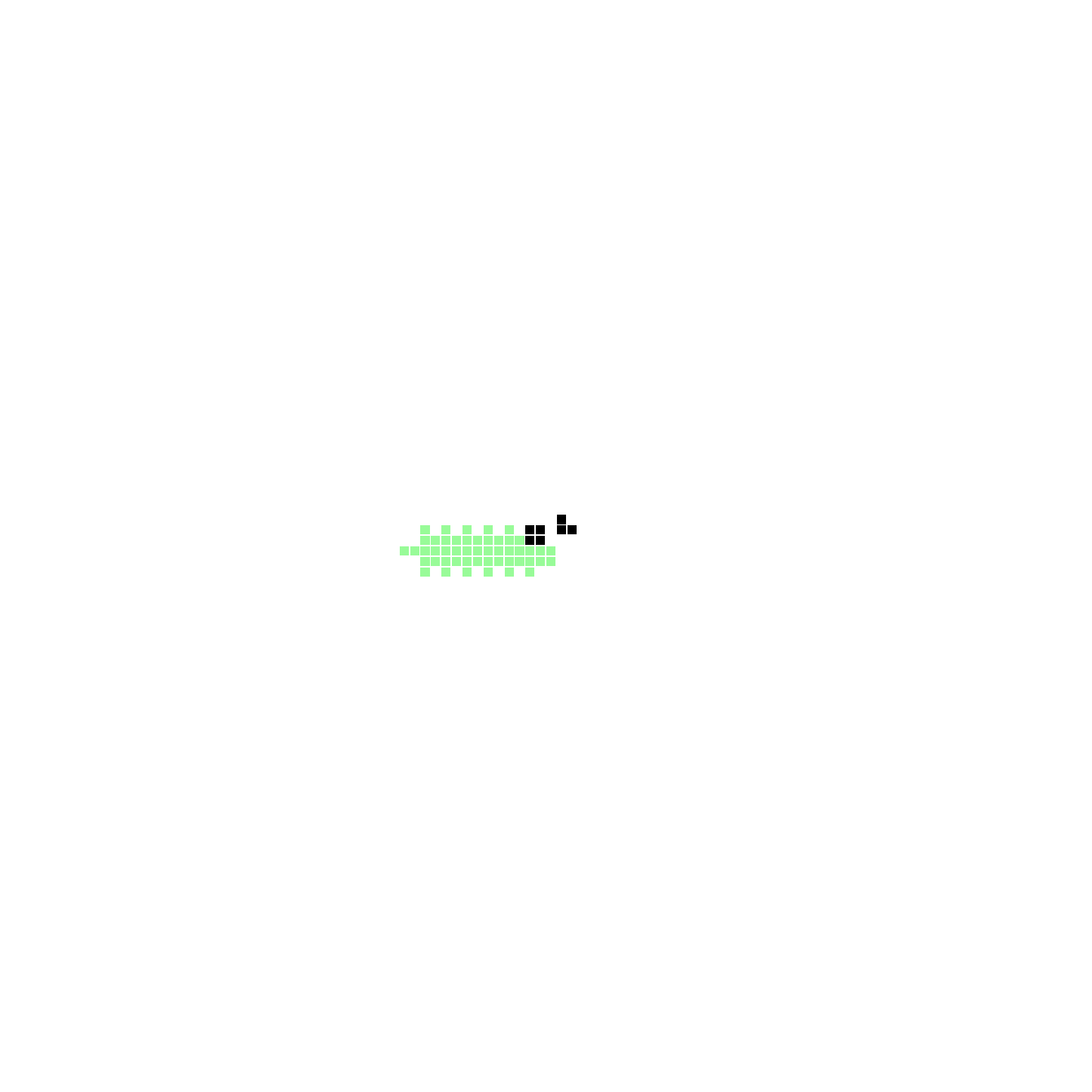}\\
\includegraphics[width=1\linewidth,bb=140 192 223 220, clip=]{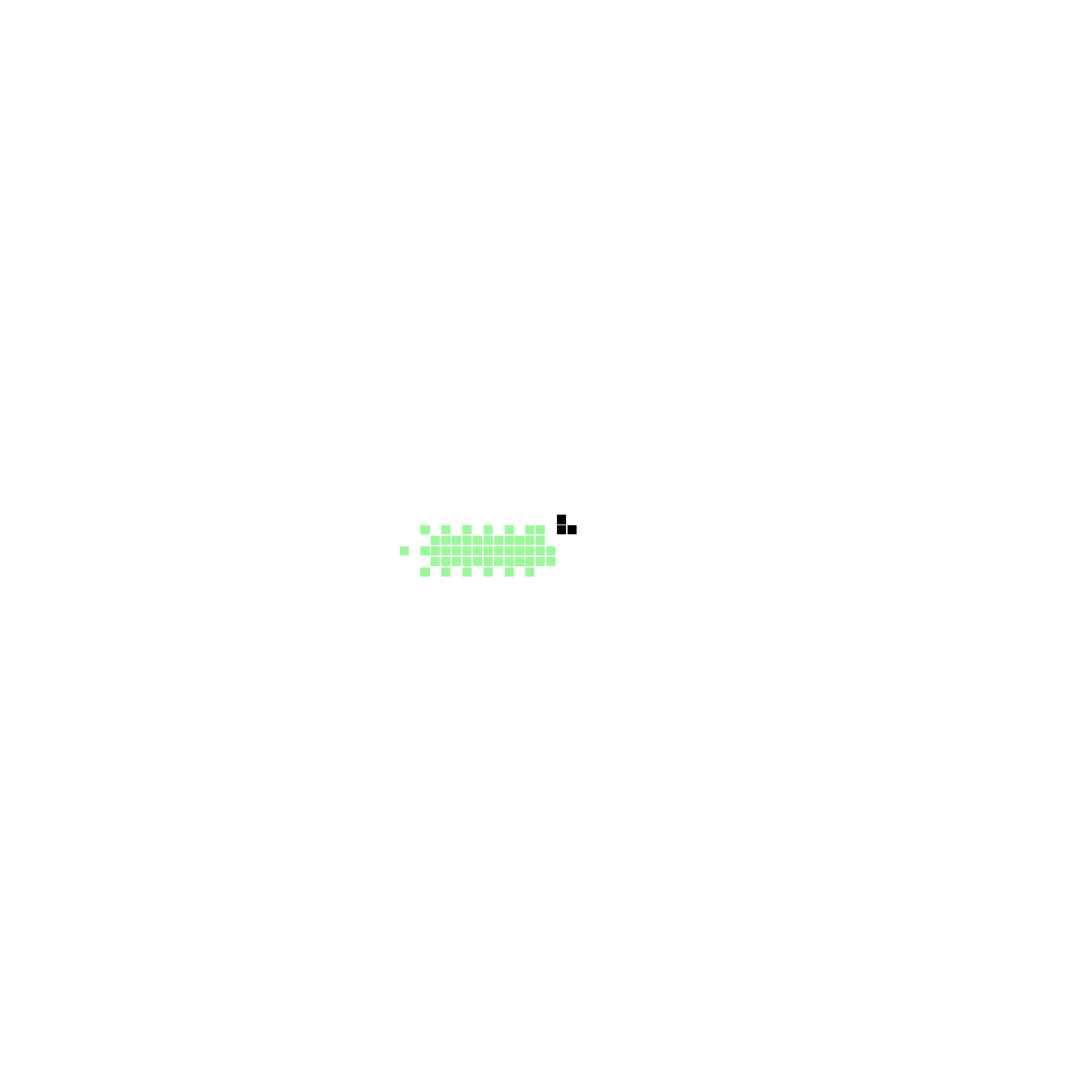}\\
\textsf{\small Gc Eater in 7 steps}
\end{minipage}}
\hfill
\fbox{\begin{minipage}[c]{.65\linewidth} 
\vspace{-4ex}
{\color{white}xx}\begin{minipage}[c]{.3\linewidth}
\includegraphics[width=1\linewidth,bb=160 182 215 244, clip=]{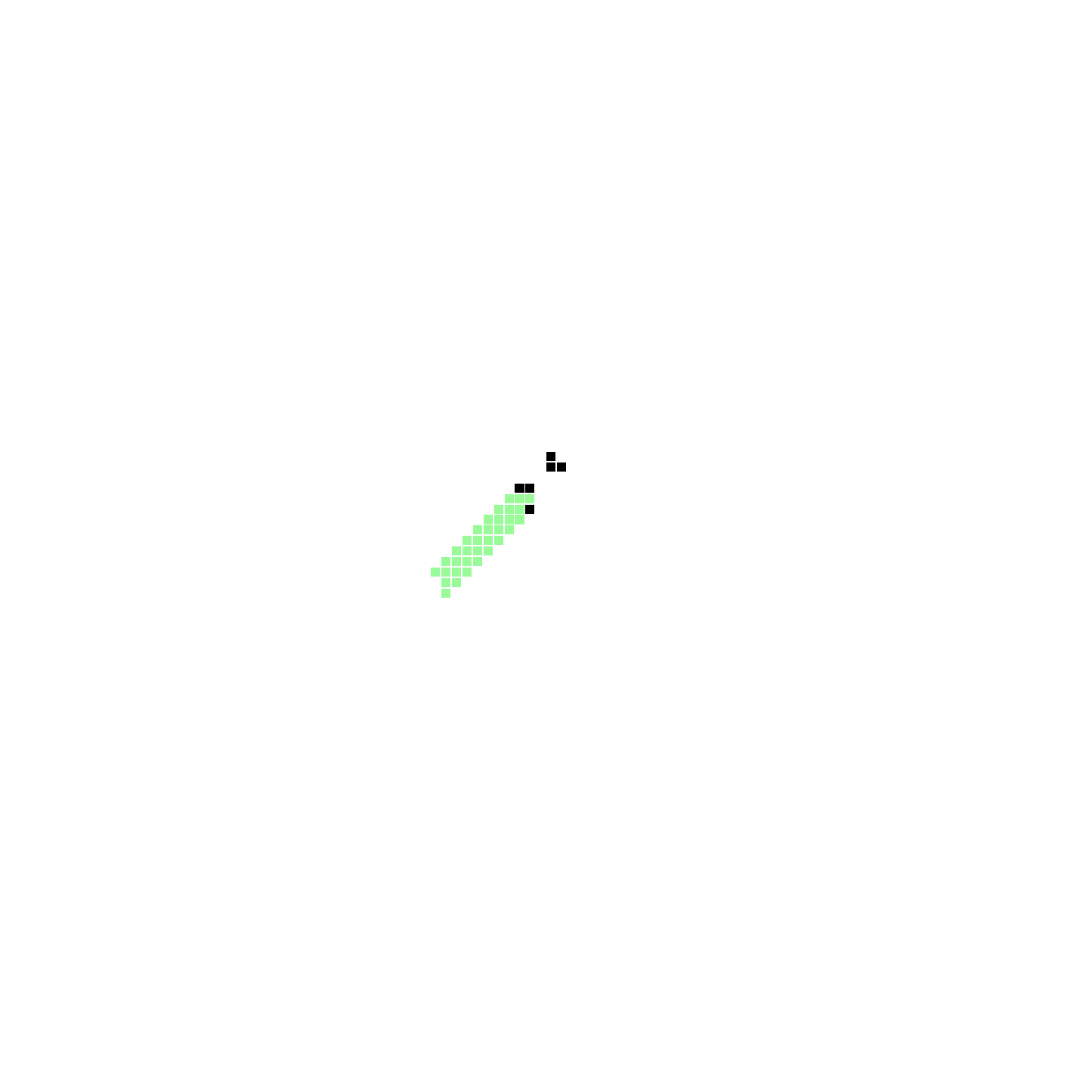}
\end{minipage}
\begin{minipage}[c]{.3\linewidth}
\vspace{8ex}\hspace{-2ex}
\includegraphics[width=1\linewidth,bb=160 182 215 244, clip=]{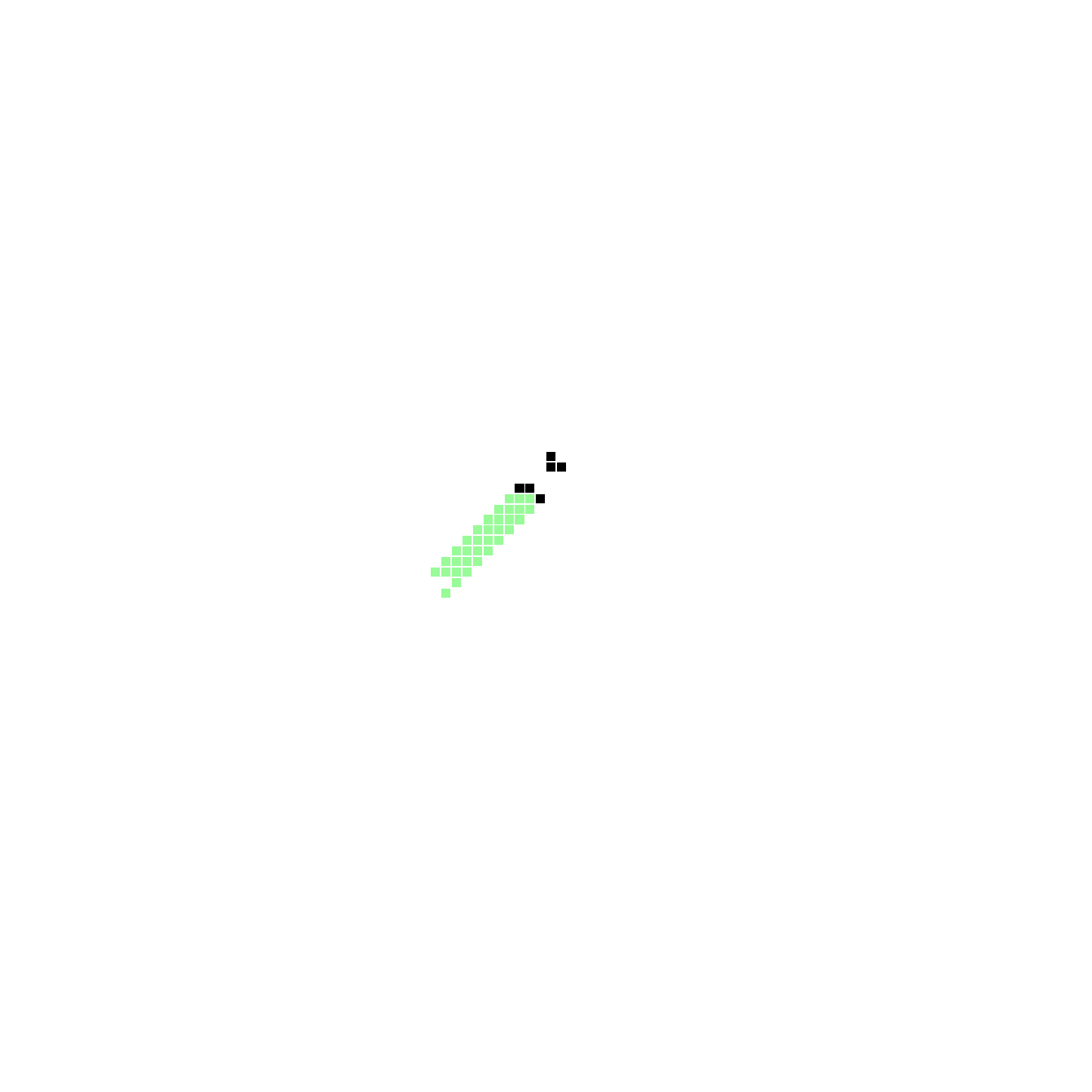}
\end{minipage}\vspace{8ex}
\begin{minipage}[c]{.3\linewidth}
\vspace{16ex}\hspace{-3.5ex}
\includegraphics[width=1\linewidth,bb=160 182 215 244, clip=]{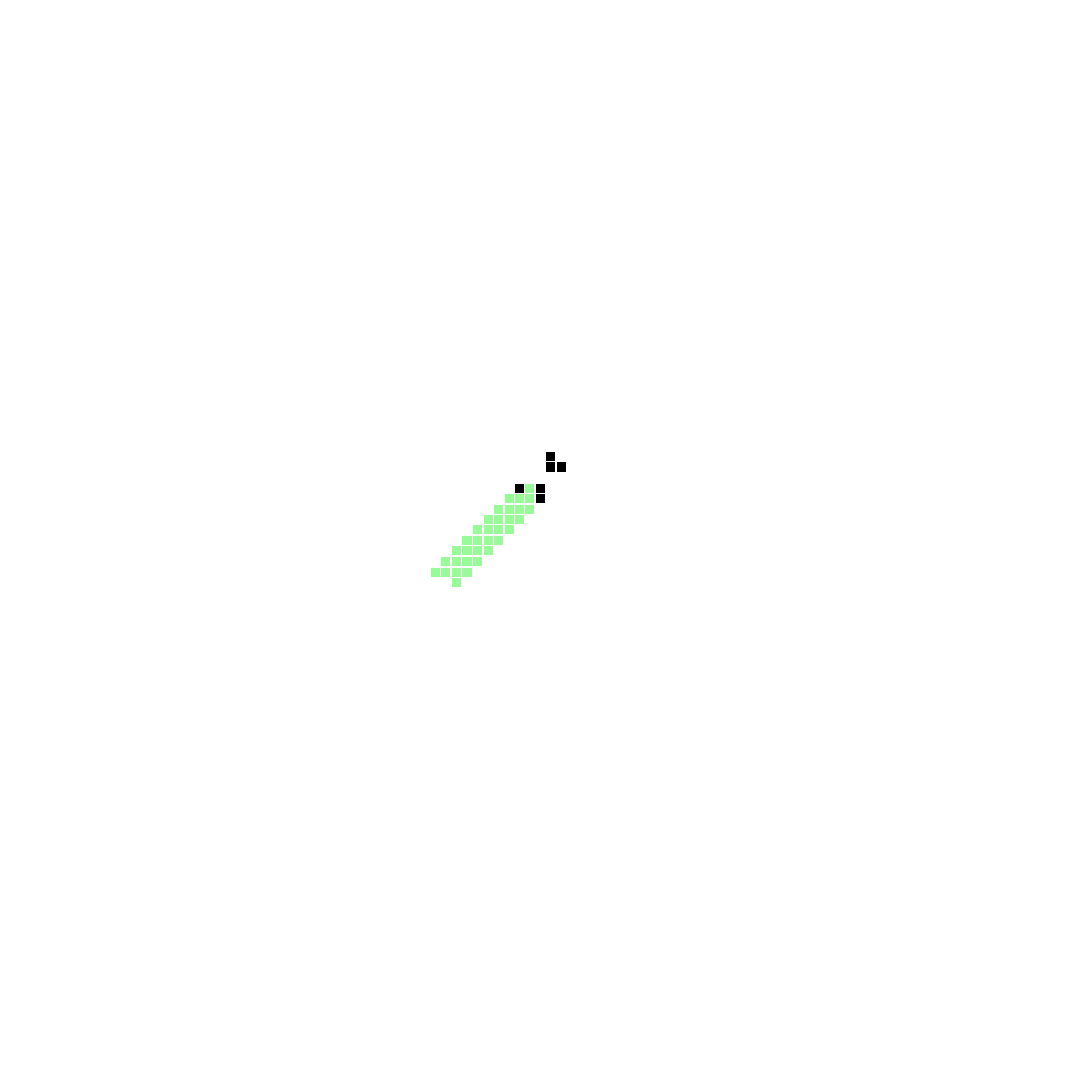}
\end{minipage}\\[-18ex]
{\color{white}xx}\begin{minipage}[c]{.3\linewidth}
\includegraphics[width=1\linewidth,bb=160 182 215 244, clip=]{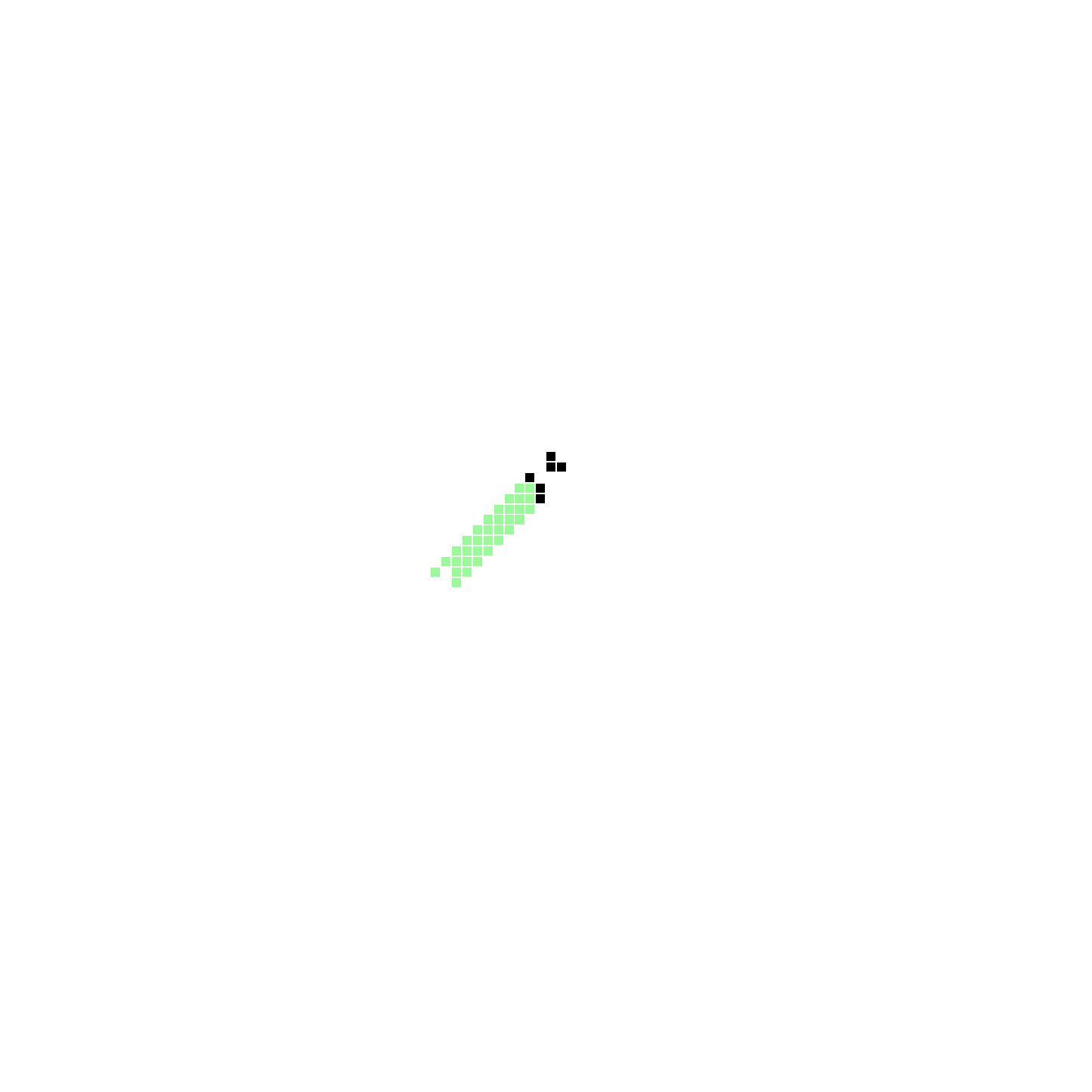}
\end{minipage}
\begin{minipage}[c]{.3\linewidth}
\vspace{8ex}\hspace{-2ex}
\includegraphics[width=1\linewidth,bb=160 182 215 244, clip=]{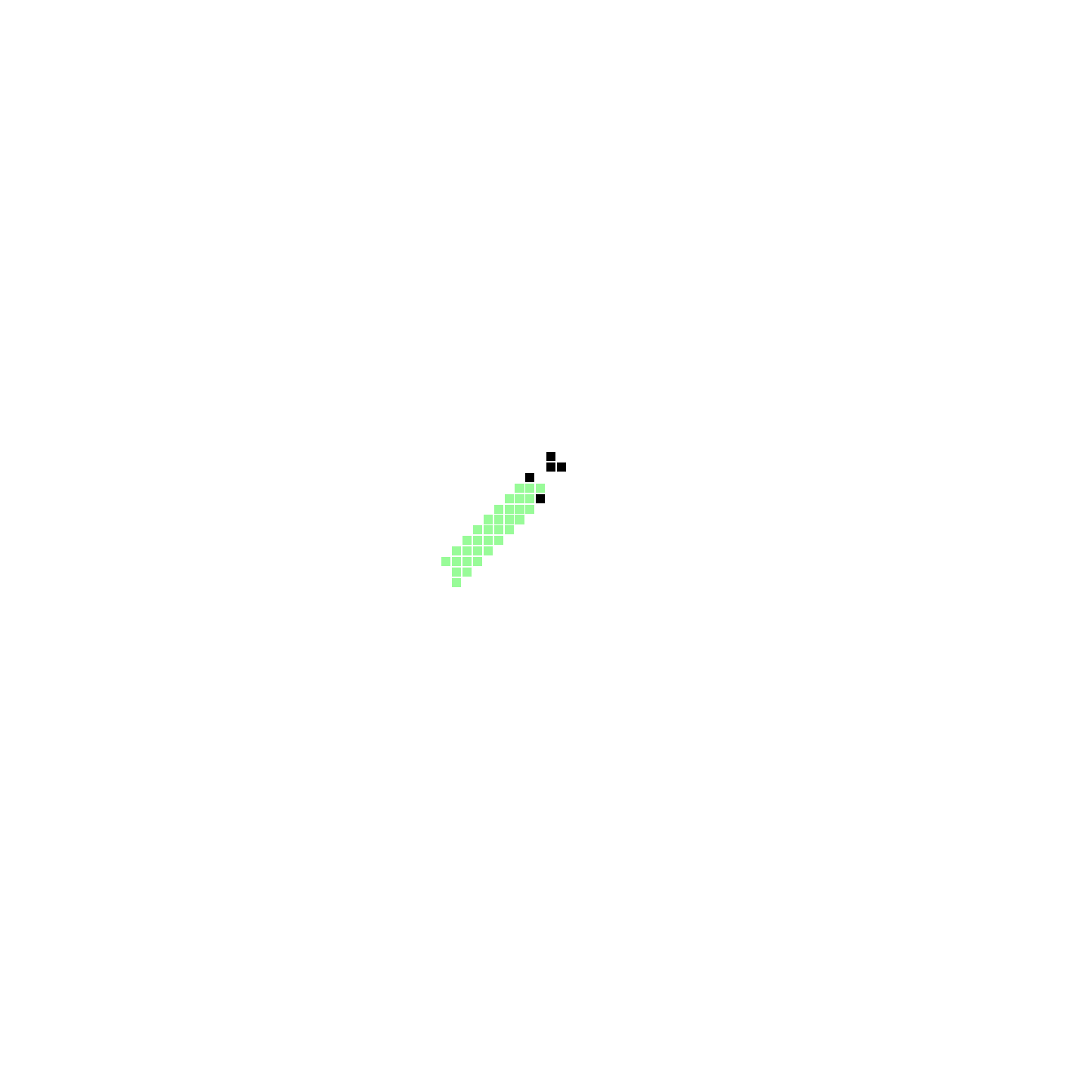}
\end{minipage}\vspace{8ex}
\begin{minipage}[c]{.3\linewidth}
\vspace{16ex}\hspace{-3.5ex}
\includegraphics[width=1\linewidth,bb=160 182 215 244, clip=]{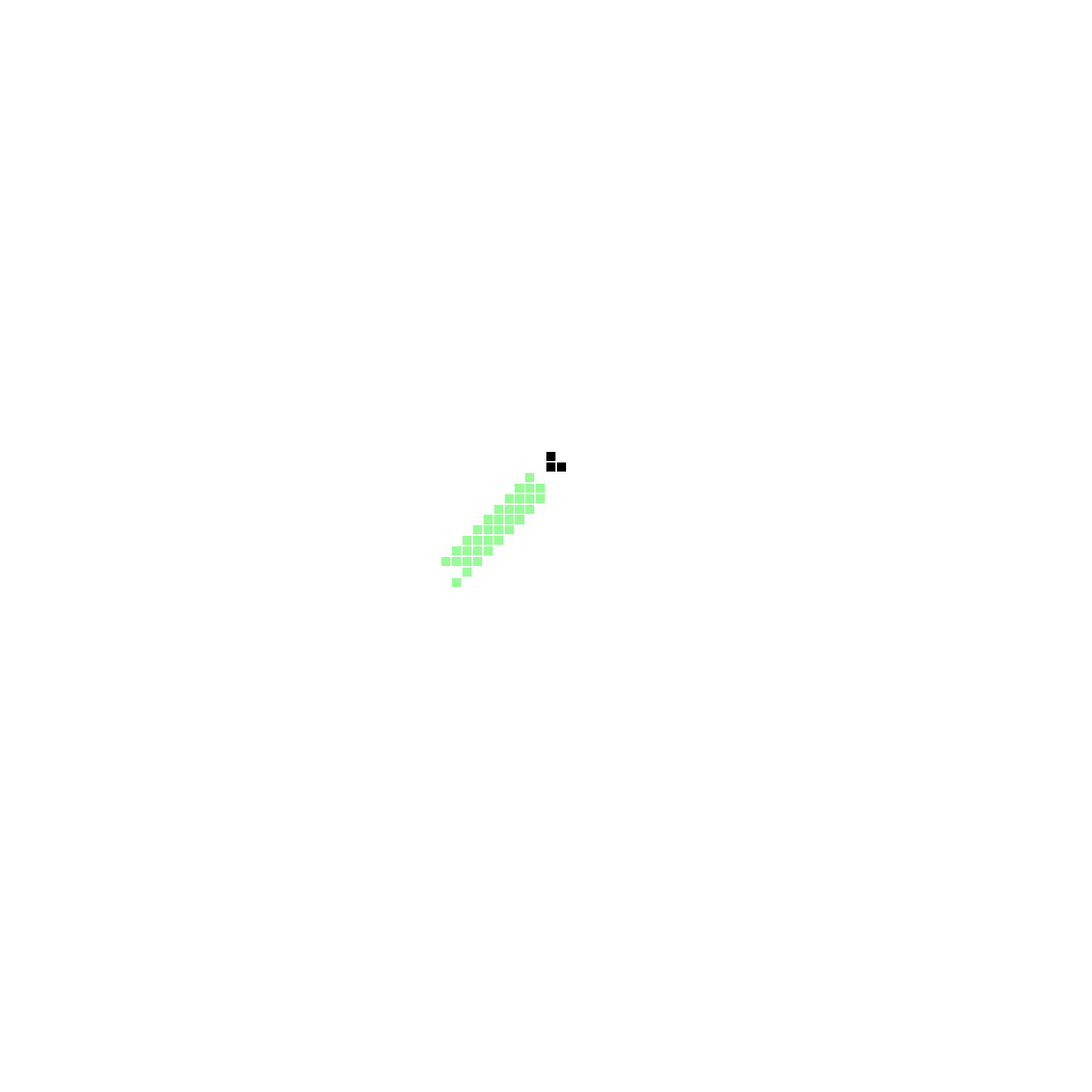}
\end{minipage}\\[-8ex]
\textsf{\small Ga Eater in 6 steps}
\end{minipage}}
\end{minipage}
\end{center}
\vspace{-2ex}
\caption[Ga and Gc Eaters]
{\textsf{Destruction of gliders Ga and Gc by an Eater stable block, which survives
the collision, showing phases/time-steps approaching the Eater as well as the impact.
These dynamics apply to both the Variant and Precursor rules.\\[-1.5ex]
The Gc eater shape can also be 
~\raisebox{-5.5ex}{\includegraphics[width=.11\linewidth]{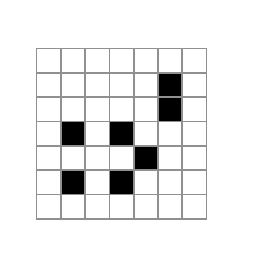}} 
and the Ga eater 
~\raisebox{-5.5ex}{\includegraphics[width=.11\linewidth]{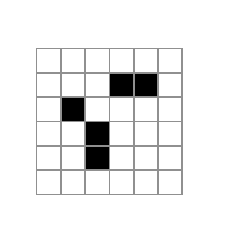}}\hspace{-2ex}
~\raisebox{-7ex}{\includegraphics[width=.11\linewidth]{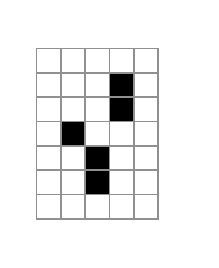}}
 \label{Ga and Gc Eaters}
}}
\vspace{-3ex}
\end{figure}

\begin{figure}[htb]
\begin{center}
\fbox{\begin{minipage}[c]{1\linewidth} 
\vspace{2ex}
\includegraphics[width=.13\linewidth,bb=93 190 139 268, clip=]{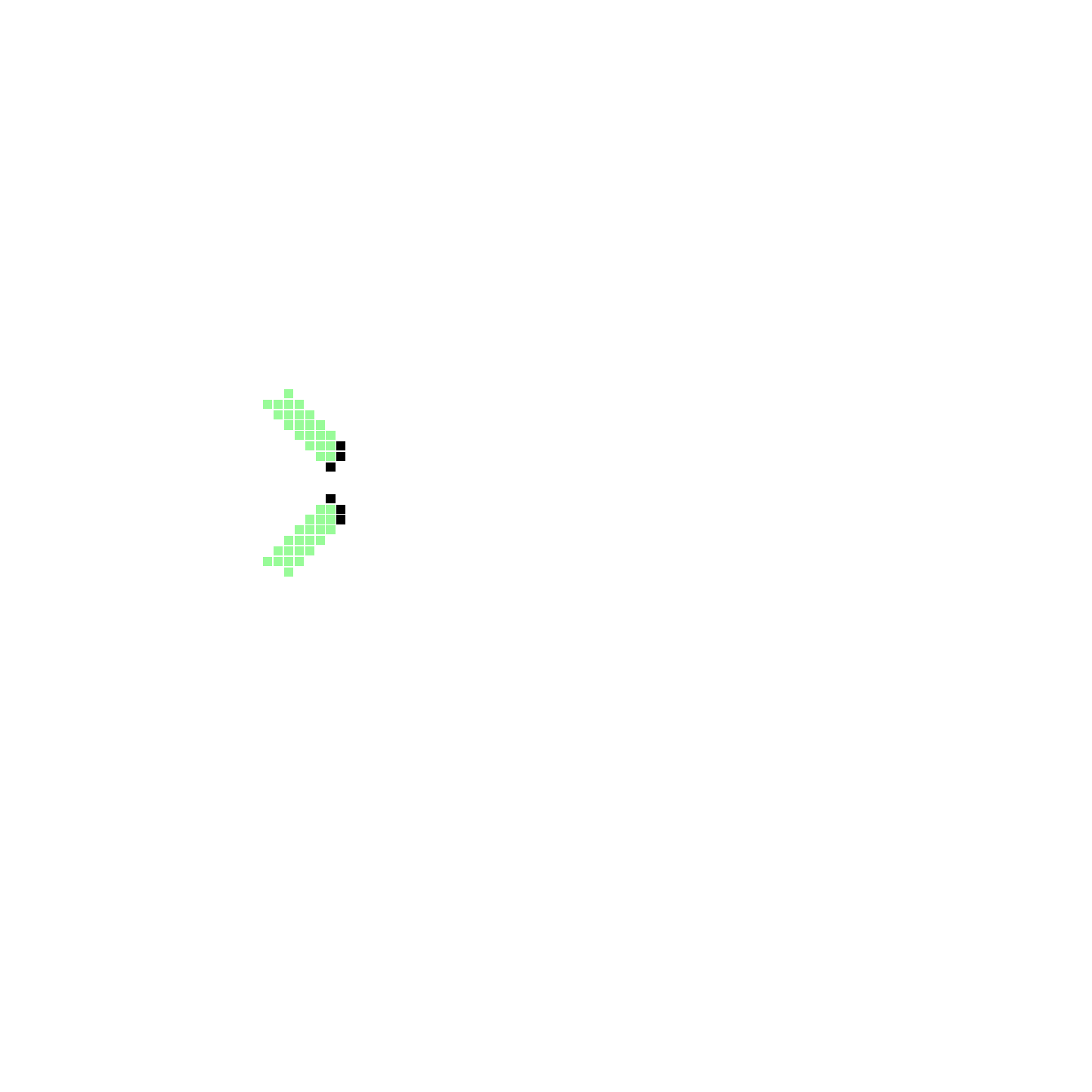} \hfill
\includegraphics[width=.13\linewidth,bb=93 190 139 268, clip=]{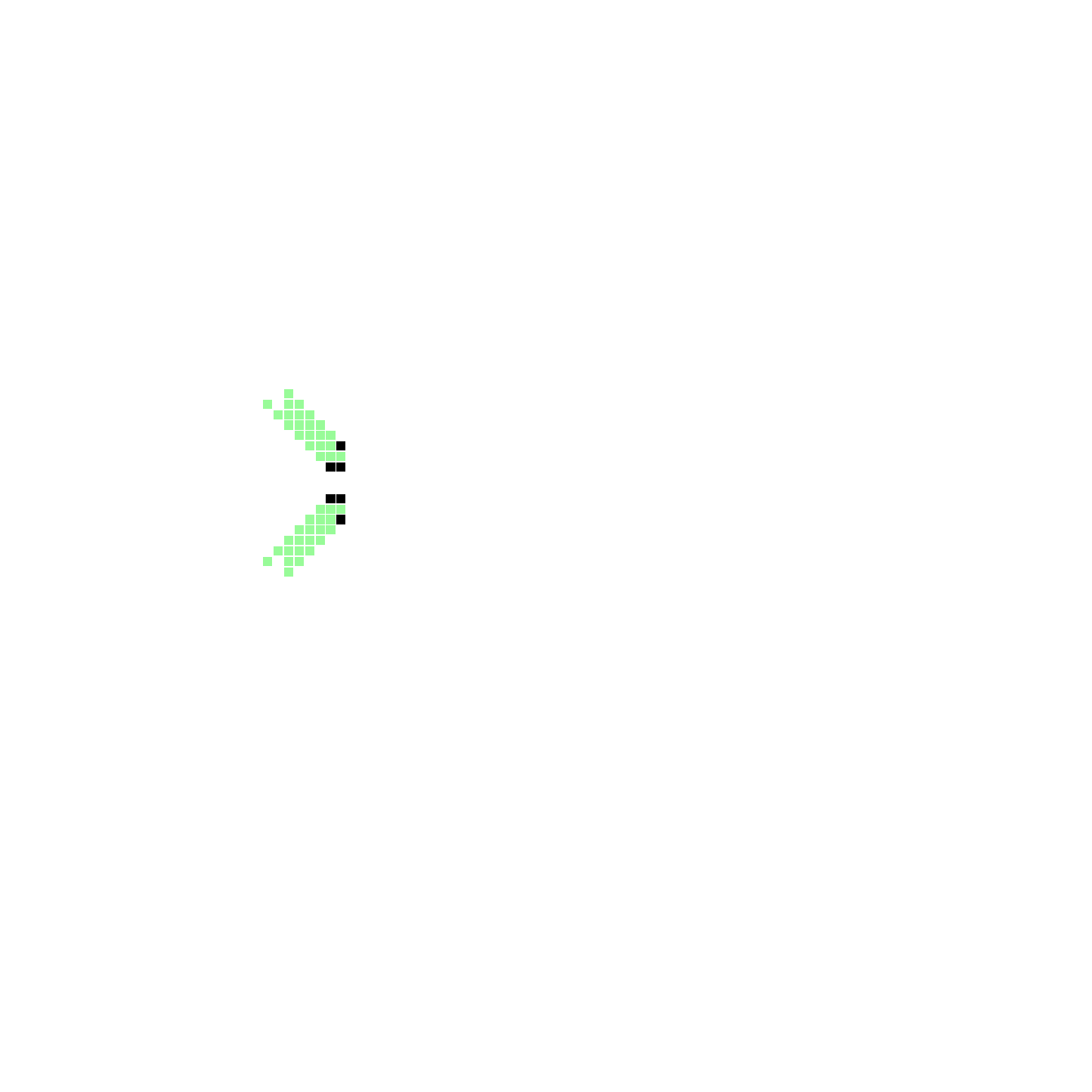} \hfill
\includegraphics[width=.13\linewidth,bb=93 190 139 268, clip=]{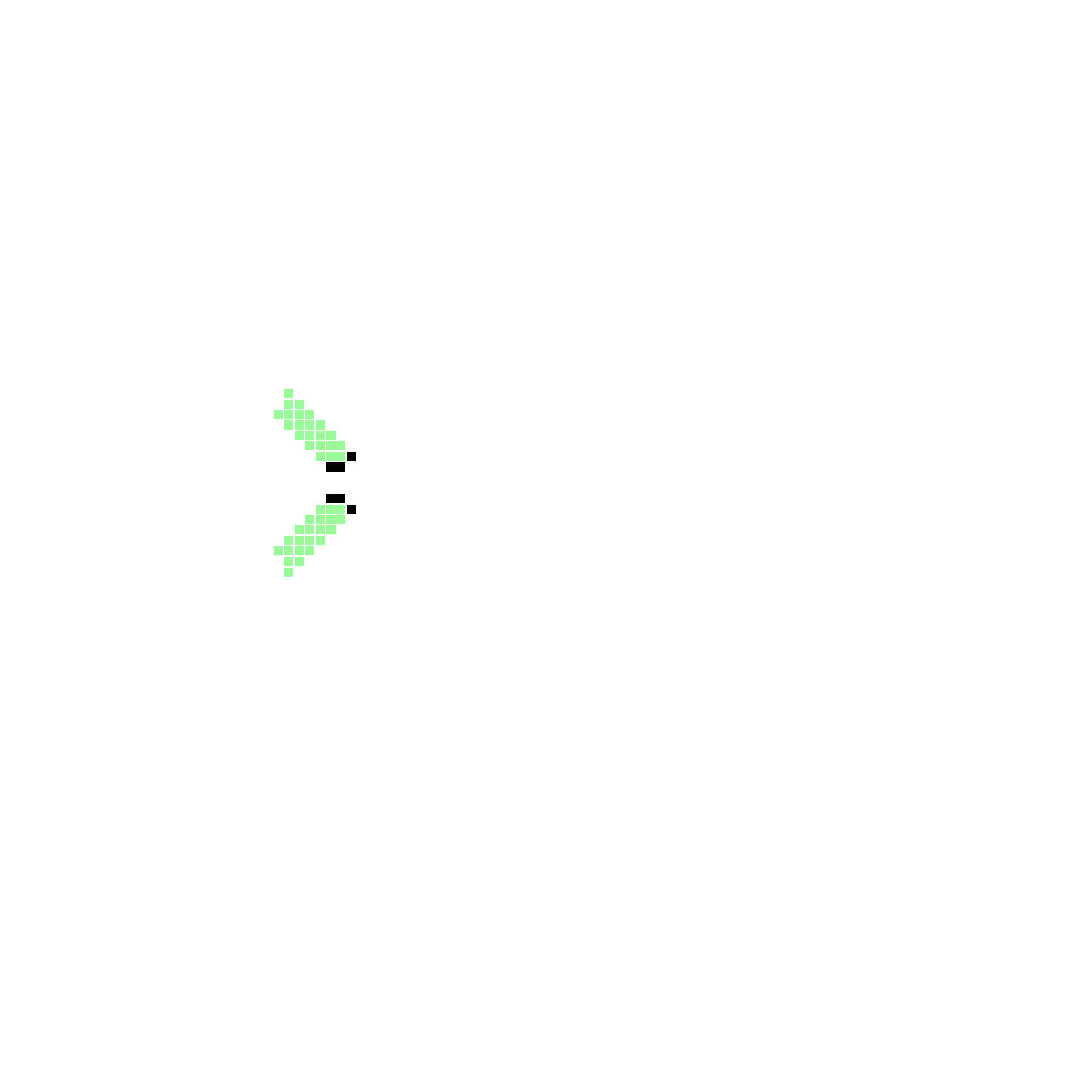} \hfill
\includegraphics[width=.13\linewidth,bb=93 190 139 268, clip=]{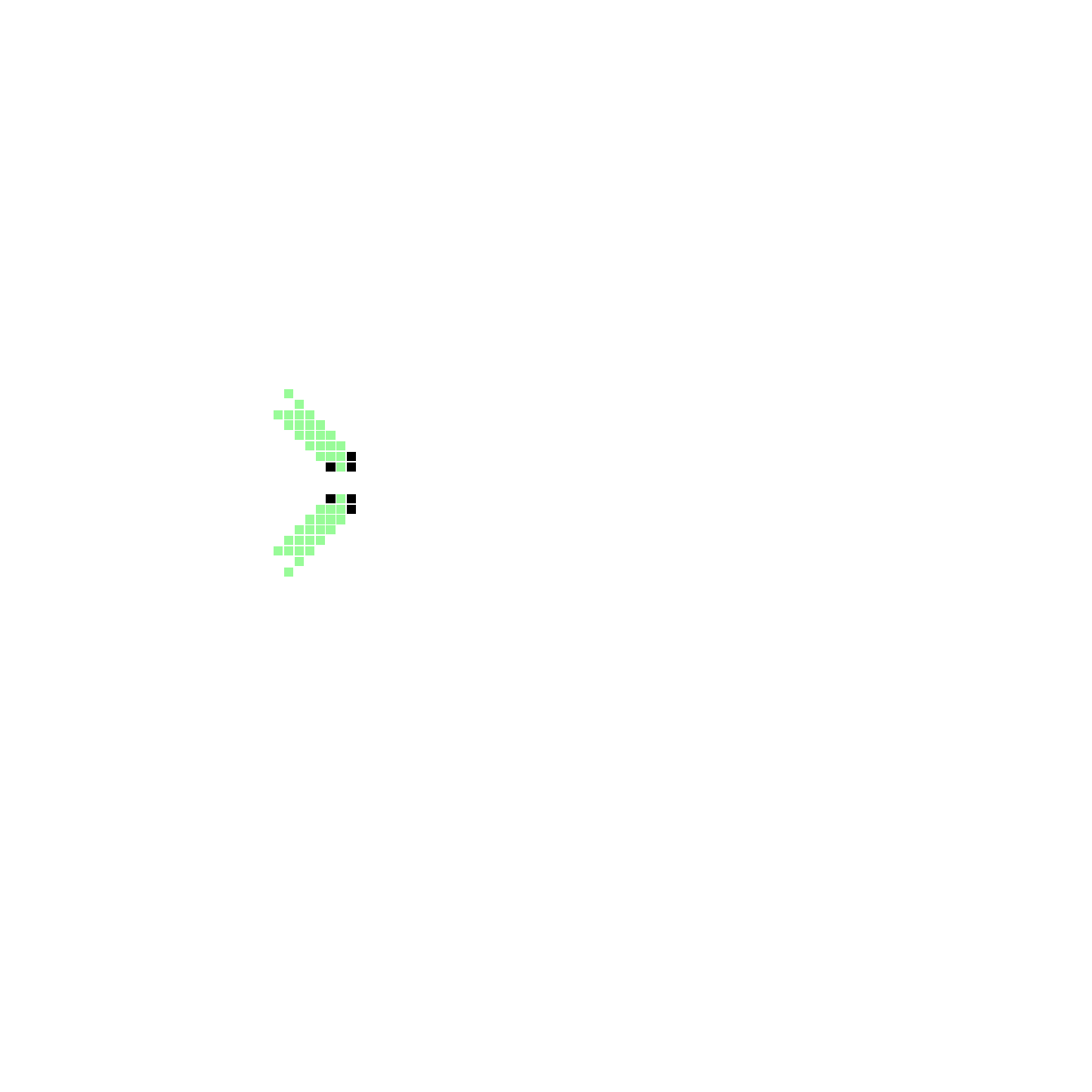} \hfill
\includegraphics[width=.13\linewidth,bb=93 190 139 268, clip=]{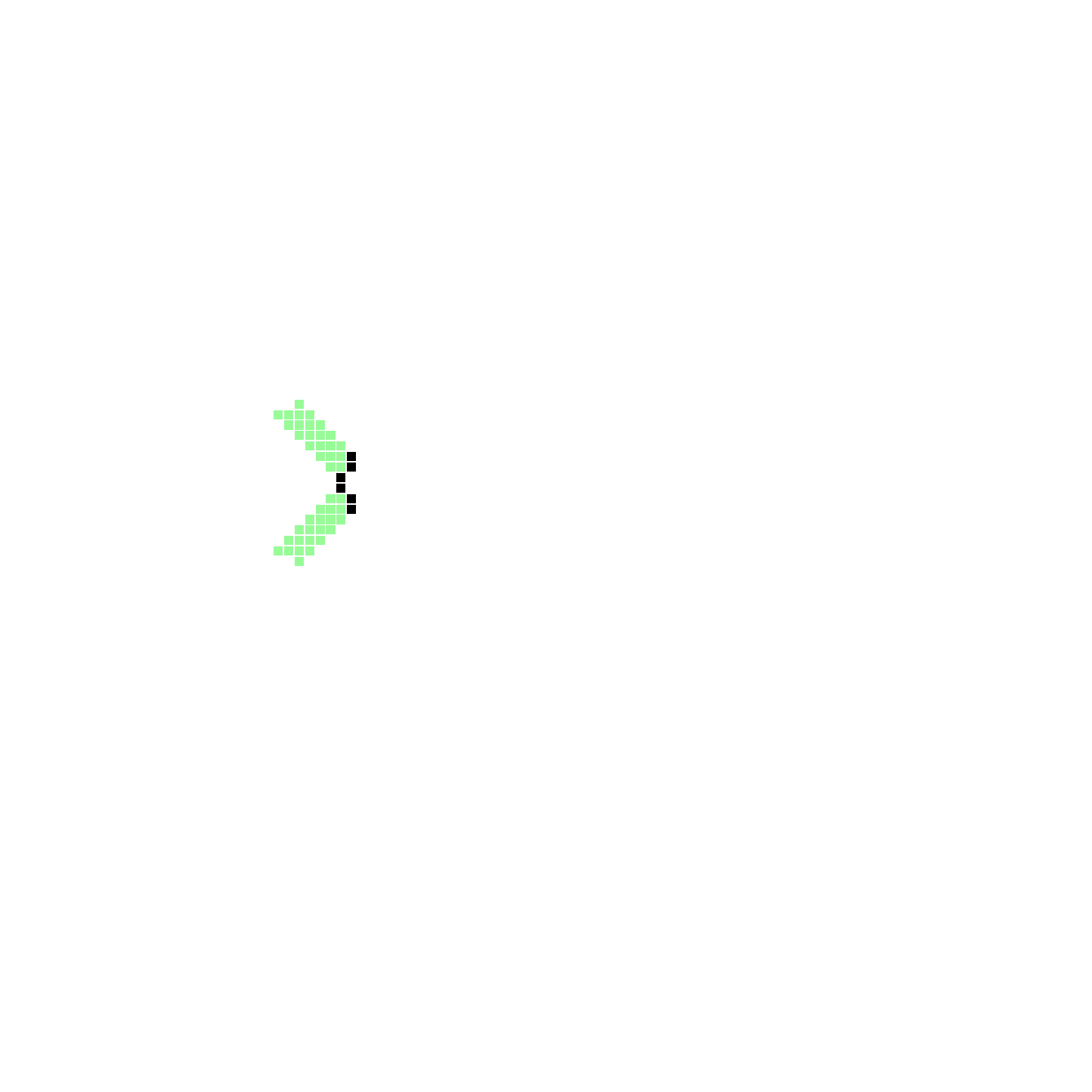} \hfill
\includegraphics[width=.13\linewidth,bb=93 190 139 268, clip=]{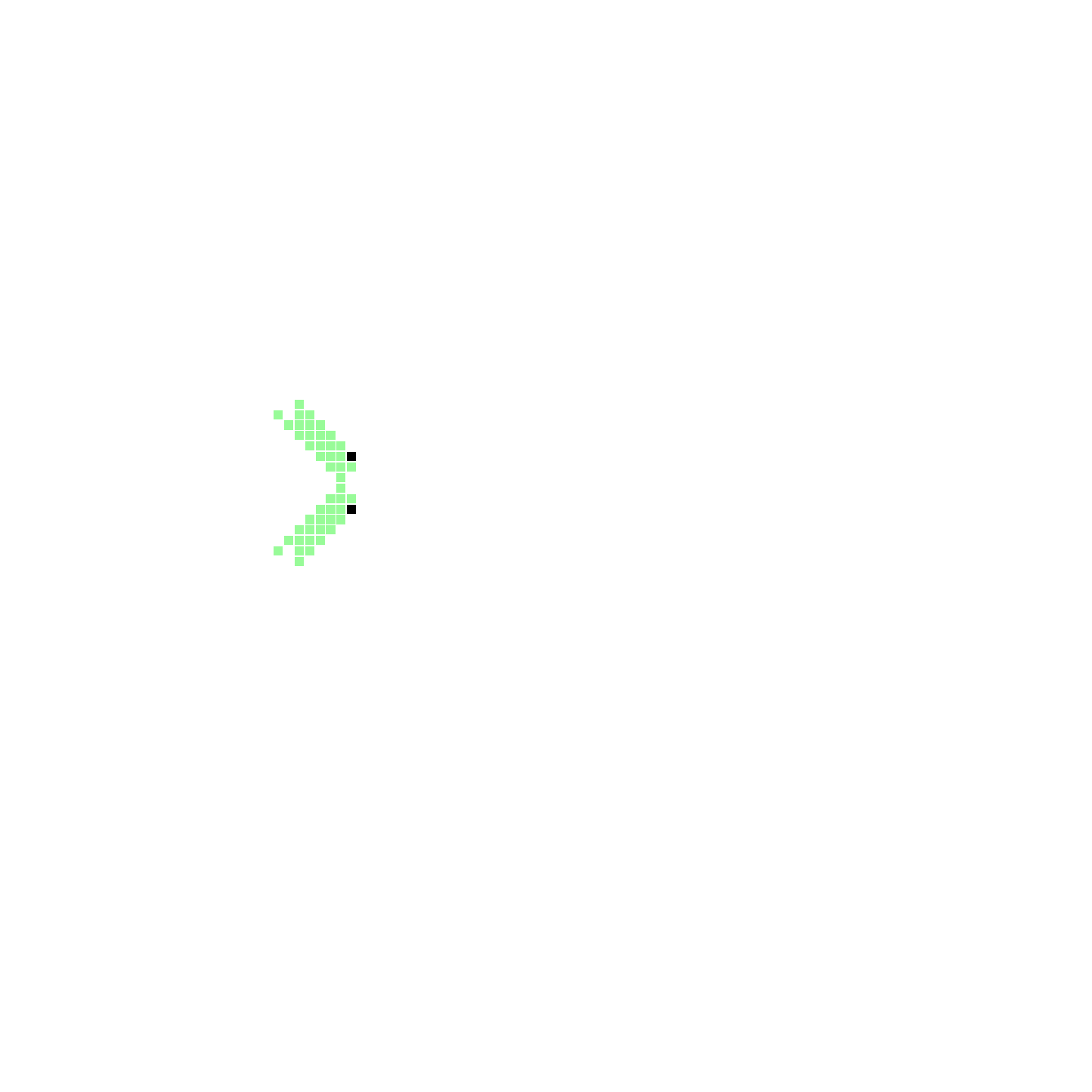} \hfill
\includegraphics[width=.13\linewidth,bb=93 190 139 268, clip=]{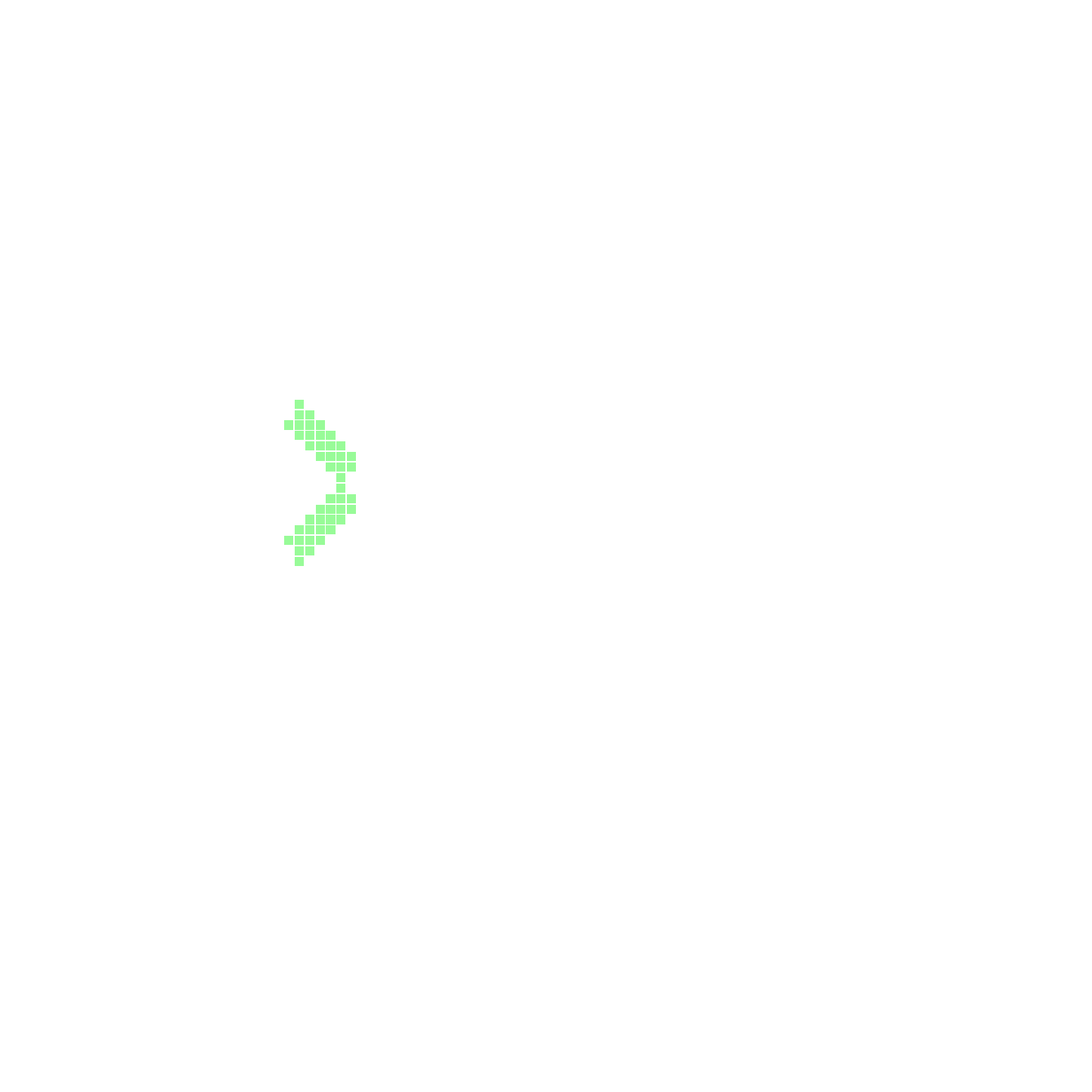} 
\textsf{\small Ga mutual destruction in 7 steps}
\end{minipage}}\\[-.3ex]
\fbox{\begin{minipage}[c]{1\linewidth} 
\vspace{2ex}
\includegraphics[width=.135\linewidth,bb=186 191 289 300, clip=]{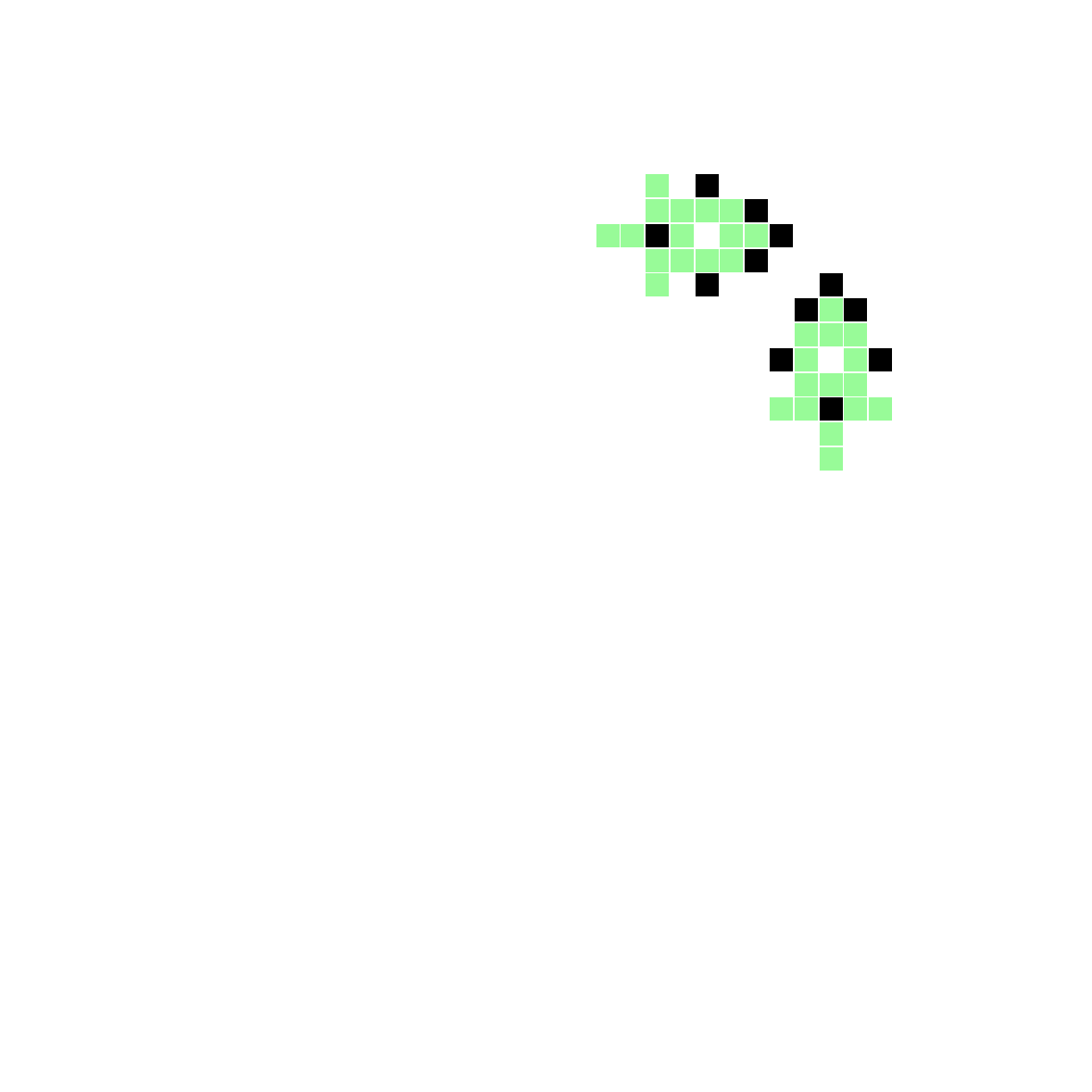} \hfill
\includegraphics[width=.135\linewidth,bb=186 191 289 300, clip=]{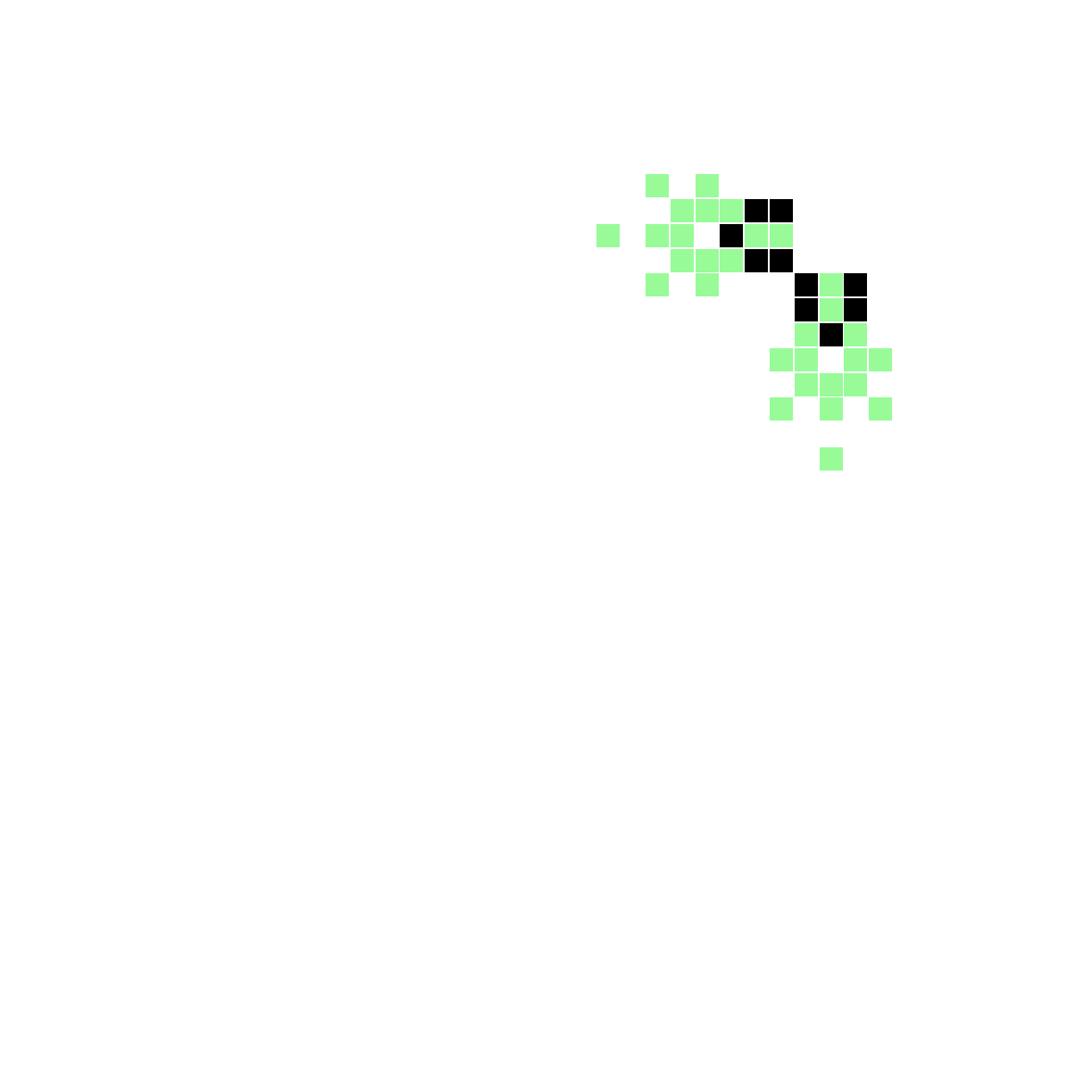} \hfill
\includegraphics[width=.135\linewidth,bb=186 191 289 300, clip=]{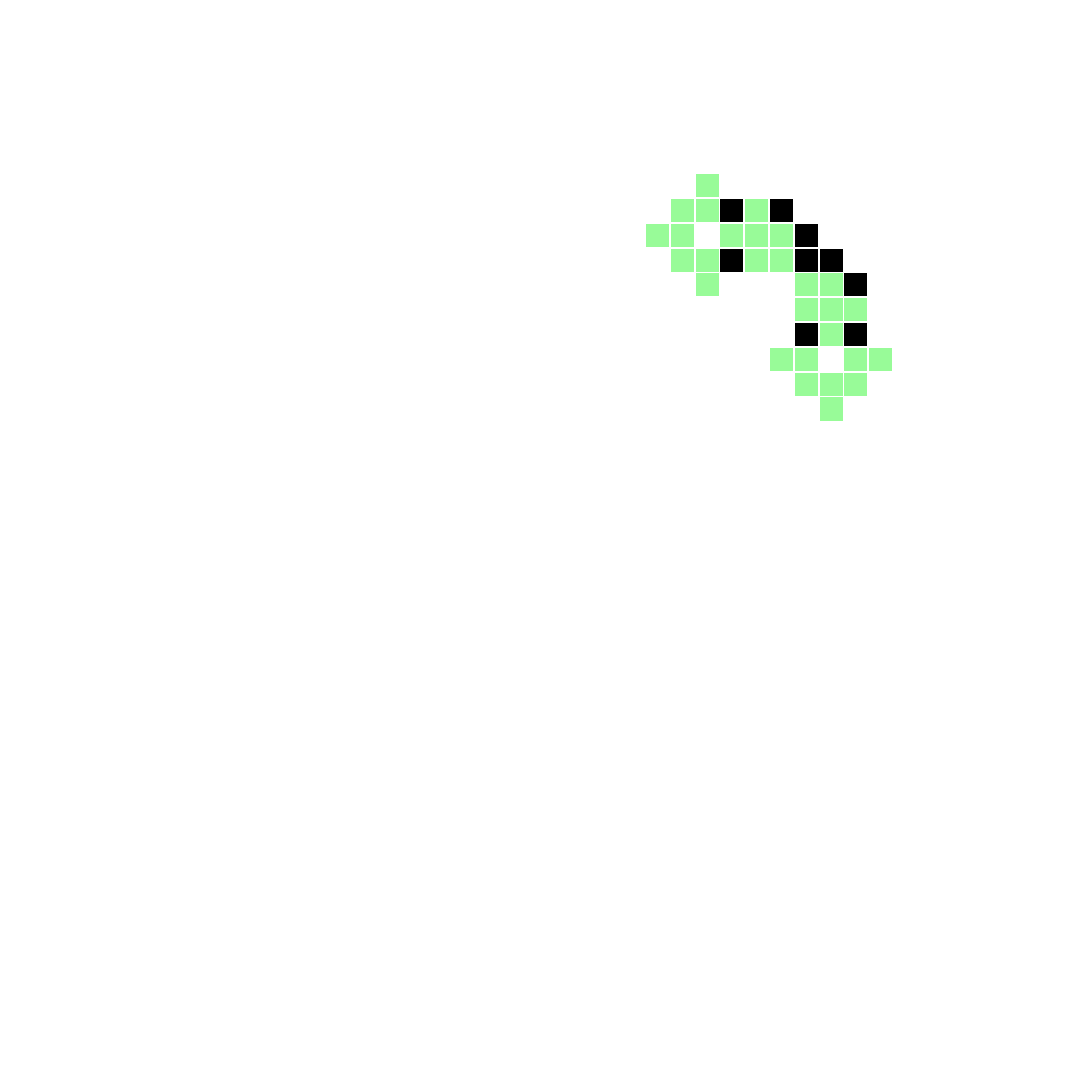} \hfill
\includegraphics[width=.135\linewidth,bb=186 191 289 300, clip=]{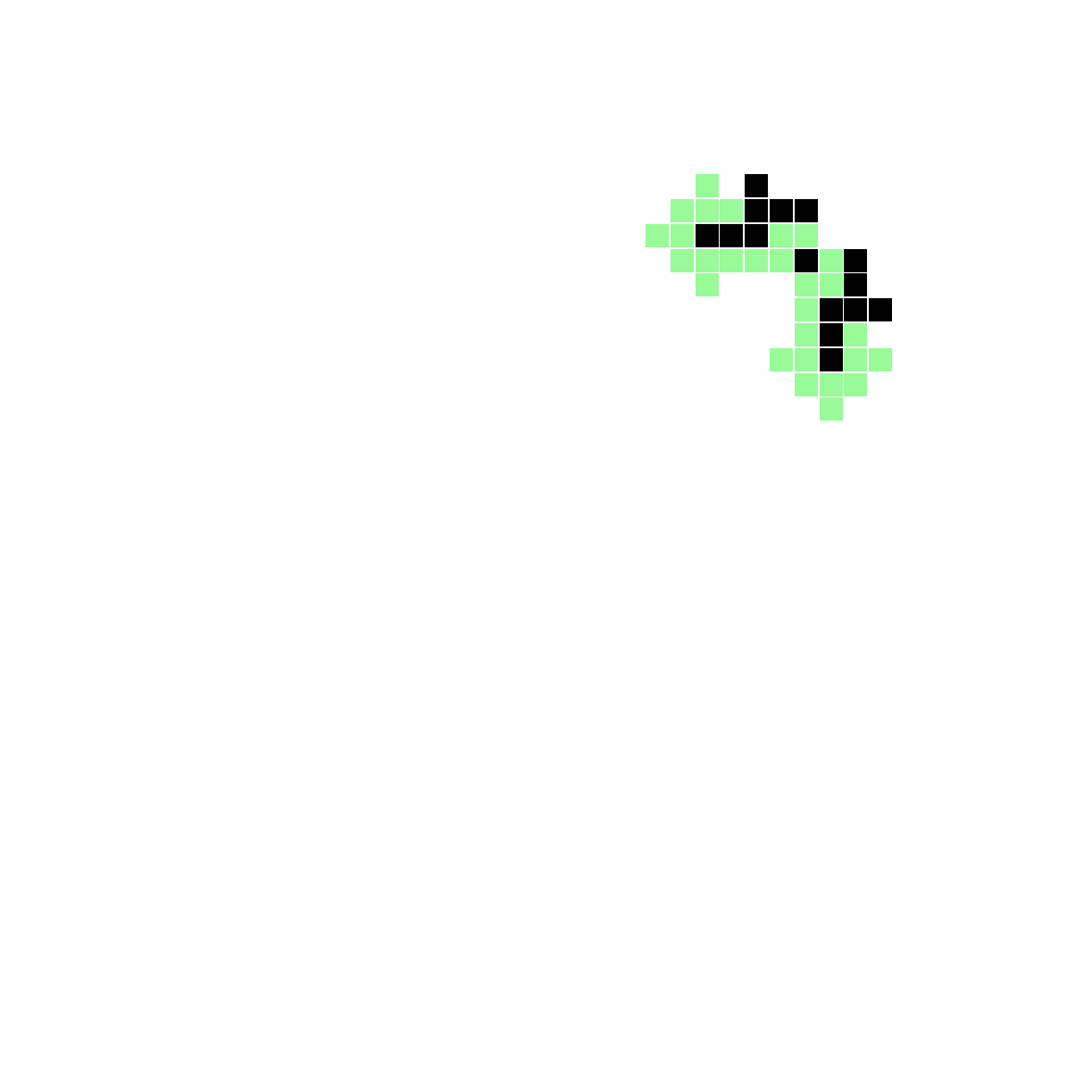} \hfill
\includegraphics[width=.135\linewidth,bb=186 191 289 300, clip=]{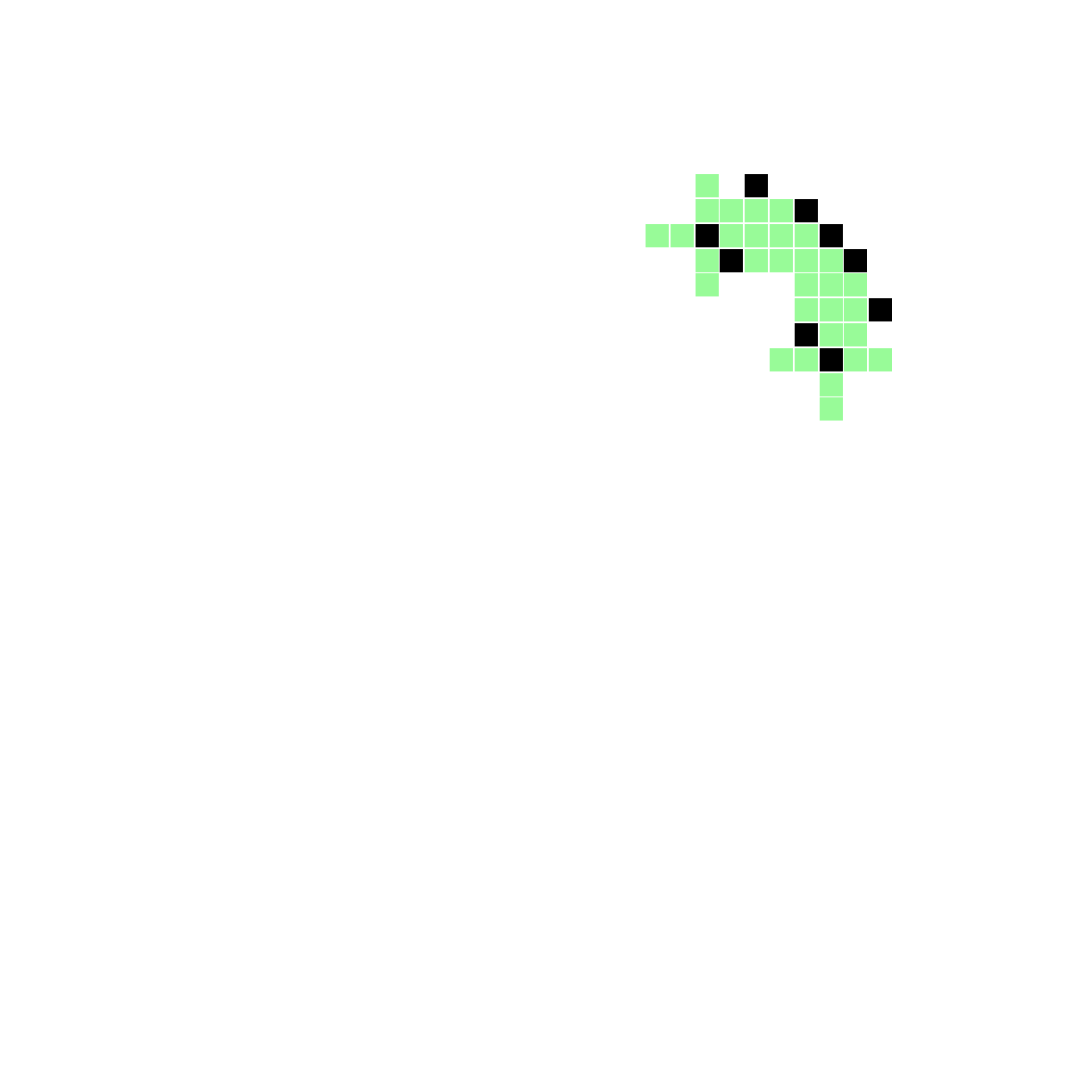} \hfill
\includegraphics[width=.135\linewidth,bb=186 191 289 300, clip=]{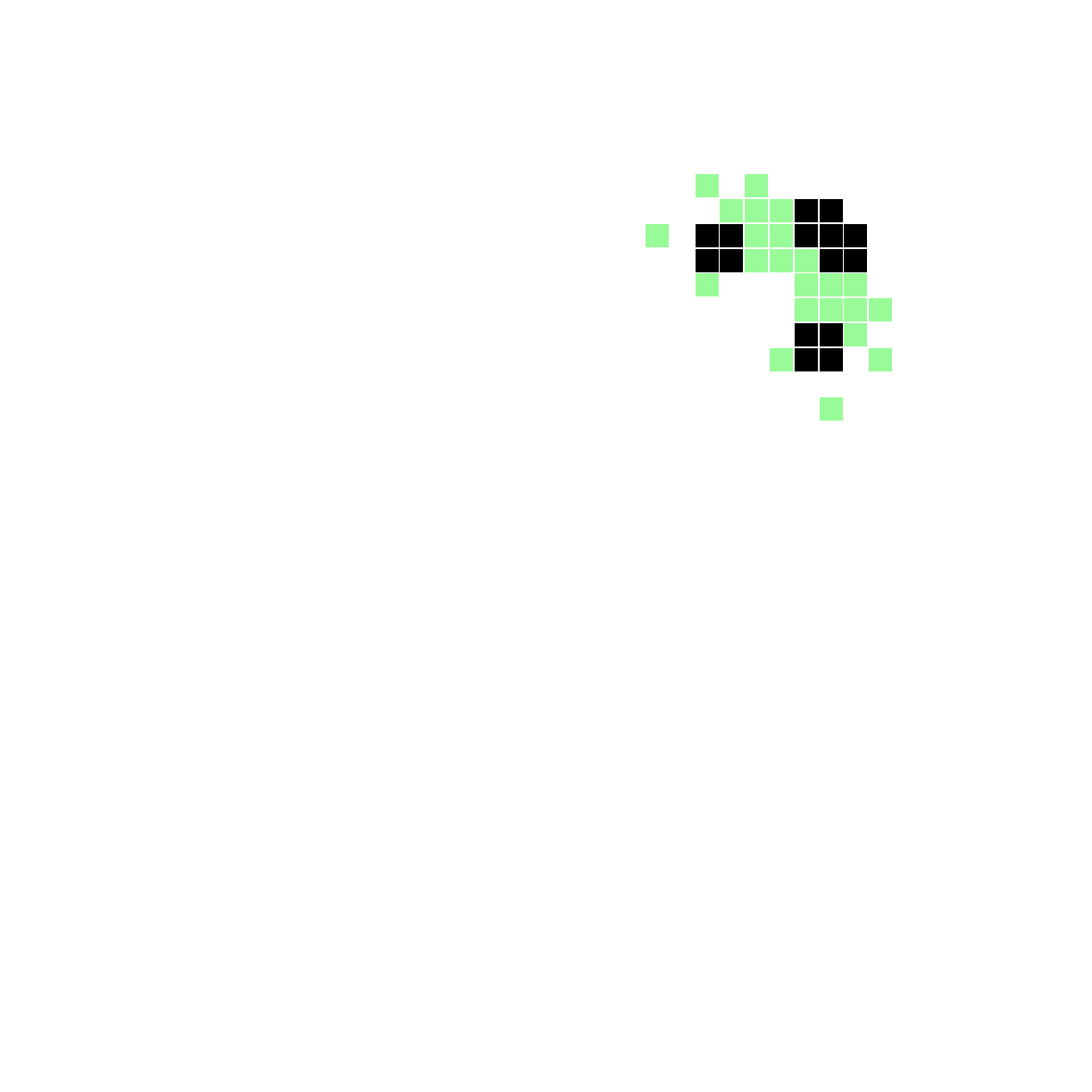} \hfill
\includegraphics[width=.135\linewidth,bb=186 191 289 300, clip=]{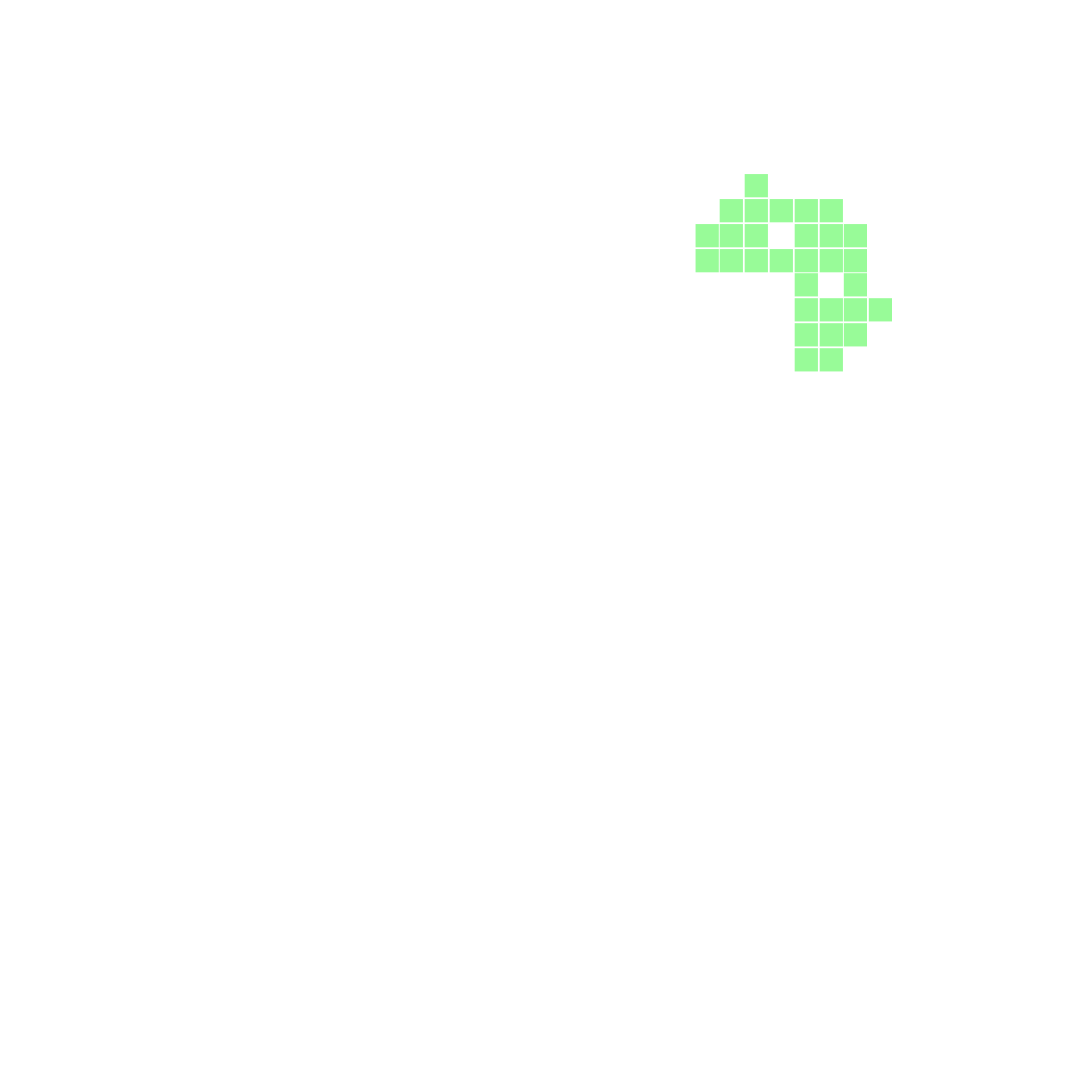} 
\textsf{\small Gc mutual destruction in 7 steps}
\end{minipage}}
\end{center}
\vspace{-3ex}
\caption[Ga and Gc mutual destruction]
{\textsf{Mutual destruction by colliding gliders, Ga and Gc, showing phases/time-steps on the approach
as well as the impact. These dynamics apply to both the Variant and Precursor rules.
\label{Ga and Gc mutual destruction}
}}
\vspace{-3ex}
\end{figure}

Among necessary collision behaviors for logical universality are,
\begin{s-itemize}
\item The destruction of a glider by a stable block (an ``eater'') which
survives intact to destroy subsequent gliders in a glider stream
(figures \ref{Ga and Gc Eaters}, \ref{GG2a glider-guns}, \ref{GG2a reflector glider-guns}).
\item  Mutual destruction when two gliders collide
(figure \ref{Ga and Gc mutual destruction}).
\end{s-itemize}

\begin{figure}[htb]
\begin{center}
\begin{minipage}[c]{1\linewidth}

\begin{minipage}[c]{.4\linewidth} 
\fbox{\includegraphics[width=1\linewidth,bb=68 246 255 321, clip=]{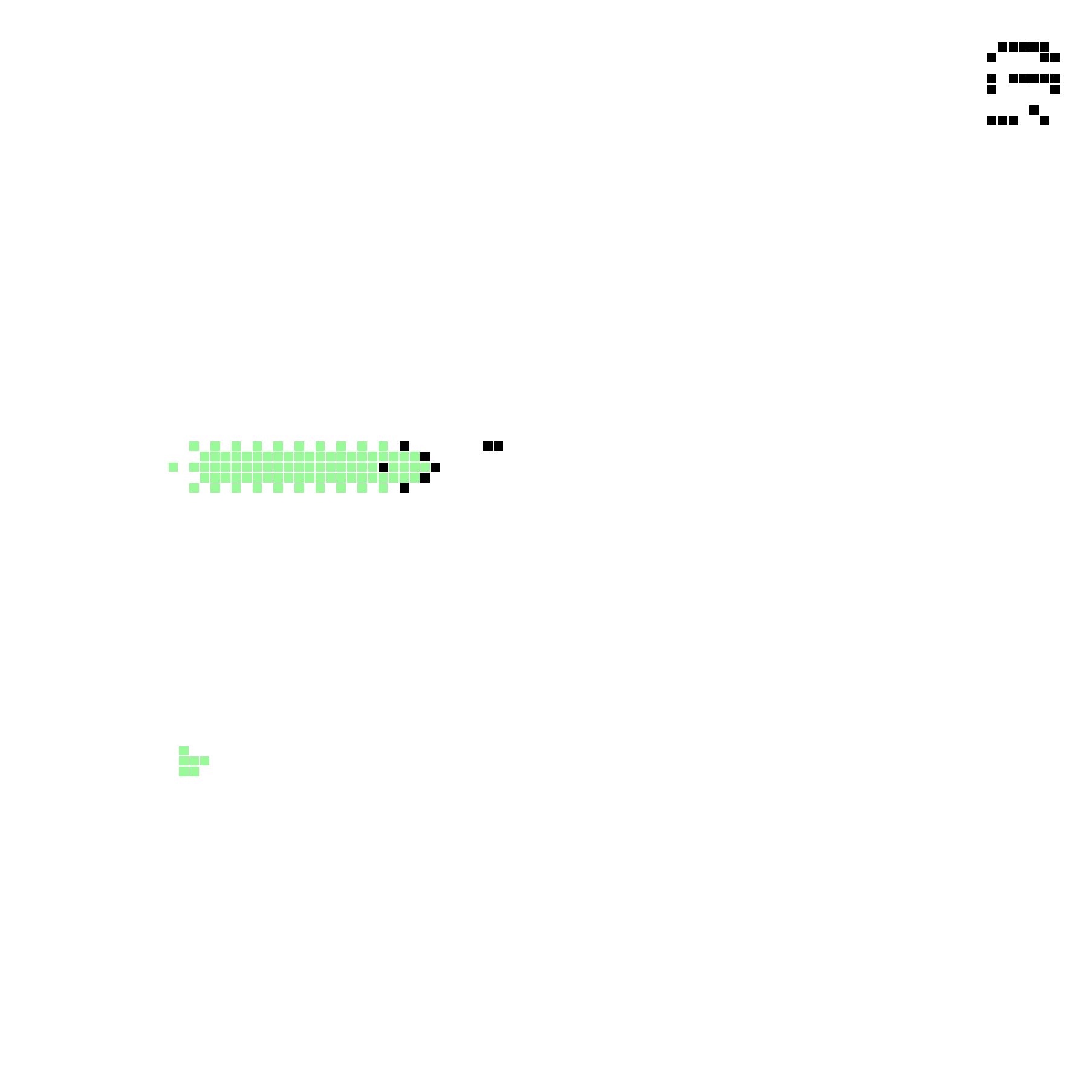}}
\end{minipage}
\hfill
\begin{minipage}[c]{.15\linewidth}
\begin{center}
\includegraphics[width=.6\linewidth,bb=10 9 32 26, clip=]{pdf-figs/ArrowE}\\
\textsf{\small{Gc to Ga}\\via block}
\end{center}
\end{minipage}
\hfill
\begin{minipage}[c]{.4\linewidth}
\fbox{\includegraphics[width=1\linewidth,bb=68 246 255 321, clip=]{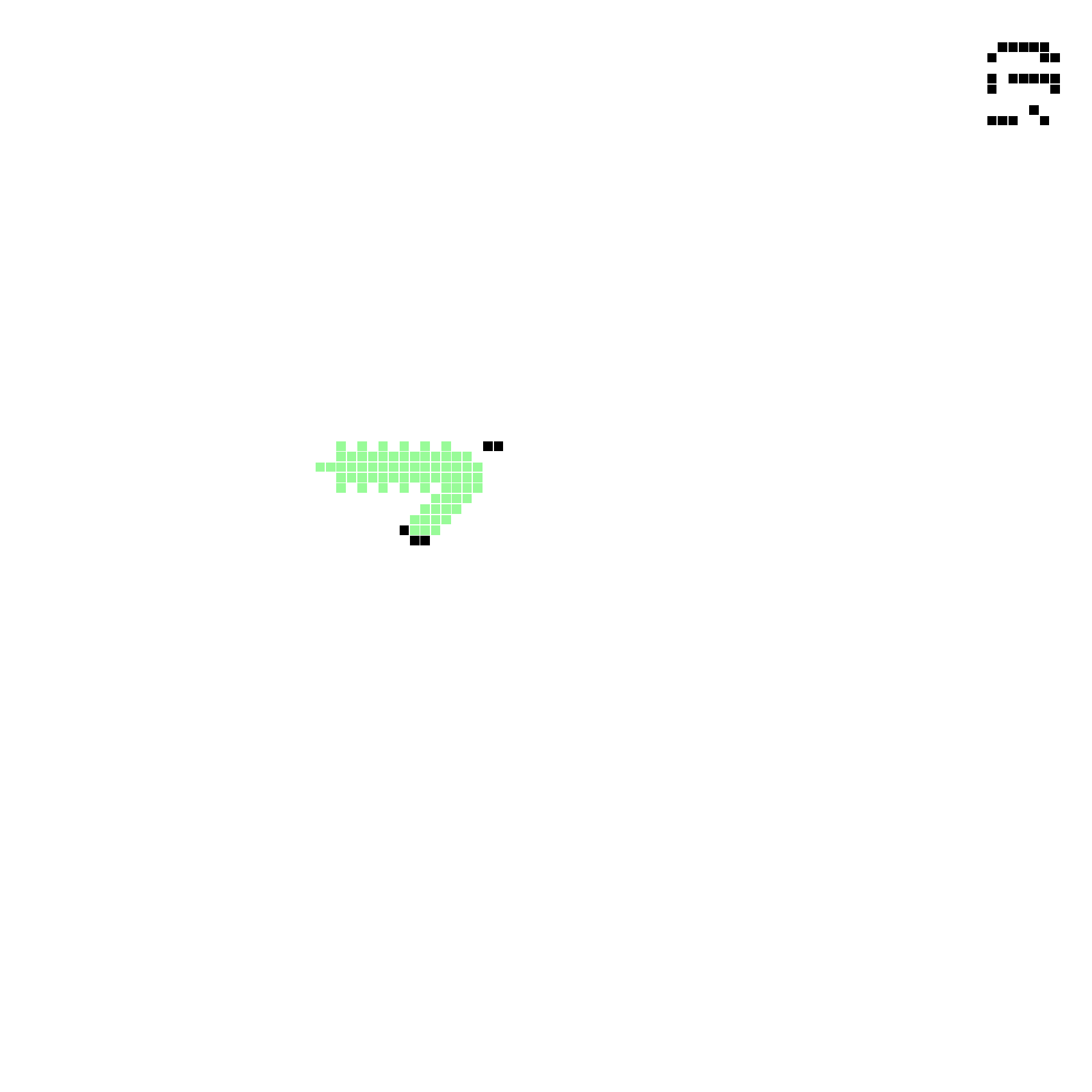}}
\end{minipage}\\[1ex]

\begin{minipage}[c]{.4\linewidth} 
\fbox{\includegraphics[width=1\linewidth,bb=170 245 355 355, clip=]{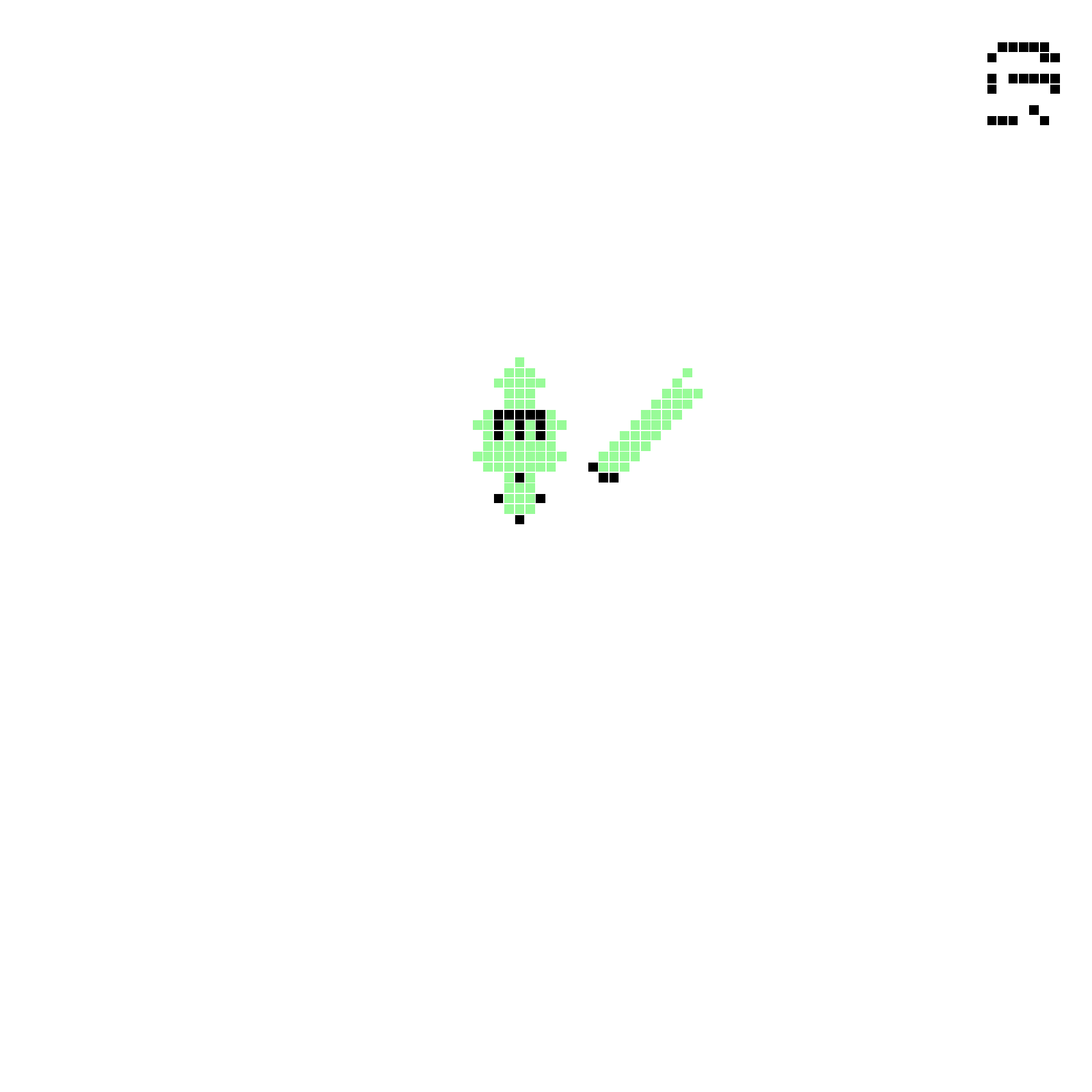}}
\end{minipage}
\hfill
\begin{minipage}[c]{.15\linewidth}
\begin{center}
\includegraphics[width=.6\linewidth,bb=10 9 32 26, clip=]{pdf-figs/ArrowE}\\
\textsf{\small{Ga to Gc}\\via P22}
\end{center}
\end{minipage}
\hfill
\begin{minipage}[c]{.4\linewidth}
\fbox{\includegraphics[width=1\linewidth,bb=170 245 355 355, clip=]{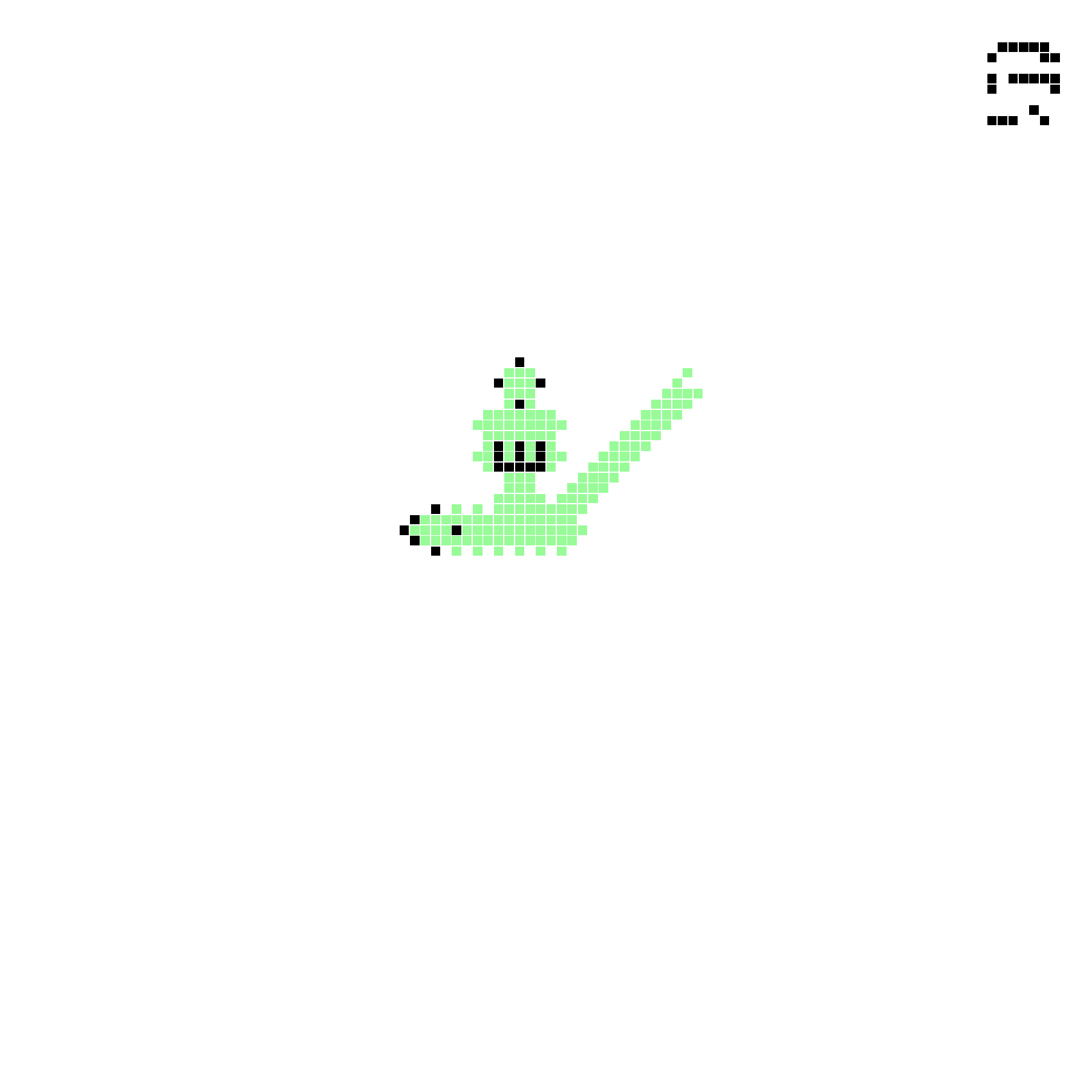}}
\end{minipage}\\[1ex]

\begin{minipage}[c]{.4\linewidth} 
\fbox{\includegraphics[width=1\linewidth,bb=186 229 365 350, clip=]{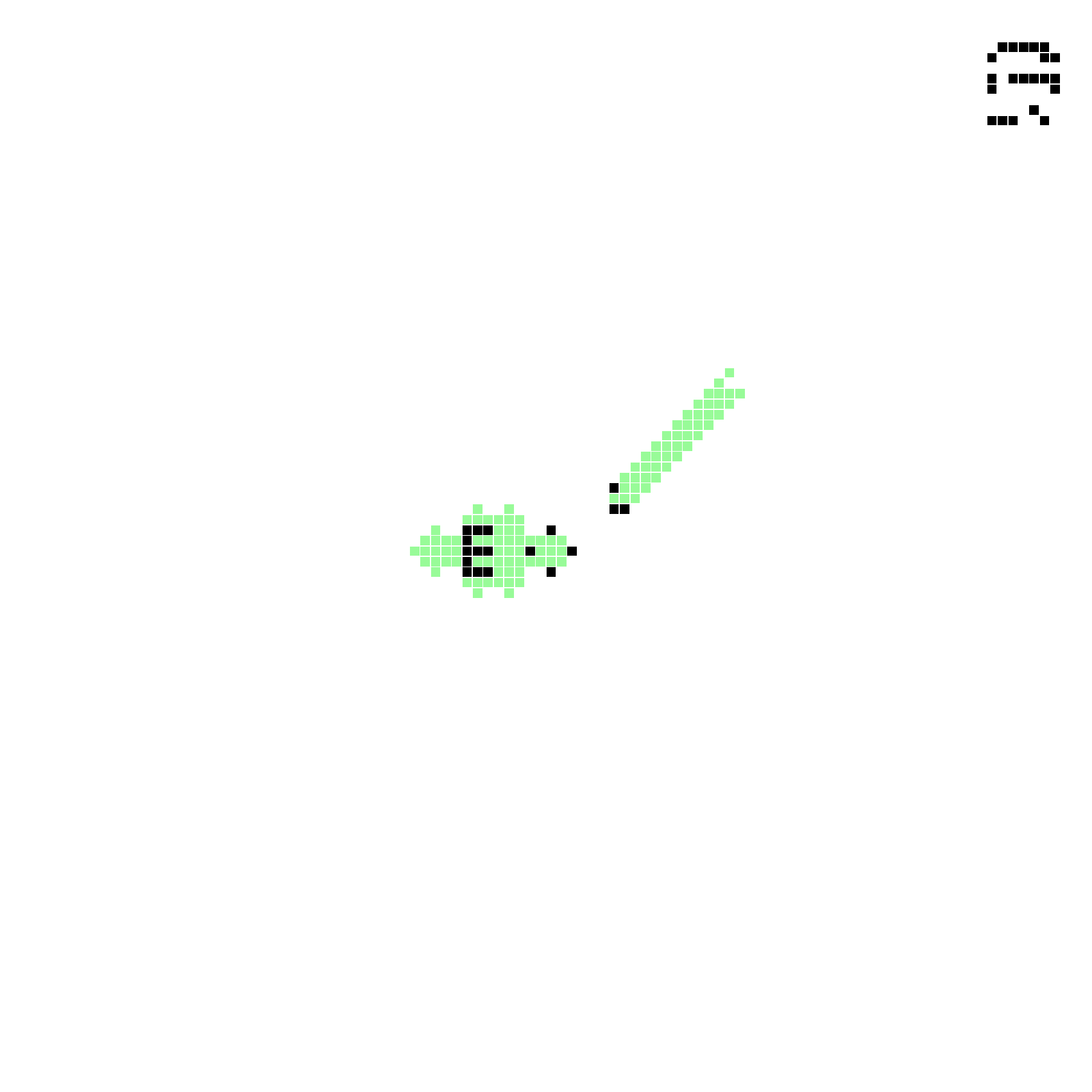}}
\end{minipage}
\hfill
\begin{minipage}[c]{.15\linewidth}
\begin{center}
\includegraphics[width=.6\linewidth,bb=10 9 32 26, clip=]{pdf-figs/ArrowE}\\
\textsf{\small{Ga to Gc}\\via P22\cite{BlinkerSpawn}}
\end{center}
\end{minipage}
\hfill
\begin{minipage}[c]{.4\linewidth}
\fbox{\includegraphics[width=1\linewidth,bb=186 229 365 350, clip=]{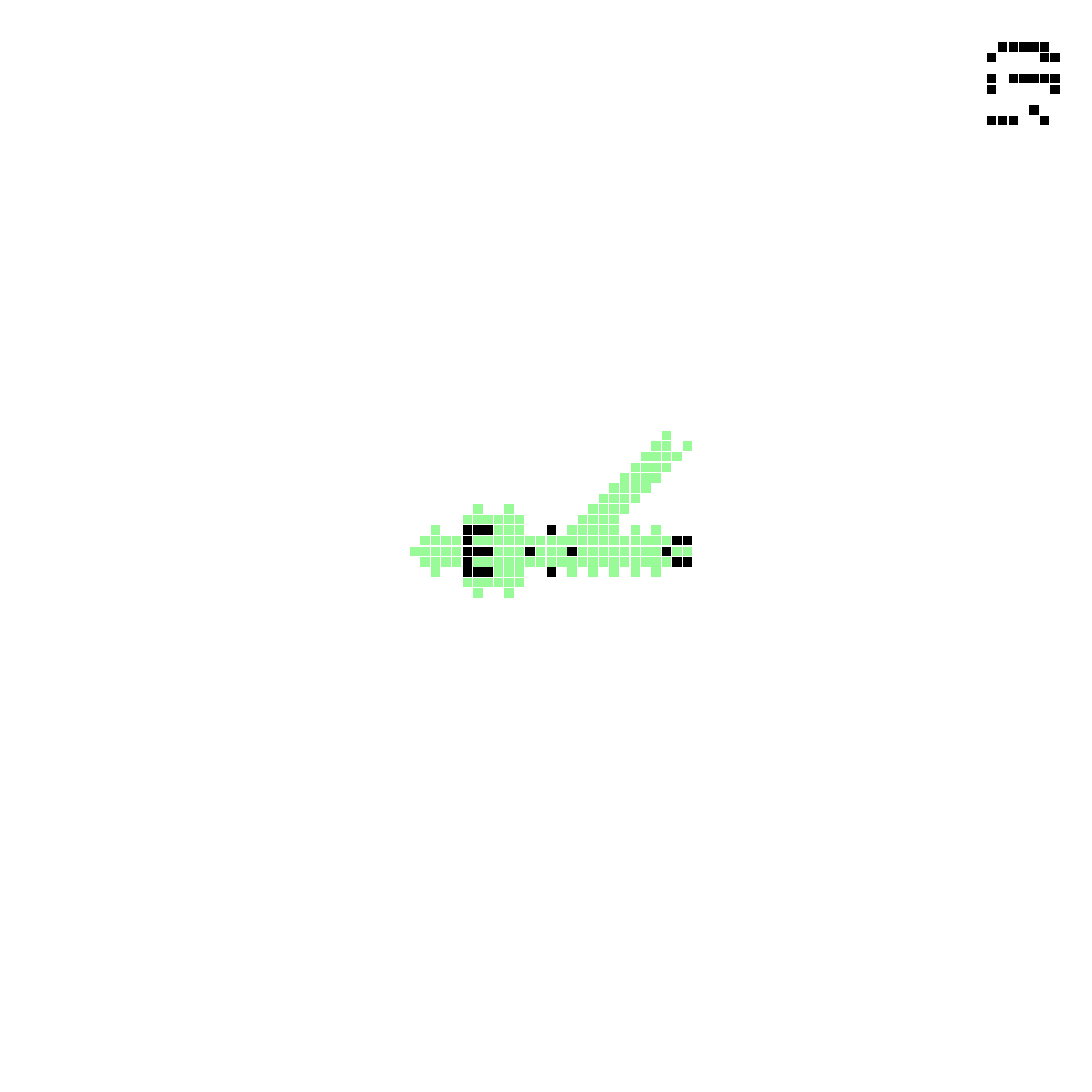}}
\end{minipage}\\[1ex]

\begin{minipage}[c]{.4\linewidth} 
\fbox{\includegraphics[width=1\linewidth,bb=148 184 322 348, clip=]{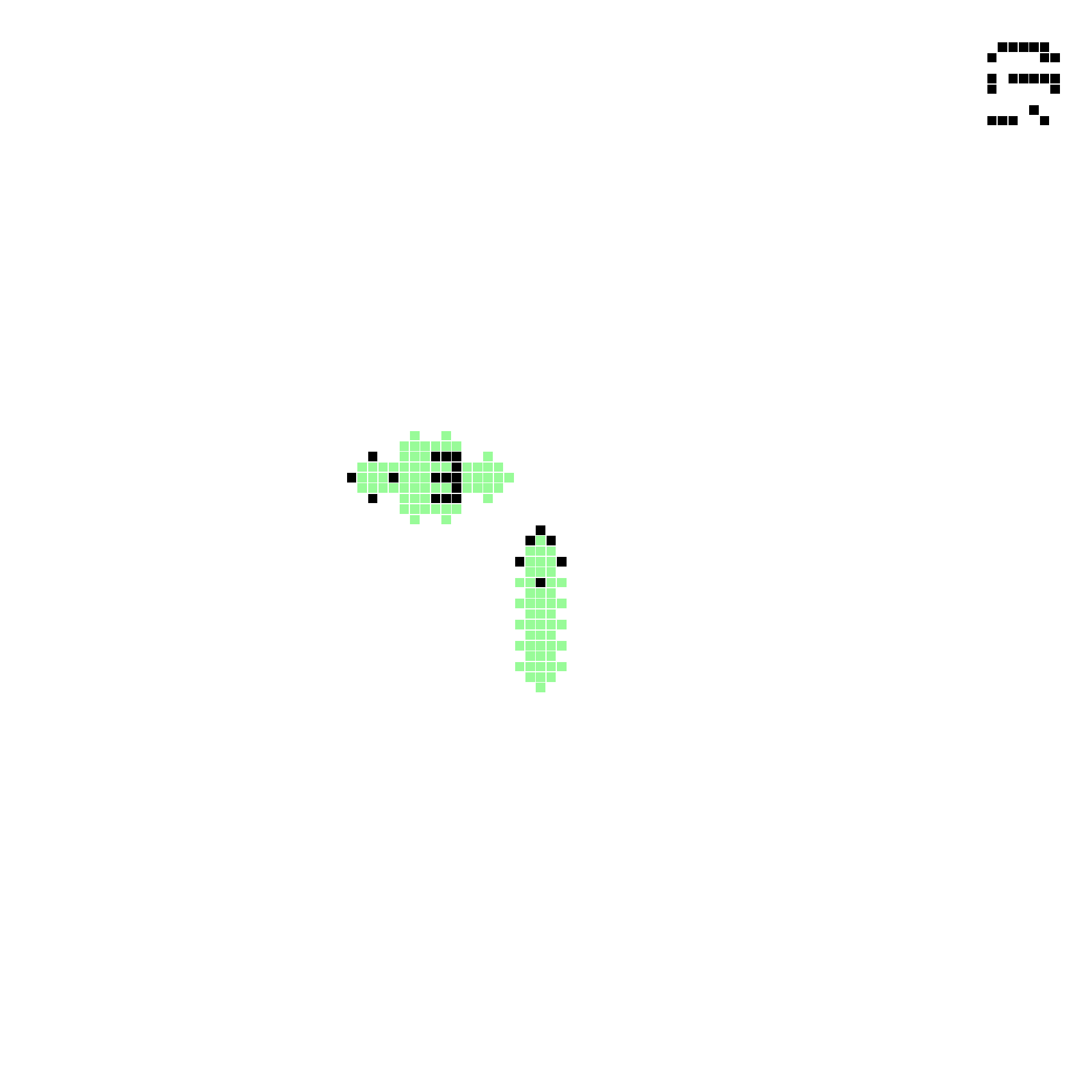}}
\end{minipage}
\hfill
\begin{minipage}[c]{.15\linewidth}
\begin{center}
\includegraphics[width=.6\linewidth,bb=10 9 32 26, clip=]{pdf-figs/ArrowE}\\
\textsf{\small{Gc to Ga}\\via P22\cite{BlinkerSpawn}}
\end{center}
\end{minipage}
\hfill
\begin{minipage}[c]{.4\linewidth}
\fbox{\includegraphics[width=1\linewidth,bb=148 184 322 348, clip=]{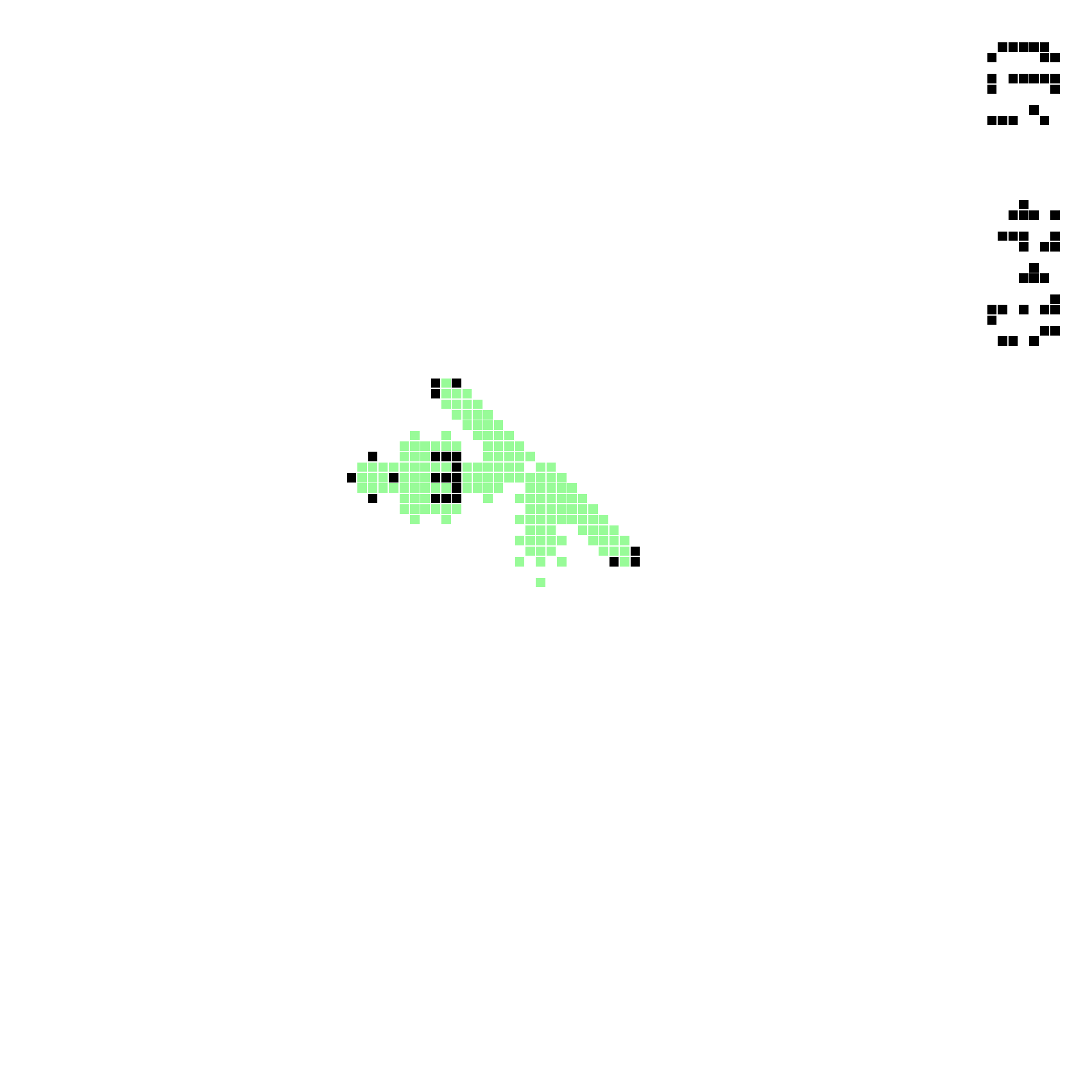}} 
\end{minipage}\\[1ex]

\end{minipage}
\end{center}
\vspace{-1ex}
\caption[collisions with P22]
{\textsf{
Examples of collisions outcomes which transform and change the direction of gliders.
$Top$: collision with a stable block. $Below$: collisions with the oscillator P22 described
in figure~\ref{P22 oscillator attractor}.
\label{collisions with P22}
}}

\end{figure}

\begin{figure}[htb]
\begin{center}
\begin{minipage}[c]{.9\linewidth}

\begin{minipage}[c]{.37\linewidth} 
\fbox{\includegraphics[width=1\linewidth,bb=121 196 270 295, clip=]{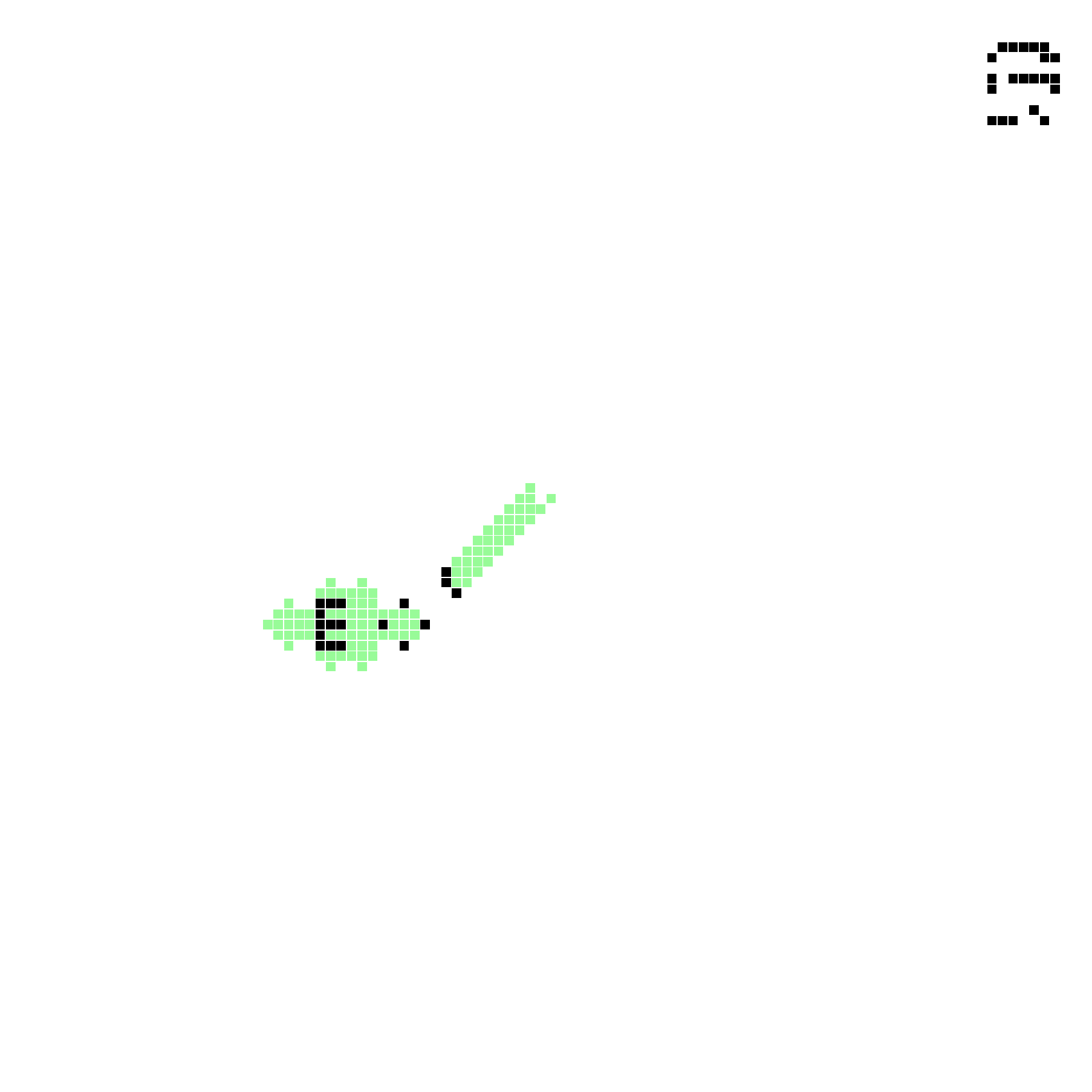}}
\end{minipage}
\hfill
\begin{minipage}[c]{.15\linewidth}
\begin{center}
\includegraphics[width=.6\linewidth,bb=10 9 32 26, clip=]{pdf-figs/ArrowE}\\
\textsf{\small{Ga$\rightarrow$P22$\rightarrow$P15}}
\end{center}
\end{minipage}
\hfill
\begin{minipage}[c]{.37\linewidth} 
\fbox{\includegraphics[width=1\linewidth,bb=121 196 270 295, clip=]{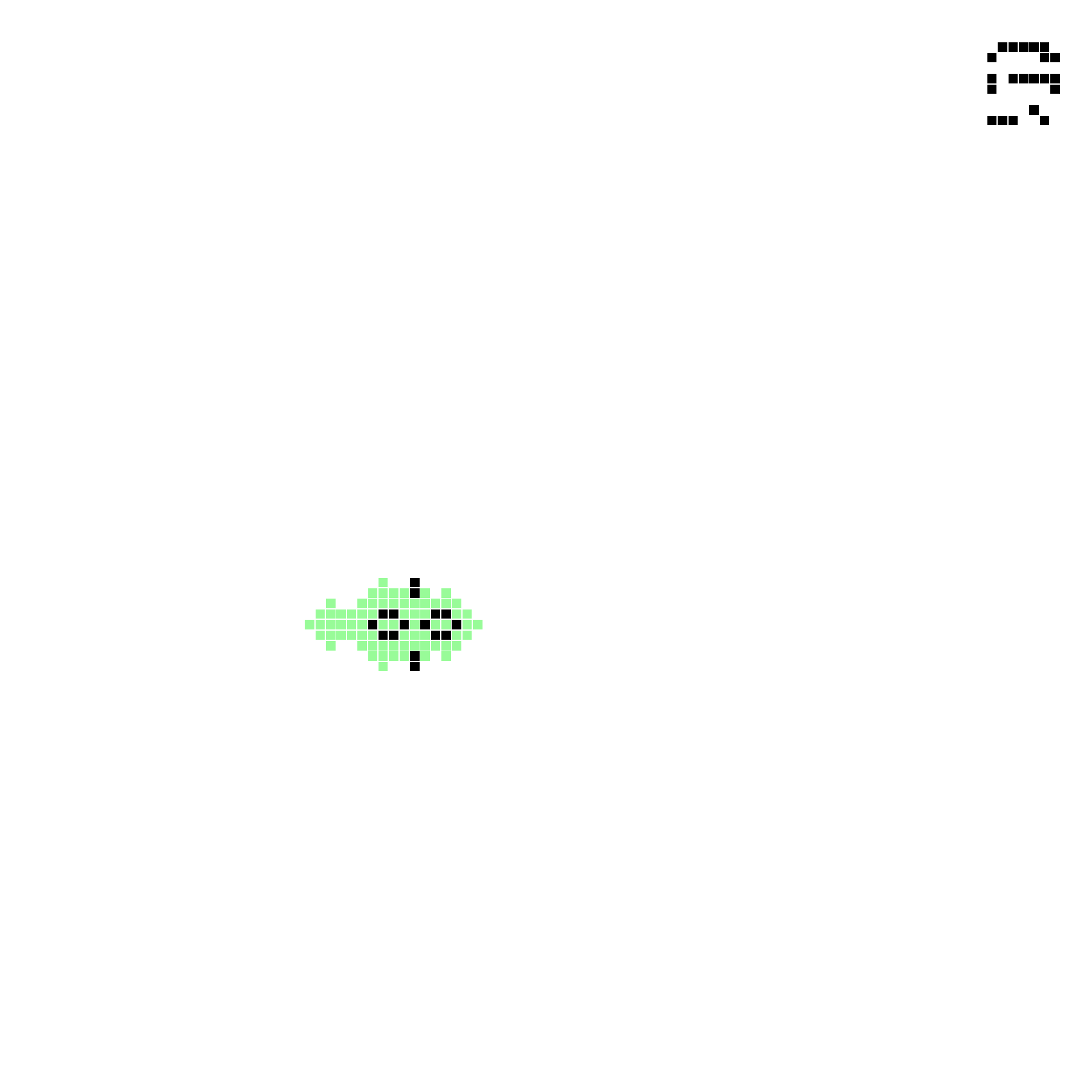}}
\end{minipage}\\[1ex]

\begin{minipage}[c]{.37\linewidth} 
\fbox{\includegraphics[width=1\linewidth,bb=131 247 281 299, clip=]{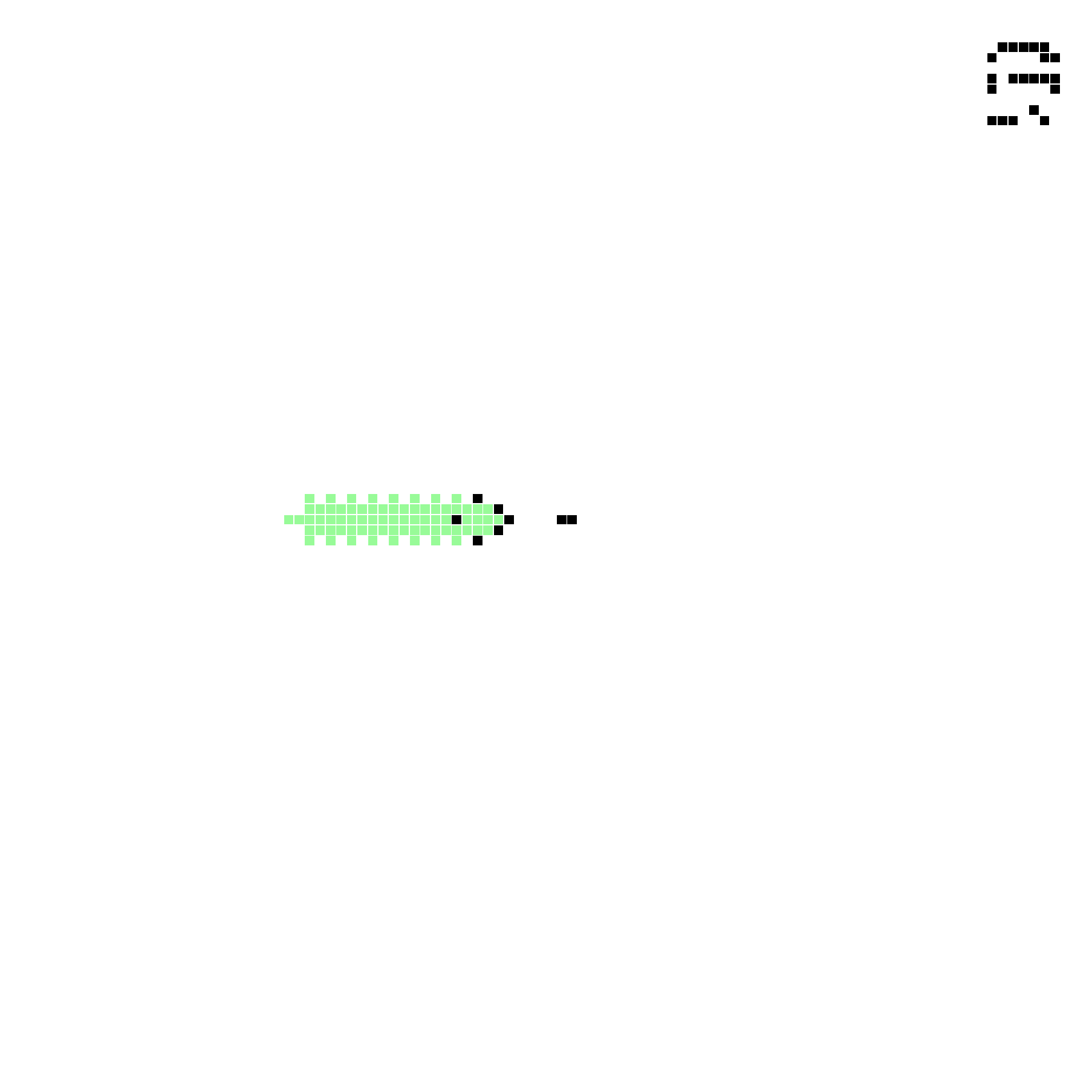}}
\end{minipage}
\hfill
\begin{minipage}[c]{.15\linewidth}
\begin{center}
\includegraphics[width=.6\linewidth,bb=10 9 32 26, clip=]{pdf-figs/ArrowE}\\
\textsf{\small{Gc$\rightarrow$Bk$\rightarrow$P15}}
\end{center}
\end{minipage}
\hfill
\begin{minipage}[c]{.37\linewidth}
\fbox{\includegraphics[width=1\linewidth,bb=131 247 281 299, clip=]{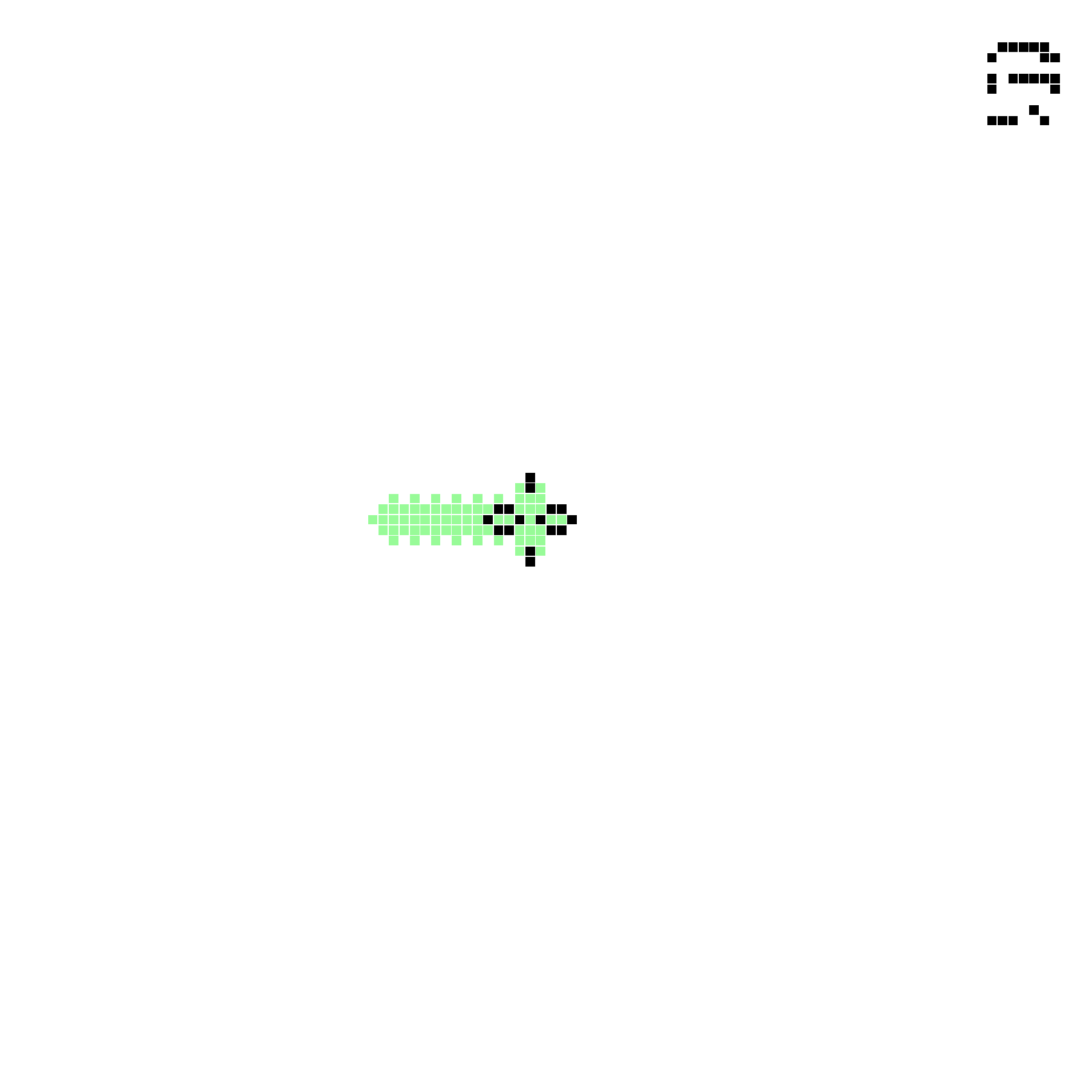}}
\end{minipage}\\[1ex]

\end{minipage}
\end{center}
\vspace{-4ex}
\caption[Collisions resulting in P15]
{\textsf{
Collisions resulting in oscillator P15.
$Top$: Ga collides with oscillator P22 and transforms it to oscillator P15
 --- Ga is destroyed\cite{BlinkerSpawn}. 
$Below$: Gc collides with a stable block resulting
in oscillator P15 --- both Gc and the block are destroyed\cite{Thunk}.
\label{Collisions resulting in P15}
}}
\vspace{-2ex}
\end{figure}

These are present in the Variant rule, but other collisions, also
involving oscillators and reflectors, enrich the behavior of the dynamical system in
unexpected ways --- transforming glider types, transforming
oscillators, changing gilder direction --- which would be significant
to achieve universality in its full sense, to include other
functionality such as memory by data storage.  
Figures \ref{Ga and Gc Eaters} -- \ref{Collisions resulting in P15}
provide a selection of collision examples.

\clearpage

\section{Glider-Guns, oscillators and reflectors}
\label{Glider-Guns, oscillators and reflectors}

Glider-guns can be built from two types of oscillator, P22 and P15,
and also from related interacting reflectors. If these sub-components
are juxtaposed to interact precisely, gliders are ejected with a
rhythm related to the oscillation or reflection period.  Several
different glider-guns are built by these methods, shooting glider
types Ga, Gc, and G2a (a double Ga). Its quite possible that other
glider-guns are out there in the Variant rule, to
be discovered.

\subsection{Glider-Guns from oscillator P22}
\label{Glider-Guns from oscillator P22}

\begin{figure}[htb]
\begin{center}
\begin{minipage}[c]{.75\linewidth}

\begin{minipage}[c]{.45\linewidth} 
\fbox{\includegraphics[width=1\linewidth,bb=110 198 347 357, clip=]{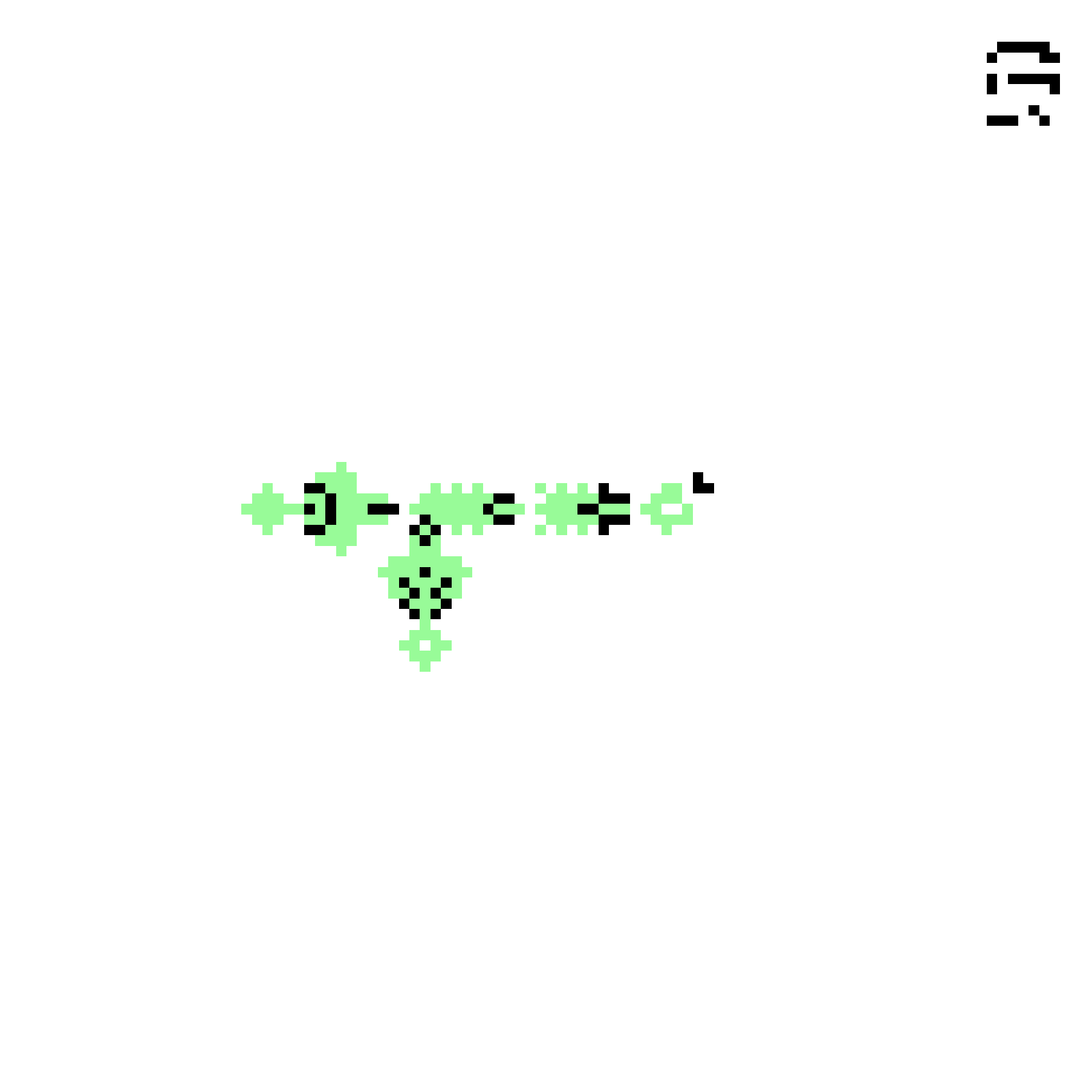}}
\end{minipage}
\hfill
\begin{minipage}[c]{.45\linewidth} 
\fbox{\includegraphics[width=1\linewidth,bb=110 198 347 357, clip=]{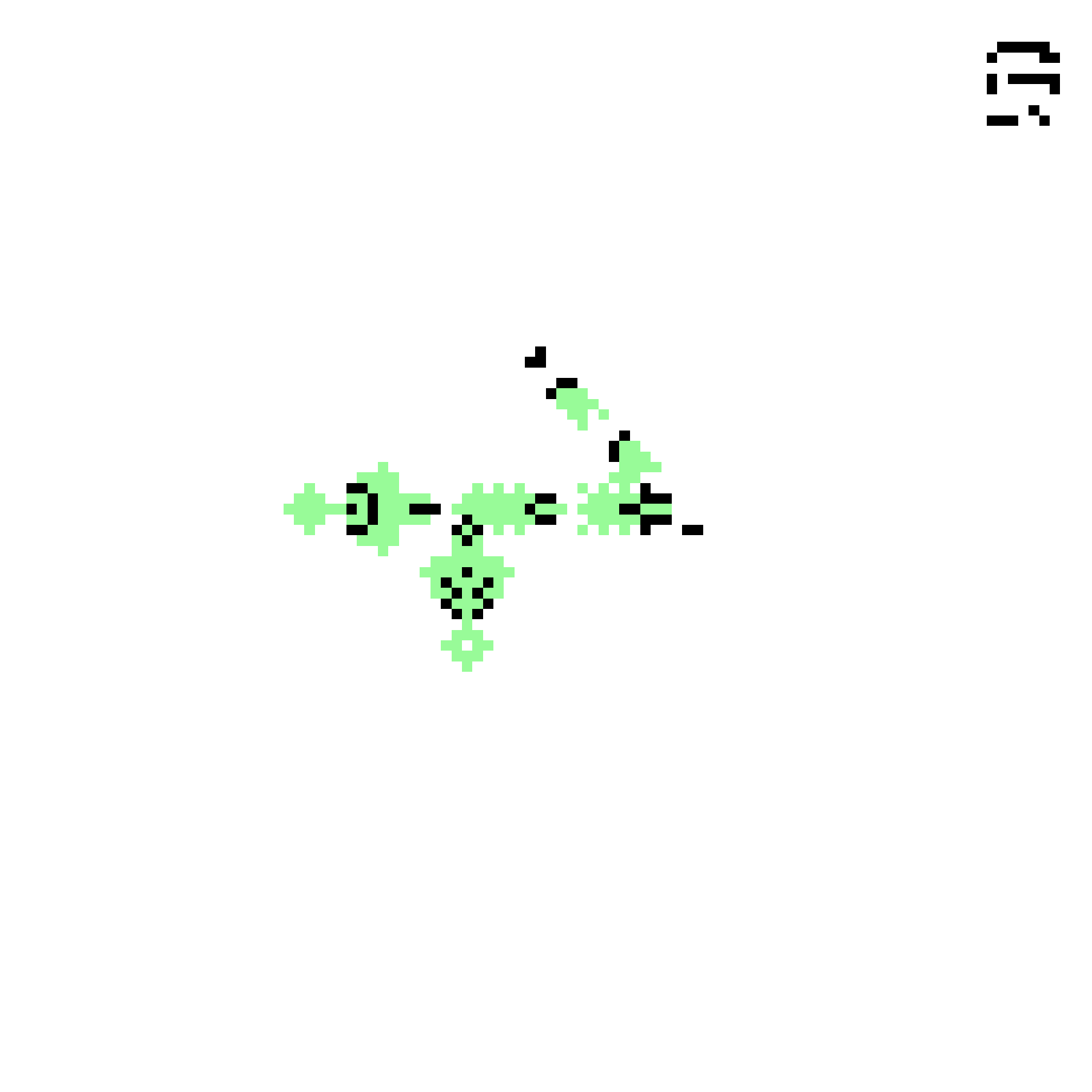}}
\end{minipage}\\[1ex]

\end{minipage}
\end{center}
\vspace{-3ex}
\caption[GGc and GGa glider-guns]
{\textsf{
$Left$: Two phases of the P22 oscillator, precisely juxtaposed at $90^\circ$, interact
to form a GGc glider-gun\cite{Wildmyron}. The two phases are successive time-steps
(figure \ref{P22 oscillator attractor}).
The Gc glider stream is stopped by an eater.
$Right$: The same structure but with a stable block that transforms
Gc to Ga creating a GGa glider-gun\cite{Wright}. 
The Ga glider stream is stopped by an eater.
\label{GGc and GGa glider-guns}
}}
\end{figure}

\begin{figure}
\begin{center}
\begin{minipage}[c]{.9\linewidth} 
\includegraphics[width=1\textwidth,bb=168 15 836 643, clip=]{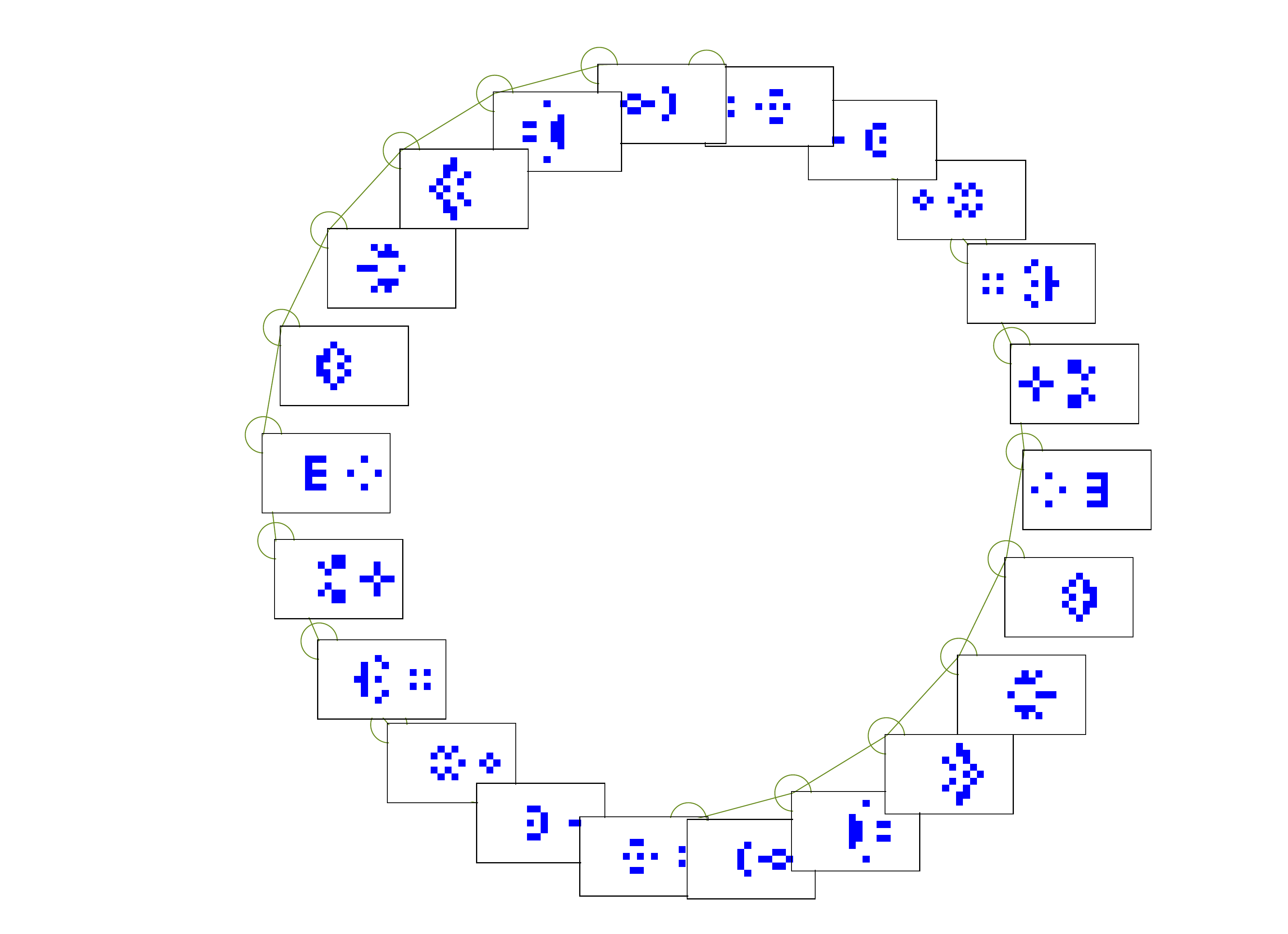}
\end{minipage}
\begin{minipage}[c]{1\linewidth} 
\vspace{-59ex}\hspace{37ex}\includegraphics[width=.3\textwidth,bb=8 12 296 176]{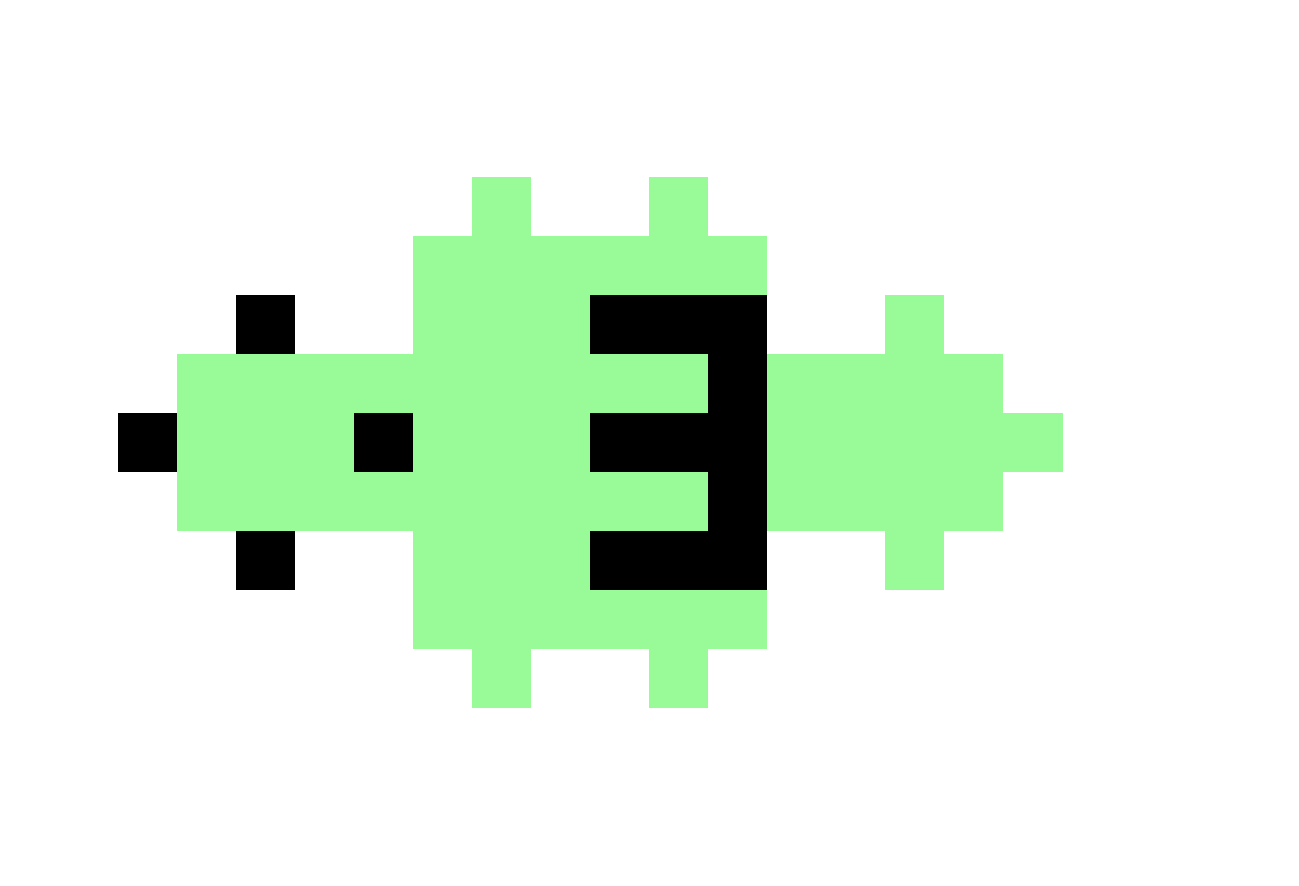}
\end{minipage}
\end{center}
\vspace{-3ex}
\caption[P22 oscillator attractor]
{\textsf{The P22 oscillator showing all 22 phases (time-steps) as an attractor 
cycle\cite{Wuensche92,Wuensche2016} where the direction of time is clockwise.
$Inset$: An oscillator phase shown at a larger scale
alongside the same phase on the attractor cycle.
}}
\label{P22 oscillator attractor}
\end{figure}

\begin{figure}[htb]
\begin{center}
\begin{minipage}[c]{.6\linewidth} 
\fbox{\includegraphics[width=1\linewidth,bb=46 115 360 307, clip=]{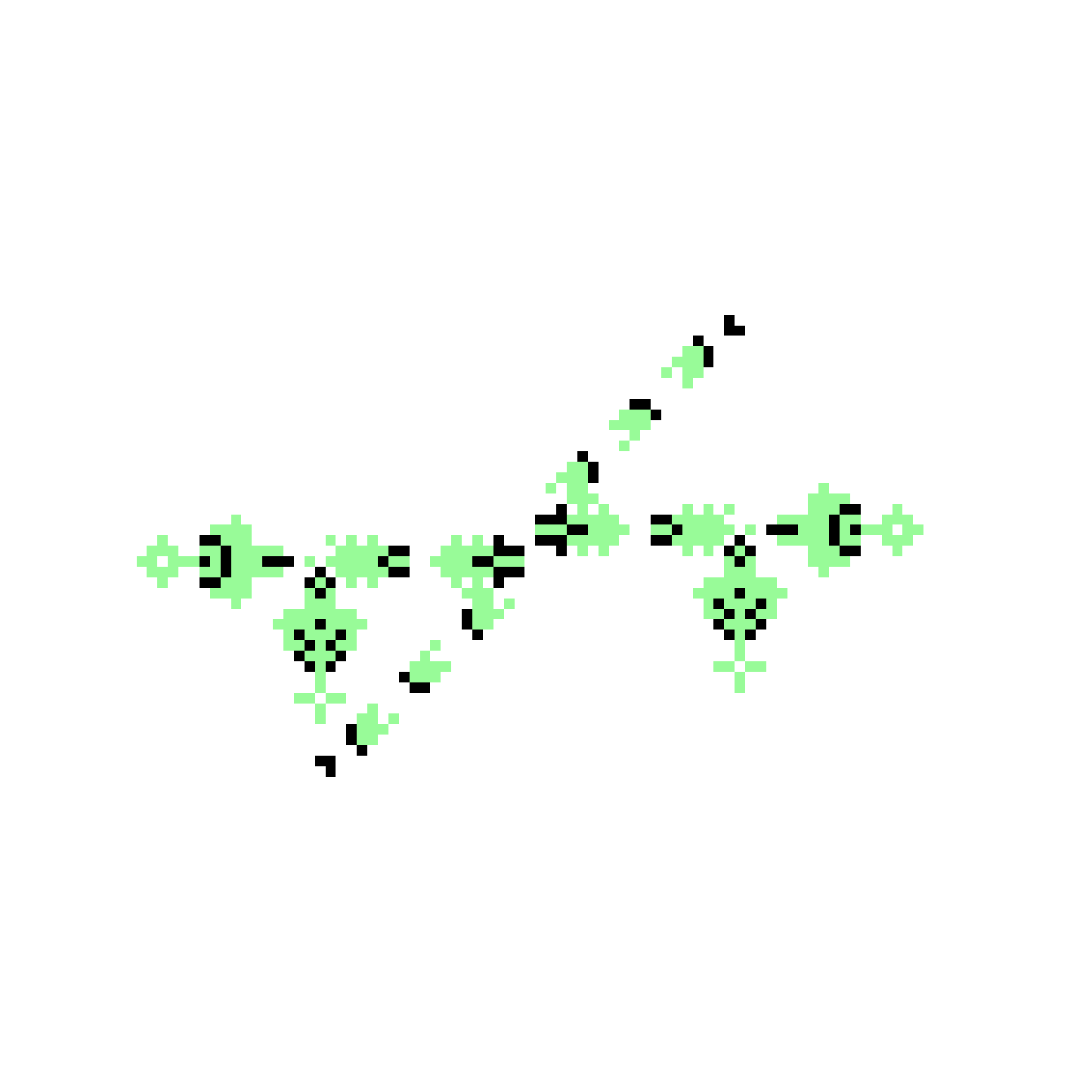}}
\end{minipage}
\end{center}
\vspace{-3ex}
\caption[GGc x GGc to GGa glider-gun]
{\textsf{
Two GGc glider-guns, their centers offset, shoot Gc gliders at each other.
The Gc collisions create two Ga gliders, thus the combination creates
a double GGa glider-gun.
The Ga glider streams are stopped by eaters.
\label{GGc x GGc to GGa glider-gun}
}}
\end{figure}

The P22 oscillator, named for its 22 time-step frequency, but which divides
into two sets of 11 reflected patterns, is detailed in
figure~\ref{P22 oscillator attractor}, and is used to build Gc and Ga
glider-guns in figures~\ref{GGc and GGa glider-guns} and 
\ref{GGc x GGc to GGa glider-gun}.

\enlargethispage{5ex} 
\clearpage

\clearpage

\subsection{Glider-Guns from oscillator P15}
\label{Glider-Guns from oscillator P15}

\begin{figure}[h]
\begin{center}
\begin{minipage}[c]{.75\linewidth} 
\includegraphics[width=1\textwidth,bb=272 132 701 540, clip=]{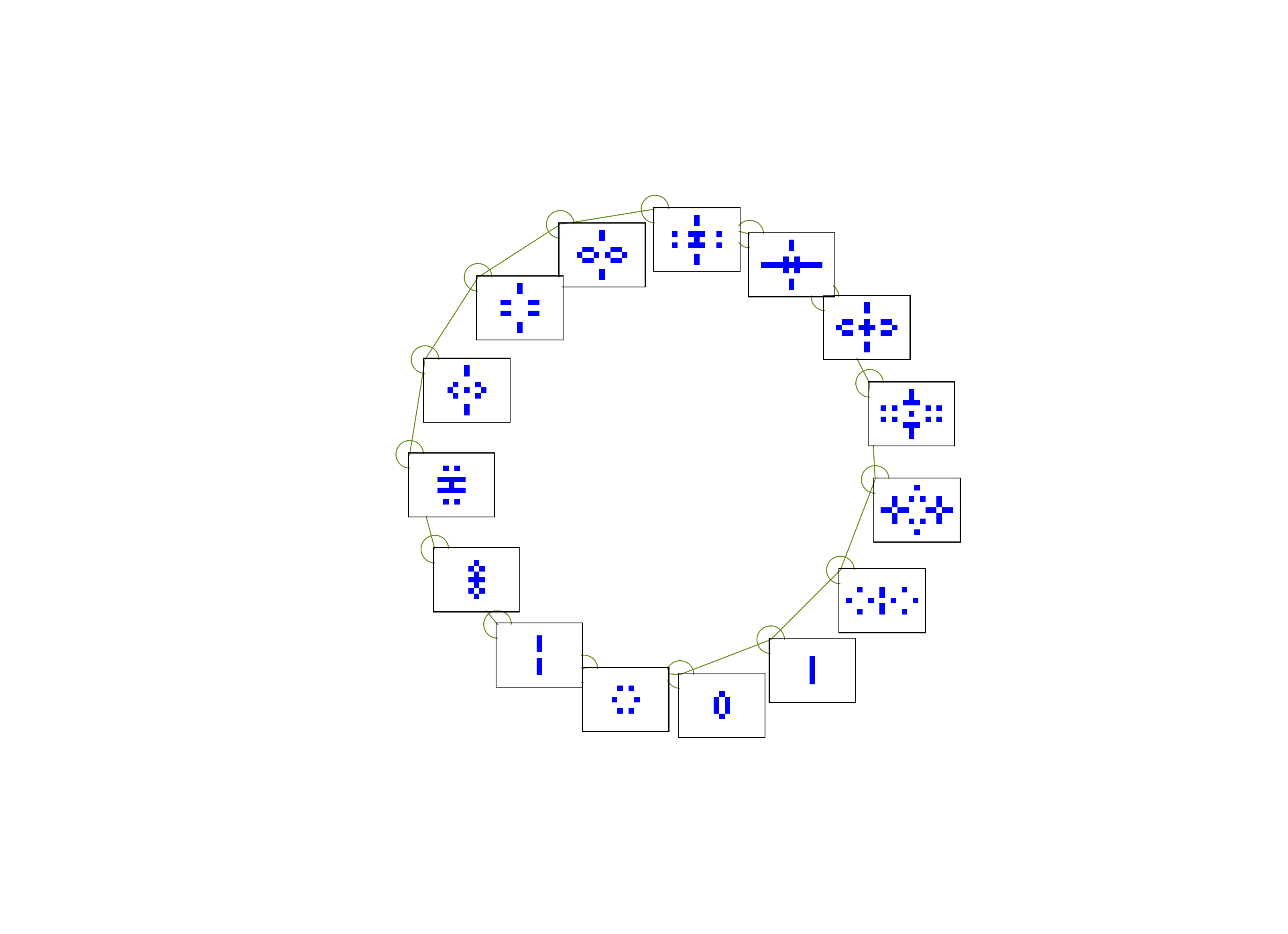}
\end{minipage}
\begin{minipage}[c]{1\linewidth} 
\vspace{-51ex}\hspace{37ex}\includegraphics[width=.24\textwidth,bb=14 21 265 198]{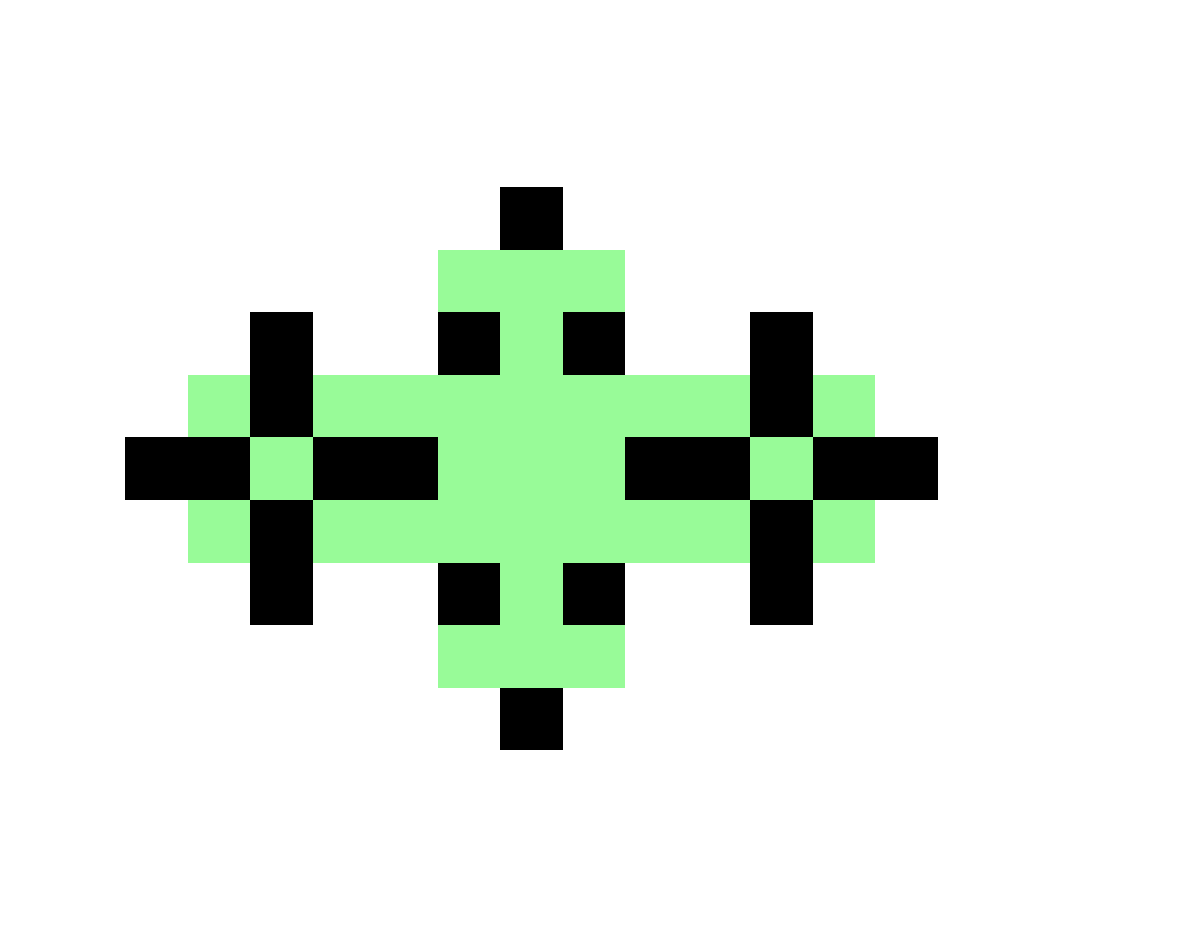}
\end{minipage}
\end{center}
\vspace{-5ex}
\caption[P15 oscillator attractor]
{\textsf{The P15 oscillator\cite{Wildmyron} showing all 15 phases (time-steps) as an attractor 
cycle\cite{Wuensche92,Wuensche2016} where the direction of time is clockwise.
$Inset$: An oscillator phase shown at a larger scale
alongside the same phase on the attractor cycle.
}}
\label{P15 oscillator attractor}
\vspace{-2ex}
\end{figure}

\begin{figure}[htb]
\begin{center}
\begin{minipage}[c]{.9\linewidth} 
\fbox{\includegraphics[height=.45\linewidth,bb=124 152 362 387, clip=]{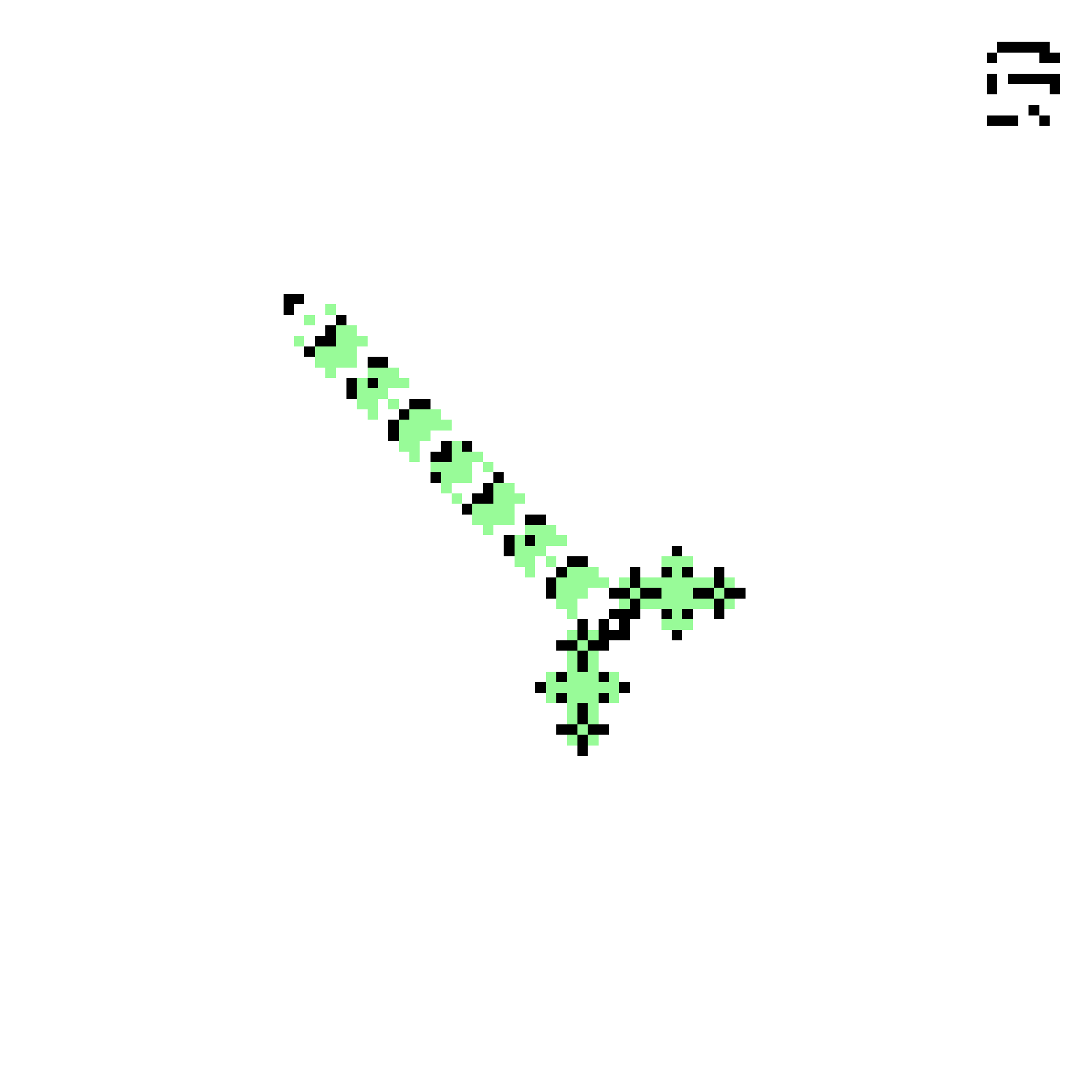}}
\hfill 
\fbox{\includegraphics[height=.45\linewidth,bb=10 18 295 291, clip=]{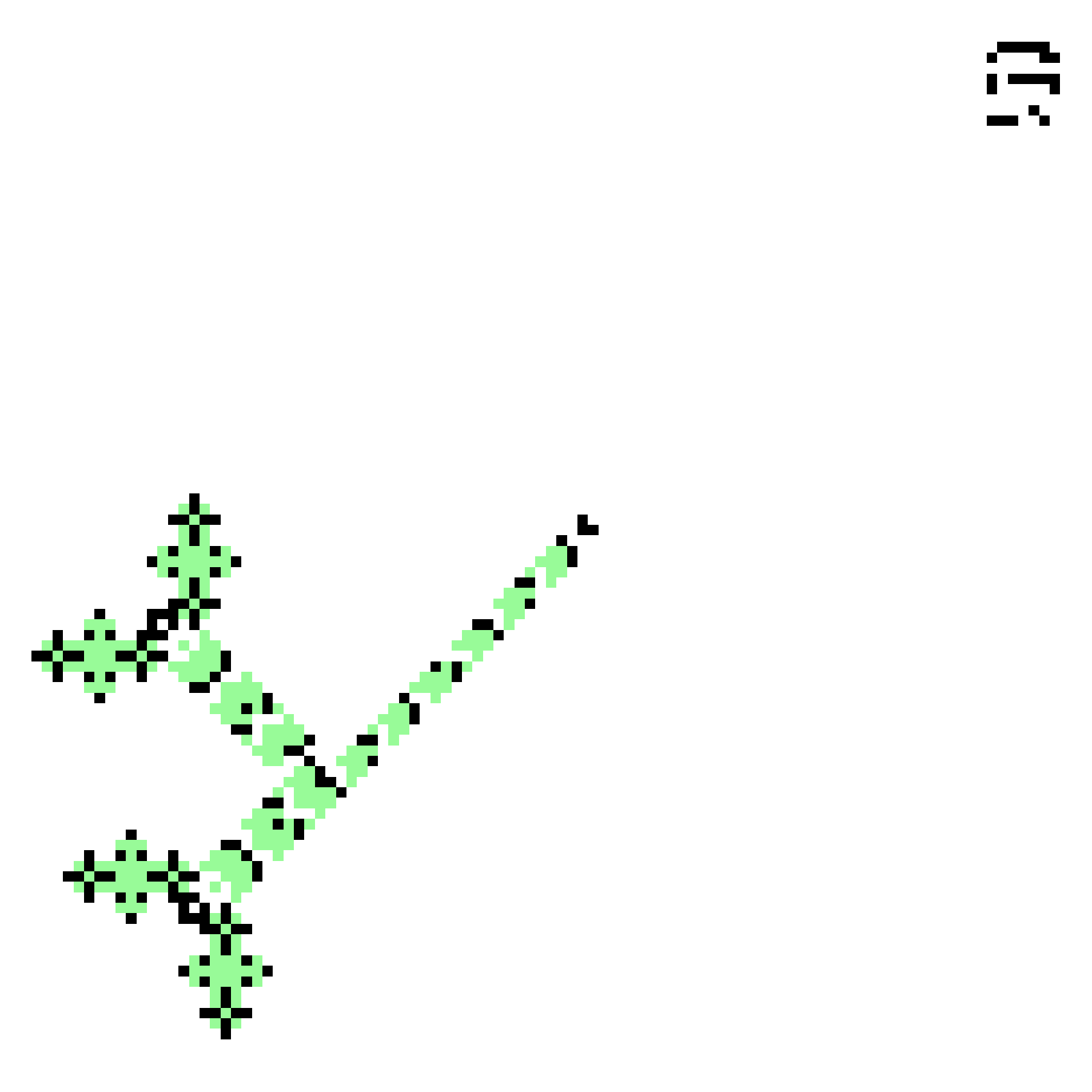}}
\end{minipage}
\end{center}
\vspace{-4ex}
\caption[GG2a glider-guns]
{\textsf{
Double Ga (G2a) gliders can be shot by a GG2a glider-gun constructed from
two (same phase) P15 oscillators correctly juxtaposed at $90^\circ$\cite{Wildmyron}.
$Left$: The glider stream is stopped by a G2a eater.
$Right$: Two GG2a glider-guns at $90^\circ$ create a Ga glider stream, stopped by an eater.
\label{GG2a glider-guns}
}}
\end{figure}

The P15 oscillator, named for its 15 time-step frequency, is detailed in
figure~\ref{P15 oscillator attractor}, and is used to build G2a and Ga
glider-guns in figure~\ref{GG2a glider-guns}.

\enlargethispage{7ex}
\clearpage

\subsection{Glider-Guns from reflector}
\label{Glider-Guns from reflector}

\begin{figure}[htb]
\begin{center}
\begin{minipage}[c]{1\linewidth} 
\includegraphics[width=.07\linewidth,bb=177 191 212 270, clip=]{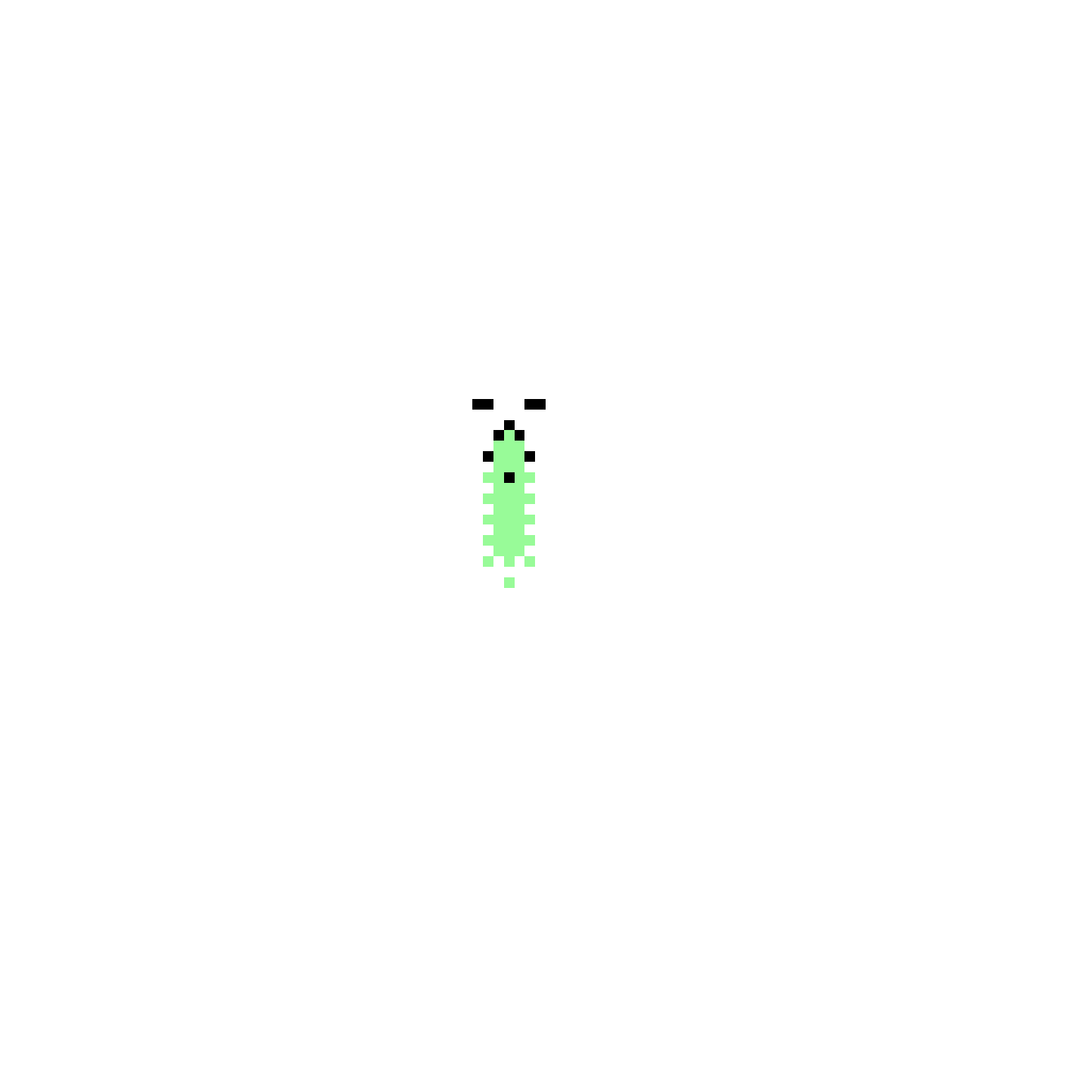}
\includegraphics[width=.07\linewidth,bb=177 191 212 270, clip=]{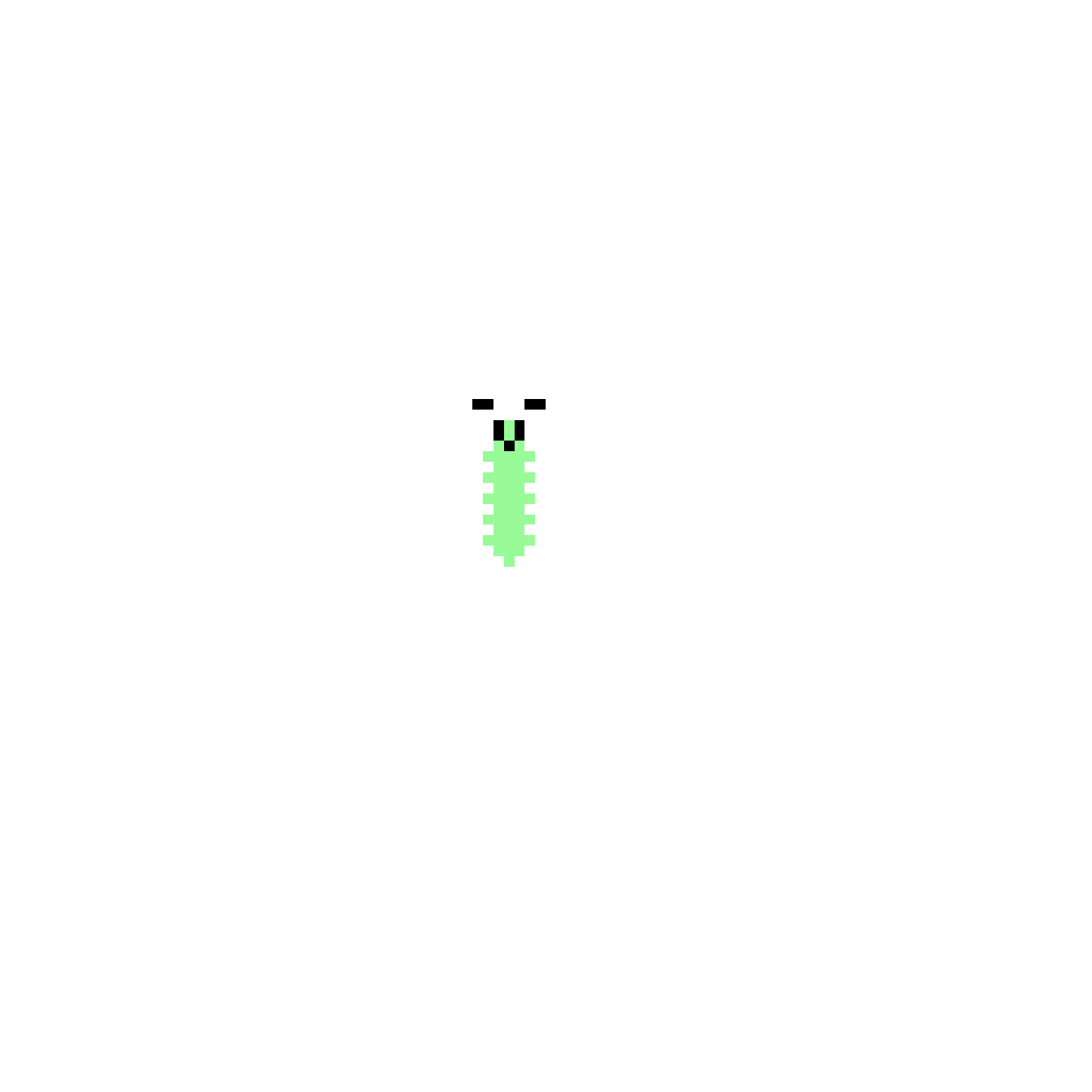}
\includegraphics[width=.07\linewidth,bb=177 191 212 270, clip=]{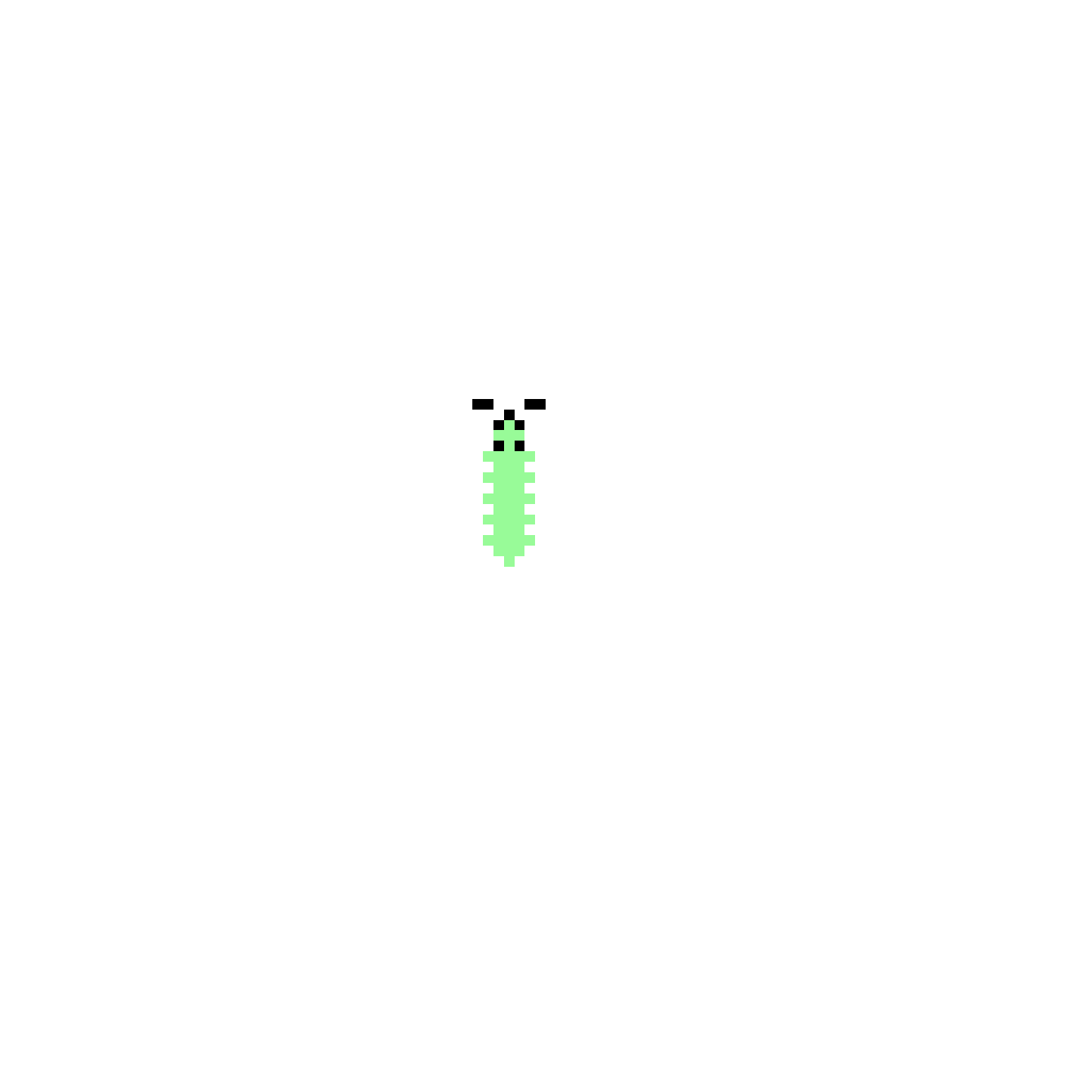}
\includegraphics[width=.07\linewidth,bb=177 191 212 270, clip=]{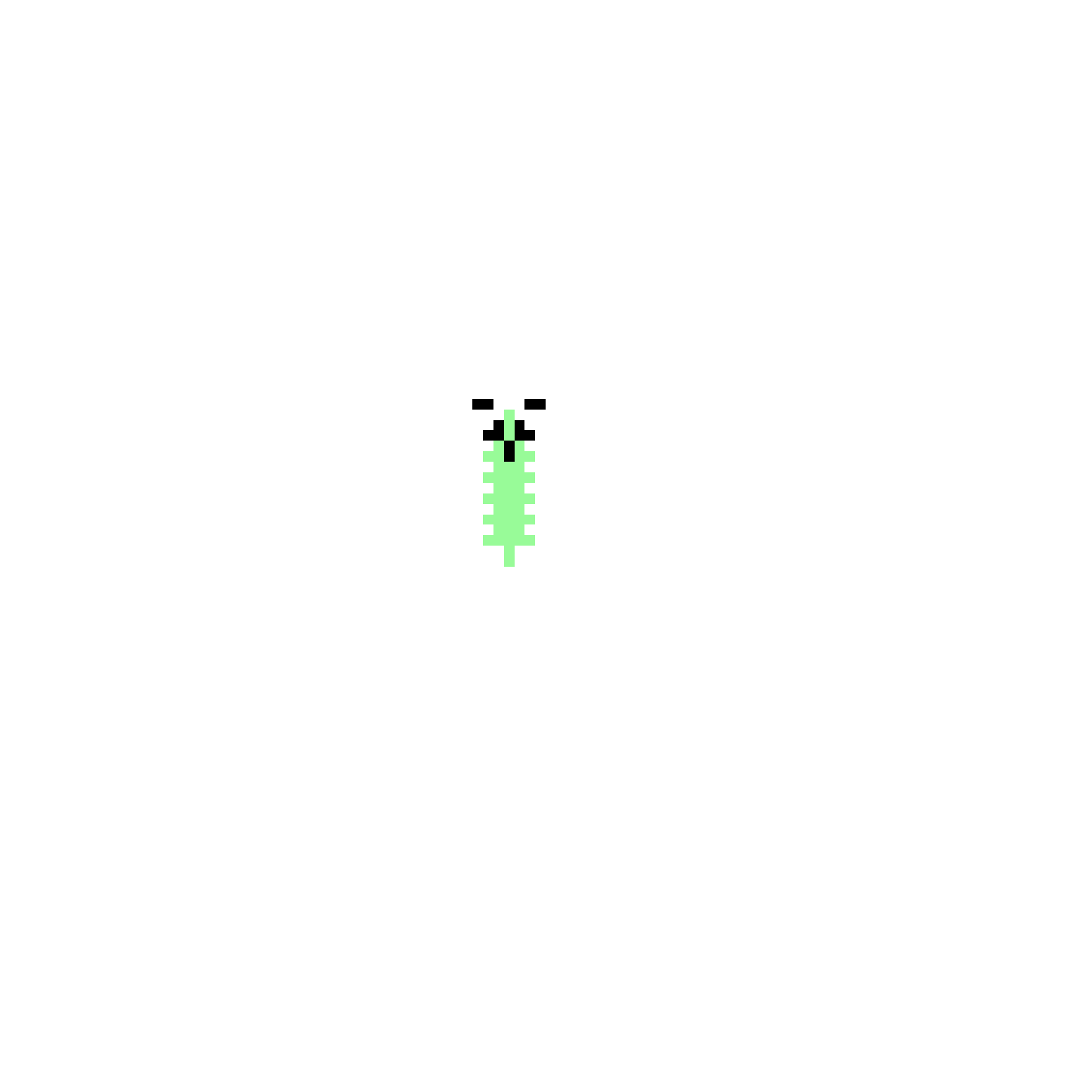}
\includegraphics[width=.07\linewidth,bb=177 191 212 270, clip=]{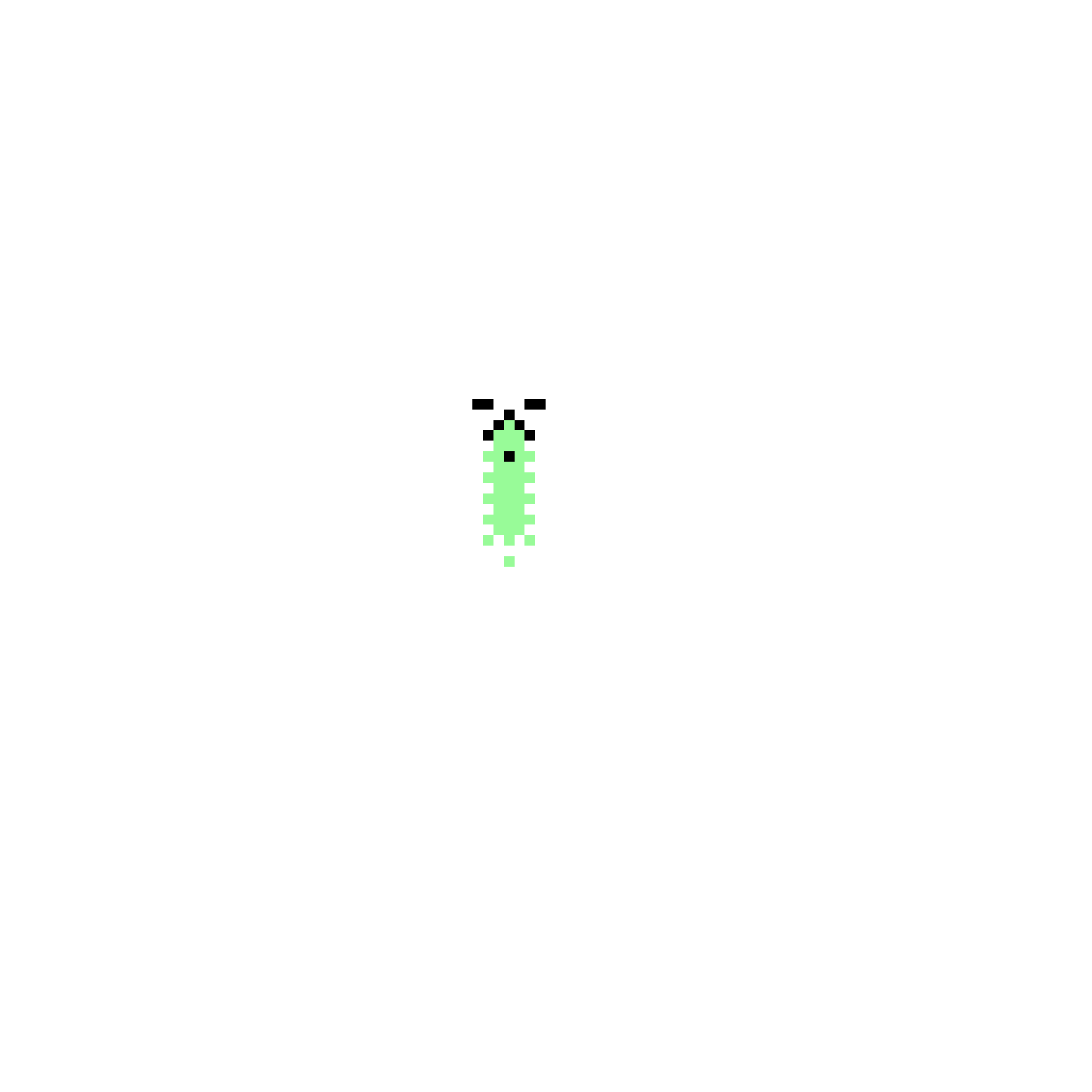}
\includegraphics[width=.07\linewidth,bb=177 191 212 270, clip=]{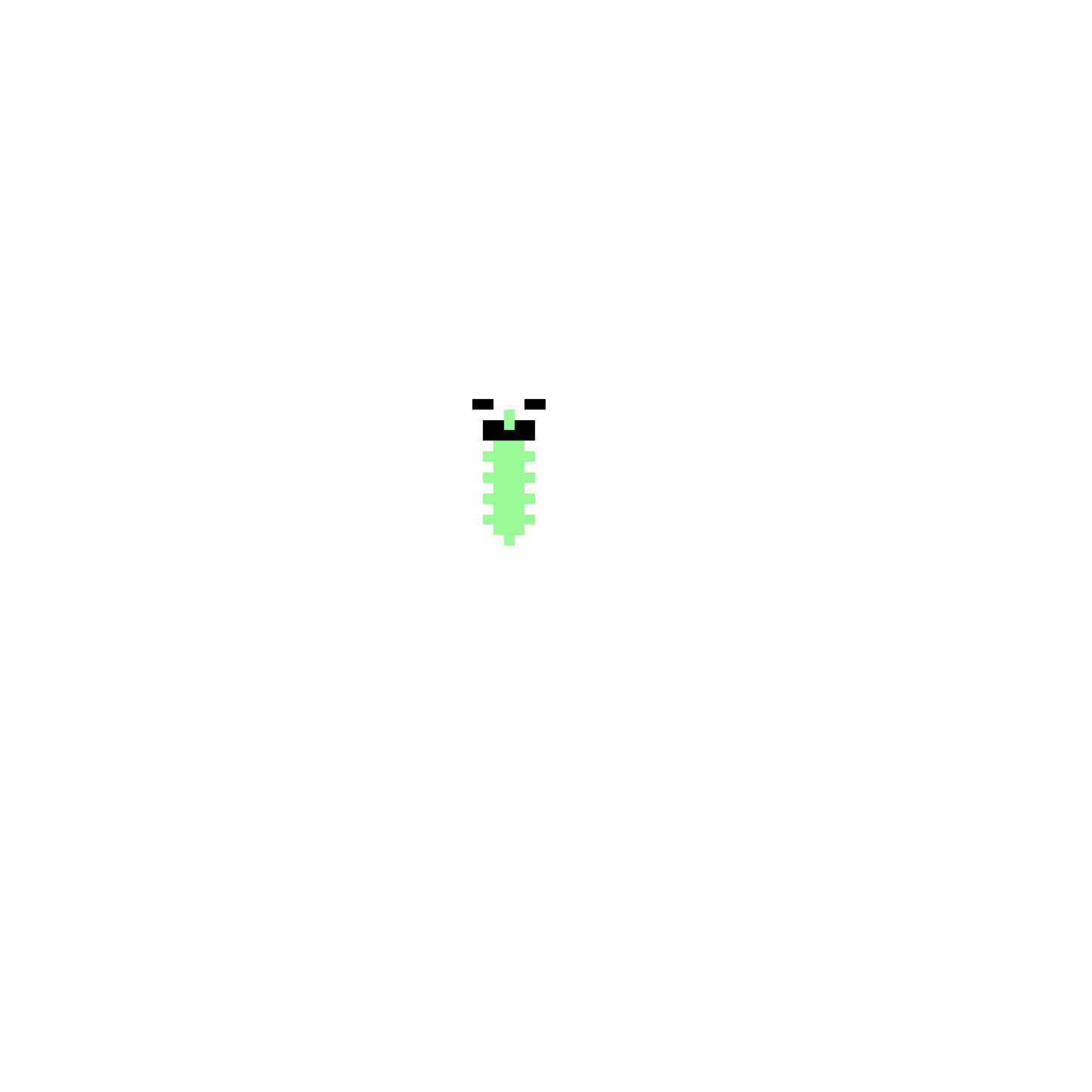}
\includegraphics[width=.07\linewidth,bb=177 191 212 270, clip=]{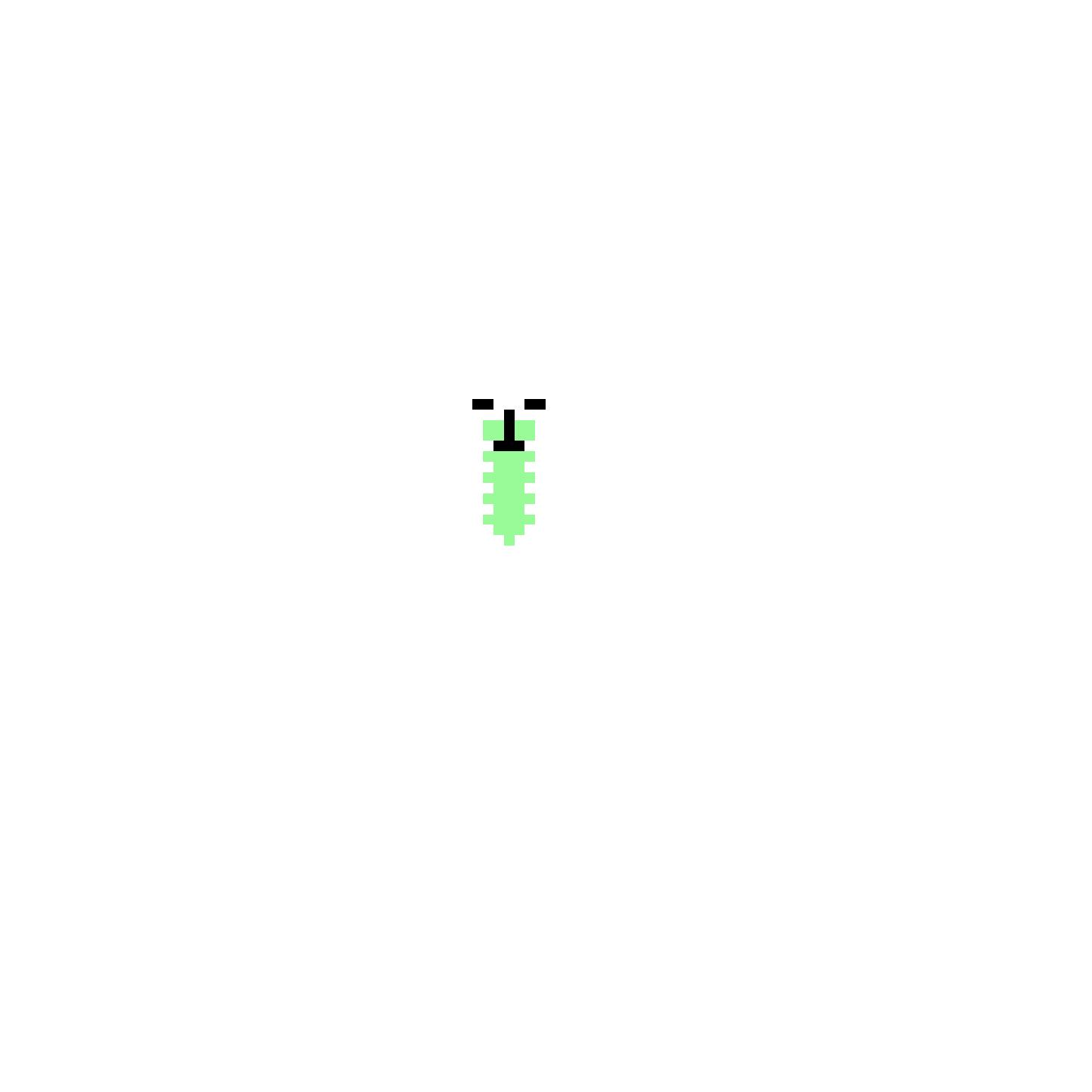}
\includegraphics[width=.07\linewidth,bb=177 191 212 270, clip=]{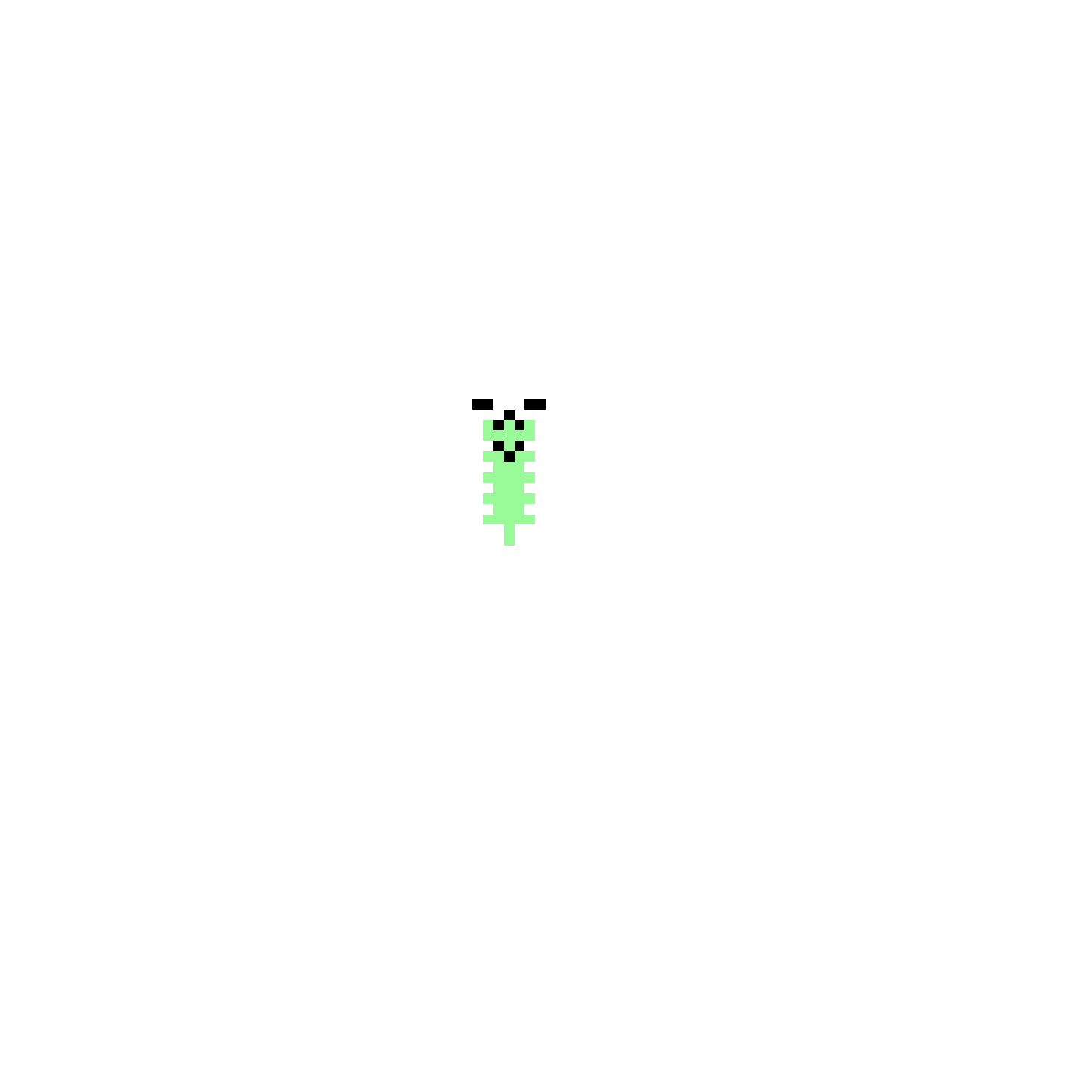}
\includegraphics[width=.07\linewidth,bb=177 191 212 270, clip=]{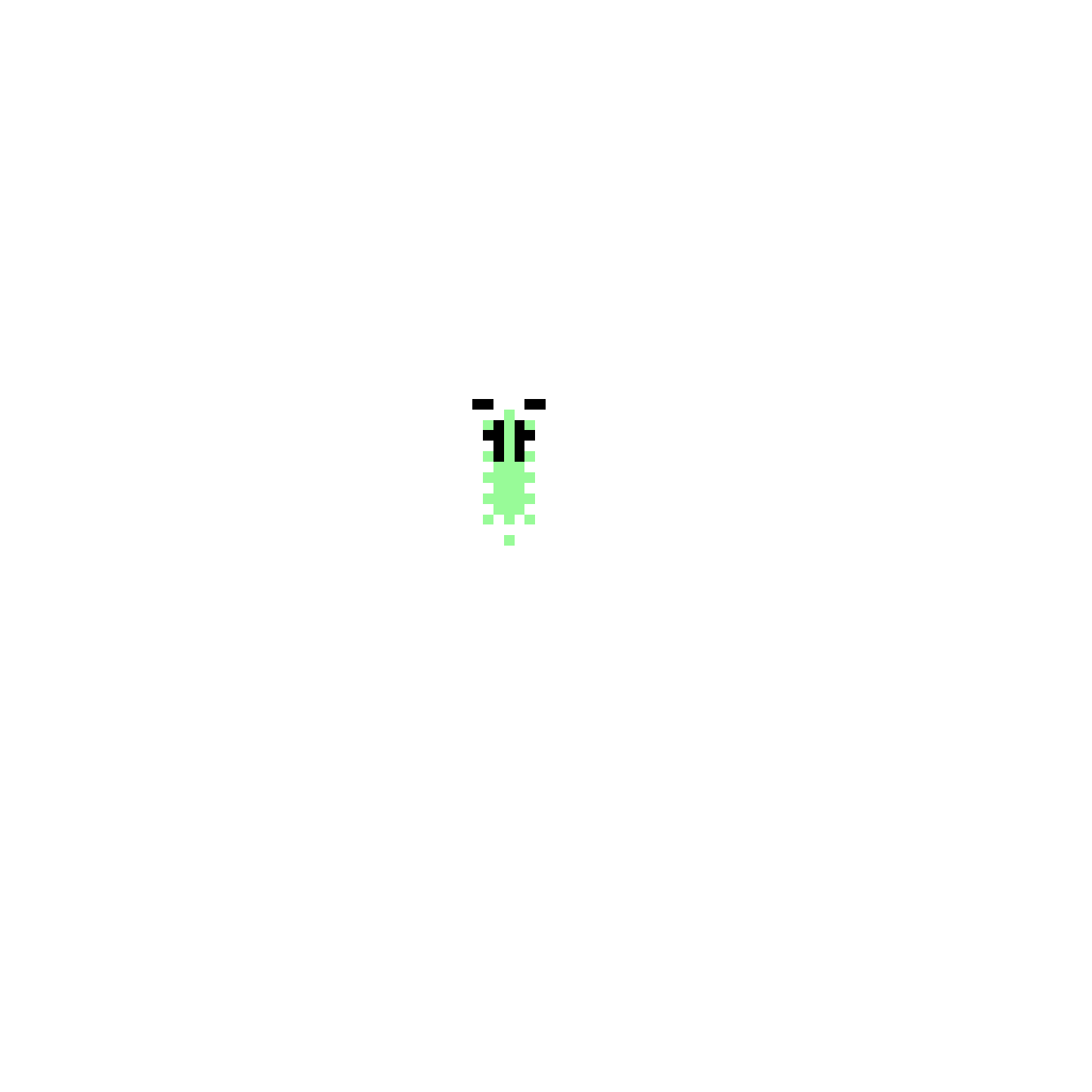}
\includegraphics[width=.07\linewidth,bb=177 191 212 270, clip=]{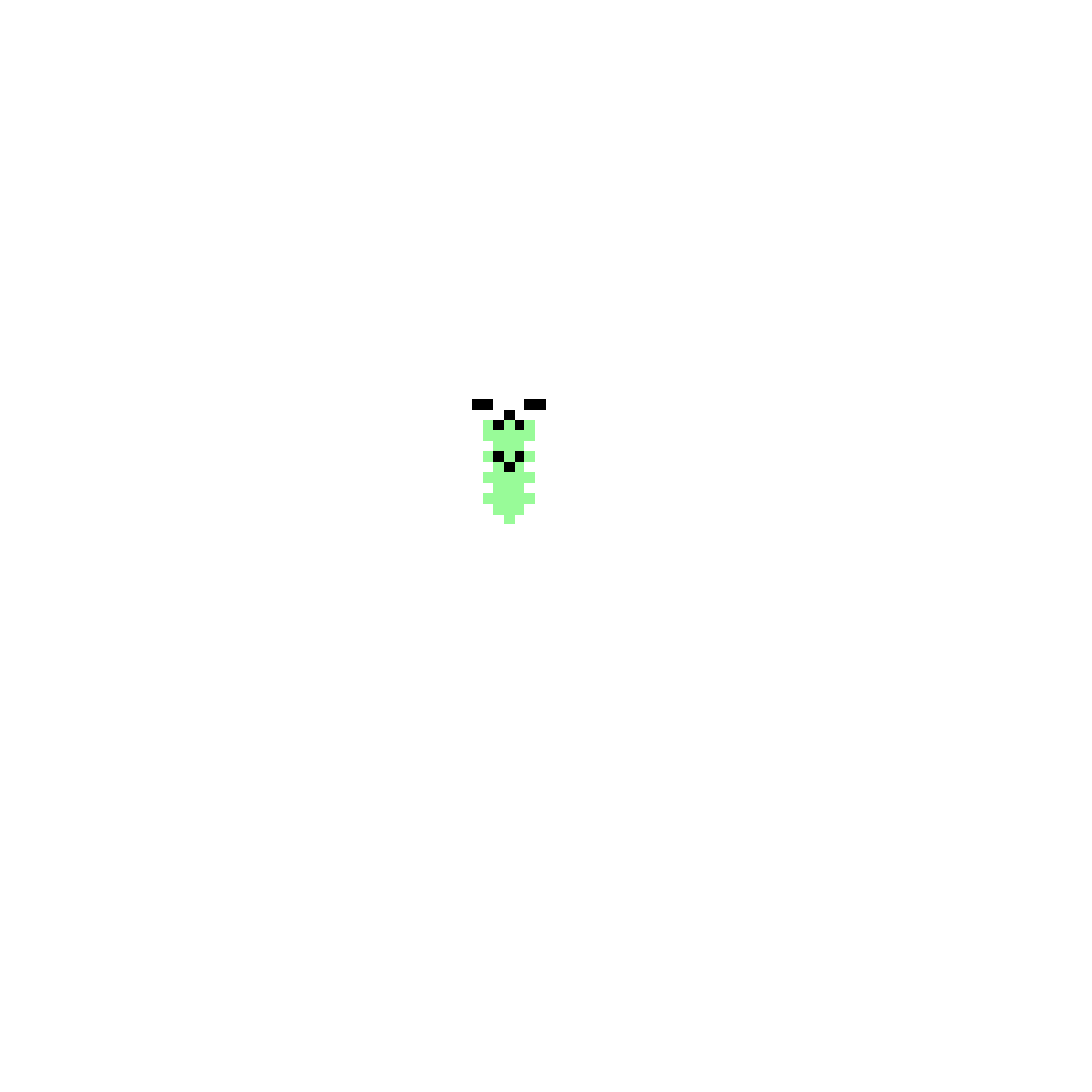}
\includegraphics[width=.07\linewidth,bb=177 191 212 270, clip=]{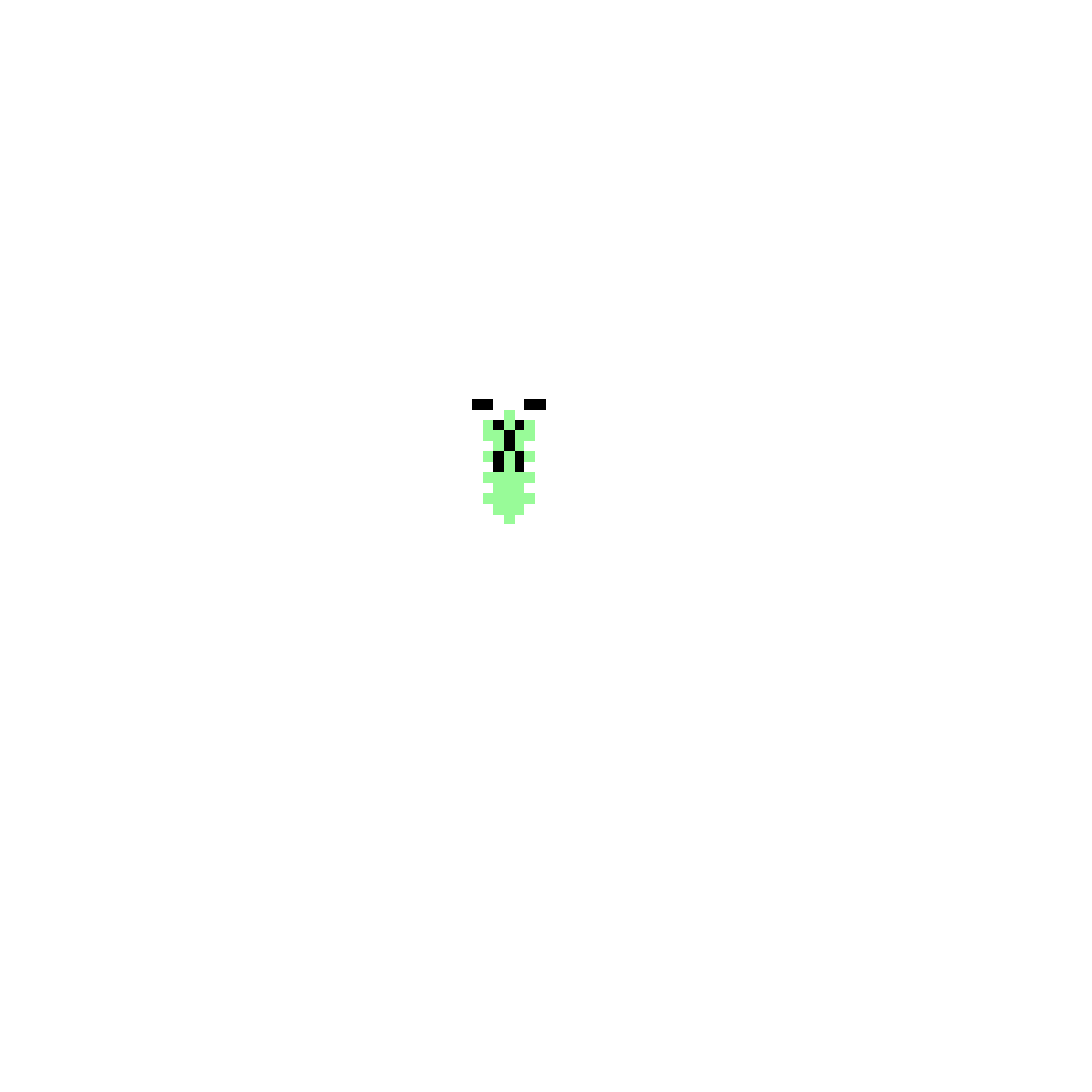}
\includegraphics[width=.07\linewidth,bb=177 191 212 270, clip=]{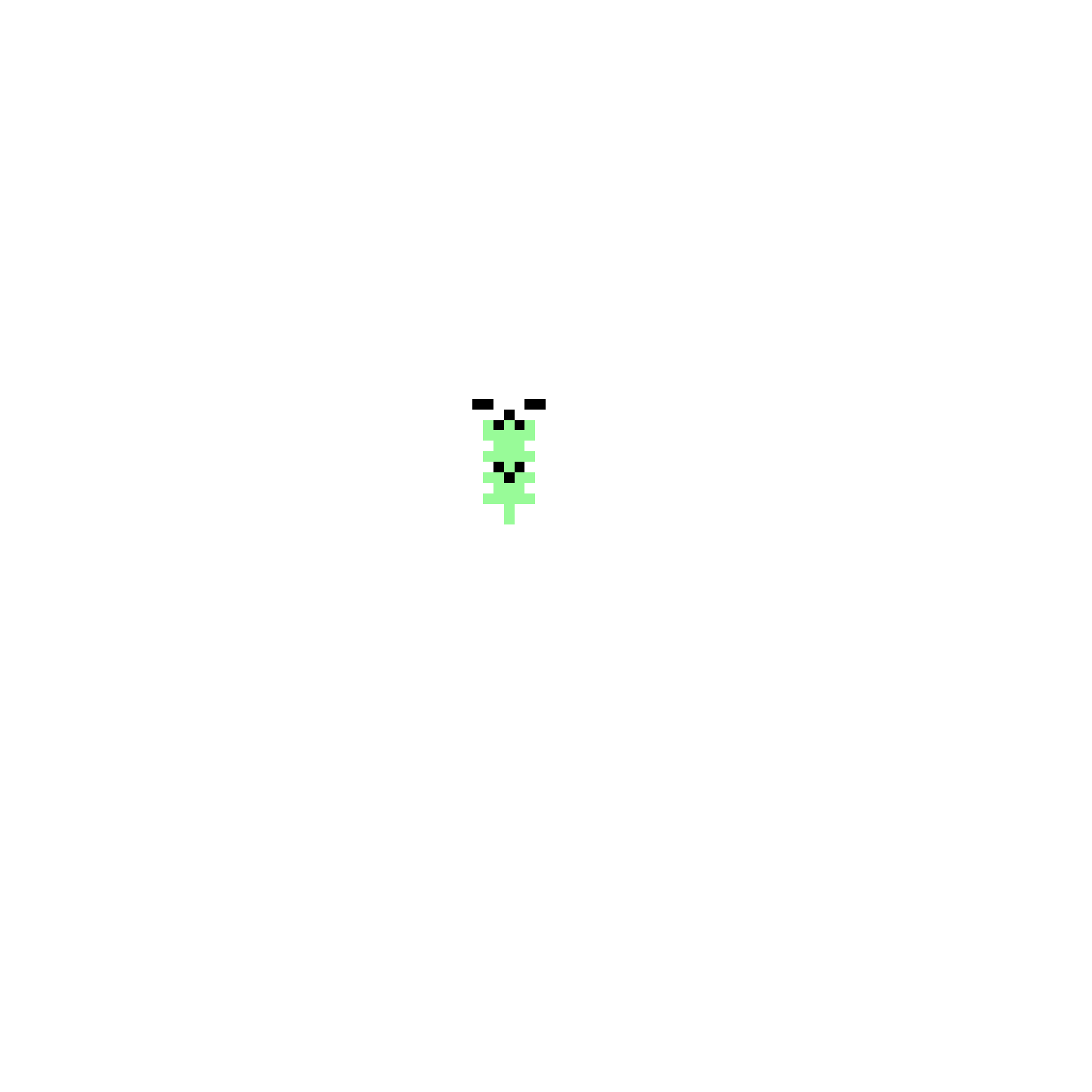}
\includegraphics[width=.07\linewidth,bb=177 191 212 270, clip=]{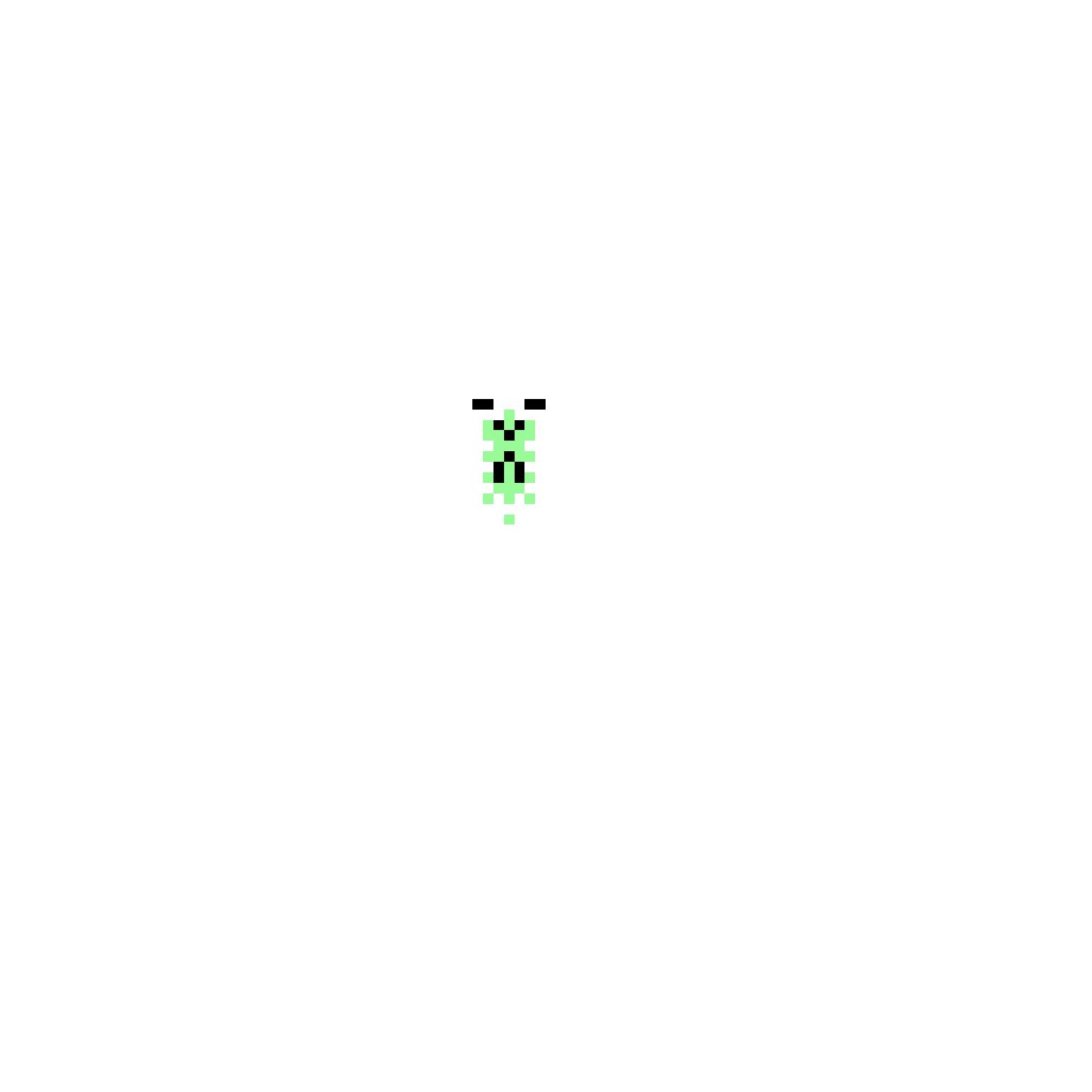}
\includegraphics[width=.07\linewidth,bb=177 191 212 270, clip=]{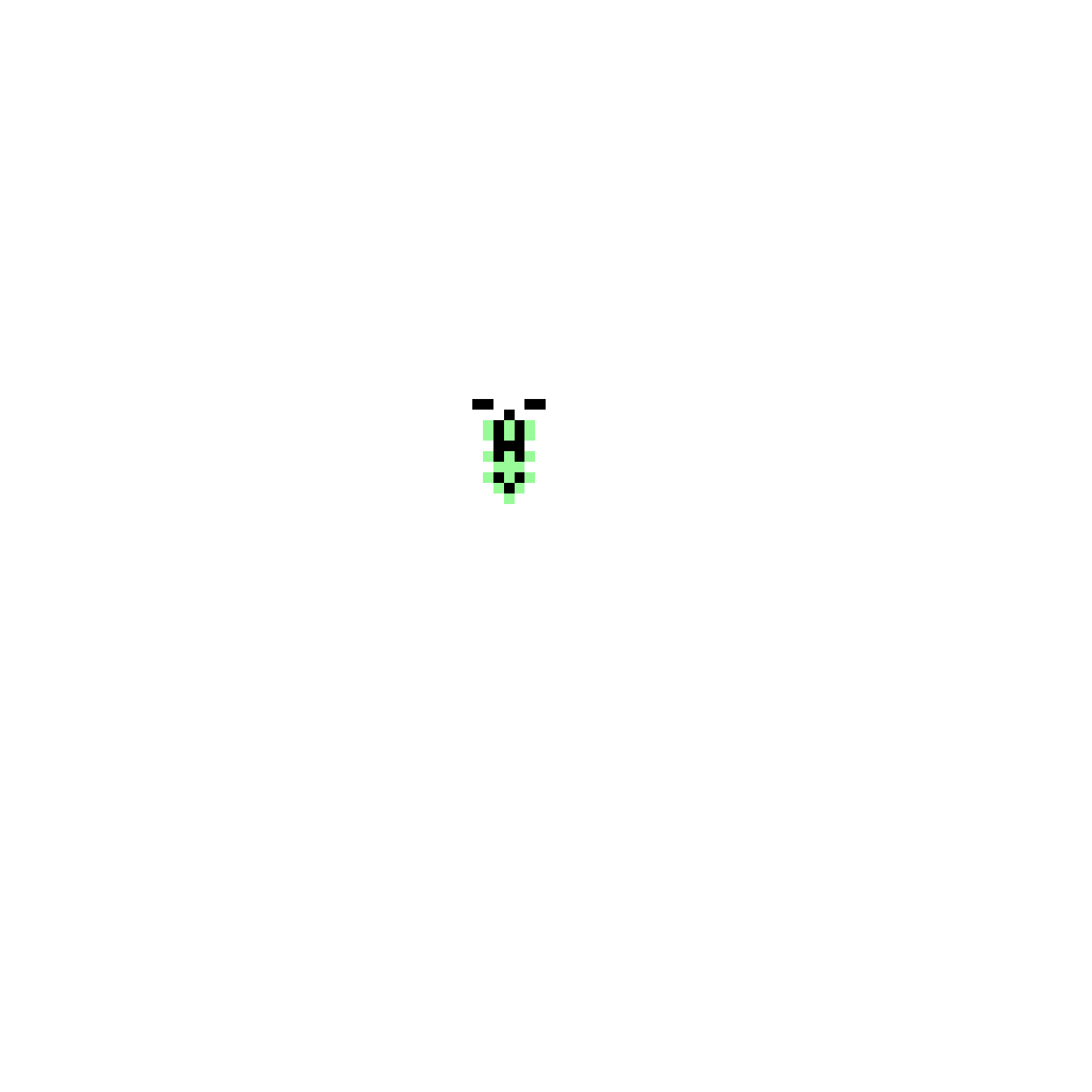}
\includegraphics[width=.07\linewidth,bb=177 191 212 270, clip=]{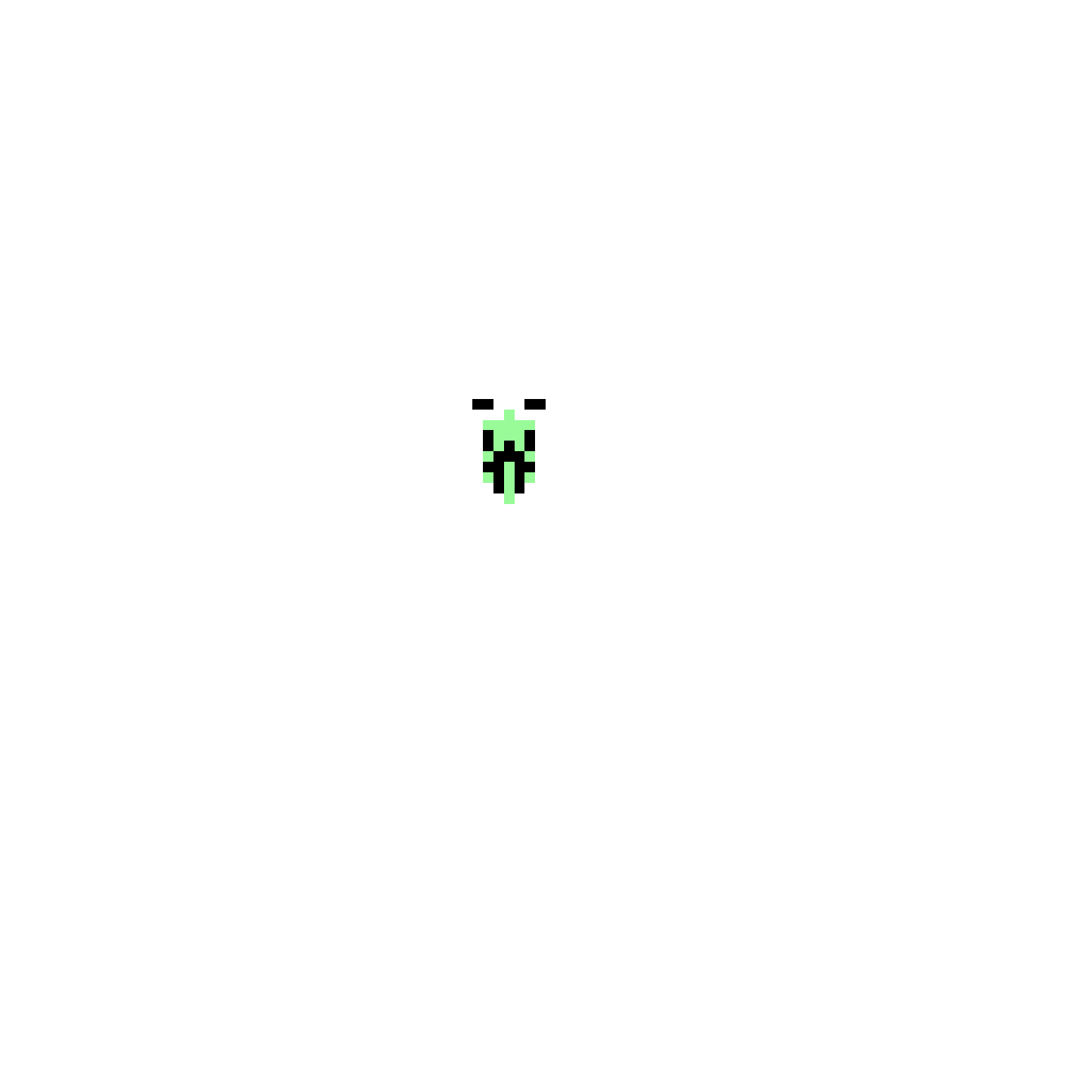}
\includegraphics[width=.07\linewidth,bb=177 191 212 270, clip=]{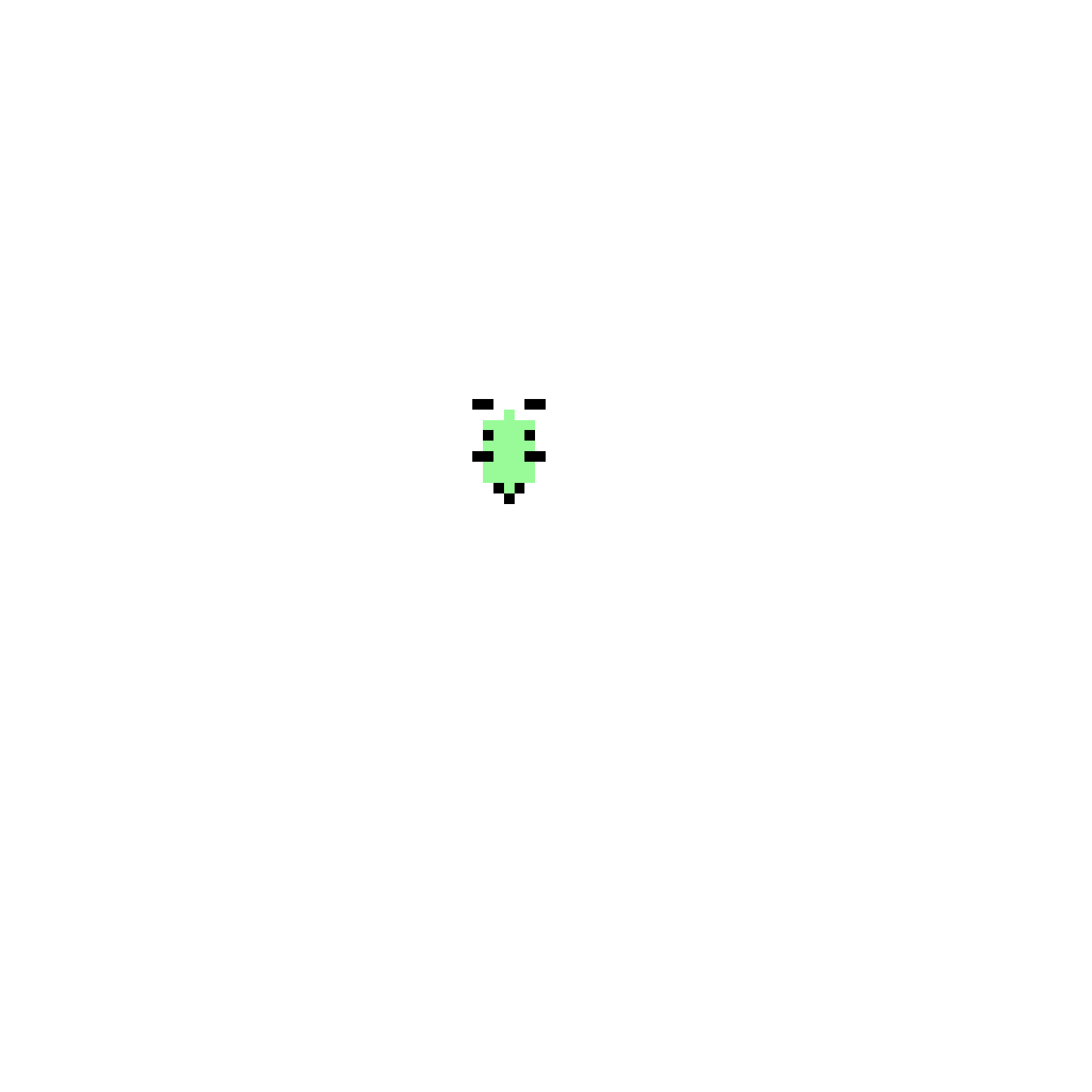}
\includegraphics[width=.07\linewidth,bb=177 191 212 270, clip=]{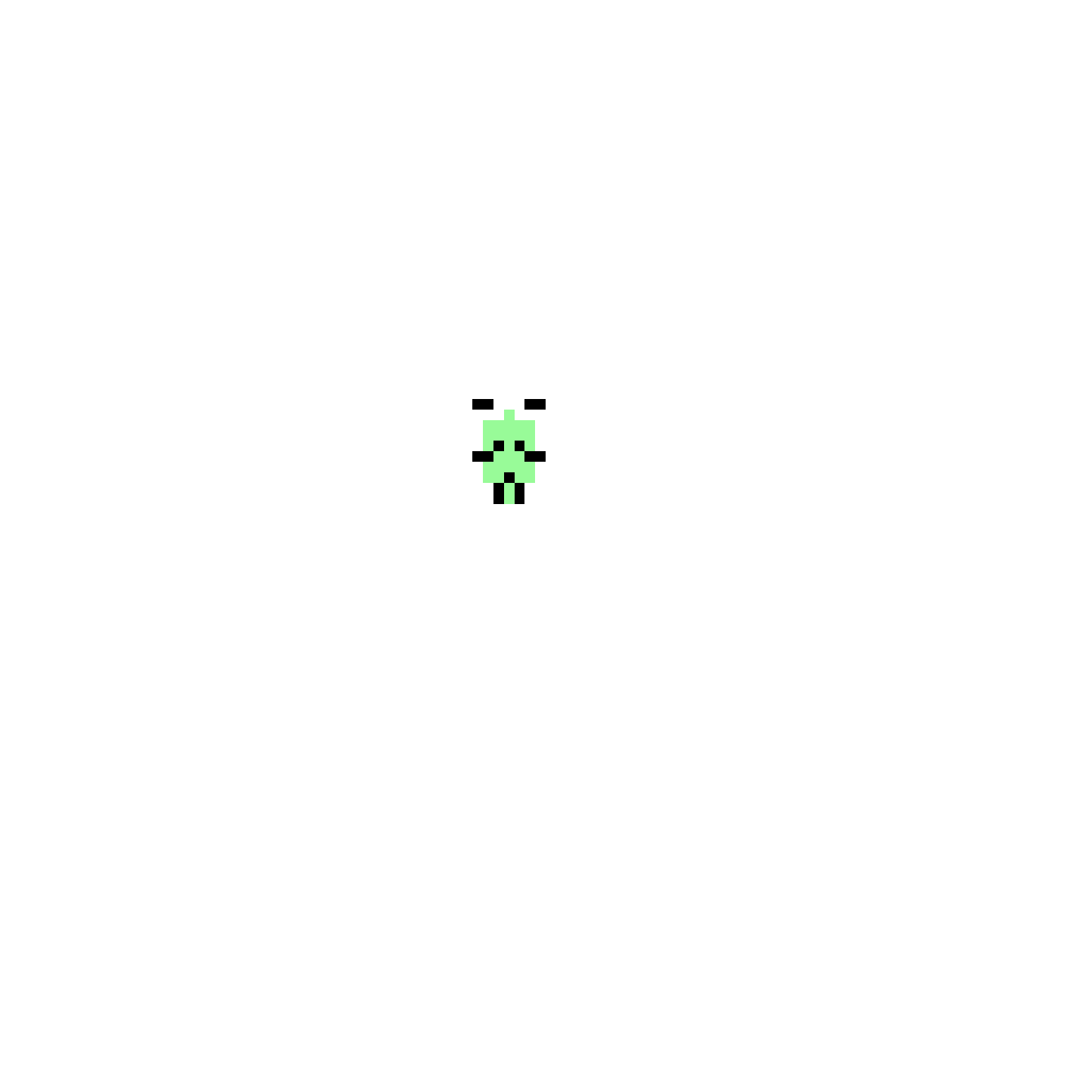}
\includegraphics[width=.07\linewidth,bb=177 191 212 270, clip=]{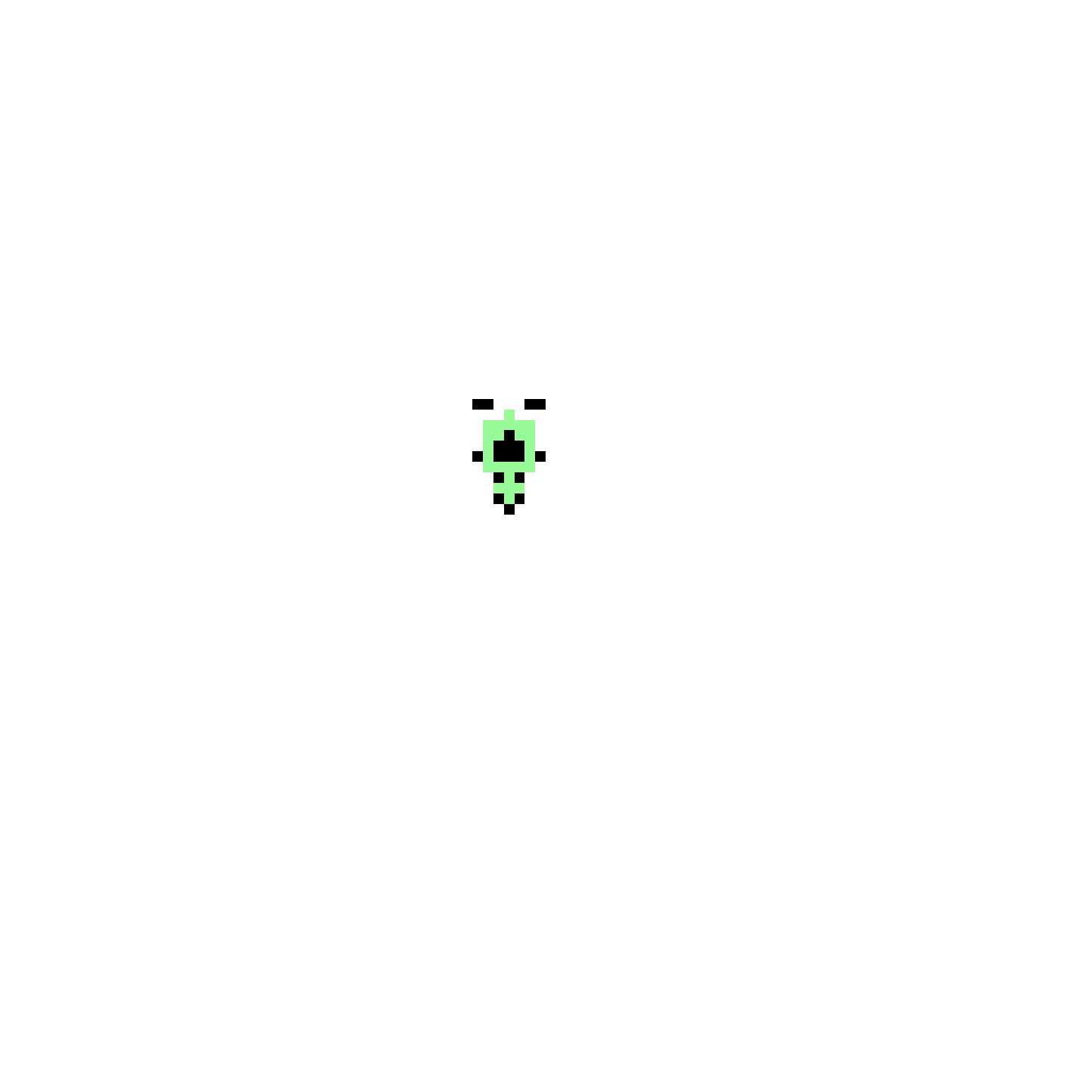}
\includegraphics[width=.07\linewidth,bb=177 191 212 270, clip=]{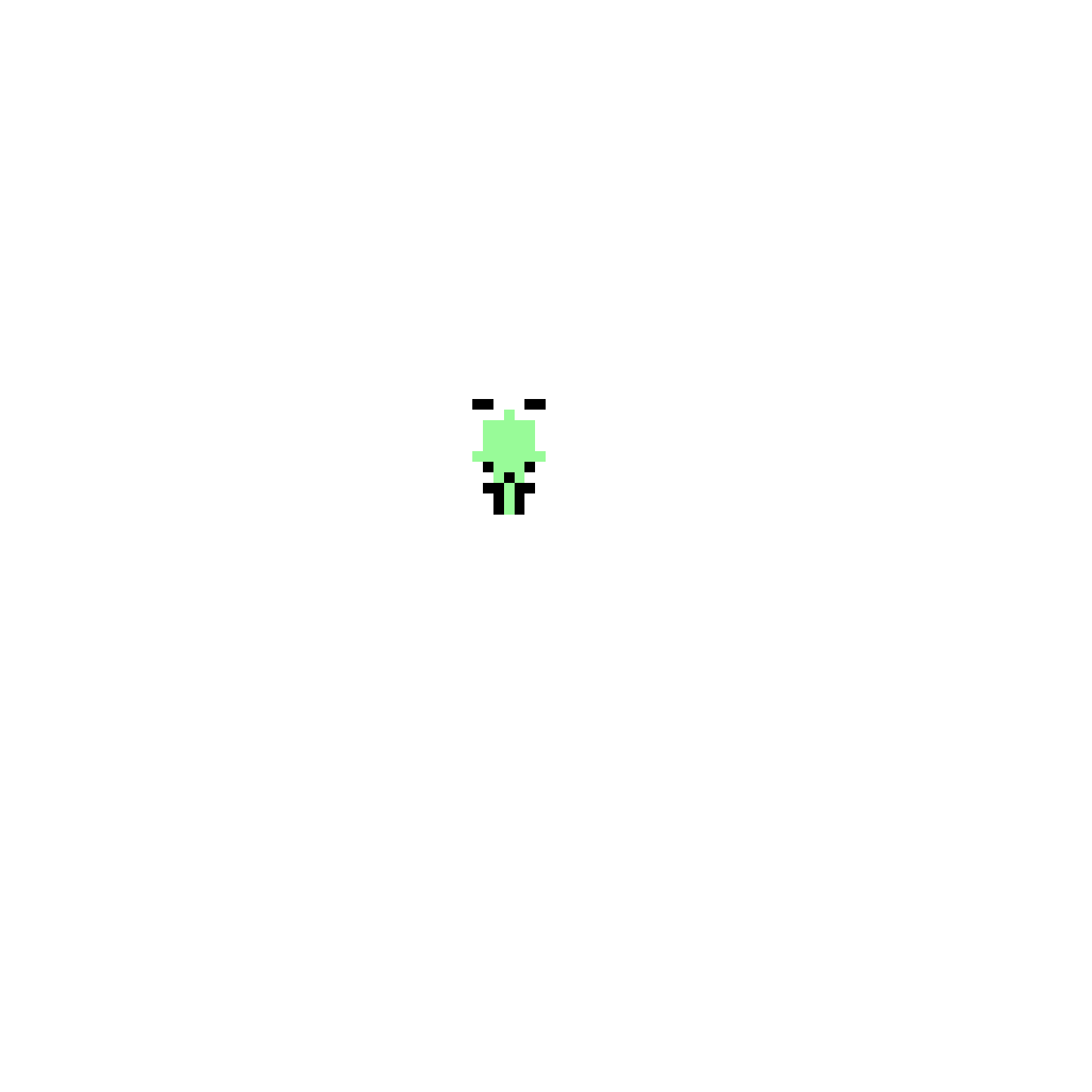}
\includegraphics[width=.07\linewidth,bb=177 191 212 270, clip=]{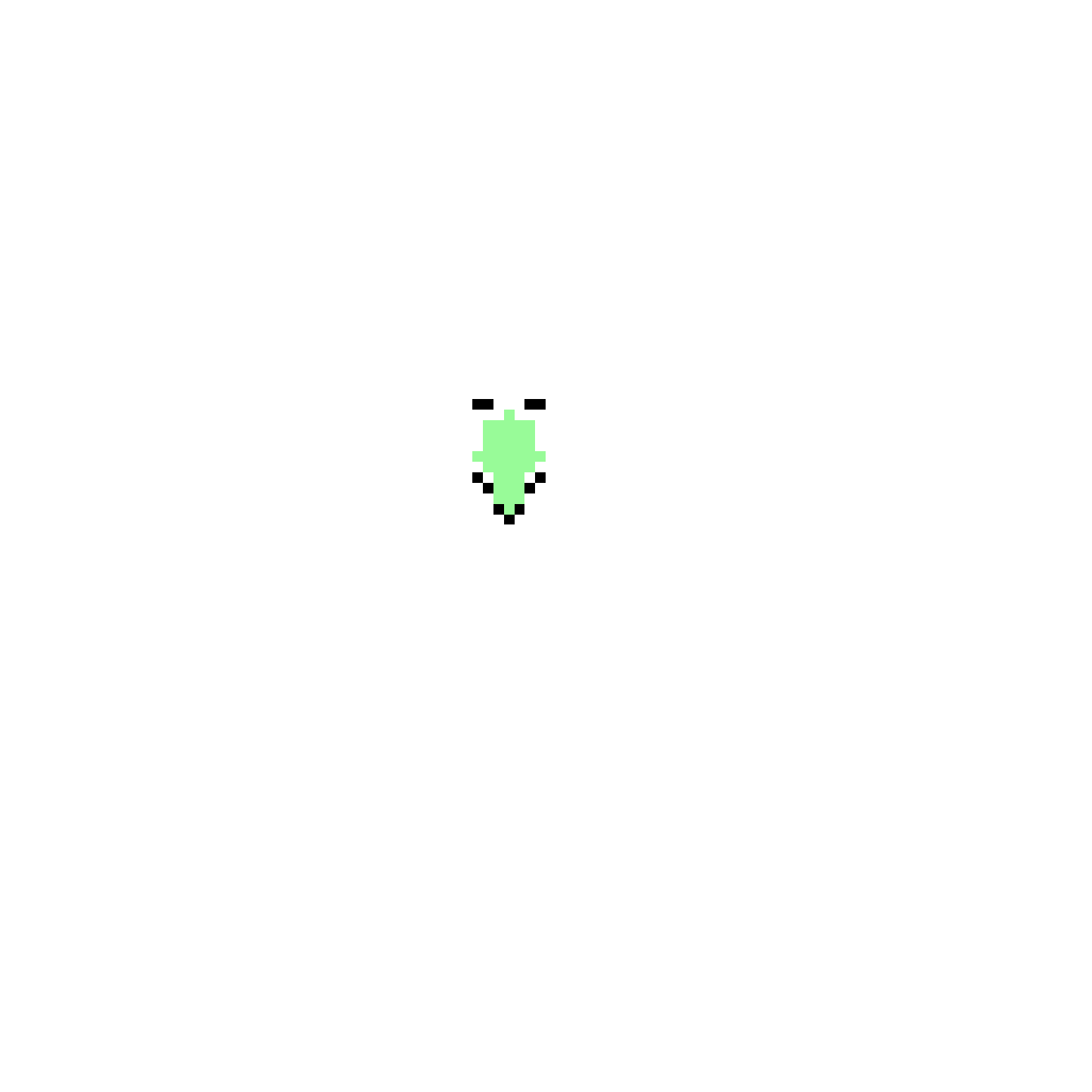}
\includegraphics[width=.07\linewidth,bb=177 191 212 270, clip=]{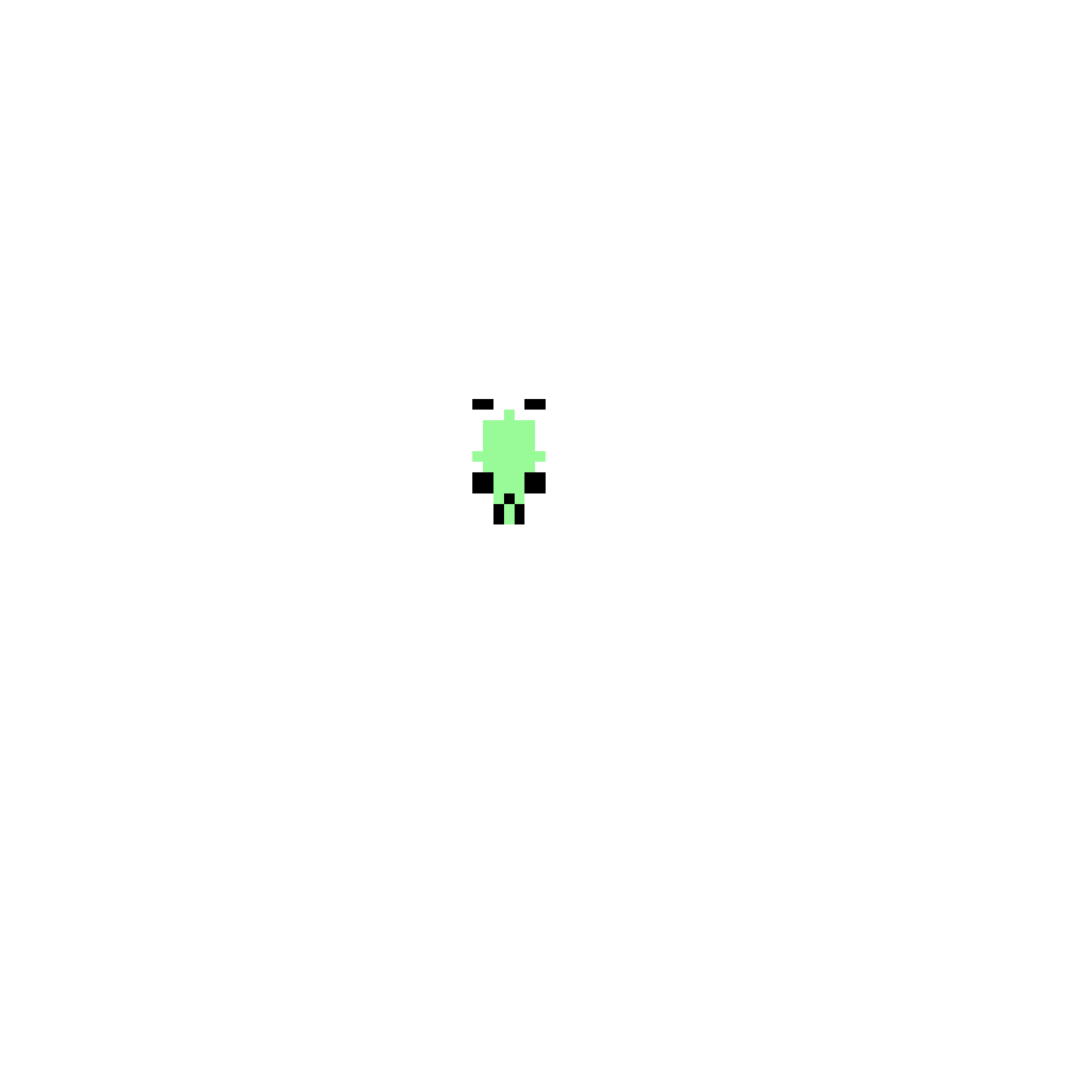}
\includegraphics[width=.07\linewidth,bb=177 191 212 270, clip=]{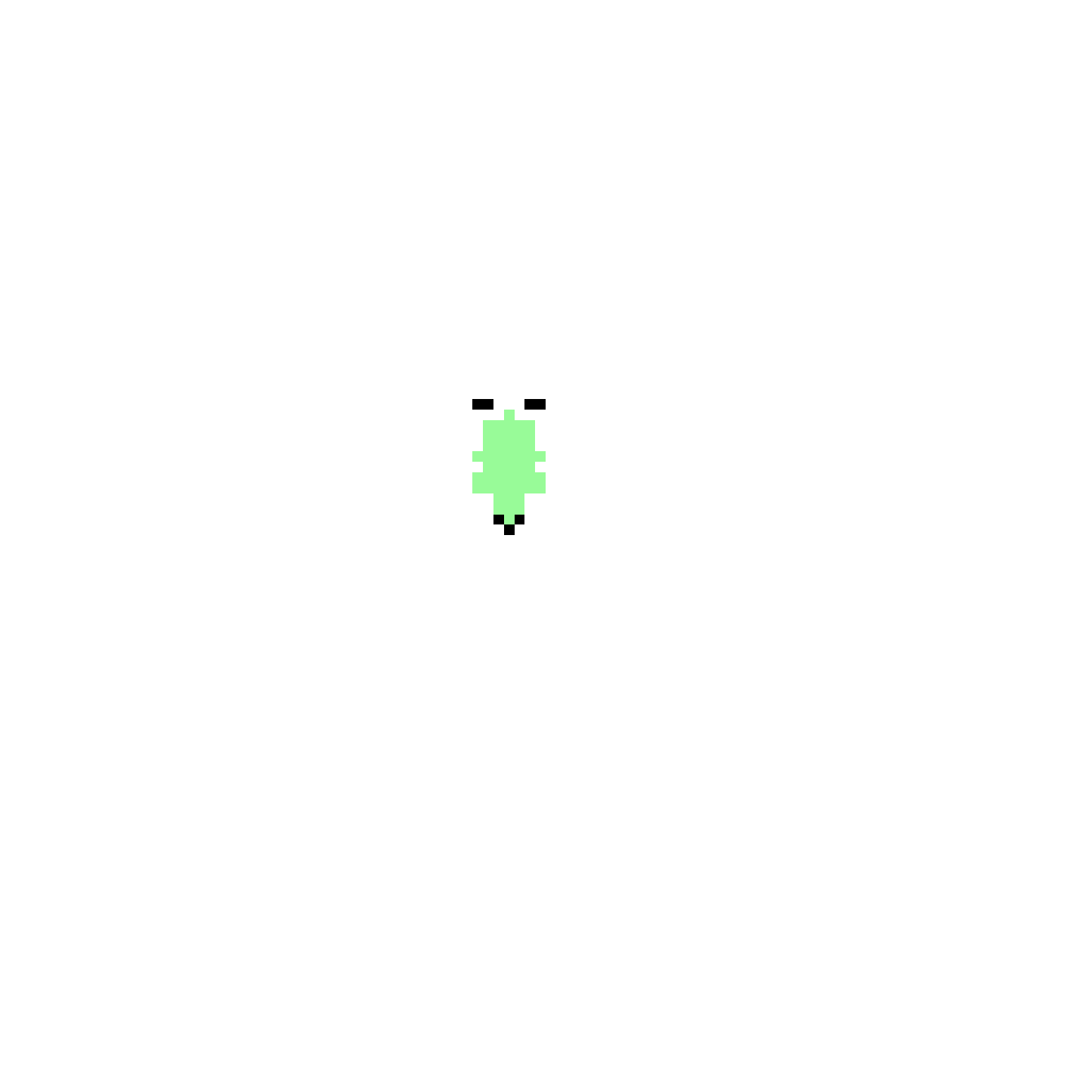}
\includegraphics[width=.07\linewidth,bb=177 191 212 270, clip=]{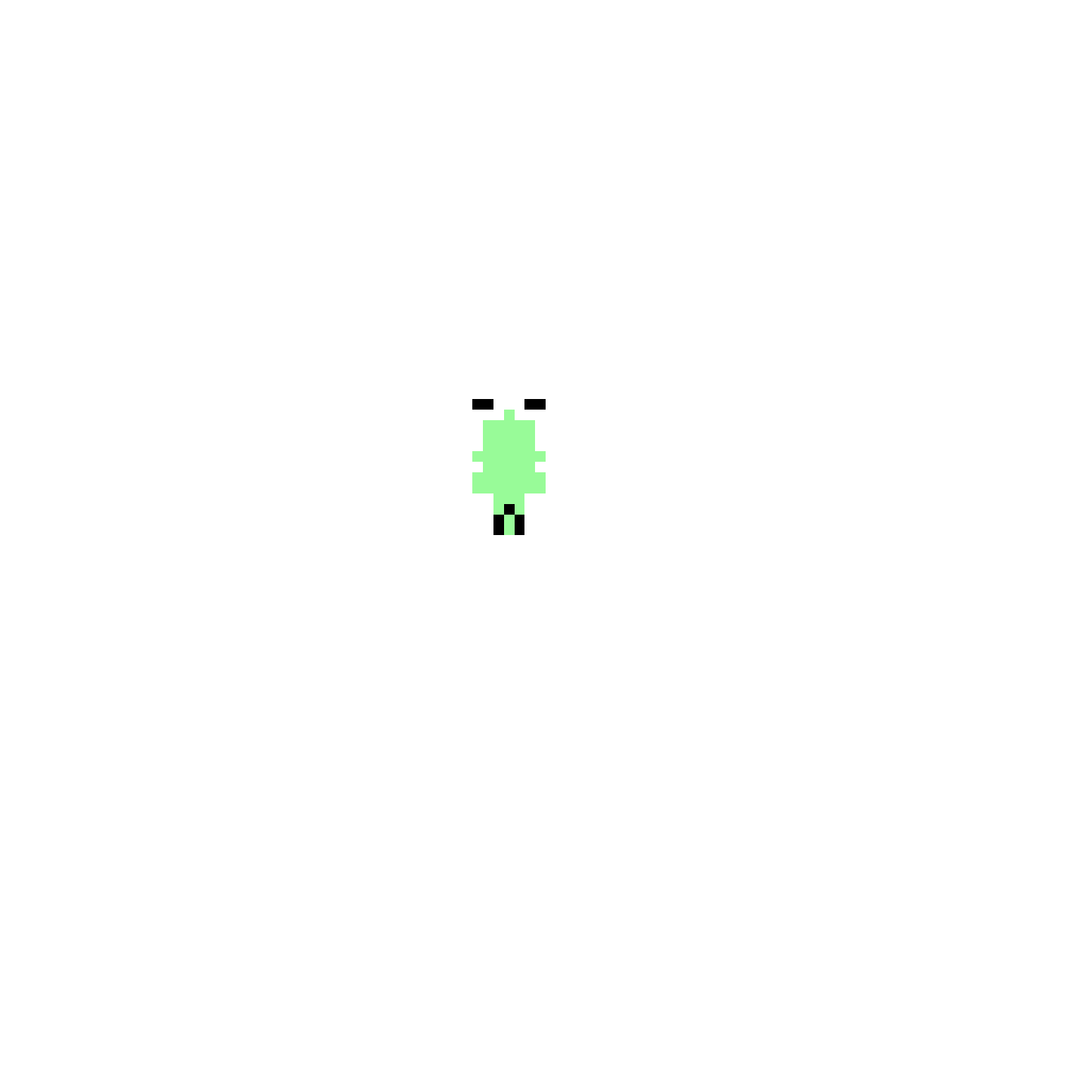}
\includegraphics[width=.07\linewidth,bb=177 191 212 270, clip=]{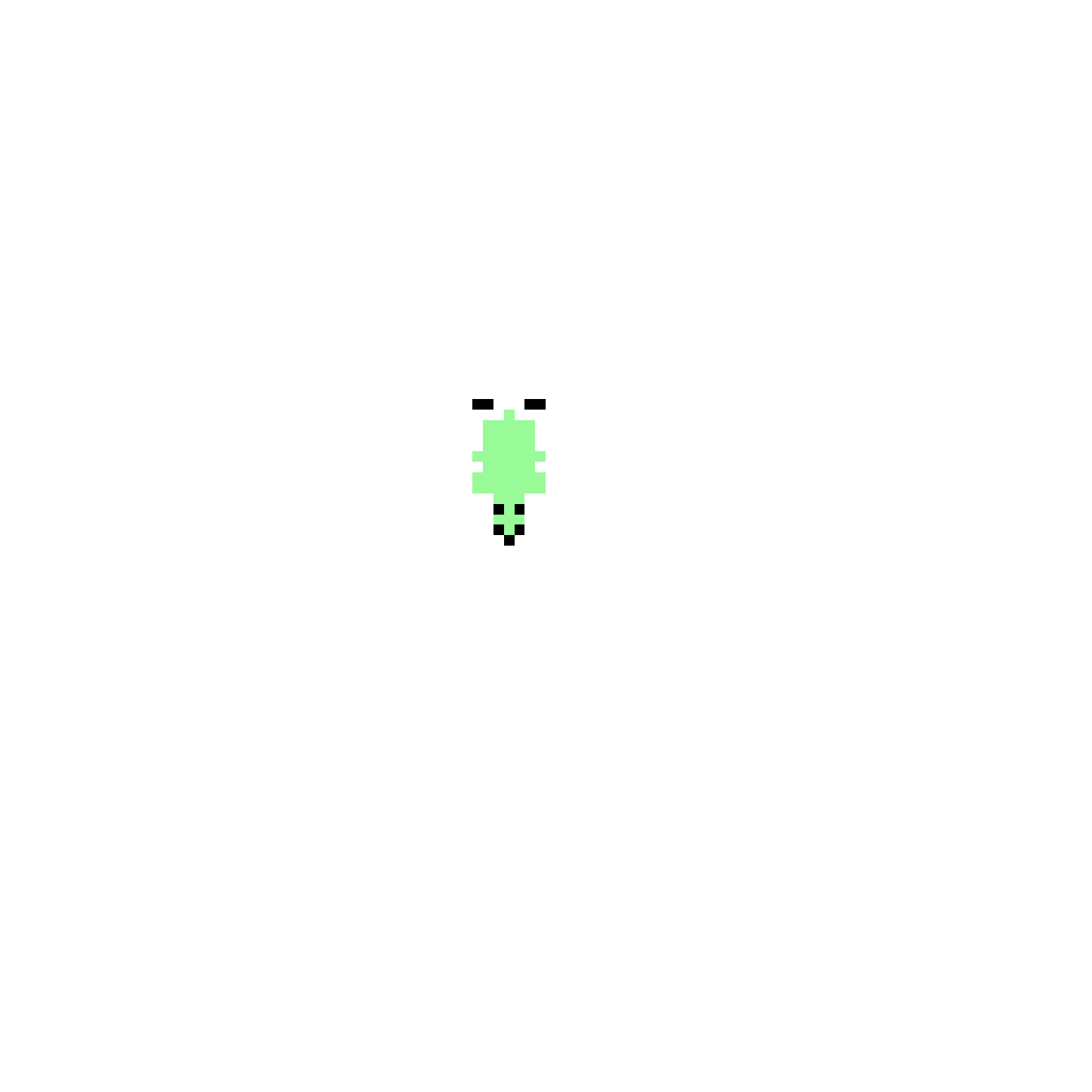}
\includegraphics[width=.07\linewidth,bb=177 191 212 270, clip=]{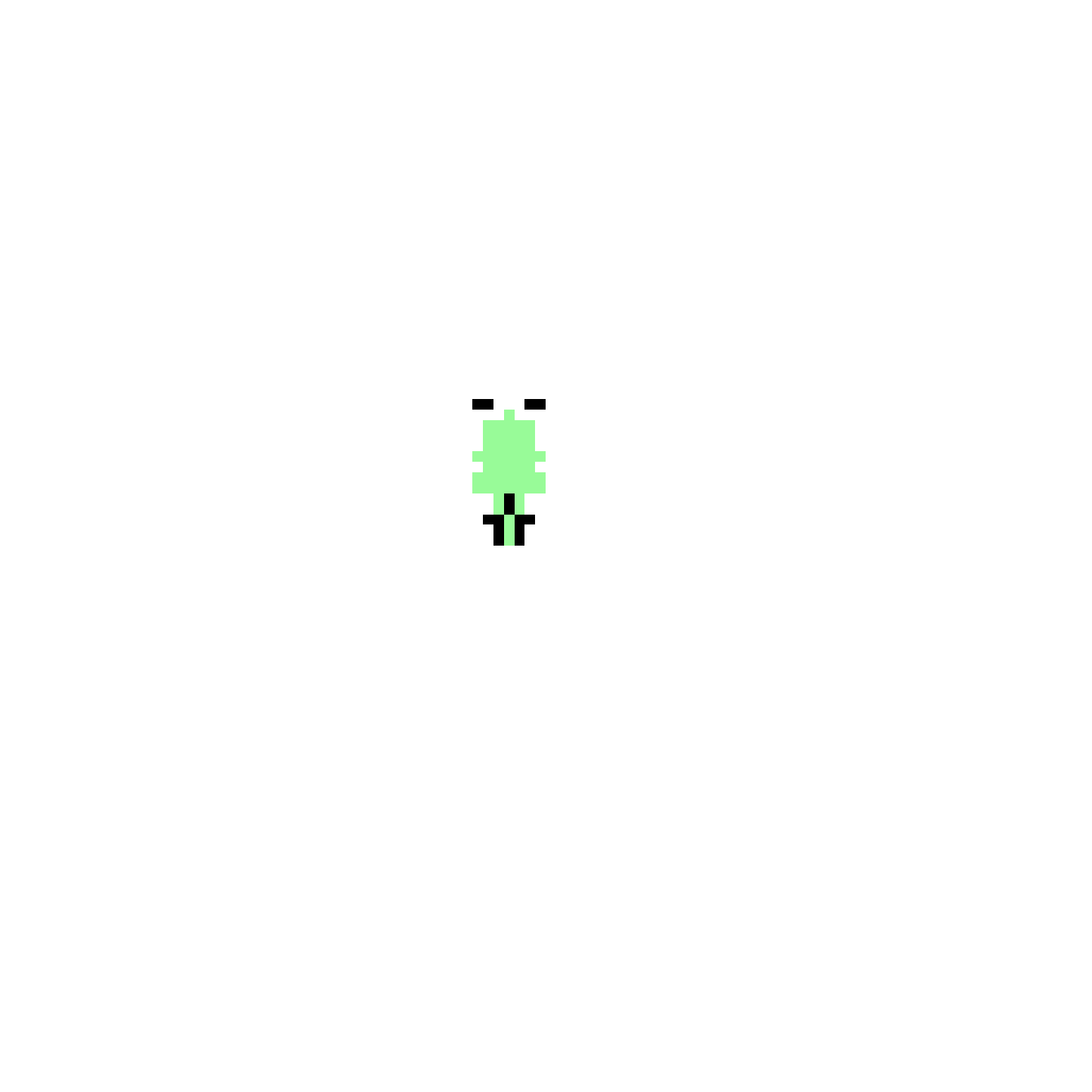}
\includegraphics[width=.07\linewidth,bb=177 191 212 270, clip=]{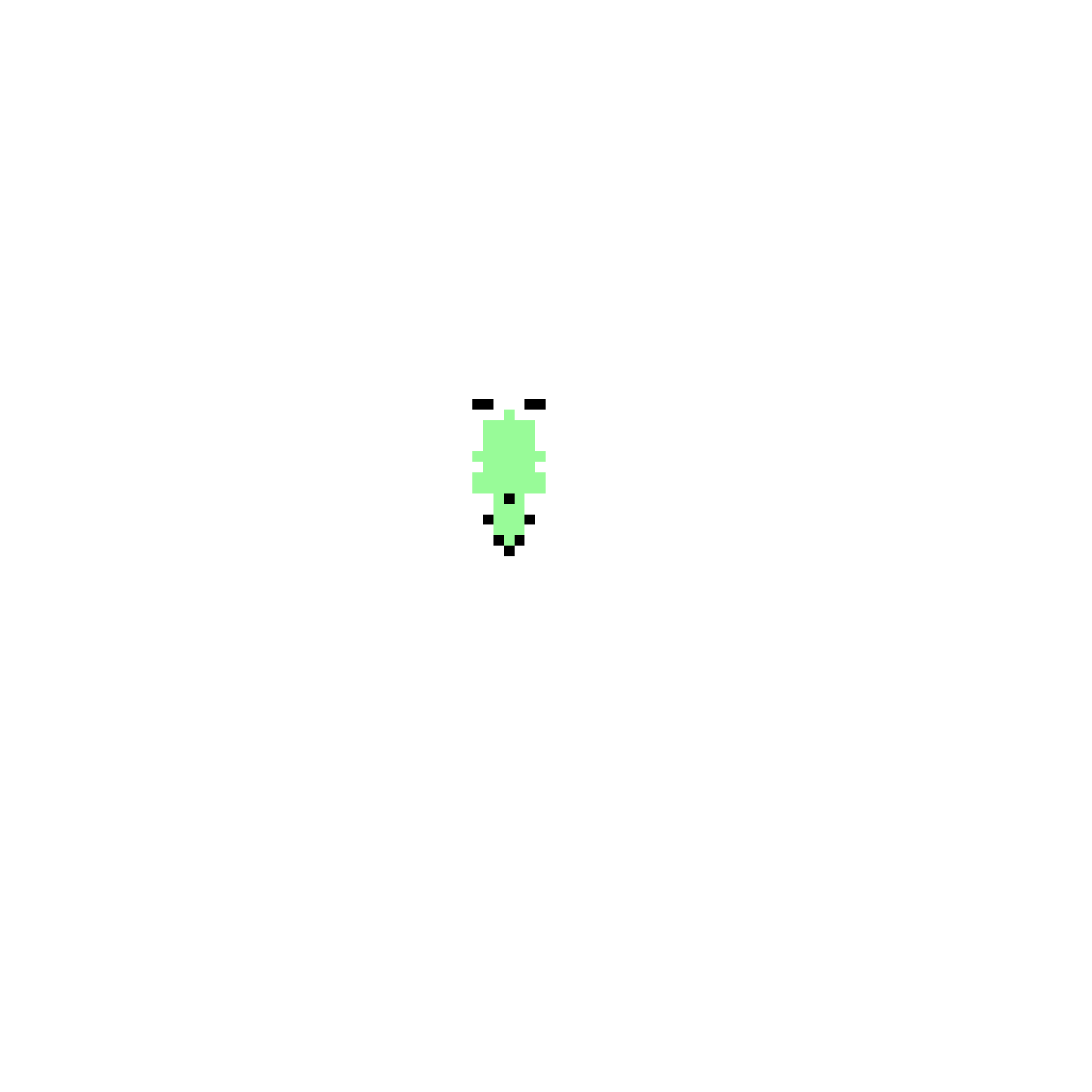}
\end{minipage}
\end{center}
\vspace{-5ex}
\caption[Gc reflection]
{\textsf{
A Gc glider bounces of a stable reflector, showing 26 consecutive time-steps.
The reflector can take up any of these shapes,
\raisebox{-4ex}
{\includegraphics[width=.13\linewidth]{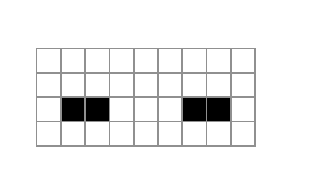}\hspace{-2ex}
\includegraphics[width=.13\linewidth]{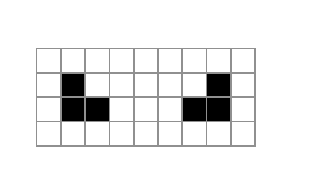}\hspace{-2ex}
\includegraphics[width=.13\linewidth]{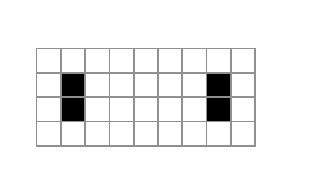}}\\[-3.5ex]
and the  corner shapes can be mixed,
\raisebox{-4ex}{\includegraphics[width=.13\linewidth]{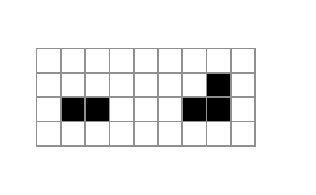}}
\label{Gc reflection}
}}
\end{figure}

\begin{figure}[htb]
\begin{center}
\begin{minipage}[c]{1\linewidth} 
\fbox{\includegraphics[height=.45\linewidth,bb=182 54 452 321, clip=]{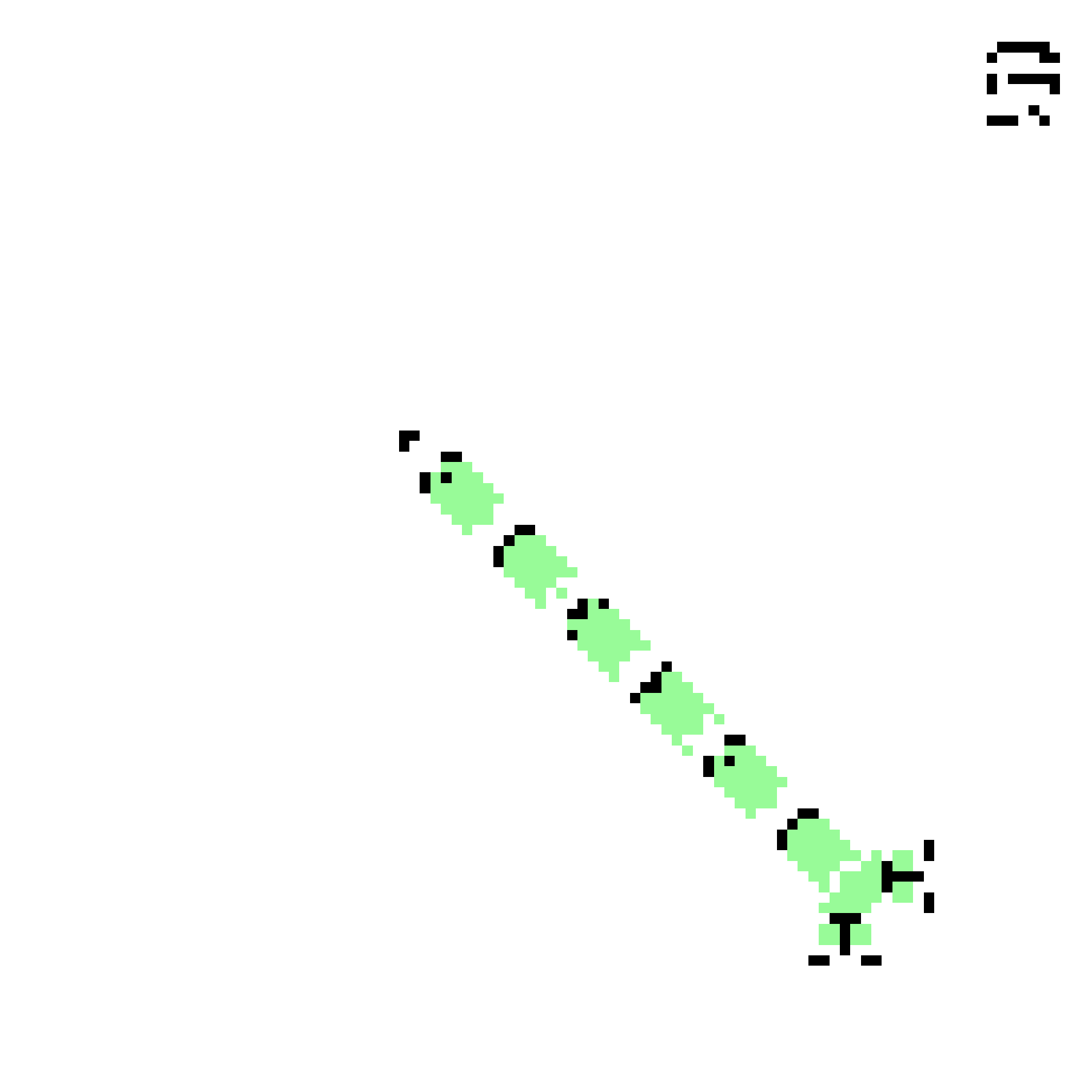}}
\hfill 
\fbox{\includegraphics[height=.45\linewidth,bb=24 36 255 249, clip=]{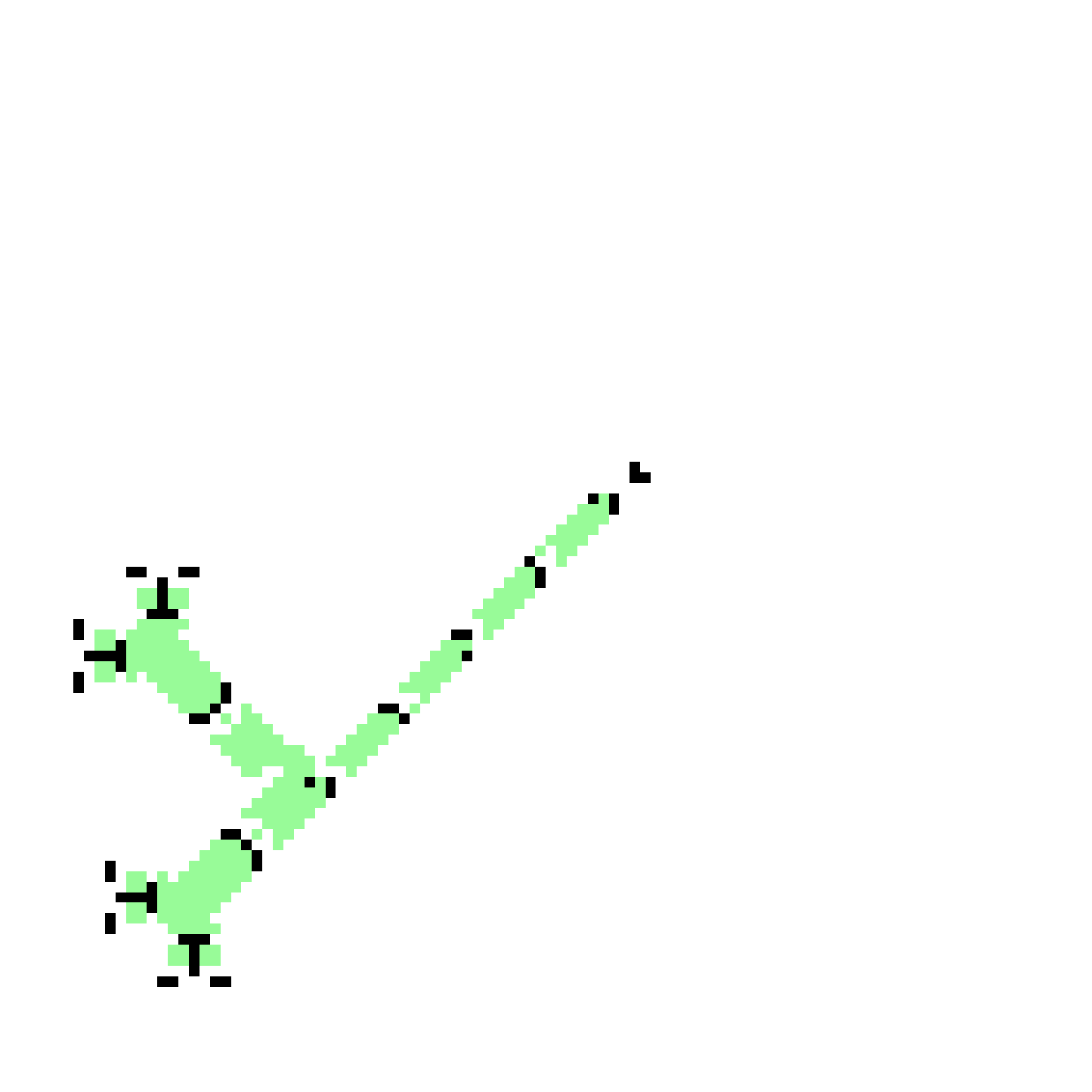}}
\end{minipage}
\end{center}
\vspace{-3ex}
\caption[GG2a reflector glider-guns]
{\textsf{
Using the the Gc reflection property in figure \ref{Gc reflection}, 
double Ga (G2a) gliders, with a frequency of 27 time-steps, 
can be shot by a GG2aR glider-gun constructed from two Gc reflectors
(same phase) correctly juxtaposed at $90^\circ$\cite{Wildmyron}.
$Left$: The glider stream is stopped by a G2a eater.
$Right$: Two GG2a glider-guns at $90^\circ$ create a Ga glider stream, stopped by an eater.
\label{GG2a reflector glider-guns}
}}
\end{figure}

A Gc glider is able to bounce off a stable reflector as in
figure~\ref{Gc reflection}. Two such reflectors correctly juxtaposed at $90^\circ$
create a G2a glider-gun with a frequency of 27 time-steps, and two of 
these correctly juxtaposed at $90^\circ$ build a Ga glider-gun 
(figure~\ref{GG2a reflector glider-guns}). The glider spacing is wider
than the glider-guns in sections~\ref{Glider-Guns from oscillator P22}
and \ref{Glider-Guns from oscillator P15}.\\

\clearpage

\subsection{Small oscillators}
\label{Small oscillators}

\begin{figure}[htb]
\begin{center}
\fbox{\includegraphics[width=.9\linewidth,bb=55 275 371 339, clip=]{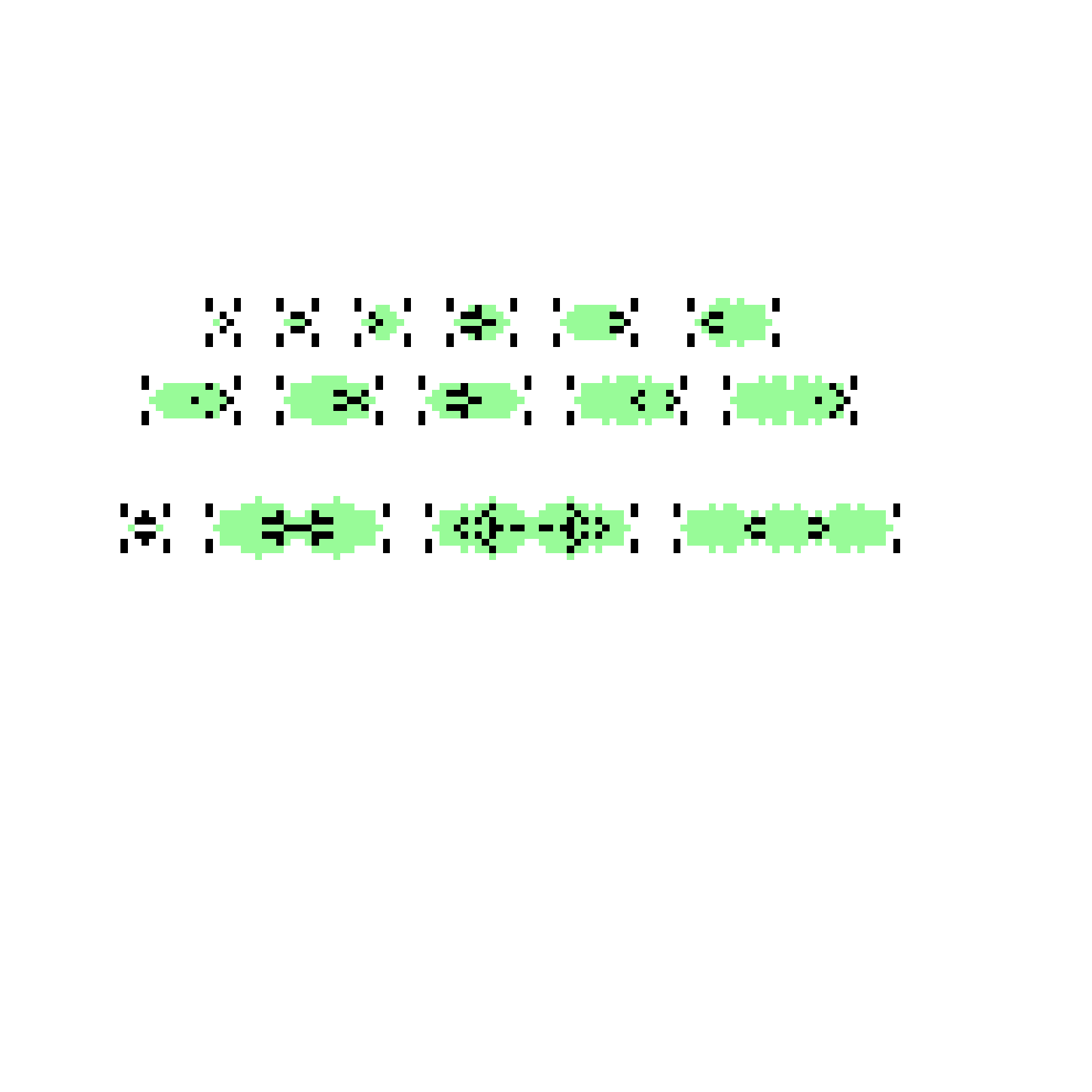}} 
\end{center}
\vspace{-2ex}
\caption[simple reflecting oscillators (SROs)]
{\textsf{Simple reflecting oscillators (SROs) --- 
a Gc glider bouncing between stable reflectors.
The period depends on the gap between reflectors.
These SROs are also present in the Precursor rule. 
\label{SROs}
}}  
\end{figure}

\begin{figure}[htb]
\hspace{-.05\linewidth}\begin{minipage}[c]{1.1\linewidth}
\fbox{\includegraphics[width=1\linewidth,bb= 64 236 634 314, clip=]{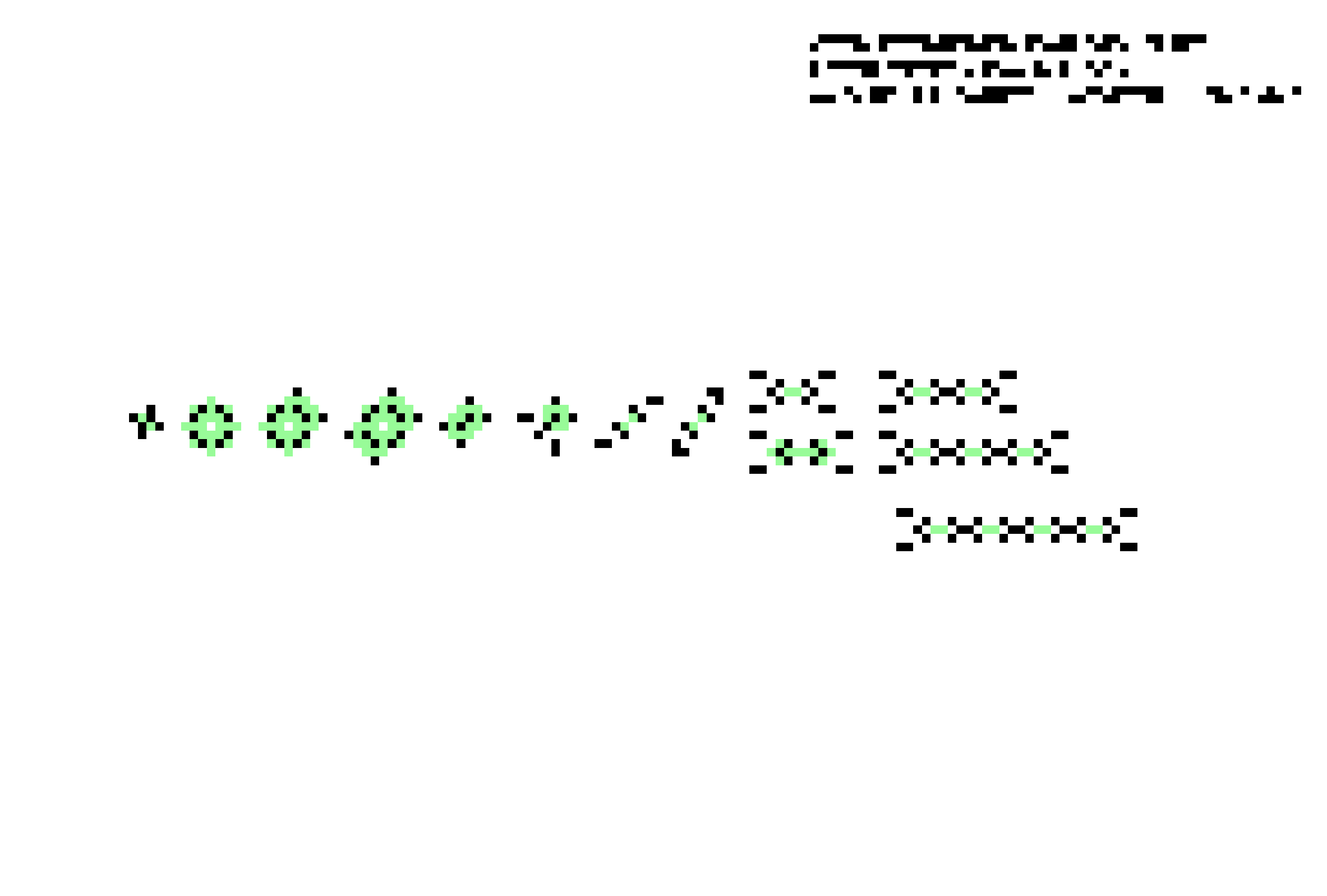}}\\ 
\textsf{P2\color{white}-----------------------------------------------------------------------------\color{black}$g$\color{white}-------------------\color{black}$h$}
\end{minipage}\\[2ex]
\begin{center}
\begin{minipage}[c]{.95\linewidth}
\begin{minipage}[c]{.35\linewidth}
\fbox{\includegraphics[width=1\linewidth,bb=60 40 193 94, clip=]{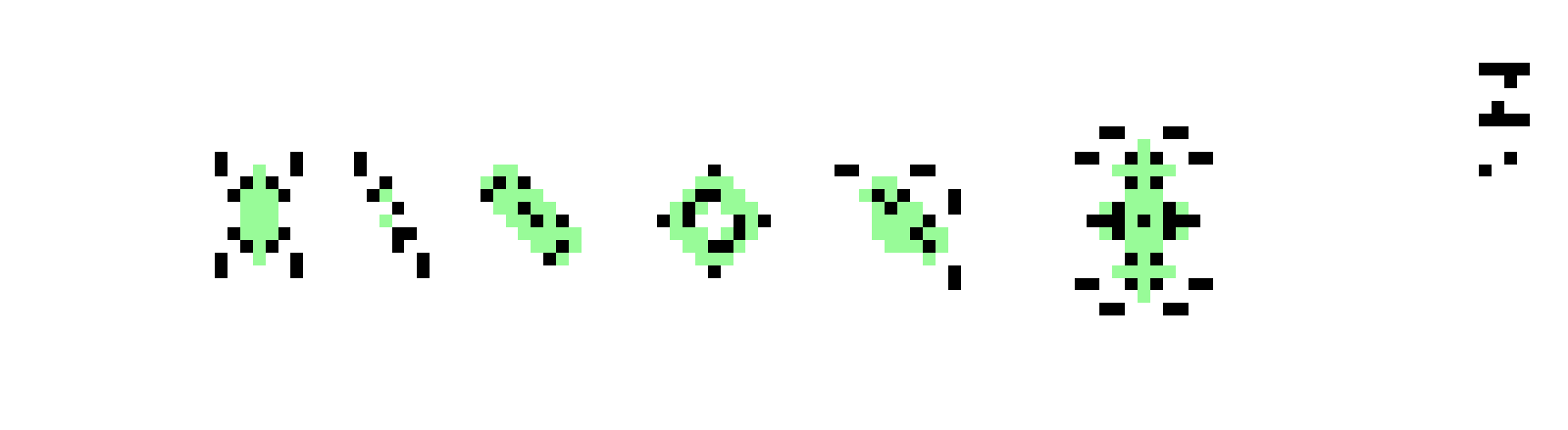}}\\ 
\textsf{P3}
\end{minipage}
\hfill
\begin{minipage}[c]{.13\linewidth}
\fbox{\includegraphics[width=1\linewidth,bb=201 42 251 88, clip=]{pdf-figs/oscP3-15}}\\ 
\textsf{P6}
\end{minipage}
\hfill
\begin{minipage}[c]{.13\linewidth}
\fbox{\includegraphics[width=1\linewidth,bb=259 37 309 89, clip=]{pdf-figs/oscP3-15}}\\ 
\textsf{P8}
\end{minipage}
\hfill
\begin{minipage}[c]{.13\linewidth}
\fbox{\includegraphics[width=1\linewidth,bb=336 31 391 100, clip=]{pdf-figs/oscP3-15}}\\ 
\textsf{P15}
\end{minipage}
\end{minipage}
\end{center}
\vspace{-3ex}
\caption[Period 2-15 oscillators]
{\textsf{$Top$: Oscillators with periods P2 ($g$ and $h$ are extendable).
$Below$: Oscillators with periods P3, P6, P8, and P15.
\label{Period 2-15 oscillators}
}}
\end{figure}

A variety of small oscillators exist in the Variant rule, some where the
period is related to the size of an extendable
pattern between reflectors with a bouncing interior.  
There are significant overlaps with small oscillators in the
Precursor rule\cite{Gomez2017}. Figures \ref{SROs}, \ref{ETOs}
and \ref{Period 2-15 oscillators} give examples.\\

\begin{figure}[htb]
\begin{center}
\fbox{\includegraphics[width=.9\linewidth,bb=48 223 387 252, clip=]{pdf-figs/oscEX}}
\end{center}
\vspace{-2ex}
\caption[Extendable trapped oscillators (ETOs)]
{\textsf{Examples of extendable trapped oscillators (ETOs) between reflectors
where the dynamics is more complicated than simple bouncing.
The periods are 2, 43, 46 and 50 respectively.
\label{ETOs}
}}  
\end{figure}

\section{Glider stream circuits}
\label{Glider stream circuits}

\begin{figure}[htb]
\begin{center}
\begin{minipage}[c]{.8\linewidth} 
\fbox{\includegraphics[width=1\linewidth,bb=43 108 353 282, clip=]{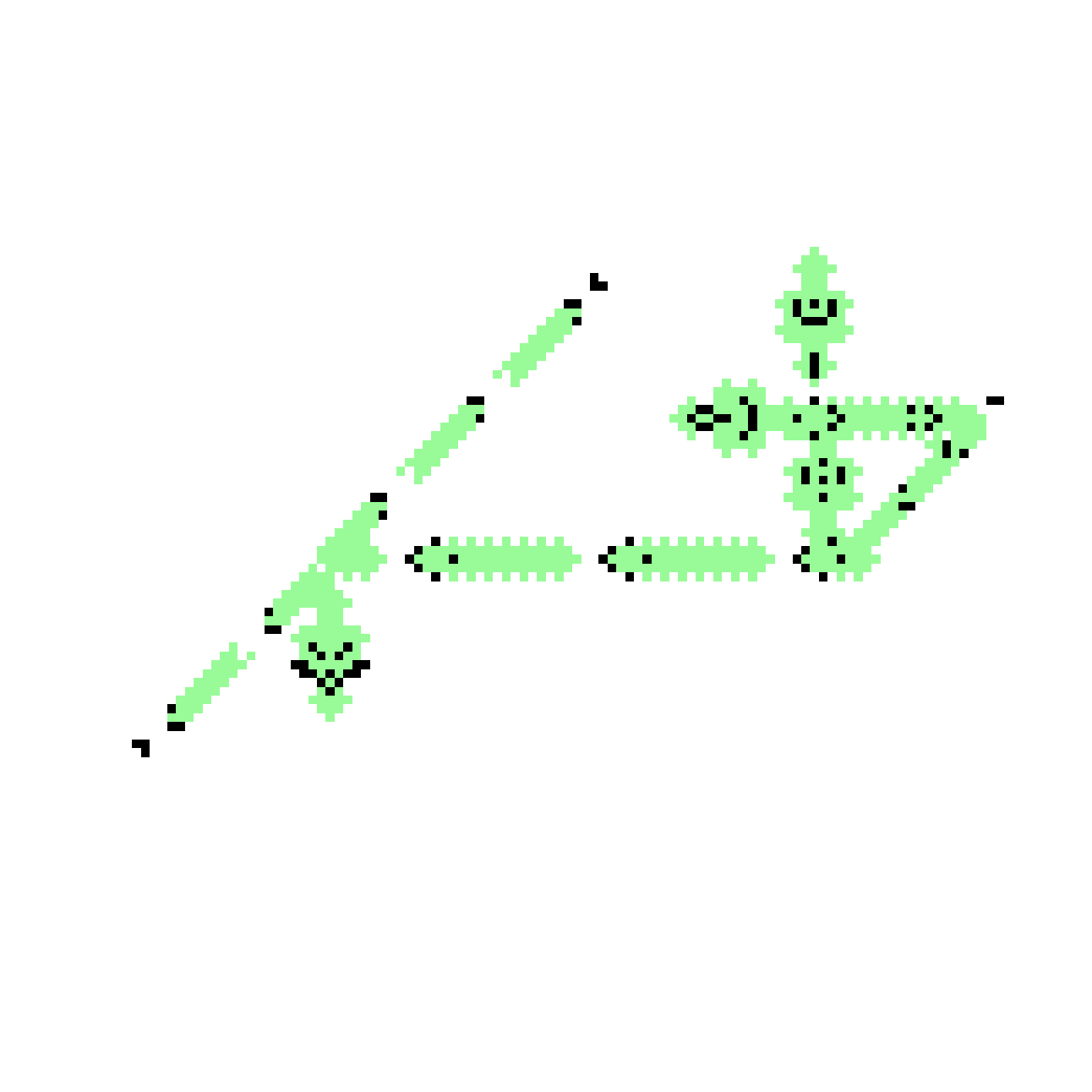}}
\end{minipage}
\end{center}
\vspace{-3ex}
\caption[glider stream circuit double spaced Ga-GS]
{\textsf{
A glider stream (GS) circuit where a Gc-GS is transformed to double spaced Ga-GS
by interaction with a block and two P22 oscillators.
The circuit sub-components: 1) GGc glider-gun, 2) Gc-GS collision with a block, 
3) transform to Ga-GS, interaction with P22 oscillator, 
4) transform to double spaced Gc-GS by eliminating
alternate gliders, 5) interaction with P22 oscillator to create 
two doubly spaced Ga-GS, stopped by eaters.
\label{GcP22Ga}
}}
\end{figure}

\begin{figure}[htb]
\begin{center}
\begin{minipage}[c]{.77\linewidth} 
\fbox{\includegraphics[width=1\linewidth,bb=23 138 469 355, clip=]{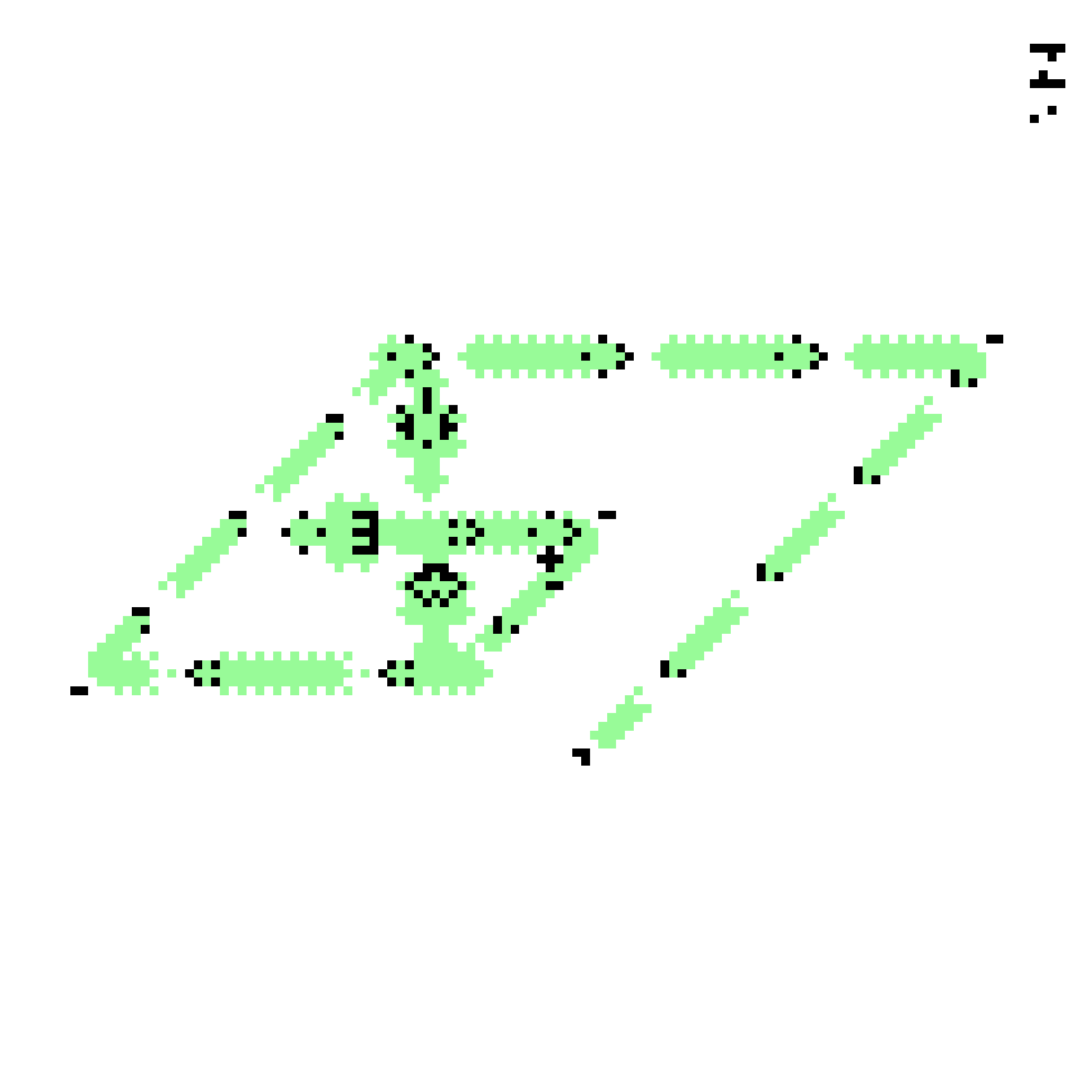}}
\end{minipage}
\end{center}
\vspace{-3ex}
\caption[glider stream circuit spiral]
{\textsf{
A glider stream (GS) circuit folds into a spiral, where a Gc-GS is transformed to double spaced Ga-GS
by interaction with two P22 oscillators and three blocks.
The circuit repeats steps 1) to 4),  then 5) collision with a block, 6) transform to Ga-GS, 
7) interaction with P22 oscillator, 8) transform to Gc-GS, 9)~collision with a block,
10) transform to Ga-GS, stopped by an eater.
\label{spiral1}
}}
\end{figure}

The various glider-guns already described can themselves become
sub-components, that together with eaters, collsions, blocks,
reflectors and oscillators, can build super-glider-guns,
super-oscillators, and glider stream circuits of ever increasing complexity 
Figures~\ref{GcP22Ga} and \ref{spiral1} give examples.

\enlargethispage{4ex}
\clearpage

\section{Variable period glider-gun}
\label{Variable period glider-guns}

\begin{figure}[htb]
\begin{center}
\begin{minipage}[c]{1\linewidth}

\begin{minipage}[c]{.37\linewidth} 
\fbox{\includegraphics[width=1\linewidth,bb=52 167 217 252, clip=]{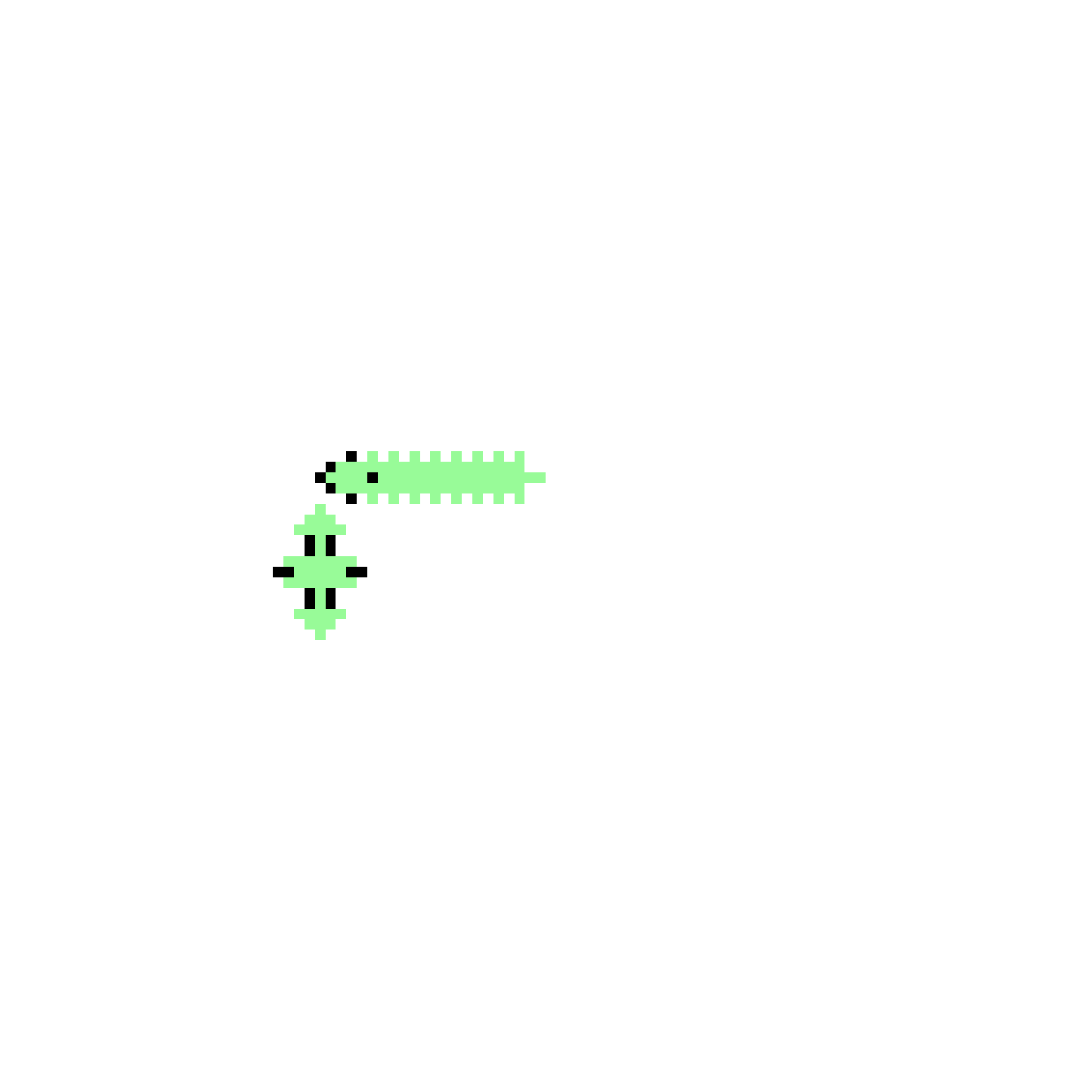}}
\end{minipage}
\hfill
\begin{minipage}[c]{.15\linewidth}
\begin{center}
\includegraphics[width=.6\linewidth,bb=10 9 32 26, clip=]{pdf-figs/ArrowE}\\
\textsf{\small{Gc$\rightarrow$P15$\rightarrow$2$\times$Gc}}
\end{center}
\end{minipage}
\hfill
\begin{minipage}[c]{.37\linewidth} 
\fbox{\includegraphics[width=1\linewidth,bb=52 167 217 252, clip=]{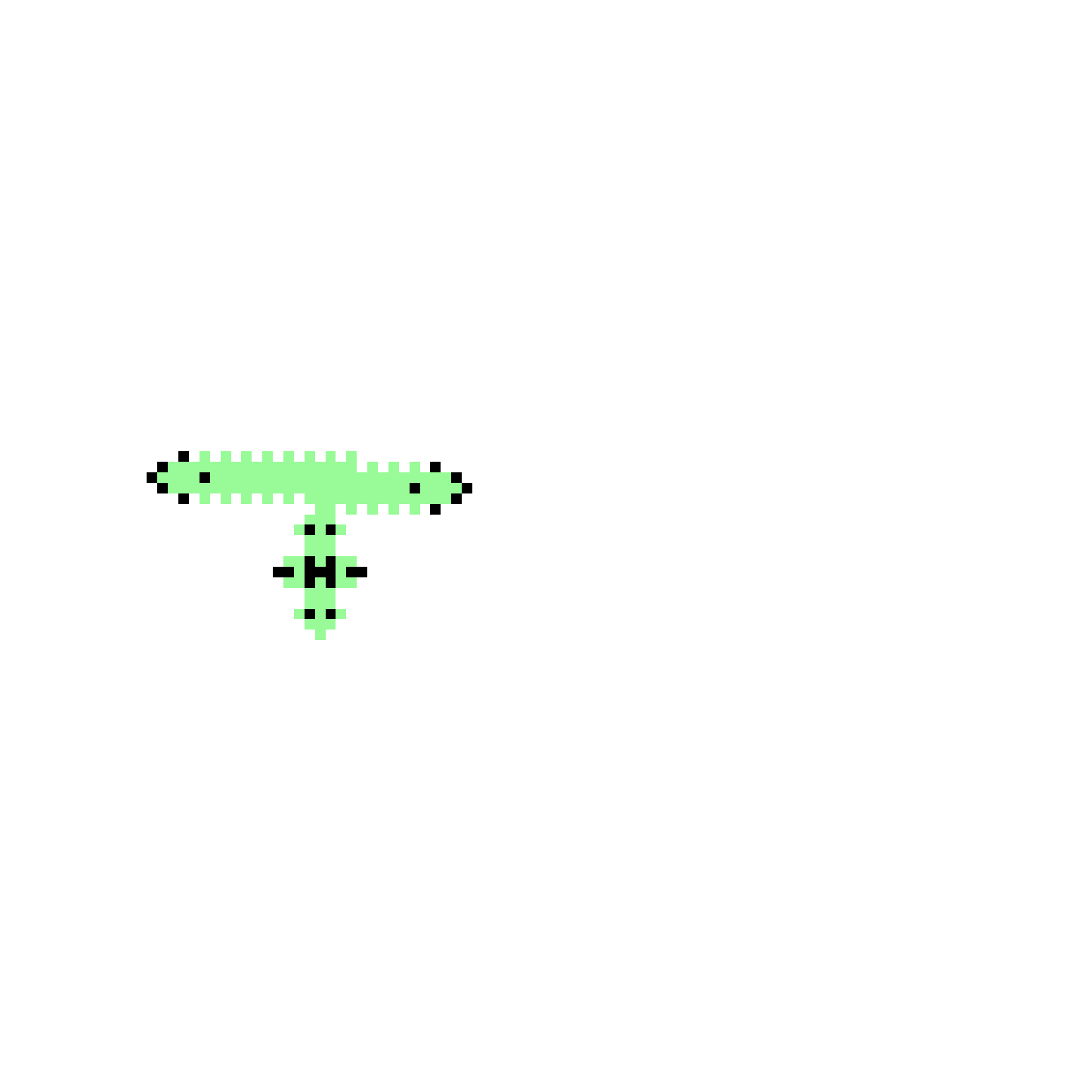}}
\end{minipage}

\end{minipage}
\end{center}
\caption[Gc P15 to 2xGc]
{\textsf{
A Gc glider precisely brushes past a P12 oscillator resulting in two Gc gliders
in opposite directions. In the sequence above, the Westward glider continues, the
new Eastward glider is displaced South by one cell.
\label{Gc P15 to 2xGc}
}}q
\end{figure}

\begin{figure}[htb]
\begin{center}
\begin{minipage}[c]{1\linewidth} 
\fbox{\includegraphics[width=1\linewidth,bb=41 128 427 230, clip=]{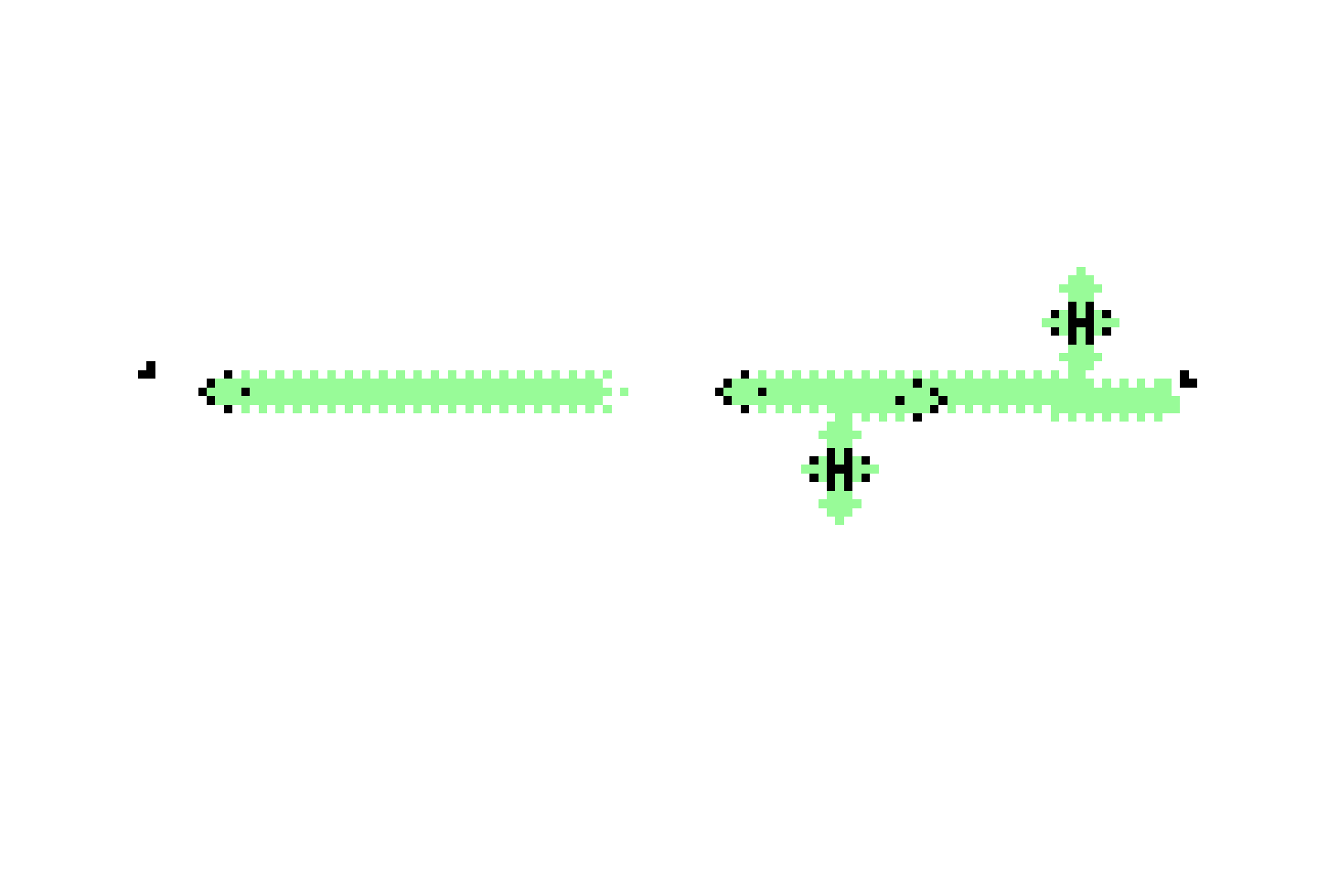}}
\end{minipage}
\end{center}
\vspace{-3ex}
\caption[GGcV27 glider-gun]
{\textsf{
A GGcV27 variable glider-gun constructed from two P15 oscillators separated by 27 time-steps,
shooting gliders with a frequency of 120 time-steps. Increasing the separation by
30 increases the frequency by 120. The Gc glider streams are stopped by eaters.
\label{GGcV27 glider-gun}
}}
\end{figure}

Variant-rule features an interesting collision where a Gc glider brushes past
a P15 oscillator resulting in two Gc gliders moving in opposite directions
(figure~\ref{Gc P15 to 2xGc}),
from which a variable period Gc glider-gun (GGcV) is built by introducing
a second P15 separated from the first by 27 cells\cite{BlinkerSpawn} 
(figure~\ref{GGcV27 glider-gun}). 
Gc gliders are shot in opposite
directions at a frequency of 120 time-steps, and can be stopped by eaters.
The P15 separation (S) can be increases by intervals of 30, which increases the period (P)
by intervals of 120, giving S/P of 57/240, 87/360, and so on.

\begin{figure}[htb]
\begin{center}
\begin{minipage}[c]{.6\linewidth} 
\fbox{\includegraphics[width=1\linewidth,bb=21 5 304 305, clip=]{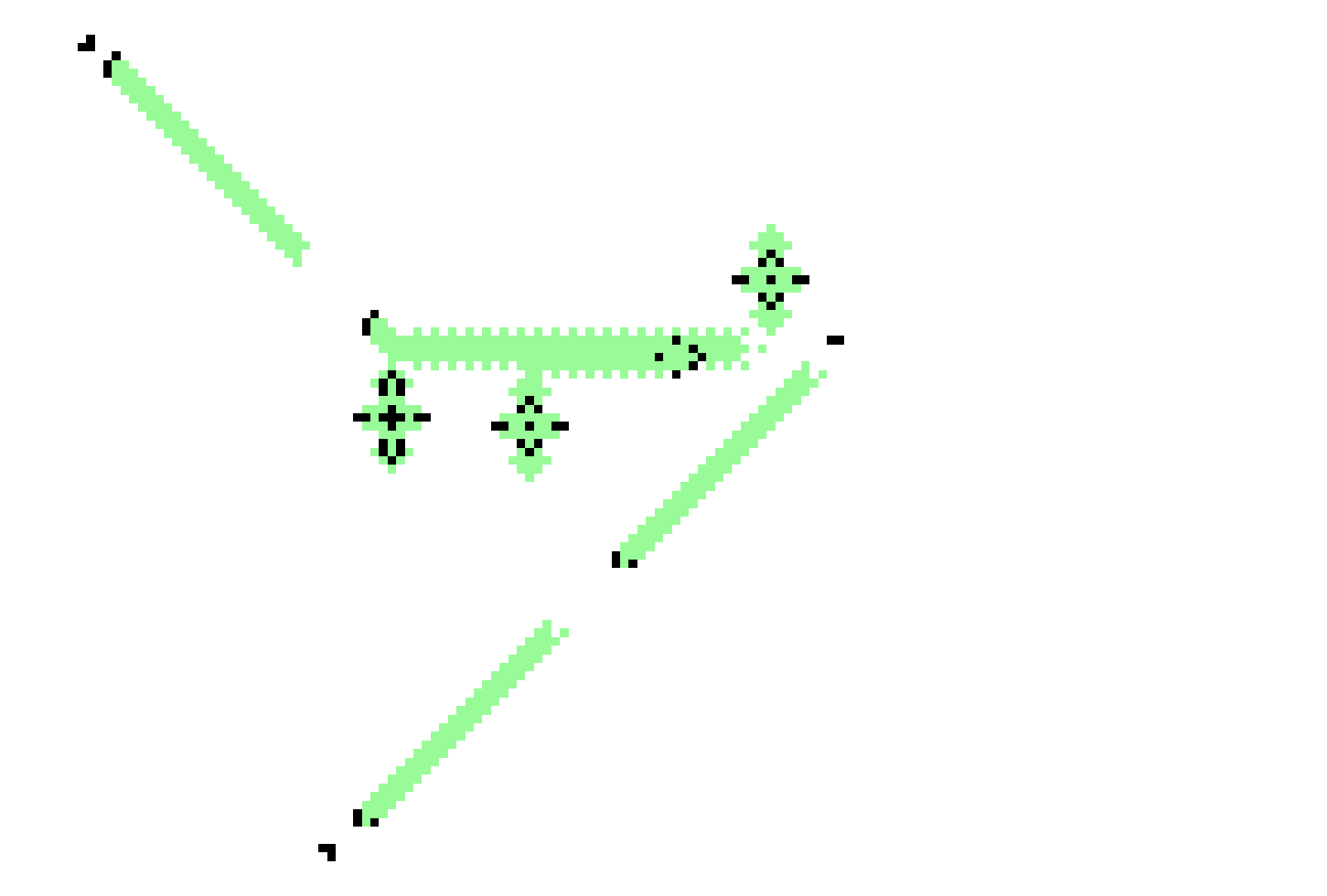}}
\end{minipage}
\end{center}
\vspace{-3ex}
\caption[GGaV27 glider-gun]
{\textsf{
A GGaV27 variable glider-gun made from a modified GGcV27 in figure~\ref{spiral1}
by introducing two transformations from Gc to Ga --- interaction with
a P15 oscillator, and collision with a stable block. 
The Ga glider streams are stopped by eaters.
\label{GGaV27 glider-gun}
}}
\end{figure}

The Gc glider stream can be transformed to a Ga glider stream
by an appropriate block or another P15 oscillator as in figure~\ref{GGaV27 glider-gun}, 
making a variable Ga glider-gun (GGaV), with the same variability as GGcV.
 
Its easy see that other permutations are possible by adapting the same
mechanism, for example a combined Ga and Gc glider-gun.\\

\section{Logical Universality}
\label{Logical Universality}

Post's Functional Completeness Theorem\cite{Post41,Francis90} established 
a disjunctive (or conjunctive) normal form formula using the
logical gates NOT, AND and OR to satisfy negation, conjunction and disjunction,
and we apply the term ``logical universality'' to a CA if these gates can be demonstrated.

However, for a CA to be universal in the full sense according to Conway\cite{Berlekamp1982},
two further conditions (1 and 2 below), are required, giving a full list as follows,

\begin{s-enumerate}
\item Data storage or memory.
\item Data transmission requiring the equivalent of wires and an internal clock.
\item Data processing requiring a universal set of logic gates NOT, AND, OR.
\end{s-enumerate} 

The Variant rule probably has a sufficient variety of logical components to
establish all three conditions following similar methods for
the \mbox{Game-of-Life\cite{{Berlekamp1982,Gardner1970}}}, but we will
postpone that investigation
and confine our demonstration to item 3, logical universality only.
To achieve this we will need the following basic ingredients\cite{Gomez2017}:

\begin{s-enumerate}

\item A glider-gun or ``pulse generator'', sending a stream of gliders
  into space.  So far 11 glider-guns have been discovered in the
  Variant-rule, shooting Gc, Ga and G2a gliders, all moving in 4
  phases.

\begin{s-enumerate}

\item Three GGc glider-guns (figures
\ref{GGc and GGa glider-guns}$L$, \ref{GcP22Ga}, \ref{GGcV27 glider-gun}).

\item Six GGa glider-guns (figures 
\ref{GGc and GGa glider-guns}$R$, 
\ref{GGc x GGc to GGa glider-gun},
\ref{GG2a glider-guns}$R$,
\ref{GcP22Ga},
\ref{spiral1},
\ref{GGaV27 glider-gun}).

\item Two GG2a glider-guns (figures
\ref{GG2a glider-guns}, \ref{GG2a reflector glider-guns}).
\end{s-enumerate}

\item A stable eater, based on a block, oscillator or another glider-gun.
The eater must destroy each incoming glider and
survive the collision to destroy the next, so capable of stopping a glider stream.

\item Complete self-destruction when two gliders collide at an angle. 
Any debris must quickly dissipate, and the gap between gliders must be sufficient
so as not to interfere with the next incoming glider.

\end{s-enumerate}

Both GGa and GGc glider-guns meet these conditions, and possibly any of the
other glider-guns. In sections \ref{GGa NOT, AND, OR}
and \ref{GGc NOT, AND, OR}
we demonstrate the logical gates NOT, AND, OR for GGa and GGc,
following Conway's method\cite{Berlekamp1982} where a data stream of 1s/0s
is implemented by gliders/gaps. Here the gaps are marked as grey discs.

\subsection{GGa NOT, AND, OR}
\label{GGa NOT, AND, OR}

\enlargethispage{5ex}
\begin{figure}[htb]
\begin{center} 
\begin{minipage}[c]{.95\linewidth}
\fbox{\includegraphics[height=.68\linewidth,bb=13 7 149 275, clip=]{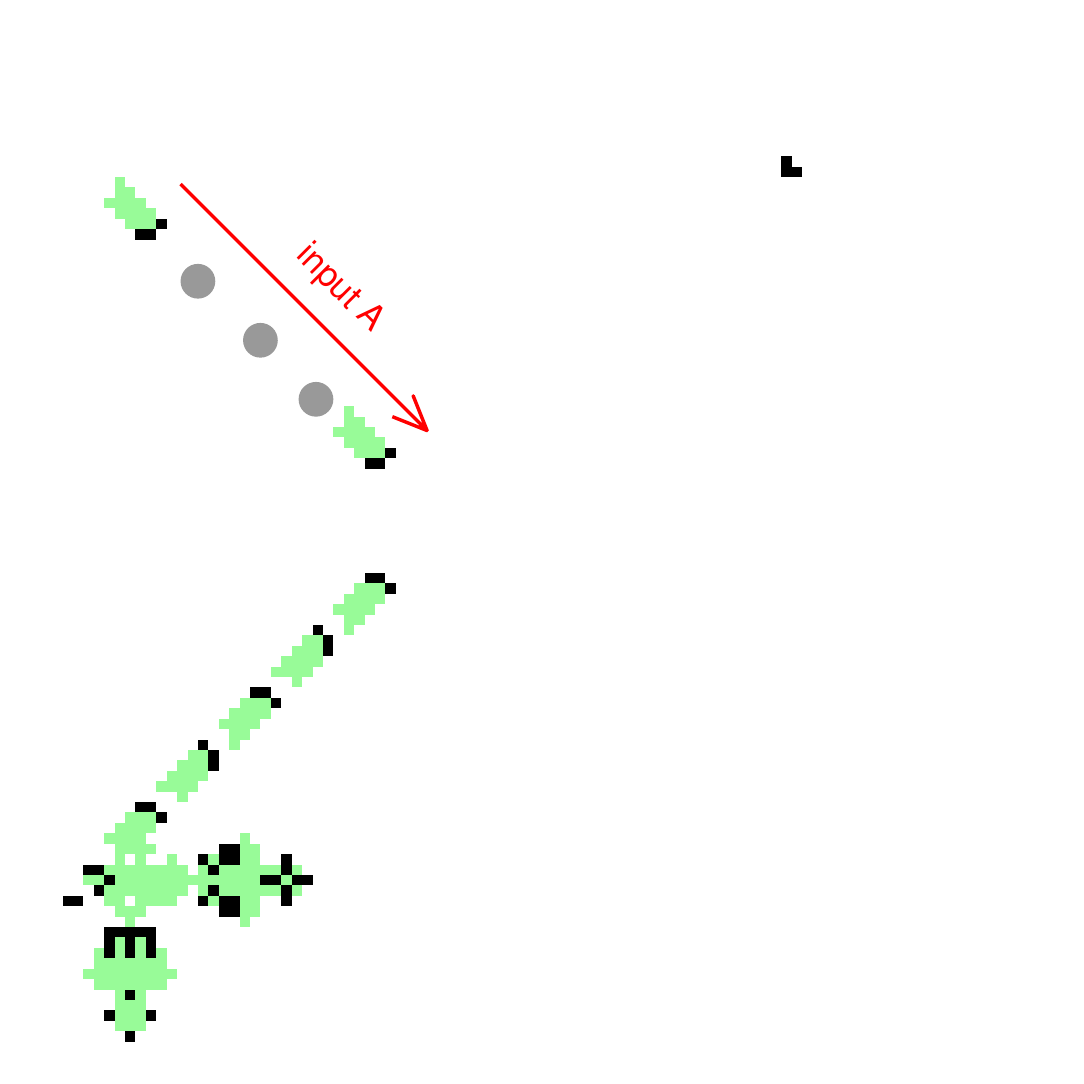}}
\hfill
\fbox{\includegraphics[height=.68\linewidth,bb=13 7 245 275, clip=]{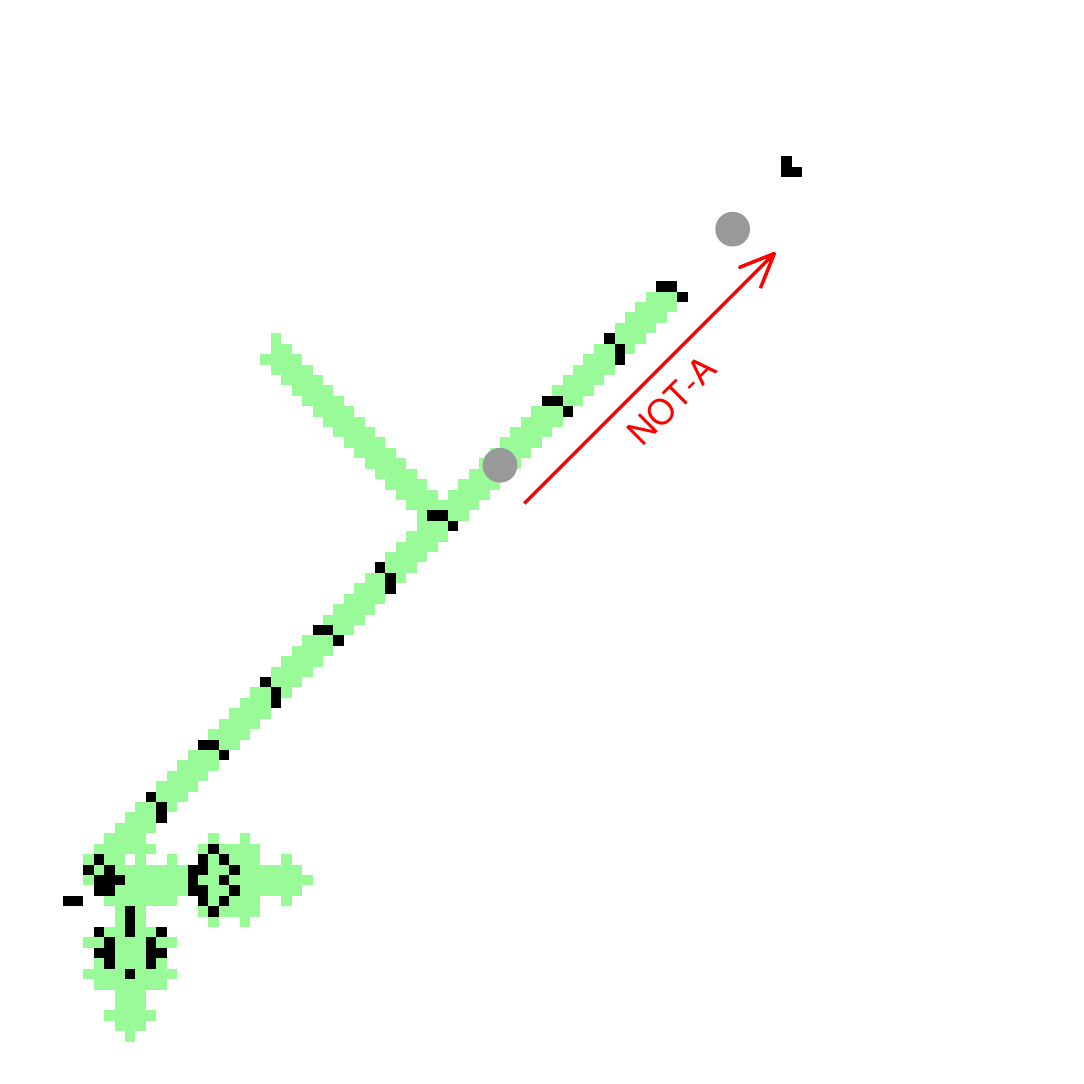}}
\end{minipage}
\end{center}
\vspace{-3ex}
\caption[GGa NOT gate]
{\textsf{
An example of the GGa NOT gate: ($\neg 1, 1 \rightarrow$ 0 and $0
    \rightarrow$ 1) or inverter, which transforms a stream of data to
    its complement, represented by gliders and gaps (grey discs). 
$Left$: The 5-bit input string A (10001) moving SE is about to interact 
with a GGa glider-stream moving NE. $Right$: The outcome is NOT-A (01110) moving
NE, shown after 134 time-steps.
\label{GGa NOT gate}
}}
\end{figure}

\begin{figure}[htb]
\begin{center} 
\begin{minipage}[c]{1\linewidth}
\fbox{\includegraphics[height=.83\linewidth,bb=30 27 160 371, clip=]{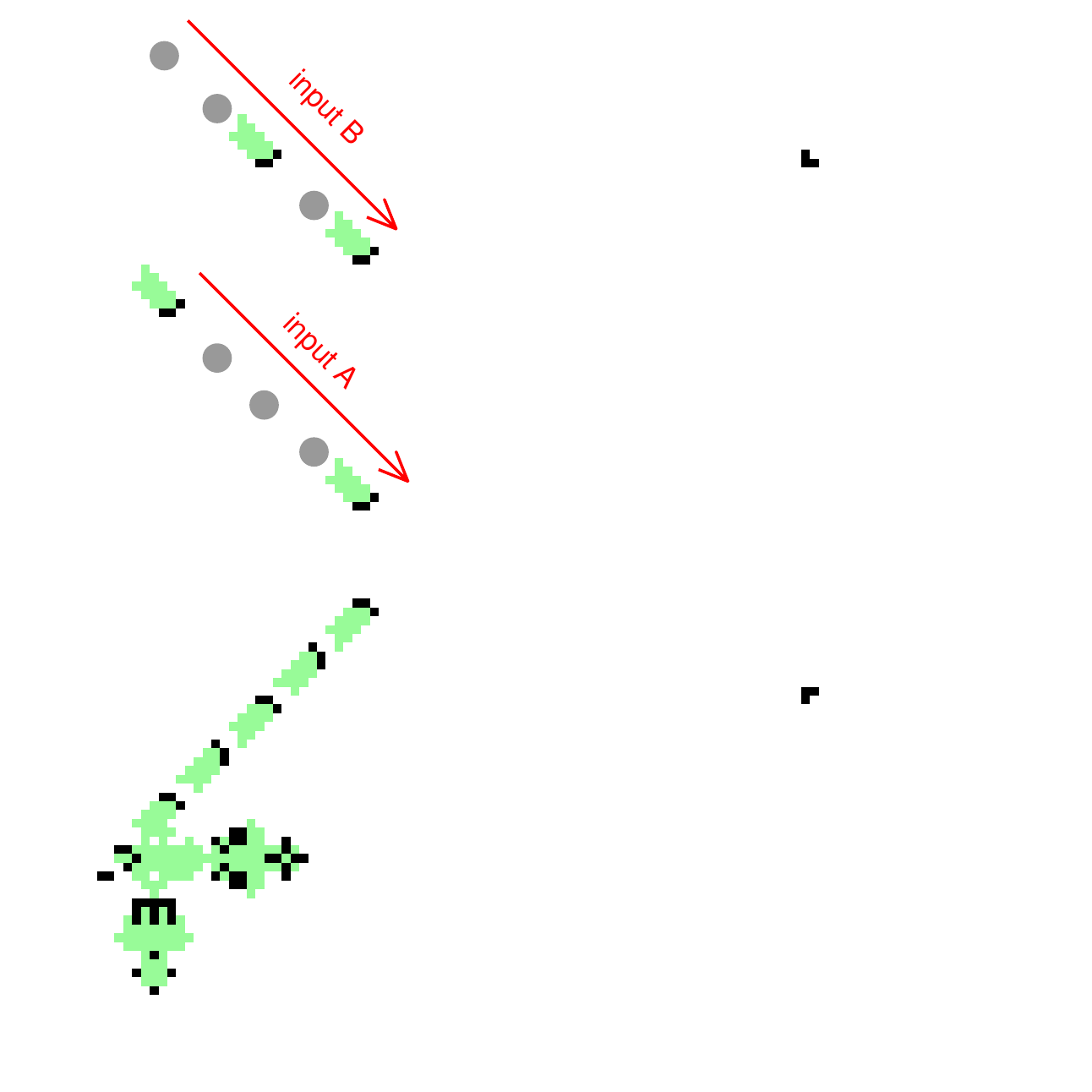}}
\hfill
\fbox{\includegraphics[height=.83\linewidth,bb=30 27 290 371, clip=]{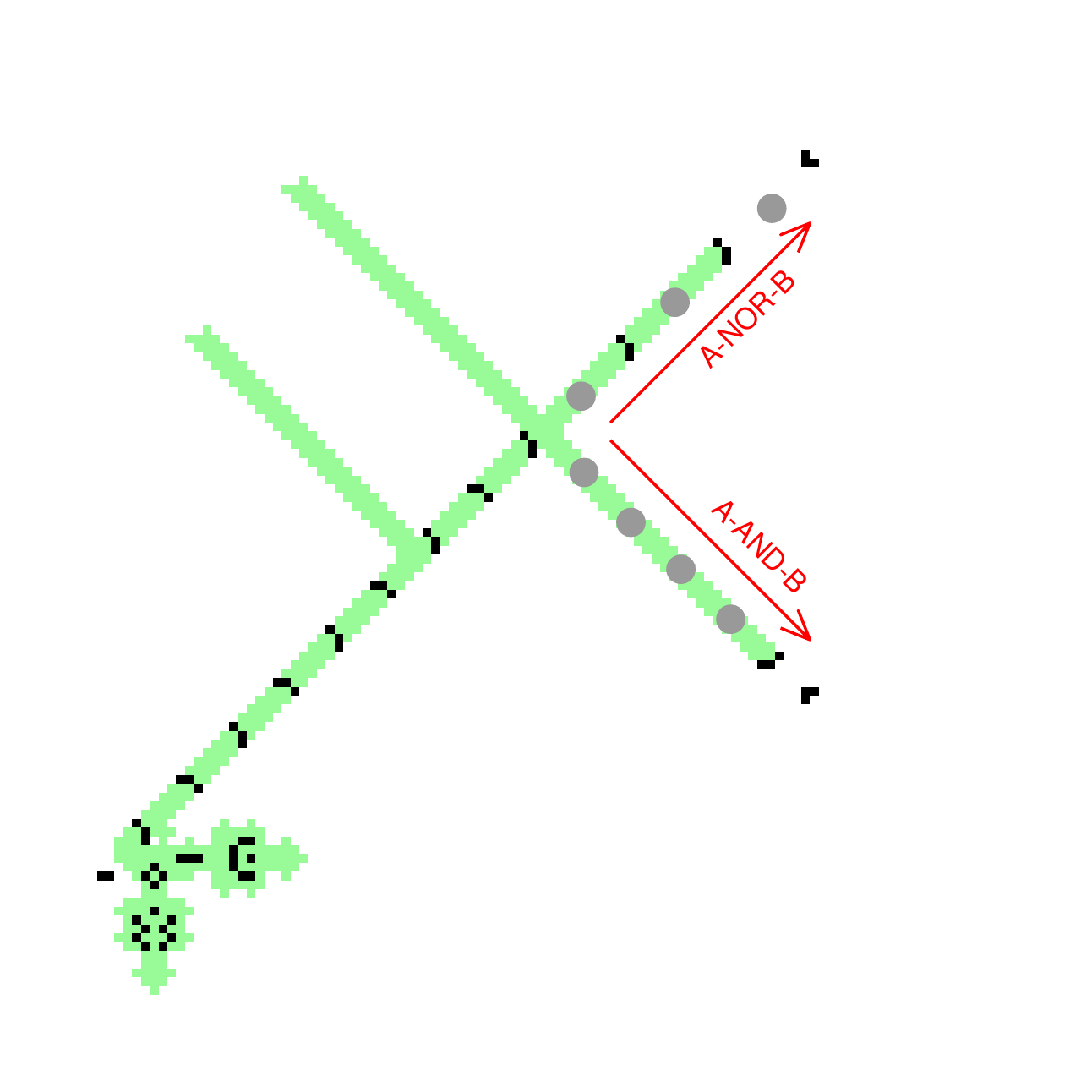}}
\end{minipage}
\end{center}
\vspace{-3ex}
\caption[GGa AND gate]
{\textsf{
An example of the AND gate (1 $\wedge$ 1 $\rightarrow$ 1,
    else $\rightarrow$ 0) making a conjunction between two streams of
    data, represented by gliders and gaps (grey discs).
$Left$: The 5-bit input strings A (10001) and B (10100) both moving SE
are about to interact with a GGa glider-stream moving NE. $Right$: The outcome is
\mbox{A-AND-B} (10000) moving SE shown after 184 time-steps.\\
{\color{white}xxx}
The dynamics making this AND gate first makes an intermediate NOT-A
(NE 01110 -- figure~\ref{GGa NOT gate}) which interacts with
input B to simultaneously produce both \mbox{A-AND-B} (SE 10000),
and the A-NOR-B (NE 01010) which will
be required to make the OR gate in figure~\ref{GGa OR gate}.
\label{GGa AND gate}
}}
\end{figure}

\clearpage
  
\begin{figure}[htb]
\begin{center} 
\begin{minipage}[c]{1\linewidth}
\fbox{\includegraphics[height=.98\linewidth,bb=20 7 203 649, clip=]{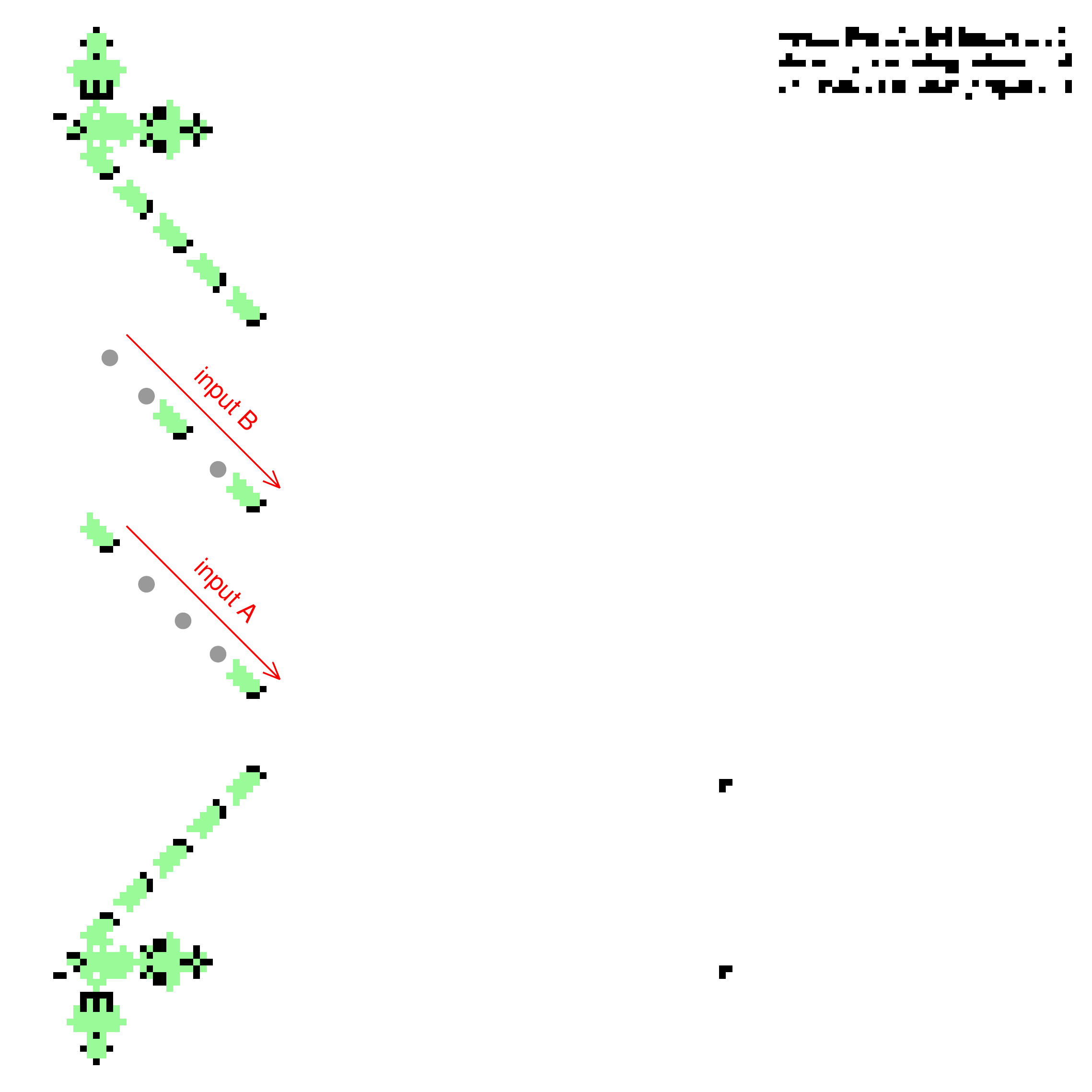}}
\hfill
\fbox{\includegraphics[height=.98\linewidth,bb=20 7 453 649, clip=]{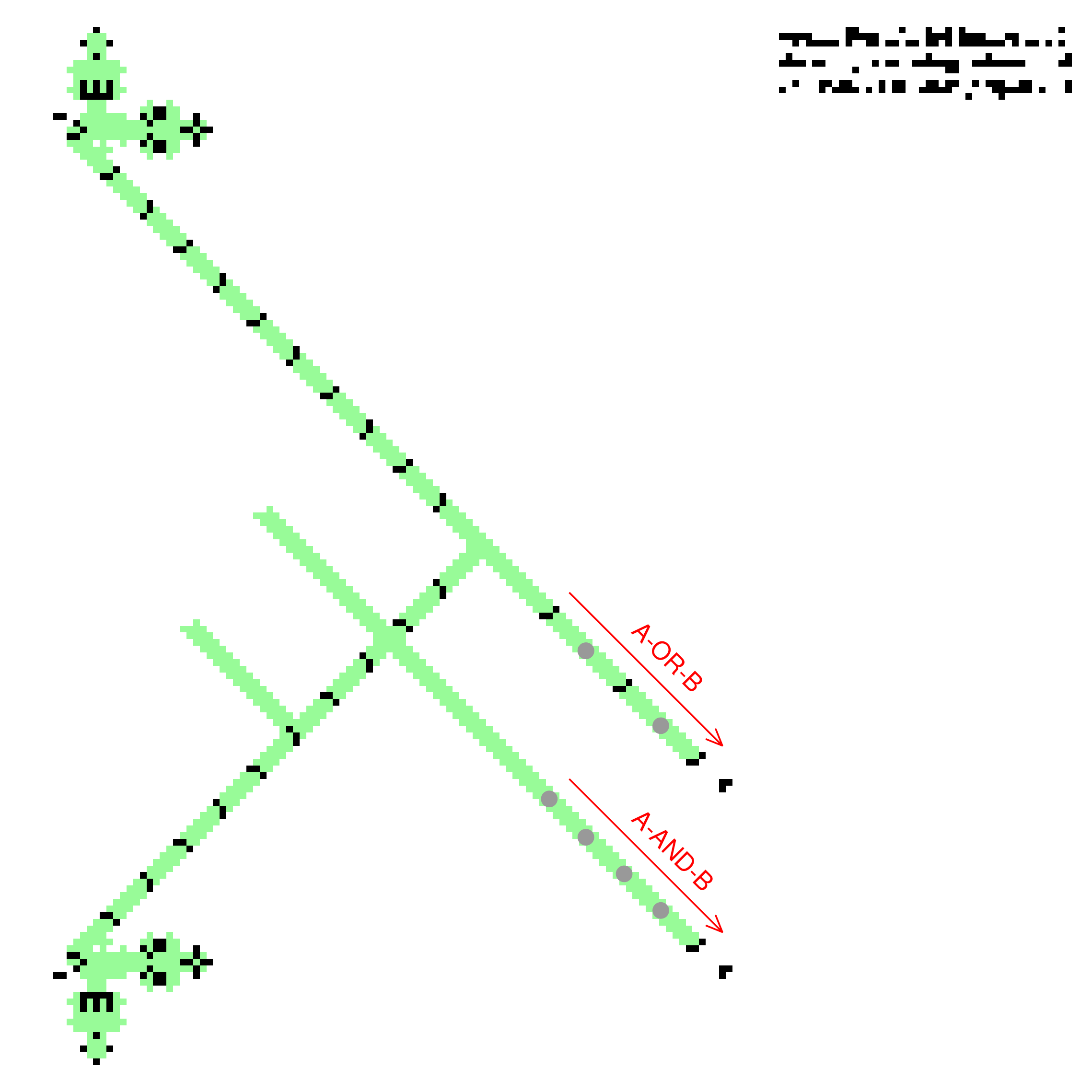}}
\end{minipage}
\end{center}
\vspace{-3ex}
\caption[GGa OR gate]
{\textsf{
An example of the OR gate (1 $\vee$ 1 $\rightarrow$ 1, else
$\rightarrow$ 0) making a disjunction between two stream of data
represented by two streams of gliders and gaps (grey discs).
$Left$: The 5-bit input strings A (10001) and B (10100) both moving SE
are about to interact with two GGa glider-streams, the lower GGa shooting NE, and
the upper GGa shooting SE. $Right$: The outcome is A-OR-B
(10101) moving SE shown after 264 time-steps.\\
{\color{white}xxx}
The dynamics making this OR gate first makes an intermediate NOT-A
(NE 01110 -- figure~\ref{GGa NOT gate}) which interacts with
input B to make A-NOR-B (NE 01010, as in figure \ref{GGa AND gate})
which interacts with
the upper GGa shooting SE to make A-OR-B (SE 10101).
A residual bi-product is A-AND-B (SE 10000 -- figure \ref{GGa AND gate}).
\label{GGa OR gate}
}}
\end{figure}
\clearpage

\subsection{GGc NOT, AND, OR}
\label{GGc NOT, AND, OR}

\begin{figure}[htb]
\begin{center} 
\begin{minipage}[c]{1\linewidth}
\fbox{\includegraphics[height=.9\linewidth,bb=106 14 354 511, clip=]{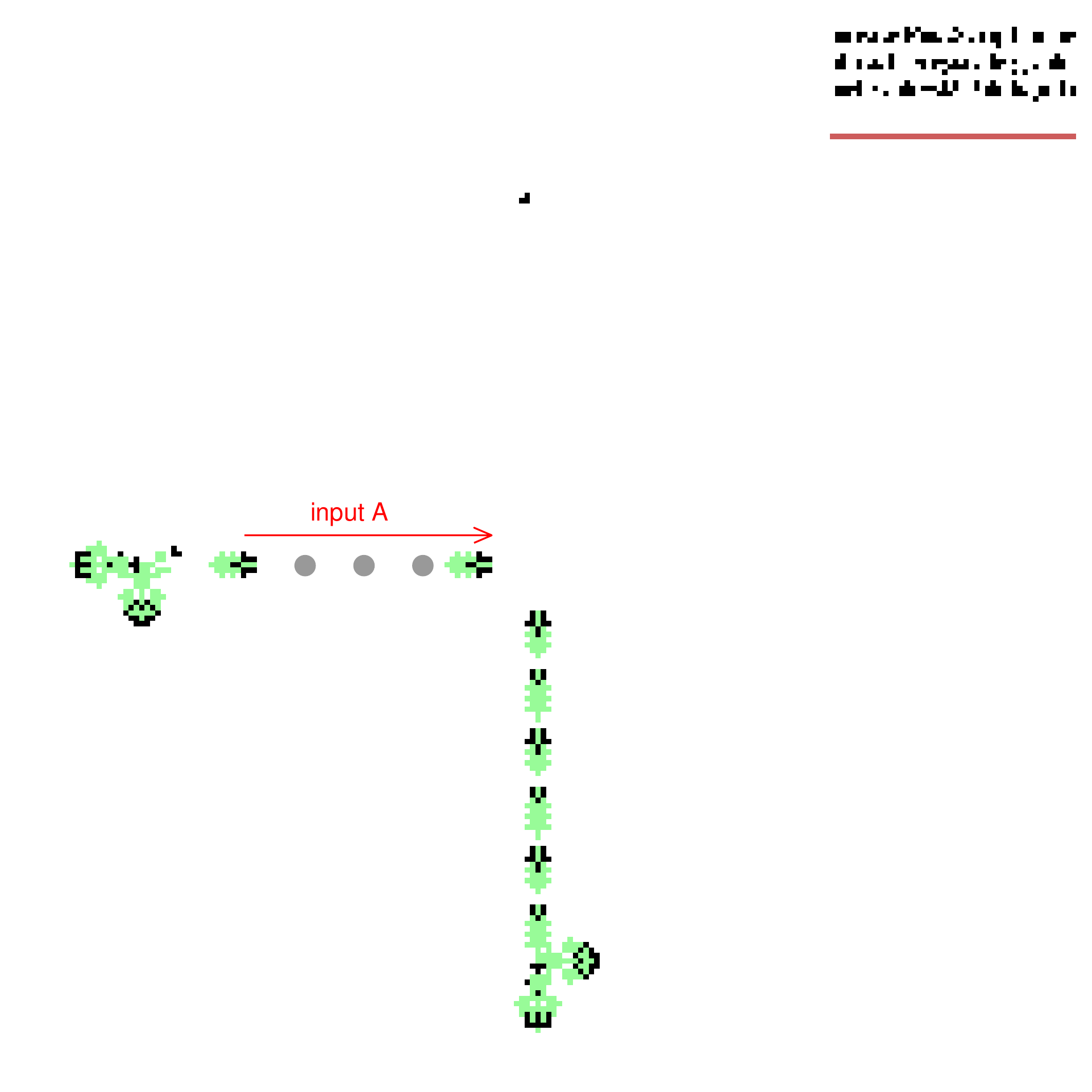}}
\hfill
\fbox{\includegraphics[angle=90,width=.45\linewidth,bb=14 256 511 504, clip=]{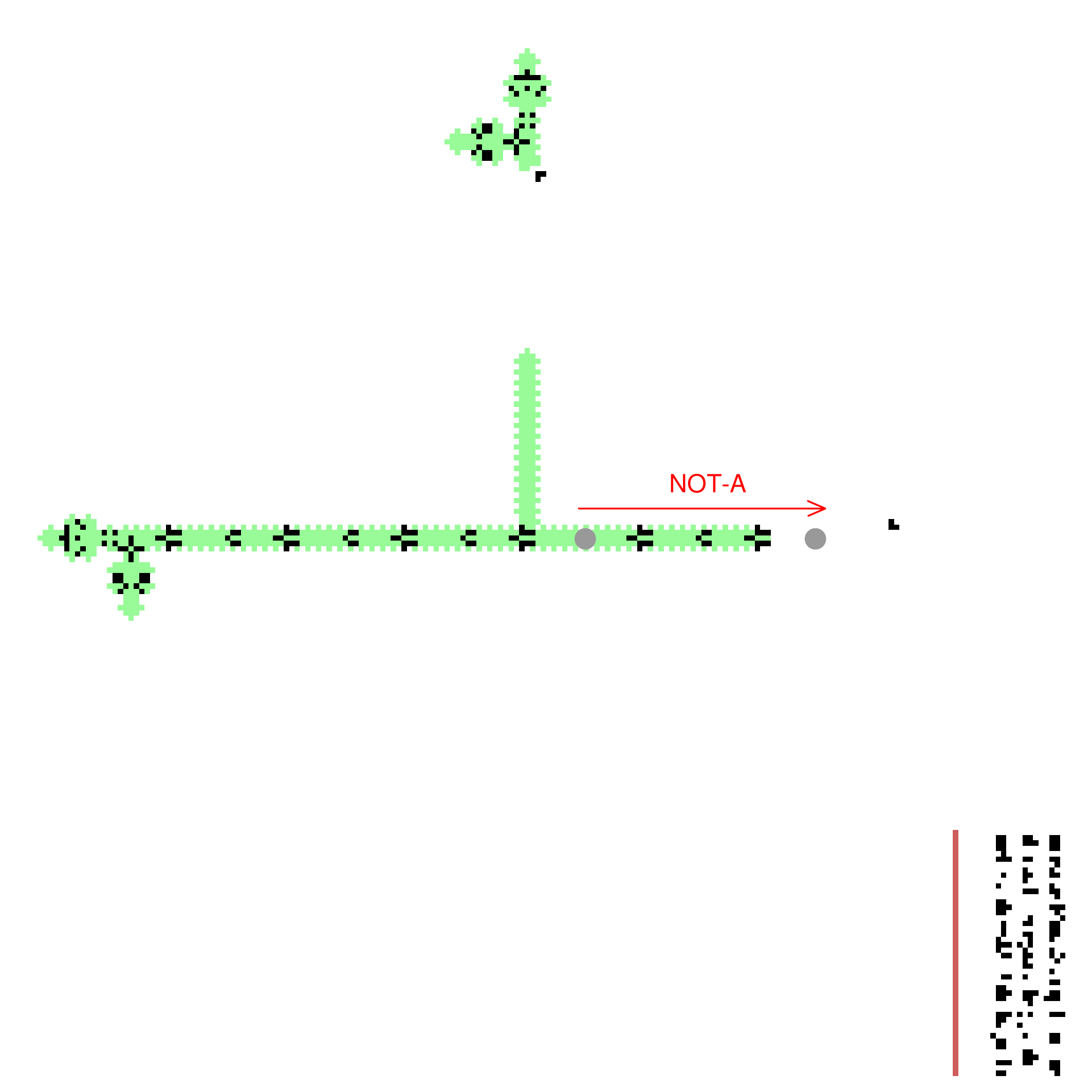}} 
\end{minipage}
\end{center}
\vspace{-3ex}
\caption[GGc NOT gate]
{\textsf{
An example of the GGc NOT gate: ($\neg 1, 1 \rightarrow$ 0 and $0
    \rightarrow$ 1) or inverter, which transforms a stream of data to
    its complement, represented by gliders and gaps (grey discs). 
$Left$: The 5-bit input string A (10001) moving East is about to interact 
with a GGc glider-stream moving North. $Right$: The outcome is NOT-A (01110) moving
North, shown after 134 time-steps.
\label{GGc NOT gate}
}}
\end{figure}

\begin{figure}[htb]
\begin{center} 
\begin{minipage}[c]{1\linewidth}
\fbox{\includegraphics[height=.82\linewidth,bb=65 35 274 413, clip=]{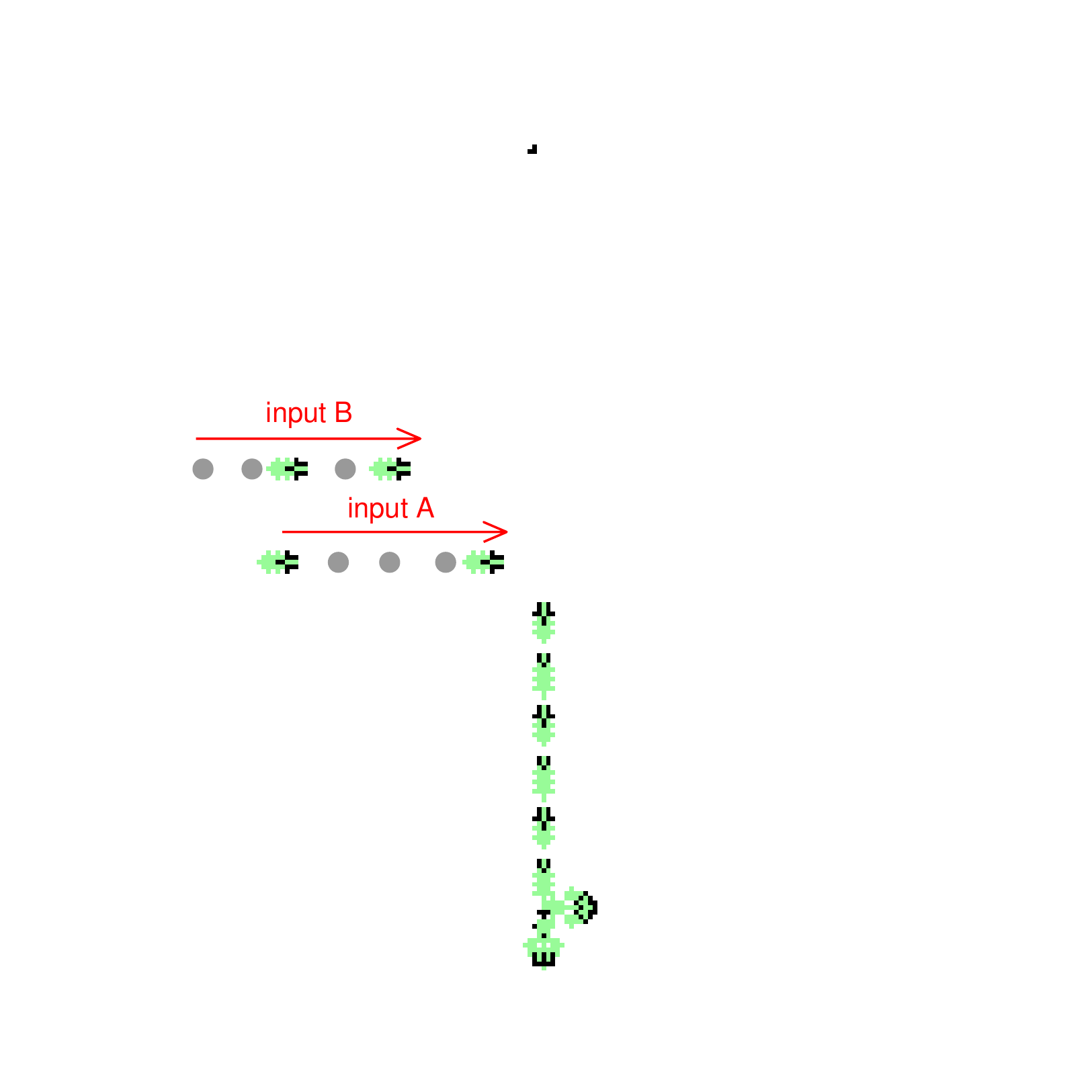}}
\hfill
\fbox{\includegraphics[height=.82\linewidth,bb=140 35 363 413, clip=]{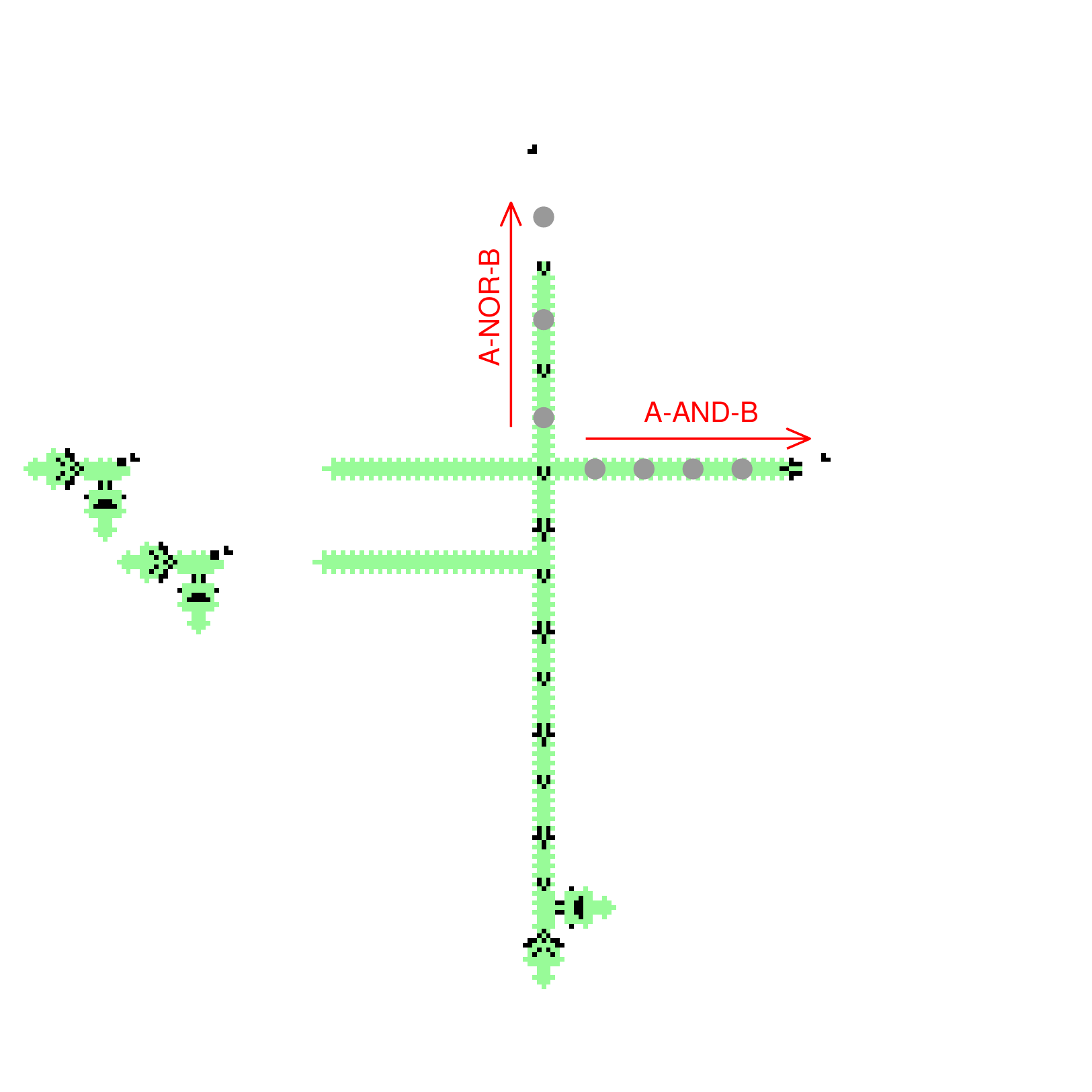}}
\end{minipage}
\end{center}
\vspace{-3ex}
\caption[GGc AND gate]
{\textsf{
An example of the AND gate (1 $\wedge$ 1 $\rightarrow$ 1,
    else $\rightarrow$ 0) making a conjunction between two streams of
    data, represented by gliders and gaps (grey discs).
$Left$: The 5-bit input strings A (10001) and B (10100) both moving East
are about to interact with a GGc glider-stream moving North.
$Right$: The outcome is \mbox{A-AND-B} (10000) moving East shown after 179 time-steps.\\
{\color{white}xxx}
The dynamics making this AND gate first makes an intermediate NOT-A
(North 01110 -- figure~\ref{GGc NOT gate}) which interacts with
input B to simultaneously produce both \mbox{A-AND-B} (East 10000),
and the A-NOR-B (North 01010) which will
be required to make the OR gate in figure~\ref{GGc OR gate}.
\label{GGc AND gate}
}}
\end{figure}
\clearpage

\begin{figure}[htb]
\begin{center} 
\begin{minipage}[c]{.85\linewidth}
\fbox{\includegraphics[width=1\linewidth,bb=11 51 418 345, clip=]{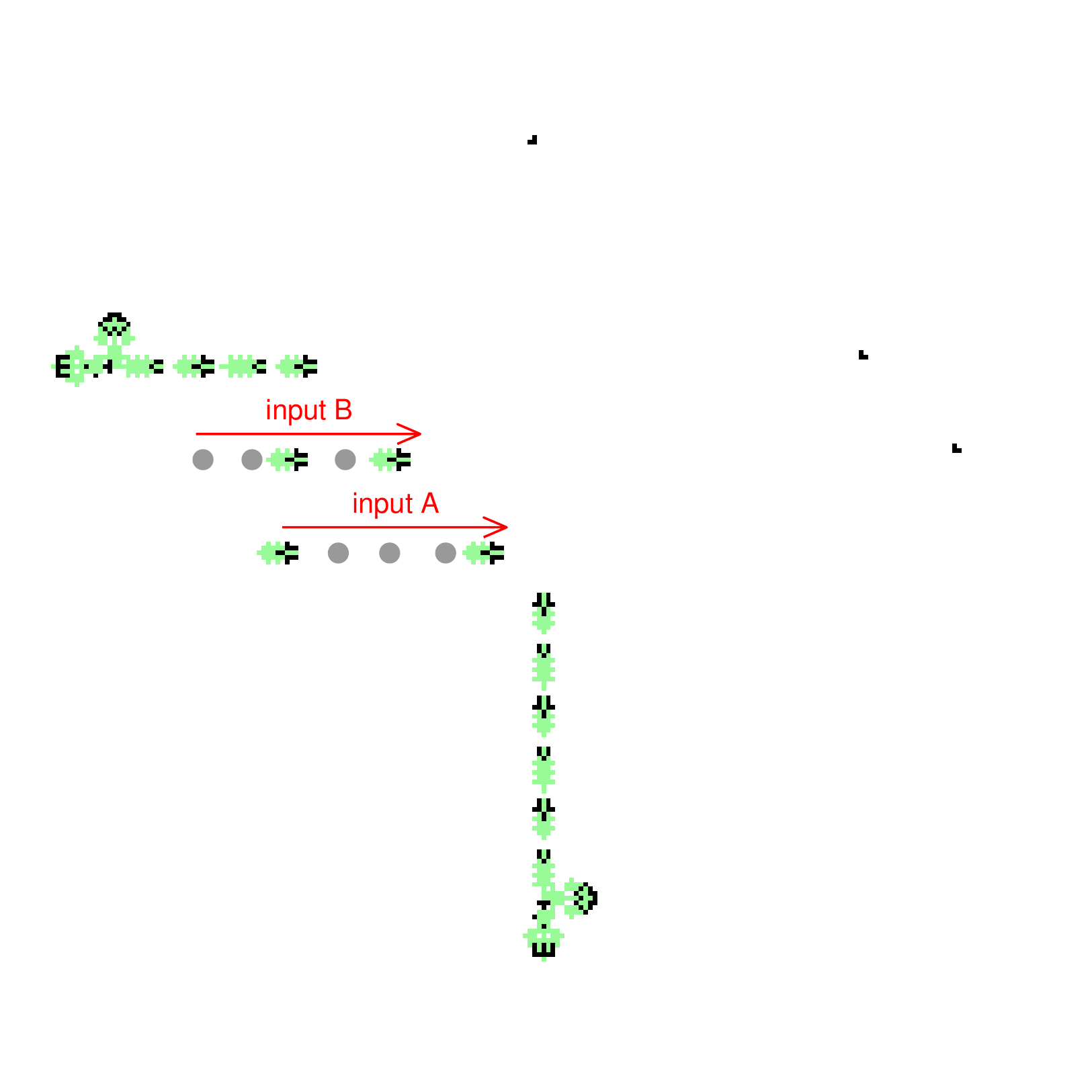}}\\[-.3ex]
\fbox{\includegraphics[width=1\linewidth,bb=11 181 418 345, clip=]{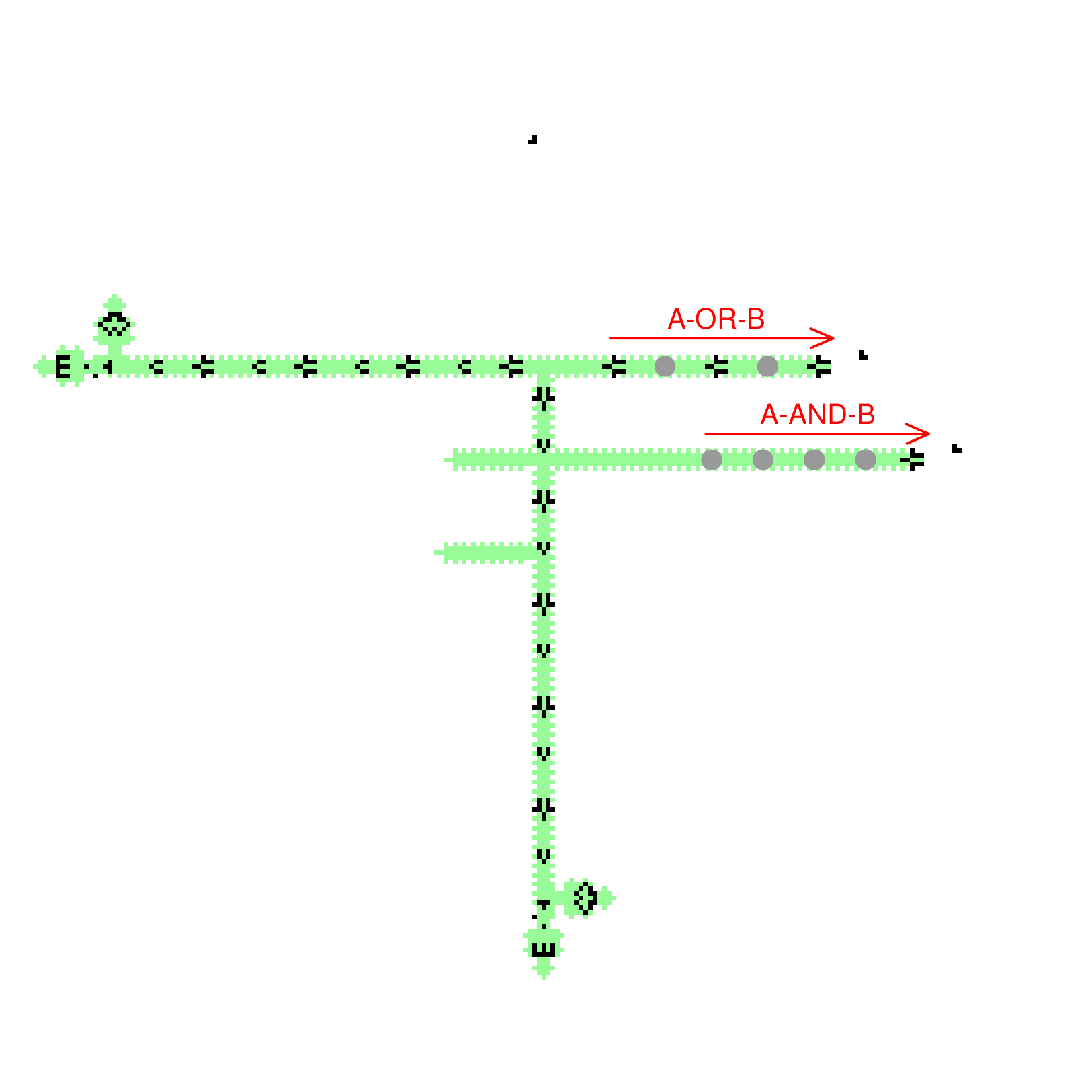}}
\end{minipage}
\end{center}
\vspace{-3ex}
\caption[GGc OR gate]
{\textsf{
An example of the OR gate (1 $\vee$ 1 $\rightarrow$ 1, else
$\rightarrow$ 0) making a disjunction between two stream of data
represented by two streams of gliders and gaps (grey discs).
$Top$: The 5-bit input strings A (10001) and B (10100) both moving East
are about to interact with two GGc  glider-streams, the lower GGc shooting North, and
the upper GGc shooting East. $Below$: The outcome is A-OR-B
(10101) moving East shown after 220 time-steps.\\
{\color{white}xxx}
The dynamics making this OR gate first makes an intermediate NOT-A
(North 01110 -- figure~\ref{GGc NOT gate}) which interacts with
input B to make A-NOR-B (North 01010 -- figure \ref{GGc AND gate})
which interacts with
the upper GGc shooting East to make A-OR-B (SE 10101).
A residual bi-product is A-AND-B (East 10000 -- figure \ref{GGc AND gate}).
\label{GGc OR gate}
}}
\vspace{-2ex}
\end{figure}

\enlargethispage{4ex}
\section{Spaceships, puffers and rakes}
\label{Spaceships, puffers and rakes}

\begin{figure}[htb]
\begin{center} 
\begin{minipage}[c]{.8\linewidth}
\fbox{\includegraphics[width=.46\linewidth,bb=141 216 269 281, clip=]{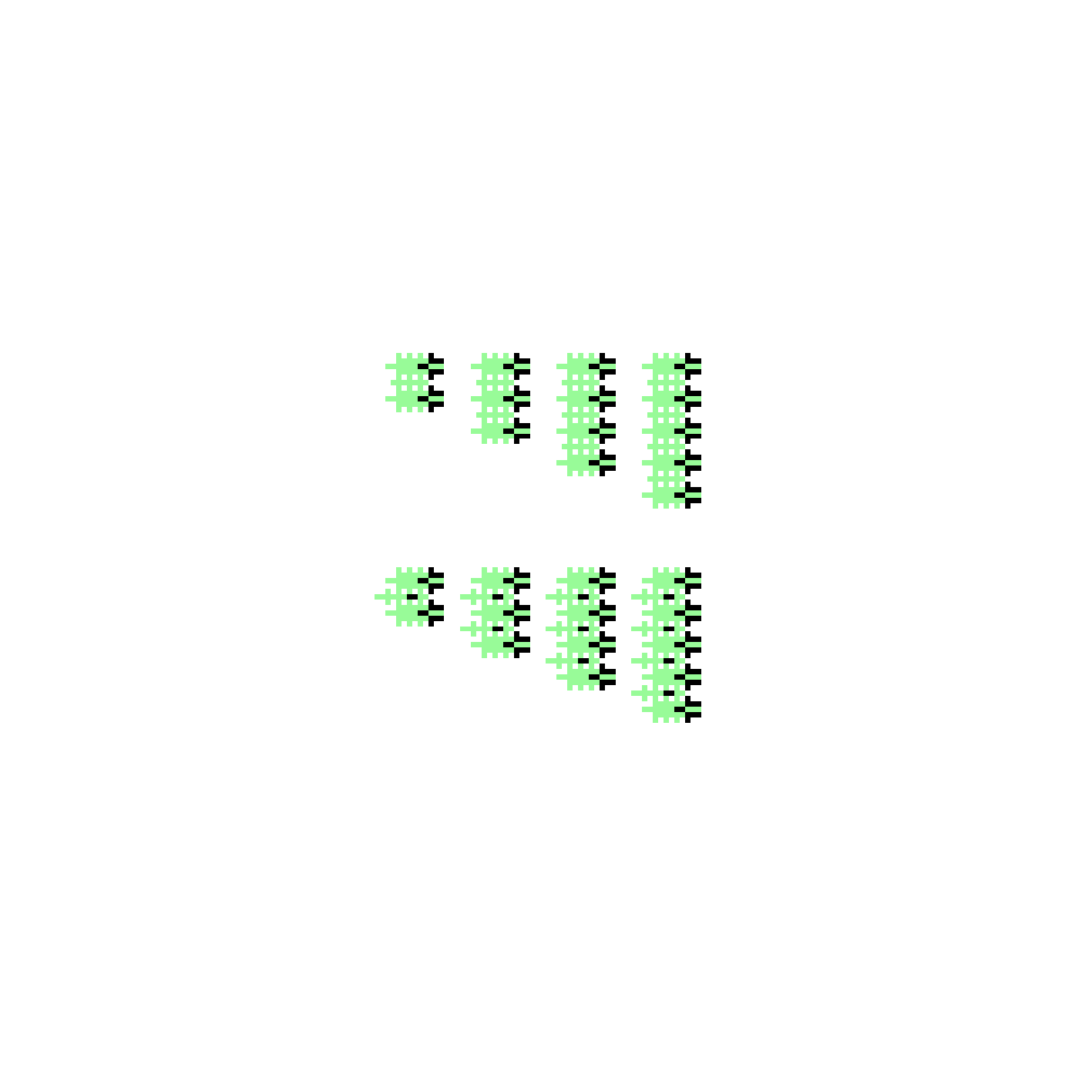}}
\hfill
\fbox{\includegraphics[width=.46\linewidth,bb=141 135 271 201, clip=]{pdf-figs/ships1+2}}
\end{minipage}
\end{center}
\vspace{-3ex}
\caption[ships]
{\textsf{$Left$: Spaceships built from Gc glider subunits separated by one cell,
    and $Right$: dragging a periodic tag\cite{Danieldb}. More units can be added.
\label{ships}
}}
\end{figure}

\begin{figure}[htb]
\begin{center} 
\textsf{\small
\begin{minipage}[c]{1\linewidth}
\fbox{\includegraphics[width=.85\linewidth,bb=44 364 351 390, clip=]{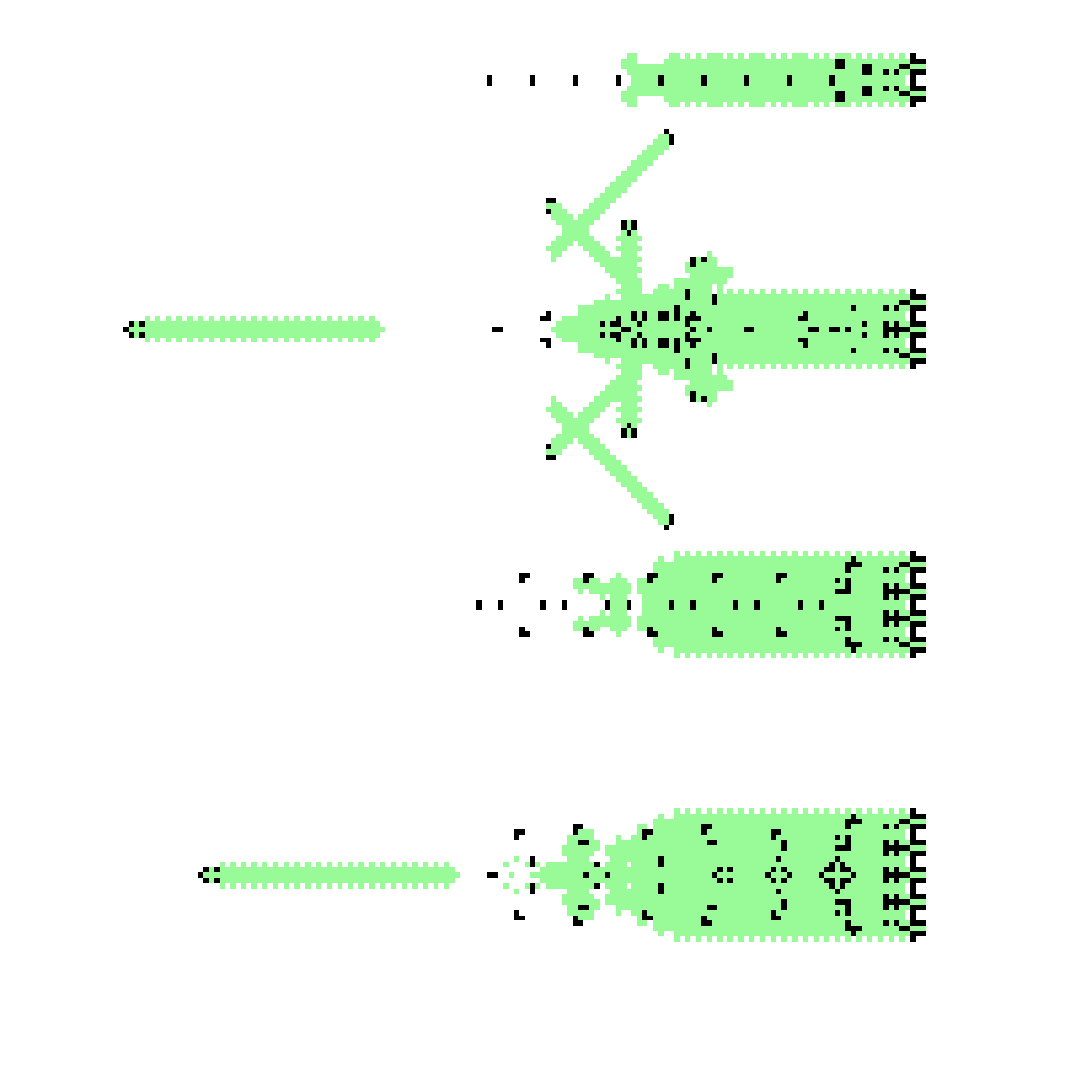}} 
\raisebox{.4ex}{\includegraphics[width=.048\linewidth,bb=185 365 203 391, clip=]{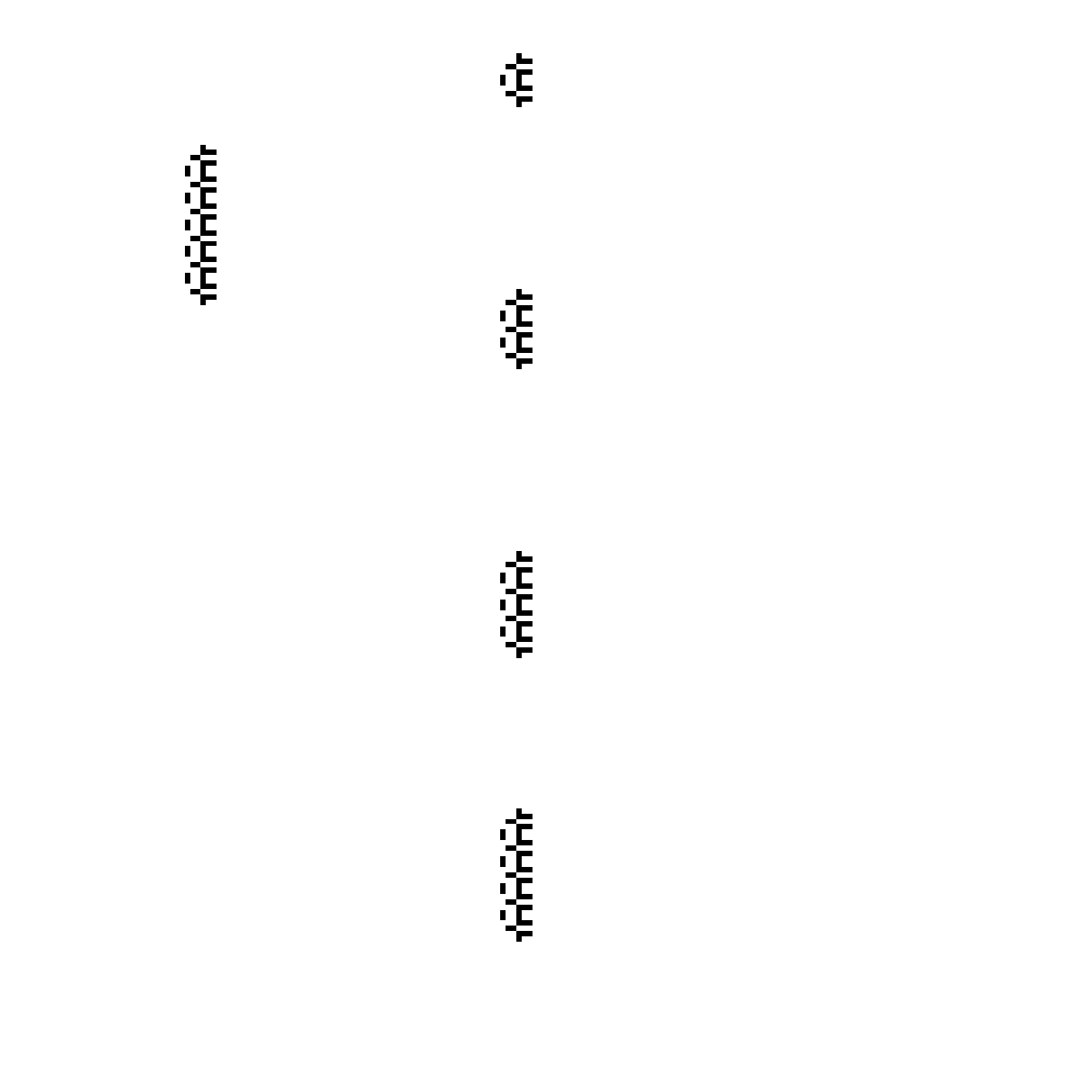}}
\raisebox{3ex}{ \begin{minipage}[c]{4ex}puf2\\[-.5ex]$p$=16\end{minipage} }   \\[-.3ex]
\fbox{\includegraphics[width=.85\linewidth,bb=42 207 349 362, clip=]{pdf-figs/puf2-5a}}
\raisebox{14.2ex}{\includegraphics[width=.048\linewidth,bb=185 266 203 302, clip=]{pdf-figs/puf2-6sd}}
\raisebox{19ex}{  \begin{minipage}[c]{4ex}puf3\\[-.5ex]$p$=96\end{minipage} }            \\[-.3ex]
\fbox{\includegraphics[width=.85\linewidth,bb=44 158 351 204, clip=]{pdf-figs/puf2-5a}} 
\raisebox{.2ex}{\includegraphics[width=.048\linewidth,bb=185 157 203 205, clip=]{pdf-figs/puf2-6sd}}
\raisebox{5ex}{ \begin{minipage}[c]{4ex}puf4\\[-.5ex]$p$=24\end{minipage} }             \\[-.3ex]
\fbox{\includegraphics[width=.85\linewidth,bb=44 49 351 110, clip=]{pdf-figs/puf2-5a}} 
\raisebox{.8ex}{\includegraphics[width=.048\linewidth,bb=185 51 203 109, clip=]{pdf-figs/puf2-6sd}}
\raisebox{7ex}{ \begin{minipage}[c]{4ex}puf5\\[-.5ex]$p$=24\end{minipage} }           \\[-.3ex]
\end{minipage}
}
\end{center}
\vspace{-5ex}
\caption[puf2-5] 
{\textsf{
Periodic puffer-trains and a rake.
$Right$: initial states are made from increasing numbers (puf2\cite{Wildmyron} to puf5)
of touching Gc+block subunits, with period $p$ shown.
$Left$: the pattern fronts advance in 4 phases with speed $c/2$, and are
shown after 149 time-steps with trailing debris.
From the debris in puf3, which could also be called a puffer-rake,
bursts of Gc and Ga gliders emerge every 96 time-steps. These and further figures
have dynamic trails of 88 time-steps.
\label{puf2-5a}
}}
\end{figure}

\begin{figure}[htb]
\begin{center} 
\textsf{\small
\begin{minipage}[c]{1\linewidth}
\fbox{\includegraphics[width=.9\linewidth,bb=107 107 376 301, clip=]{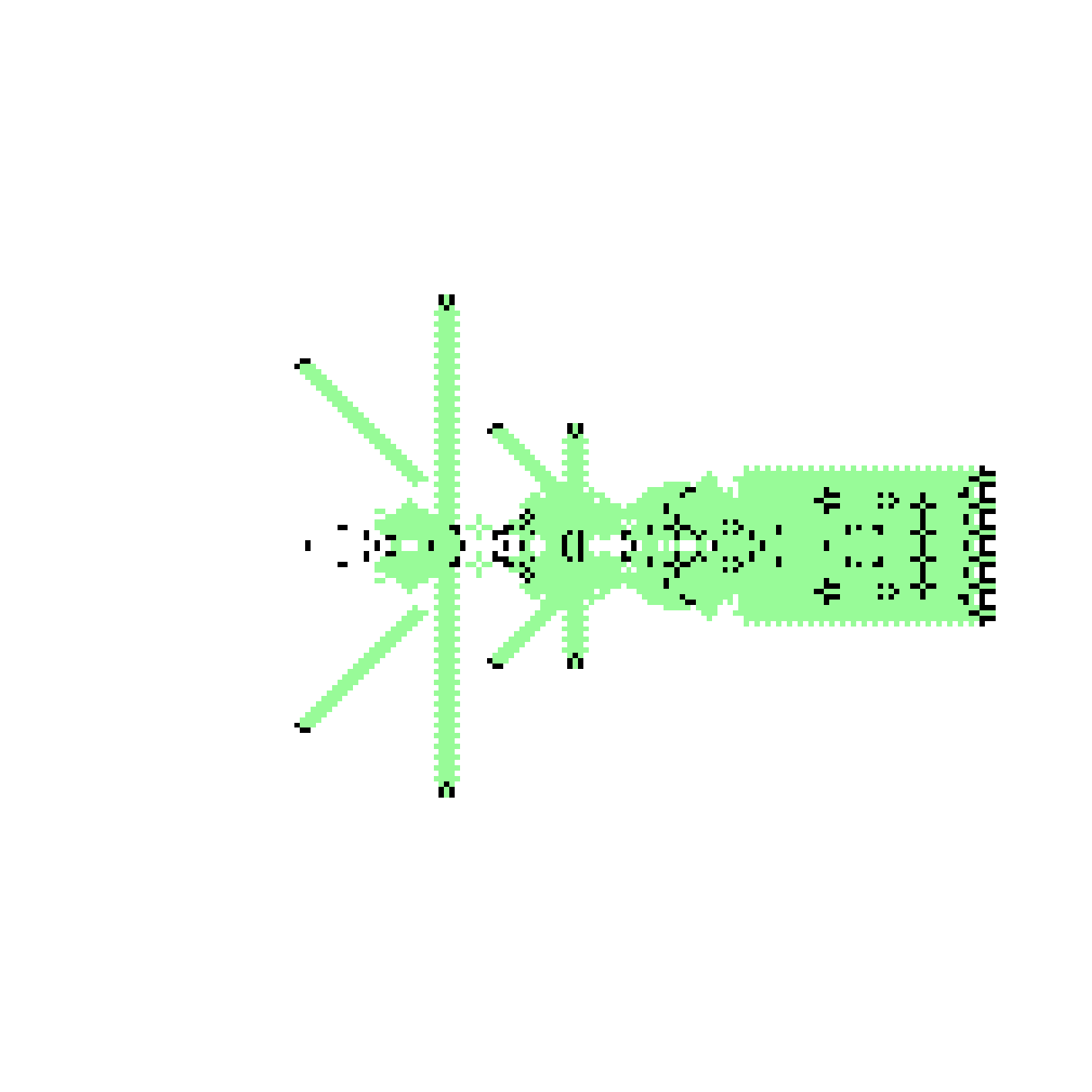}}
\raisebox{18.6ex}{\includegraphics[width=.065\linewidth,bb=66 290 86 357, clip=]{pdf-figs/puf2-6sd}}
\end{minipage}
}
\end{center}
\vspace{-3ex}
\caption[puf6a]
{\textsf{Periodic puffer-rake.
$Right$: the initial states.
$Left$: the pattern front advances in 4 phases with speed $c/2$, and is
shown after 274 time-steps with trailing debris, from which
bursts of Ga and Gc gliders emerge every 24 time-steps.
\label{puf6a}
}}
\end{figure}

\begin{figure}[htb]
\begin{center} 
\textsf{\small
\begin{minipage}[c]{1\linewidth}
\fbox{\includegraphics[width=.9\linewidth,bb=45 351 355 382, clip=]{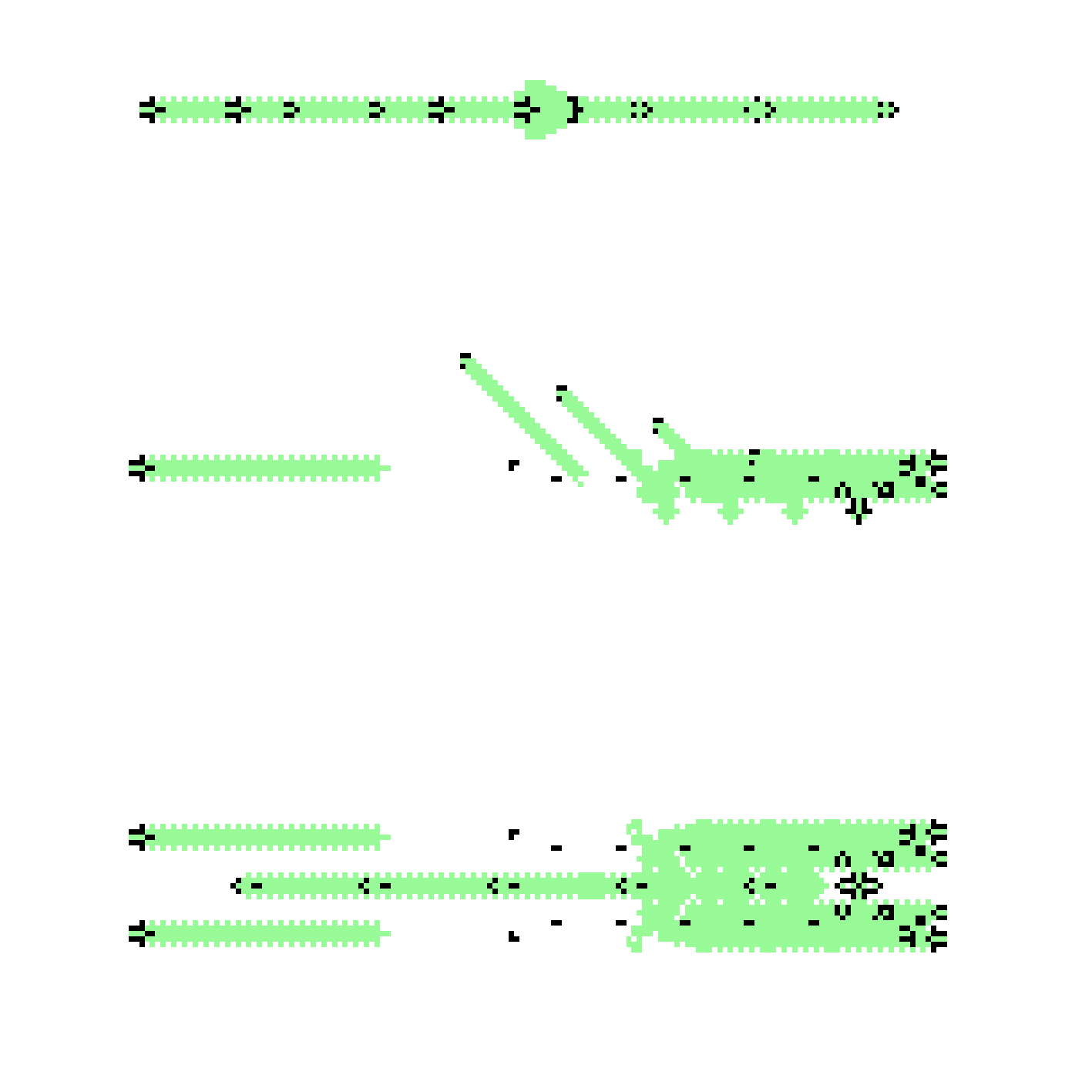}} 
\raisebox{1ex}{\includegraphics[width=.035\linewidth,bb=198 358 209 375, clip=]{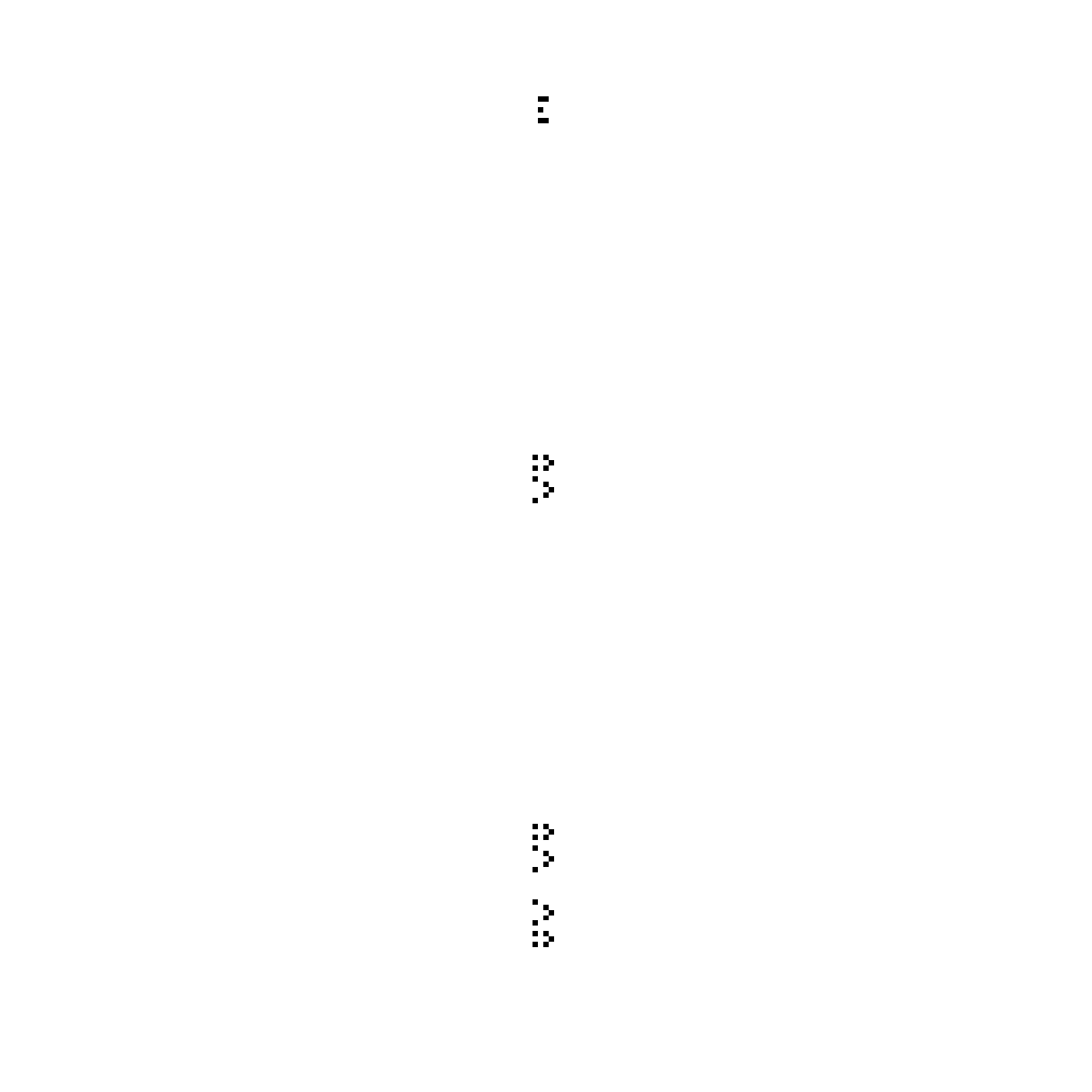}}
\raisebox{2.5ex}{a}\\[-.3ex]
\fbox{\includegraphics[width=.9\linewidth,bb=45 209 355 280, clip=]{pdf-figs/rk124}}
\raisebox{1ex}{\includegraphics[width=.035\linewidth,bb=198 216 209 241, clip=]{pdf-figs/rk124sd}}
\raisebox{3.5ex}{b}\\[-.3ex]
\fbox{\includegraphics[width=.9\linewidth,bb=45 47 355 105, clip=]{pdf-figs/rk124}}
\raisebox{0ex}{\includegraphics[width=.035\linewidth,bb=198 49 209 103, clip=]{pdf-figs/rk124sd}}
\raisebox{7ex}{c}
\end{minipage}}
\end{center}
\vspace{-3ex}
\caption[rk124]
{\textsf{
Three periodic rakes\cite{Wildmyron}.
$Right$: initial states.
$Left$: Gc gliders/patterns move West/East in 4 phases with speed $c/2$,
shown after 149 time-steps. (a) the central zone has a period of 50. 
(b) Ga gliders emerge every 24 time-steps. (c) Gb gliders (figure~\ref{glider types})
are shot East every 48 time-steps.
\label{rk124}
}}
\end{figure}

\begin{figure}[htb]
\begin{center} 
\textsf{\small
\begin{minipage}[c]{1\linewidth}
\fbox{\includegraphics[width=.83\linewidth,bb=115 12 614 384, clip=]{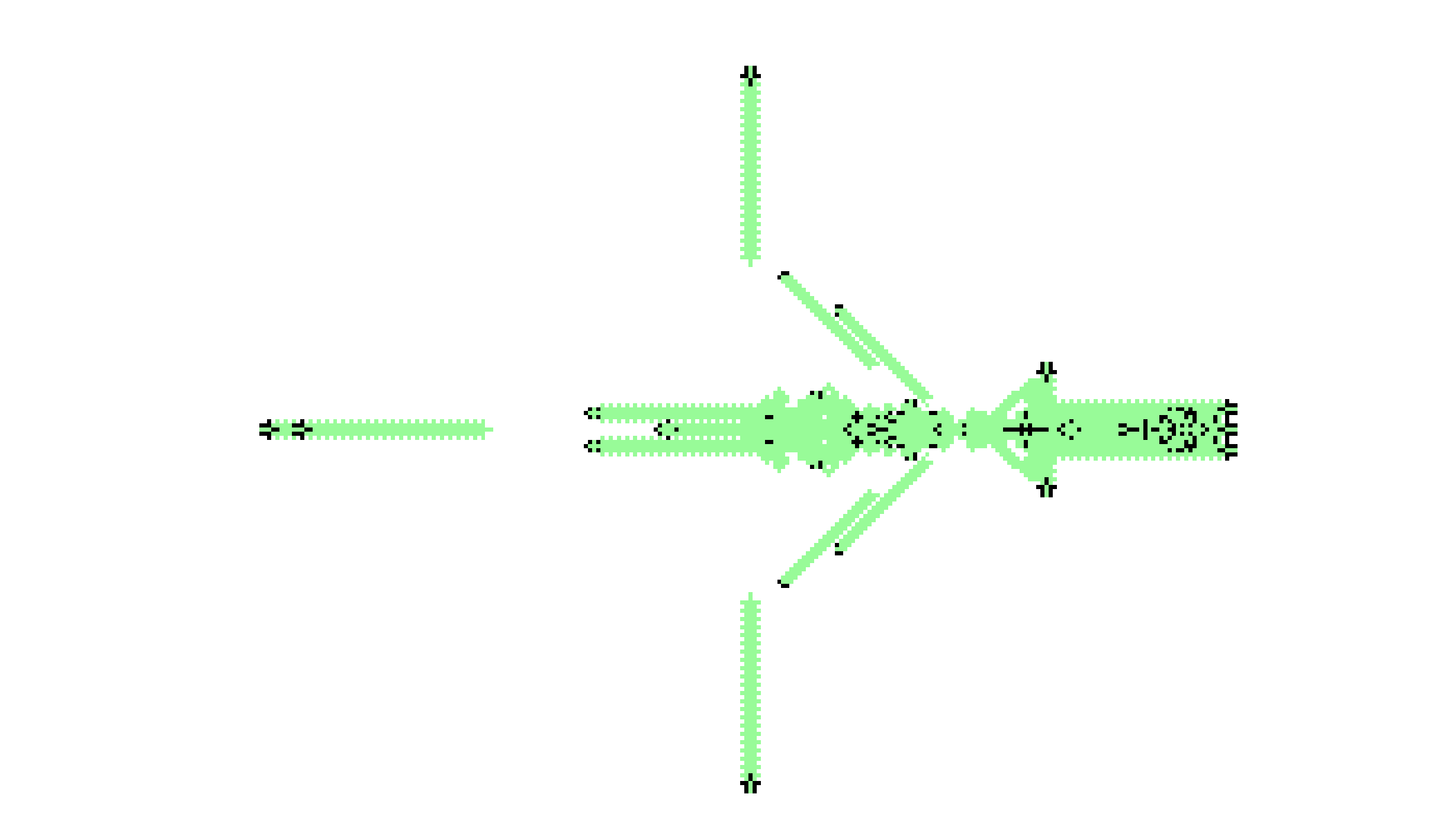}} 
\raisebox{22ex}{\includegraphics[width=.12\linewidth,bb=177 280 222 324, clip=]{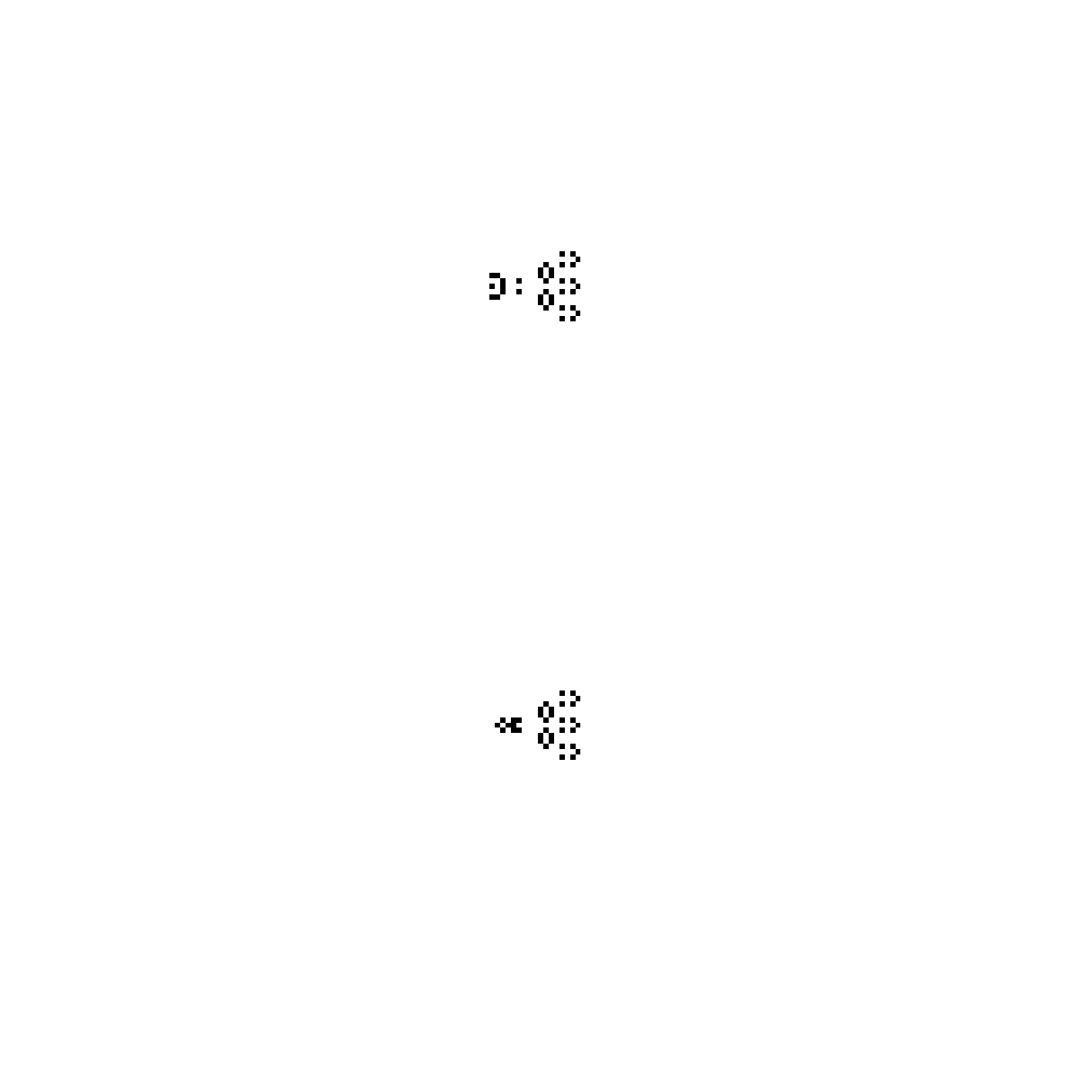}}
\raisebox{26ex}{a}\\[-.3ex]
\fbox{\includegraphics[width=.83\linewidth,bb=115 50 614 343, clip=]{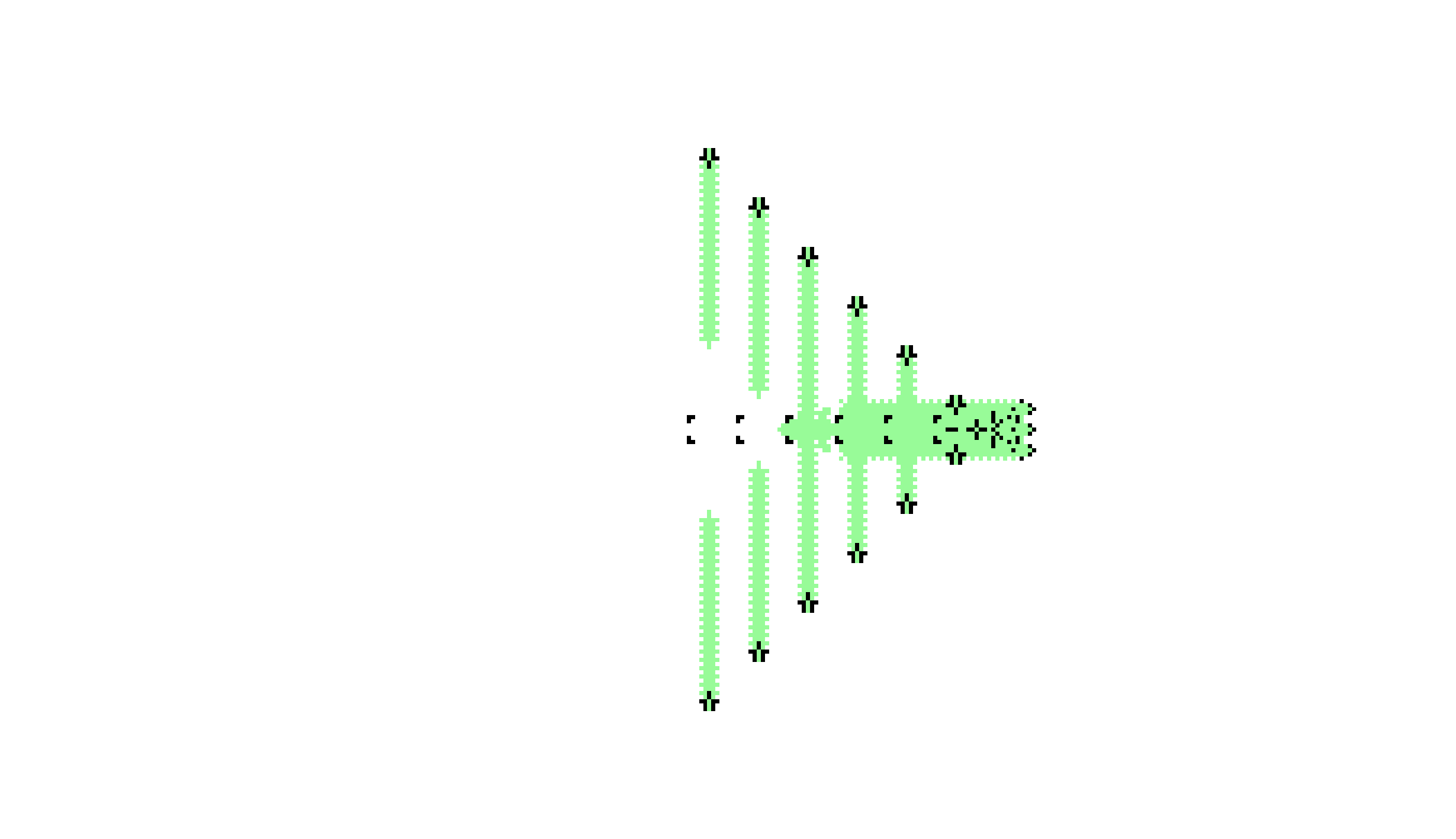}} 
\raisebox{16ex}{\includegraphics[width=.12\linewidth,bb=177 114 222 158, clip=]{pdf-figs/rk3a3b}}
\raisebox{20ex}{b}\\[-.3ex]
\end{minipage}}
\end{center}
\vspace{-3ex}
\caption[Two periodic rakes]
{\textsf{
Two periodic rakes\cite{Wildmyron}.
$Right$: initial states based on 3 leading Gc's separated by 2 cells, and other structures.
$Left$(a): Gc travel West, and North/South every 144 time-steps. Ga gliders travel NW.
The pattern front moves East in 4 phases with speed $c/2$.
Shown after 243 time-steps. 
$Left$(b) Gc gliders travellig North/South emerge every 24 time-steps,
as the  pattern front moves East in 4 phases with speed $c/2$. Shown after 138 time-steps. 
\label{Two periodic rakes}
}}
\end{figure}

The Variant rule features many other interesting larger scale moving
patterns, named from the Game-of-Life lexicon.  A spaceship is a
compound-glider built from subunits, a puffer-train is a spaceship
leaving debris in its wake, a rake ejects gliders as it moves, and
there are intermediate or ambiguous structures such as puffer-rakes.
Most of these patterns were discovered by members of the ConwayLife
forum\cite{ConwayLife-forum}.  Figures~\ref{ships} to \ref{Two periodic rakes}
give examples.

\enlargethispage{4ex}

\clearpage

\section{Concluding remarks}
\label{Concluding remarks}

A chance mutation while the ConwayLife Forum\cite{ConwayLife-forum}
scrutinised the Precursor rule\cite{Gomez2017}, revealed a new
``Variant'' rule with an interesting diversity of different patterns,
despite the small divergence from the Precursor.  Members of the forum
applied their considerable know-how in Game-of-Life pattern search to
discover glider-guns and other patterns in the Variant rule ---
a selection are now documented and elaborated in this paper. It's very
possible that other significant patterns exist, to be discovered.

The Variant-rule has gliders, glider-guns, eaters and convenient
collision from which we are able to construct the logical gates in at
least two distinct ways, and demonstrate universality in the logical
sense.  More work would be necessary to show universality in the
Turing sense\cite{Randall2002} and in terms of
Conway\cite{Berlekamp1982}.  The Variant-rule enriches the family of
cellular automata with glider-gun complex properties that are not
based on Life-like birth/survival schemes.

It would be interesting in the future to study which patterns are
common or not, between the Variant,  Precursor\cite{Gomez2017} and
X-rule\cite{Gomez2015}, and also the Sayab-rule\cite{Gomez2018}, as
well as the Game-of-Life.  The Variant and Precursor rules have very
different glider-guns and larger scale pattern behaviors despite their
genetic closeness, though they share glider types and small scale
features, illustrating both the robustness and fragility of evolution.
The discovery of new complex glider-gun rules based on small mutations
would be a promising approach towards a general theory of
glider-gun dynamics.

\section{Experiments and Acknowledgements}
\label{Acknowledgements}

Experiments were done with Discrete Dynamics Lab (DDLab)\cite{Wuensche2016,Wuensche-DDLab}, 
Mathematica and Golly, and can be repeated and extended from initial states 
detailed at the ``Logical Universality in 2D Cellular Automata'' web page\cite{UC2DCA-webpage}.

The Precursor-Rule was found during a collaboration at a workshop
in June 2017 at the DDLab Complex Systems Institute in Ariege, France, and
in London, UK.  and also at the Universidad Aut\'onoma de Zacatecas,
M\'exico in 2018, 2019, Later patterns were discovered during
interactions with the ConwayLife Forum\cite{ConwayLife-forum} where
many people made important contributions.  J.M. G\'omez Soto also
acknowledges his residency at Discrete Dynamics Lab, and financial
support from the Research Council of Mexico (CONACyT).

\end{document}